%% file: MGRdoctoralTHESIS.tex
\documentclass[11pt,a4paper]{book}

\usepackage{sidecap}
\usepackage{pdflscape}

\usepackage{amsfonts}
\usepackage{amssymb}
\usepackage{amsmath}

\usepackage{dsfont}
\usepackage{cancel}
\usepackage{graphicx}
\newcommand{\ul}{\underline}

\usepackage{bm}
\usepackage{epsfig}
\usepackage{psfrag}

\usepackage{makeidx}
\makeindex

\begin{document}
\begin{titlepage}
\centering
\begin{center}
\vspace{15cm}
\huge
\textsc{Electroweak Hadron Structure \\ within a Relativistic \\ Point-Form Approach}

\vspace{12cm}
 \Large
{Mar\'ia G\'omez-Rocha}\\

\vspace{2cm}
 \Large
\Large{Dissertation}\\
\large
\vspace{0.1cm}
{zur Erlangung des Doktorgrades der Naturwissenschaften}\\
{an der
Karl-Franzens-Universit\"at Graz\\
Betreuer:
Ao.~Univ.-Prof.~Mag.~Dr.~Wolfgang Schweiger\\
\vspace{1cm}
Graz, Dezember 2012}
\end{center}

 \newpage
\thispagestyle{empty}
\phantom{:-)}
\end{titlepage}

 \newpage
\thispagestyle{empty}

\begin{center}
 \section*{Abstract}
 \end{center}
 
In this thesis a general relativistic framework for the calculation of the electroweak structure of mesons 
of arbitrary constituent-quark masses is presented. The physical processes in which the structure is measured, 
i.e. electron-meson scattering and semileptonic weak decays, are treated in a Poincar\'e invariant way by making 
use of the point-form of relativistic quantum mechanics. The electromagnetic and weak meson currents are 
extracted from the 1-photon or 1-W-exchange amplitudes that result from a Bakamjian-Thomas type mass operator 
for the respective systems. The covariant decomposition of these currents provides the electromagnetic and weak 
(transition) form factors. The formalism is first applied to the study of heavy-light systems. 
Problems with cluster separability, which are inherent in the Bakamjian-Thomas construction, are discussed and 
it is shown how to keep them under control. It is proved that the heavy-quark limit of the electroweak form 
factors leads to one universal function, the Isgur-Wise function, confirming that the requirements of 
heavy-quark symmetry are satisfied. These results are discussed and compared with analogous calculations 
in the front form of dynamics.

The formalism is further applied to the study of bound states whose binding is caused by dynamical particle 
exchange. The problem of how to take into account retardation effects in the particle-exchange potential is 
formulated and it is shown how they affect the binding energy and wave-function solution for a dynamical model 
of the deuteron.

At the end of this work an example where the Clebsch-Gordan coefficients of the Poincar\'e group are applied 
is presented. The angular momentum decomposition of chiral multiplets is given in the instant and in the front 
forms.
 
\thispagestyle{empty}
 \newpage
\thispagestyle{empty}

 \printindex 
 \pagenumbering{roman}
\tableofcontents
 \newpage
\thispagestyle{empty}

\include{Ch1}

\include{Ch2}

\include{Ch3}

\include{Ch4}

\include{Ch5}

\include{Ch6}

\include{Ch7}

\include{Ch8}

\include{Ch9}

\include{Ch10}

\appendix

\include{App1}

\include{App2}

\include{App3}

\include{App4}

\include{Ackn}

\glossary

\include{Bibl}

\end{document}

%% file: Ch1.tex
\chapter{Introduction}\label{part:intro}
\pagenumbering{arabic}

This thesis is part of a bigger project which aims at the development
 of a theoretical formalism to
 describe the structure of hadrons or, more general, of 
few-body bound states in terms of the properties of
  their constituents within the 
framework of the point form of relativistic dynamics~\cite{Dirac:1949cp}.
The observables that encode the internal structure
 of hadrons are called form factors. 
They are functions of the Lorentz invariant variables one 
can build from the four-momenta of the incoming and outgoing hadron.
The theoretical analysis of hadron form factors
 amounts to the derivation of hadron currents in terms of 
constituents' currents. 
The electroweak hadron currents we are interested in can 
be extracted from invariant one-boson-exchange amplitudes, 
which are written as the contraction of a (pointlike) lepton current with 
a hadron current times the gauge-boson ($\gamma$, $W^\pm$, $Z^0$) propagator. 

A proper relativistic formulation of the electroweak structure of
few-body bound states poses several problems. 
The hadron current cannot be a simple sum of the constituent 
currents~\cite{Siegert:1937yt}. 
Even if one has model
wave functions for the few-body bound states one is interested in,
it is not straightforward to construct electromagnetic and weak
currents with all the properties they should have. 

Two basic features
are Poincar\'e covariance and cluster
separability~\cite{Keister:1991sb,Sokolov:1977ym,Coester:1982vt}.
The latter means that the bound-state current should become a sum of
subsystem currents, if the interaction between the subsystems is
turned off. This property is closely related 
to the requirement that
the charge of the whole system should be the sum of the subsystem
charges, irrespective whether the interaction is present or
not~\cite{Lev:1994au}. Electromagnetic currents should, furthermore,
satisfy current conservation and in the case of electroweak currents
of heavy-light systems one has restrictions coming from heavy-quark
symmetry that should be satisfied if the mass of the heavy quark
goes to infinity~\cite{Isgur:1989vq,Isgur:1989ed,Neubert:1993mb}.

Quantum chromodynamics (QCD) has been established as the theory of the 
strong interaction. 
The extraction of hadron observables such as masses and electroweak form factors 
from first principles requires to solve the QCD bound-state problem. 
As long as a complete answer
is not available in the low-energy regime, 
different approximations and QCD-motivated models are required as a step towards
the understanding of hadron properties
that might be justified by the underlying theory a posteriori. 
The difficulty of deriving electromagnetic and weak hadron currents and form factors
 lies in the fact that one has to respect, at the same time, Poincar\'e covariance and 
 the non-perturbative nature of strongly bound states. 
The scale of reactions where energy and momentum 
transfers are comparable to the masses of the constituent particles 
and where particle production may occur,
requires to combine quantum theory with the principles of relativity.

Dirac formulated the problem of including interactions in relativistic classical Hamiltonian 
dynamics~\cite{Dirac:1949cp}. His formulation generalizes in a natural way to quantum 
mechanical systems by means of canonical quantization. 
He identified three particular 
\textit{forms} for which the solution of the problem simplifies.
He called them the
\textit{instant form}, the \textit{point form} and the \textit{front form}.
Each form is associated with a 
hypersurface in  Minkowski space that is left invariant
under transformations belonging to the \textit{kinematical} subgroup
of the Poincar\'e group. 
The corresponding generators (kinematical generators) are free of interactions.
Interaction terms enter the, so called, \textit{dynamical} generators.
Classically, initial conditions are posed on those hypersurfaces, quantum mechanically they serve 
as quantization surfaces. 
Although Dirac formulated the problem in classical mechanics, the three forms 
exist also in quantum mechanics and in quantum field theory. In fact, in quantum
field theory the interaction terms enter automatically the dynamical generators by integrating
the corresponding Noether currents over the respective quantization surfaces. 
An explanation how this is done within the instant, 
front and point forms, respectively,
can be found in Refs.~\cite{Rocha:2009xq,Brodsky:1997de,Biernat:2007sz}.
Reference~\cite{Rocha:2009xq}, e.g., demonstrates that it is highly non-trivial
to boost bound states in QCD using instant-form boosts.

Of the three forms of dynamics the point form is the least known one, 
despite it possesses definite virtues in applications to low- and medium-energy 
hadron problems~\cite{Biernat:2010tp}. It has the nice 
feature that the Lorentz group (rotations and boosts) is kinematical. 
This allows to boost \text{and} rotate bound-state wave functions in a simple way. 
As a price, all components of the 4-momentum operator become interaction dependent.
The framework of \textit{relativistic quantum mechanics} combines
quantum theory and the principles of special relativity.
It deals with a finite number of degrees of freedom 
and aims at the construction of dynamical models compatible with a set of 
general principles, including relativity. 
By construction, the aspired symmetries 
are thus realized exactly~\cite{Keister:1991sb}. 

The framework we will adopt is based on 
the \textit{point form of relativistic 
quantum mechanics} (PFRQM) and 
makes use of the  Bakamjian-Thomas 
construction~\cite{Keister:1991sb,Bakamjian:1953kh}
 for introducing interactions in a fully 
Poincar\'e invariant manner. As a consequence 
the 4-momentum operator factorizes into an
interacting mass operator and a free 4-velocity 
operator so that it suffices to consider
only an eigenvalue problem for the mass operator. 
We use a multichannel version of a Bakamjian-Thomas type
mass operator~\cite{Keister:1991sb,Bakamjian:1953kh} that is represented
in a velocity-state basis~\cite{Klink:1998zz}.
A strategy that certainly distinguishes our approach from other approaches
is the description of interaction vertices, 
which are motivated by quantum field theory and given by means of an appropriate relation  
to the respective interaction Lagrangian density~\cite{Klink:2000pp}.

The multichannel formalism we are going to use was first applied to calculate 
the spectrum and decay widths of vector mesons within
 the chiral constituent quark 
model~\cite{Krassnigg:2003gh,Krassnigg:2004sp}.
More recently, electromagnetic properties of spin-0 and 
spin-1 two-body bound states consisting of equal mass 
particles~\cite{Biernat:2010tp,Biernat:2009my,Biernat:2011mp} have been studied. 
These calculations
were restricted to space-like momentum transfers. For instantaneous
binding forces the results were found to be equivalent with those
obtained with a one-body ansatz for the current in the covariant
front-form approach~\cite{Carbonell:1998rj}.
The present work is an extension of this foregoing 
work to unequal-mass constituents and to
weak decay form factors in the time-like momentum transfer region. 
A great part of the work presented here can also be found in 
Refs.~\cite{GomezRocha:2012zd,Rocha:2010wm,GomezRocha:2011qs,GomezRocha:2012hc}.

An additional requirement for the description of 
systems with unequal constituent masses
is to respect the heavy-quark symmetry predictions 
in the limit in which one of the masses is 
infinitely heavy~\cite{Neubert:1993mb}. This work is also intended as
a check whether the additional restrictions imposed by heavy-quark symmetry are 
respected if one lets one of the masses go to infinity. 

The literature on point-form calculations of the electroweak structure of 
heavy-light systems is very sparse, although the point form seems
to be particularly suited for the treatment of this kind of
systems. We are aware of two papers by
Keister~\cite{Keister:1992wq,Keister:1997je}. 
It is possible to formulate
a covariant one-body current in the point form by imposing the general constraints 
that such a current 
should have~\cite{Klink:1998qf,Melde:2004qu}. However, following 
Refs.~\cite{Biernat:2010tp,Biernat:2009my,Biernat:2011mp},
our purpose is to derive these currents in such a way 
that they are compatible with the binding forces,
avoiding to make a particular ansatz that imposes the conditions that the current
should have.
There is a long list of papers in which relativistic
constituent-quark models serve as a starting point for the
calculation of the electroweak structure of heavy-light mesons
 in front form. To mention a few, see 
those in
Refs.~\cite{Jaus:1996np,Simula:1996pk,Demchuk:1995zx,Cheng:1996if,Choi:1999nu}.
In these papers the electromagnetic and weak meson
currents are usually approximated by one-body currents, which means
that those currents are assumed to be a sum of contributions in
which the gauge boson couples only to one of the constituents,
whereas the others act as spectators. It is well known that this
approximation leads to problems with covariance of the currents in
front form and in instant form~\cite{Lev:1994au}. The form factors
extracted from such a one-body approximation of a current depend, in
general, on the frame in which the approximation is made. 
In the
covariant front-form formulation suggested in
Ref.~\cite{Carbonell:1998rj} this problem is circumvented by
introducing additional, spurious covariants and form factors that
are associated with the chosen orientation of the light front.
One way to (partly) cure this problem is the introduction of a
non-valence contribution leading to a, so called,
$Z$-graph~\cite{Simula:2002vm,Bakker:2003up}. 
This is necessary, in particular, when one considers weak decays, 
where the momentum transfer
is time like and it is thus not possible to use the very convenient $q^+=0$ frame
in the front form.
Such a non-valence
contribution to the currents is also included in an effective way in
the instant-form approach of Ebert et al.~\cite{Ebert:2006nz}. 
In connection with instant-form constituent-quark models
one should also mention the papers of Le Yaouanc et al. (see, e.g.,
Ref.~\cite{Morenas:1997nk} and references therein). They were the
first to prove that covariance of a one-body current is recovered,
if the mass of the heavy quark goes to
infinity~\cite{LeYaouanc:1995wv}. Thereby they made use of the known
boost properties of wave functions within the Bakamjian-Thomas
formulation of relativistic quantum mechanics.

Another focus of investigation of this thesis concerns the question of cluster separability. 
It is know that the Bakamjian-Thomas construction entails cluster problems~\cite{Keister:1991sb}, 
which are manifest also in our calculation of form factors and lead to
unphysical contributions in the electromagnetic 
currents~\cite{Biernat:2010tp,Biernat:2011mp}. This 
resembles the occurrence of analogous contributions within the covariant light-front
formulation of Carbonell \textit{et al.}~\cite{Carbonell:1998rj}. It is our purpose to 
investigate these nonphysical dependence in the case of electroweak
form factors of heavy-light systems. 

A further step forward
is done in Chap.~\ref{ChDeuteronExCurrents}, 
where we extend the formalism 
to the study of bound states whose binding is caused by dynamical particle exchange.
The problem how to take into account retardation effects in the particle-exchange potential is 
formulated and we show the wave-function solution for 
a simple dynamical model of the deuteron. 
Similar studies on this effects in front-form relativistic quantum mechanics
were done in Ref.~\cite{Huang:2008jd}.

Within the coupled-channel approach it is also 
possible to deal with additional dynamical degrees of freedom, such that one 
can, e.g., account for non-valence Fock-state contributions
in hadrons. Some work in this direction has already been done in Ref.~\cite{Kleinhappel:2011is}.
A long-term goal 
would be to formulate QCD in terms of 
\textit{point-form quantum field theory} (PFQFT). 
Some work on this matter can be found in 
Refs.~\cite{Biernat:2007sz,Gromes:1974yu,Fubini:1972mf,Klink:2008qt}.

\subsubsection{Structure of this document}
This dissertation is organized as follows:

\noindent
Chapter~\ref{ChPointForm} presents the basics ideas of the 
point-form framework on which our work is based and settles the prerequisites for 
the subsequent chapters. 
The coupled-channel formulation is introduced in 
Chap.~\ref{ChCoupledChannel}, where we derive the one-photon-exchange amplitude
 for electron scattering off a heavy-light meson and the one-$W$-exchange amplitude for 
the semileptonic decay of a heavy-light meson into another 
heavy-light meson. 
From these transition amplitudes we identify the 
electromagnetic and weak hadron currents. 
The Lorentz structure of these currents is studied in Chap.~\ref{ch:currents:and:ff}
which contains also a short
discussion of cluster problems. As a result of this analysis
the electromagnetic and weak (transition) form factors are
obtained. 
In  Chap.~\ref{HQ:sym} heavy-quark symmetry is checked
 by taking one of the quark masses to infinity. The heavy-quark limit of the electromagnetic 
and weak decay form factors yields a single universal function, the Isgur-Wise function.  
Cluster separability is studied in the heavy-quark limit and the
 relation with front-form results is
discussed. Numerical results for electroweak form factors of heavy-light systems 
as well as for the Isgur-Wise function are 
presented and discussed in Chap.~\ref{ChNumStudiesI}. 
A numerical study of heavy-quark symmetry breaking is made by comparisons with 
the Isgur-Wise function.
In Chap.~\ref{ChNumStudiesII} the method is applied to semileptonic heavy-to-light meson
decays.
A numerical comparison with results obtained within the light-front quark model
is given, observing the importance of considering the non-valence contributions. 
Chapter~\ref{ChDeuteronExCurrents} extends the point-form coupled-channel approach 
to the study of bound states whose binding is caused by dynamical particle exchange,
which leads to, so-called, \textit{exchange currents}. 
We formulate the coupled-channel problem for electron scattering off such a bound
state, identify again the electromagnetic current from the one-photon-exchange
amplitude, including now the exchange current, and study the effect of retardation of the exchanged
particle on the bound-state wave function for a simple Walecka-type model of the deuteron.
Finally, Chapter~\ref{ChFFChiralMultiplets} presents an example where the Clebsch-Gordan
coefficients of the Poincar\'e group defined in the context of relativistic quantum
mechanics~\cite{Keister:1991sb} are applied. The angular momentum decomposition of chiral 
multiplets is realized in the instant and in the front forms. These are  results already
published in Ref.~\cite{GomezRocha:2012yk}. 
The summary, conclusions and an outlook are given in Chap.~\ref{ChConclusions}.
For notations, conventions and details of particular calculations the reader may consult the Appendix.
 \newpage
\thispagestyle{empty}

%% file: Ch2.tex
\chapter{The point form of dynamics}\label{ChPointForm}

The most important concepts needed in the sequel
are presented in this chapter. 
The framework is the \textit{point form of relativistic quantum mechanics}.
We summarize here the most important ideas, which can
be read in much more detail in the bibliography provided
in this section. The most important references are
~\cite{Dirac:1949cp,Keister:1991sb,Bakamjian:1953kh,Klink:1998zz,Klink:2000pp,Biernat:2011mp}.

\section{Introduction}

Our point-form approach is formulated within the framework of
\textit{relativistic quantum mechanics}~\cite{Keister:1991sb,Wigner:1939cj,Bargmann:1954gh}. 
This requires to combine the principles of special relativity with the postulates of 
quantum mechanics. 
Relativity  implies that the measured probabilities are not changed by the 
action of a symmetry transformation of the Poincar\'e group. 
This can be achieved by constructing an appropriate representation of the Poincar\'e 
generators that acts on a certain Hilbert space and that satisfies the Poincar\'e algebra.
Unlike quantum field theory, relativistic quantum mechanics
describes systems with a finite number of degrees of freedom. 
A consistent way to introduce the interactions in a system with a finite number of particles 
preserving Poincar\'e invariance is provided by the Bakamjan-Thomas 
construction~\cite{Keister:1991sb,Bakamjian:1953kh}. 
We will employ its point-form version, which allows to split
the 4-momentum operator into an interacting mass and a free velocity operator. This permits
to separate the overall velocity of the system from the internal motion, so that one 
can concentrate on the study of the  dynamics of the internal variables only. 
In this framework it is convenient to define a special basis of multiparticle states that 
differs from the usual tensor-product basis. We will introduce 
\textit{velocity states}~\cite{Klink:1998zz}.
At the end of the chapter we present how to include the creation and annihilation of 
particles via vertex operators that are defined by means of a quantum-field theoretical
interaction Lagrangian densities~\cite{Klink:2000pp}. They will be necessary for the construction of 
a coupled-channel formalism that allows to describe particle-exchange interactions. 
This will complete the basic concepts and tools used in the 
next chapters. They will be, if necessary, presented in more detail for particular cases.

\section{Forms of relativistic dynamics}

The construction of a Poincar\'e-invariant quantum theory is 
equivalent to finding a representation of 
the Poincar\'e generators in terms of self-adjoint 
operators that satisfy the Poincar\'e algebra and 
that act on an appropriate Hilbert space. 
The Poincar\'e algebra in its manifest
covariant form is given by
\begin{align}
[\hat P^\mu ,\hat P^\nu]&= 0, \label{PP}\\
[\hat P^\mu,\hat J^{\nu\rho}] &= i(g^{\mu\nu}\hat P^\rho-g^{\mu\rho}\hat P^\nu), \label{PJ}\\
[\hat J^{\mu\nu},\hat J^{\rho\sigma}]  &= - i(g^{\mu\rho}\hat J^{\nu\sigma}-
g^{\mu\sigma}\hat J^{\nu\rho}+g^{\nu\sigma}\hat J^{\mu\rho}-g^{\nu\rho}\hat J^{\mu\sigma}). \label{JJ}
\end{align}
The introduction of interactions has to be made 
in such a way that the group structure is 
preserved, i.e. so that the commutation 
relations are not altered. From the 
commutation relations of the Poincar\'e algebra it
follows that the inclusion of  interaction terms
in the Hamiltonian $\hat P^0$ affects the structure of, at least,
some of the other generators. 
As an example it is instructive to consider the commutation relation:
\begin{equation}\label{irrep}
 [\hat P^j,\hat K^k]=i\hat P^0\delta^{jk}.
\end{equation}
It is straightforward to notice that adding 
interactions to $\hat P^0$ on the right-hand side requires also
adding interactions on the left-hand side,
modifying either $\hat P^j$, $\hat K^k$ or both of them~\cite{Keister:1991sb}.
The different ways how one introduces the interactions in the Poincar\'e
generators leads to the different 
\textit{forms of relativistic dynamics}.

In his seminal paper of 1949 Dirac  distinguished 
three prominent ways of combining the 
principles of relativity with the Hamiltonian 
formulation of dynamics~\cite{Dirac:1949cp}.\footnote{Although 
Dirac formulated the problem  in classical mechanics, the goal of
letting the equation of motion have a Hamiltonian 
form was to allow the transition to the quantum theory~\cite{Dirac:1949cp}. The forms of
dynamics exist also in the quantum theory and in quantum field theory. In our discussion we 
will refer only to quantum theory.}
The three prominent forms of relativistic dynamics 
are characterized by three different ways of
separating \textit{kinematical} generators -- free of interactions --,  from 
\textit{dynamical} ones -- interaction dependent --. The latter
 were called ``Hamiltonians'' by Dirac~\cite{Dirac:1949cp}.

The standard way of including the interactions 
between particles is the \textit{instant form}, 
which expresses everything in terms of dynamical variables
at one instant of time, e.g. $x^0=0$; in quantum theory this is
 the quantization surface.
The Hamiltonians in the instant form are given by the set 
$\{\hat P^0,\hat K^1,\hat K^2, \hat K^3\}$, 
which are the energy and the three
generators of the boosts. The kinematical group, consisting
of translations and rotations, leaves the equal-time surfaces invariant. 

The second form suggested by Dirac 
poses the physical conditions
on a three-dimensional hyperboloid, $x^\mu x_\mu=\tau^2$. 
In this form the Lorentz group,
i.e. the group consisting of rotations 
and boosts, is kinematical, while
the four generators of space-time translations are dynamical. The set 
$\{\hat P^0,\hat P^1,\hat P^2, \hat P^3\}$ 
are the Hamiltonians in this form,
which was called the \textit{point form}, for being characterized by the kinematic 
subgroup leaving the origin invariant. 

The third form is characterized by a 
three-dimensional
hyperplane in space-time that is tangent to the light-cone.
It was called the \textit{front form}. 
The quantization surface is customary chosen as $x^+:=x^0+x^3=0$.
The front form has the largest number of
kinematical generators, namely 7, 
 that leaves the \textit{light front} $x^+=\text{const.}$ invariant:
$\{\hat P^1,\hat P^2, \, \hat J^3, \hat K^3, \hat P^+=\hat P^0+\hat P^3,
\hat E^1=\hat K^1+ \hat J^2,\hat E^2=\hat K^2- \hat J^1\}$. 
The set of generators that play the role of Hamiltonians is given by only three dynamical operators 
$\{\hat P^-=\hat P^0-\hat P^3,\hat F^1=\hat K^1-\hat J^2,\hat F^2=\hat K^2+\hat J^1\}$.

In the point form 
dynamical and kinematical generators are clearly separated into two subgroups 
of the Poincar\'e group, namely the space-time translations $\hat P^\mu$ 
and the homogeneous Lorentz group $\hat J^{\mu\nu}$. This makes Lorentz-covariance properties of physical
quantities quite obvious.
The investigations carried out in this work therefore employ the point form.

\section{The point form}
The kinematical nature of the Lorentz group in the point-form formulation
 permits to write equations 
in a manifestly Lorentz covariant way. One can perform changes of reference frames
 in a simple fashion, since
the boost operator is not affected by interactions.
Using rotationless (canonical) boosts
 allows for the addition of 
angular momentum and spin by means of usual $SU(2)$-Clebsch-Gordan coefficients.
This is done in the 
center-of-momentum frame (of the (sub)system for which spins and orbital angular momenta should 
be added)\footnote{
The relativistic
addition of spin and orbital angular momentum away from the center-of-momentum frame
amounts to the construction of the Clebsch-Gordan coefficients
of the Poincar\'e group, which are different for every form. We will refer to the Clebsch-Gordan 
coefficients of the Poincar\'e group at the end of this work, where an example of application
to chiral multiplets 
will be shown in Chap.~\ref{ChFFChiralMultiplets}.}. 
The construction of the 4-momentum operator that satisfies the Poincar\'e algebra
is not trivial and one needs to be
particularly careful when one attempts to relate the momenta of the individual 
particles to the total momentum of 
the system.

The problem of including interaction terms in the 4-momentum
concerns the commutators (\ref{PP}) and (\ref{PJ}) 
of the Poincar\'e algebra. The latter is satisfied
provided that $\hat P^\mu$ transforms like a 4-vector. Satisfying (\ref{PP}) involves quadratic
conditions in the interaction terms. Dirac posed the latter problem as the real difficulty
in constructing a relativistic dynamical theory in the point form~\cite{Dirac:1949cp}. 
The requirements for Poincar\'e invariance in this form
can be summarized in the so-called \textit{point-form equations}~\cite{Biernat:2010tp}:
\begin{align}\label{PFequations}
  [\hat P^\mu , \hat P^\nu]&= 0, \\
\hat U_\Lambda\hat P^\mu\hat U_\Lambda^\dagger &=\left(\Lambda^{-1}\right)^\mu_\nu \hat P^\nu,
\end{align}
where $\hat U_\Lambda$ is the unitary operator representing 
the Lorentz transformation $\Lambda$
on the Hilbert space. 
Any representation of the Poincar\'e group formulated in the point form must satisfy the last two
conditions. 

In quantum field theory the derivation of the Poincar\'e generators
is given by integration of the Noether's currents associated with the Poincar\'e group
 over the respective quantization surfaces that determine
the interaction dependence in the Poincar\'e generators. 
In the point form the interactions enter
the 4-momentum operator $\hat P^\mu$ when 
one integrates over the hyperboloid $x_\mu x^\mu=\tau^2$~\cite{Biernat:2007sz}.

The problem of constructing a Poincar\'e invariant quantum theory for a restricted number 
of particles is specially involved because of the (in general) non-linear constraints imposed
by the Poincar\'e algebra on the interaction terms. 
A consistent method to construct the Poincar\'e generators that guarantees Poincar\'e
invariance is the Bakamjian-Thomas construction~\cite{Keister:1991sb,Bakamjian:1953kh}.
It requires only linear conditions for the interactions.
We will briefly summarize the procedure for the point form. A more detailed description
 of the Bakamjian-Thomas construction that includes also the
procedure in the other forms can be found in Ref.~\cite{Keister:1991sb}.

\subsection{The Bakamjian-Thomas construction}

The Bakamjian-Thomas construction is a four-step construction that allows
 for the Poncar\'e-invariant addition of interactions. We briefly summarize these steps
in its point-form version.

The first step is common to every form and is given by the construction of the generators 
of the Poincar\'e algebra for a free many-particle system by means of the Poincar\'e generators
for a free particle. The Hilbert space for a free many-particle system 
is given by the tensor product of single-particle Hilbert spaces. 
For a system of two particles, 1 and 2, the representation of the Poincar\'e
generators on the tensor-product Hilbert space is given by:
 \begin{align}
  \hat P^\mu_{\text{free}}&:=\hat P^\mu_{1}\otimes\mathds{1}_2
 +\mathds{1}_1\otimes \hat P^\mu_{2},\\
 \hat J^{\mu\nu}_{\text{free}}&:=\hat J^{\mu\nu}_{\text{1}}\otimes\mathds{1}_2
 +\mathds{1}_1\otimes \hat J^{\mu\nu}_{2}.
 \end{align}
The second step is the construction of a convenient set of auxiliary operators from the
free generators, one of them being the mass operator. In the point form, from the set of free
generators $\{\hat P^\mu_{\text{free}}, \hat J^{\mu\nu}_{\text{free}}\}$, 
one may construct mass and velocity operators defined as 
follows:
\begin{align}
 \hat M_{\text{free}}&:=\sqrt{ \hat P^\mu_{\text{free}} \hat P_{\mu}^{\text{free}}},\\
\hat V_{\text{free}}^\mu&:=\frac{\hat P^\mu_{\text{free}}}{ \hat M_{\text{free}}}.
\end{align}
The auxiliary set of operators in the point form 
is then $\{\hat M_{\text{free}}, \hat{\vec V}_{\text{free}}, \hat J^{\mu\nu}_{\text{free}}\}$.
The mass operator $ \hat M_{\text{free}}$ is the square root of a Casimir
operator for the Poincar\'e group and therefore
 commutes with all the generators or functions of them.

In the third step the interactions are included into the mass operator in the form of a potential 
 that we call $\hat M_{\text{int}}$, giving an \textit{interacting} mass operator:
\begin{equation}
 \hat M:=  \hat M_{\text{free}}+\hat M_{\text{int}}.
\end{equation}

The condition to satisfy the point-form equations (\ref{PFequations}) is that the interaction 
term $\hat M_{\text{int}}$ must be a Lorentz scalar that commutes with the free velocity 
operator,
\begin{equation}
 [\hat M_{\text{int}},\hat V^\mu_{\text{free}}]=0.
\end{equation}
This ensures that the interaction dependent mass operator $\hat M$ still commutes with the other
operators of the auxiliary set. 

The fourth and final step requires the reconstruction of the original set of generators 
from the set of auxiliary operators in which the free mass operator $\hat M_{\text{free}}$
is replaced
by the interacting one, $\hat M$. The new set of generators for interacting particles 
satisfies the Poincar\'e algebra.
In the point form the only generators that contain interactions are the components of the 
4-momentum. The (interacting) 4-momentum operator reads:

\begin{equation}\label{EqPdef}
 \hat P^\mu=\hat P^\mu_{\text{free}}+\hat P^\mu_{\text{int}}=\left( \hat M_{\text{free}}
+\hat M_{\text{int}}\right) \hat V_{\text{free}}^\mu.
\end{equation}
Note that the definition of a \textit{free} velocity operator in terms of a mass operator
is only feasible in the point form, since the requirement that the 3-momentum operator 
remains interaction independent in 
the instant or in the front forms would imply interactions in the velocity.
The possibility of defining a free 4-velocity in the point form suggests to
utilize a particular basis to define multiparticle states. This leads to the introduction
of the so-called \textit{velocity states}~\cite{Klink:1998zz}.

The Bakamjian-Thomas construction ensures the relativistic invariant treatment of a
finite number of interacting particles. 
It is known, however, that the Bakamjian-Thomas construction
causes problems with cluster separability~\cite{Keister:1991sb}.
The latter is related to the difficulty to treat properly separated 
subsystems such that their dynamics 
decouples for large space-like separations~\cite{Keister:1991sb,Foldy:1960nb}. 
We will refer to this problem later,
since it will appear along the next chapters.

\subsection{Velocity states}\label{SecVelocityStates}
Velocity states were introduced by Klink for the purpose of coupling 
 multiparticle relativistic states simultaneously~\cite{Klink:1998zz} . They
have the property that internal variables such as spin and orbital angular momenta
of relativistic multiparticle systems can be coupled together as is done in nonrelativistic
quantum mechanics. 

An $n$-particle velocity state $|v;\vec k_1,\mu_1;\vec k_2,\mu_2;...;\vec k_n,\mu_n\rangle$
is defined through an overall velocity $v$ and $n$ individual momenta and spin 
projections $\{\vec k_i,\mu_i\}$, such that $\sum_{i=1}^n \vec k_i=0$. By construction one of the
individual 3-momenta $\vec k_i's$ is redundant. A velocity state represents an
$n$-particle system in the rest frame that is boosted to a frame with
a total 4-velocity $v$ ($v^\mu v_\mu=1$) by means of a canonical boost 
$B_c(v)$ (cf.~App.~\ref{AppCanonicalBoosts}):
\begin{equation}\label{eq:velstat}
\vert v; \vec{k}_1, \mu_1; \vec{k}_2, \mu_2;\dots;
\vec{k}_n, \mu_n \rangle:= \hat{U}_{B_c(v)} \, \vert
\vec{k}_1, \mu_1; \vec{k}_2, \mu_2;\dots; \vec{k}_n, \mu_n \rangle.
\end{equation}
They satisfy the orthogonality and completeness relations:
\begin{eqnarray}\label{eq:vnorm}
&&\langle v^\prime; \vec{k}_1^\prime,
\mu_1^\prime; \vec{k}_2^\prime, \mu_2^\prime;\dots;
\vec{k}_n^\prime, \mu_n^\prime \vert \, v; \vec{k}_1, \mu_1;
\vec{k}_2, \mu_2;\dots; \vec{k}_n, \mu_n \rangle  \nonumber\\
&&\quad= v_0 \, \delta^3(\vec{v}^\prime-\vec{v})\, \frac{(2 \pi)^3 2
\omega_{k_n}}{\left( \sum_{i=1}^n \omega_{k_i}\right)^3}
\left( \prod_{i=1}^{n-1} (2\pi)^3
2 \omega_{k_i} \delta^3(\vec{k}_i^\prime-\vec{k}_i)\right) \left(
\prod_{i=1}^{n} \delta_{\mu_i^\prime \mu_i}\right)\,\nonumber\\
\end{eqnarray}
and 
\begin{eqnarray}\label{eq:vcompl}
\mathds{1}_{1,\dots,n}&=&\sum_{\mu_1=-j_1}^{j_1}
\dots \sum_{\mu_n=-j_n}^{j_n} \int
\frac{d^3 v}{(2\pi)^3 v_0}  \left[
\prod_{i=1}^{n-1}\frac{d^3k_i}{(2 \pi)^3 2 \omega_{k_i}}
\right]\frac{\left(\sum_{i=1}^n \omega_{k_i}\right)^3}{2
\omega_{k_n}}\nonumber\\ & & \times \vert v; \vec{k}_1, \mu_1;
\vec{k}_2, \mu_2;\dots; \vec{k}_n, \mu_n \rangle
 \langle v; \vec{k}_1, \mu_1; \vec{k}_2,
\mu_2;\dots; \vec{k}_n, \mu_n\vert\, ,\nonumber\\
\end{eqnarray}
with $m_i$, $\omega_{k_i}:= \sqrt{m_i^2+\vec{k}_i^2}$, and $j_i$,
being the mass, the energy, and the spin of the $i$th particle,
respectively. 

Velocity states transform under Lorentz transformations $\Lambda$ as
\begin{eqnarray}\label{eq:vstateboost}
\lefteqn{\hat{U}_\Lambda \vert v; \vec{k}_1, \mu_1; \vec{k}_2,
\mu_2;\dots; \vec{k}_n, \mu_n \rangle} \nonumber\\ &&=
\sum_{\mu_1^\prime, \mu_2^\prime,\dots,\mu_n^\prime}\, \left\{
\prod_{i=1}^n \, D^{j_i}_{\mu_i^\prime \mu_i}\left[R_{\mathrm
W}(v,\Lambda)\right] \right\}\nonumber\\&& \quad\times
 \vert \Lambda v; \overrightarrow{R_{\mathrm
W}(v,\Lambda)k}_1, \mu_1^\prime; \overrightarrow{R_{\mathrm
W}(v,\Lambda)k}_2, \mu_2^\prime;\dots; \overrightarrow{R_{\mathrm
W}(v,\Lambda)k}_n, \mu_n^\prime \rangle\, ,\nonumber\\
\end{eqnarray}
with the Wigner-rotation matrix
\begin{equation}\label{eq:wignerrot}
R_{\mathrm W}(v,\Lambda) = B_c^{-1}(\Lambda v)\Lambda B_c(v)\, .
\end{equation}
Using velocity states rather than the usual tensor-product states 
allows to perform the addition of angular
momentum in the same way as in nonrelativistic 
quantum mechanics, since all individual particle
states transform with the same Wigner rotation. 
In a 
velocity-state basis, the Bakamjian-Thomas type 4-momentum operator, Eq.~(\ref{EqPdef}), 
is diagonal in the 4-velocity $v$, which can 
be factored out as a velocity-conserving delta function, allowing to separate it from the 
internal motion in such a way that one can concentrate on studying the mass operator $\hat M$
which is a function of the internal variables only. 

Note that the momenta $k_i^\mu$ do not transform like 4-vectors under Lorentz transformations.
The effect of a Lorentz transformation on such momenta is a Wigner rotation 
(see Eq.~(\ref{eq:vstateboost})).
The relation between single particle and internal particle momenta is given by a 
canonical boost  $p_i=B_c(v)k_i$. The $p_i$ transform like 4-vectors~\cite{Biernat:2009my,Biernat:2011mp}.

Velocity states are eigenstates of the operators $\hat M:=\sqrt{\hat P^2}$, 
$\hat V^\mu:=\frac{\hat P^\mu}{\hat M}$ and also $\hat k_i^\mu$. Their eigenvalues
are given by
\begin{align}
 \hat M |v,\vec k_1,\mu_1;\vec k_2,\mu_2;...;\vec k_n,\mu_n\rangle 
&= \sum_{i=1}^n \omega_{k_i} |v,\vec k_1,\mu_1;\vec k_2,\mu_2;...;\vec k_n,\mu_n\rangle \\
 \hat V^\mu|v,\vec k_1,\mu_1;\vec k_2,\mu_2;...;\vec k_n,\mu_n\rangle 
&=  v^\mu|v,\vec k_1,\mu_1;\vec k_2,\mu_2;...;\vec k_n,\mu_n\rangle\\
&=  \frac{\sum_{i=1}^n p_i^\mu}{\sum_{i=1}^n \omega_{k_i}} |v,\vec k_1,\mu_1;\vec k_2,\mu_2;...;\vec k_n,\mu_n\rangle. \nonumber
\end{align}

\subsection{Mass operators from interaction Lagrangians}\label{SecVertexLagrangian}
The creation and annihilation of particles is introduced in this framework by means of 
\textit{vertex operators} $\hat K$ that are specified by the velocity state representation
and an appropriate relation to the pertinent field-theoretical interaction-Lagrangian 
density $\hat{\mathcal{L}}_{\text{int}}$. Due to velocity conservation that 
follows from the point-form version of the  Bakamjian-Thomas construction, one is led to 
define matrix elements of $\hat K$ by~\cite{Biernat:2010tp,Klink:2000pp}:
\begin{eqnarray}
&& \langle v,\vec k_1,\mu_1;...;\vec k_{n+1},\mu_{n+1}| 
\hat K^\dagger| v,\vec k_1,\mu_1;\vec k_2,\mu_2;...;\vec k_n,\mu_n\rangle \nonumber\\
&&\quad= \langle v,\vec k_1,\mu_1;\vec k_2,\mu_2;...;\vec k_n,\mu_n| 
\hat K|  v,\vec k_1,\mu_1;...;\vec k_{n+1},\mu_{n+1}\rangle ^*\nonumber\\
&&\quad=\mathcal{N}_{n+1,n}v^0\delta^3(\vec v-\vec v')\nonumber\\
&&\quad\quad\times \langle v,\vec k_1,\mu_1;...;\vec k_{n+1},\mu_{n+1}| 
\hat{\mathcal L}_\text{int}(0)f(\Delta m)| v,\vec k_1,\mu_1;\vec k_2,\mu_2;...;\vec k_n,\mu_n\rangle,\nonumber\\
\end{eqnarray}
where $\mathcal{N}_{n+1,n}=(2\pi)^3/\sqrt{\mathcal M'_{n+1}\mathcal M'_{n}}$, 
$\mathcal M'_{n}=\sum_{i=1}^k \omega_{i}$ 
and $f(\Delta m=\mathcal M'_{n+1}-\mathcal M'_{n})$ denotes a vertex form factor that can be
introduced in order to account for (part of) the neglected off-diagonal velocity contributions and to regulate 
integrals.

In the following all this concepts will be used and explained in more detail for 
particular cases.

%% file: Ch3.tex
\chapter{Coupled-channel approach}\label{ChCoupledChannel}

The Bakamjian-Thomas approach permits a Poincar\'e invariant 
formulation of reactions
that involve particle production. 
In order to describe particle-exchange 
interactions in such a
way that retardation effects are fully taken into 
account we use a multichannel framework that 
allows for the creation and annihilation of a finite 
number of additional particles. 

We will start with the simplest case,
namely the two-channels problem. 
We illustrate the generic mass eigenvalue problem for an
$N$-particle channel coupled to an $(N+1)$-particle channel. 
The eigenvalue problem reads:
\begin{equation}\label{twochannelgeneral}
 \left(\begin{array}{cc} \hat M_N & \hat K \\
 \hat K^\dagger & \hat M_{N+1}\end{array}\right)
 \left(\begin{array}{l}  |\psi_N\rangle \\
 |\psi_{N+1}\rangle \end{array}\right)   =
 m \left(\begin{array}{l} |\psi_{N}\rangle \\
 |\psi_{N+1}\rangle \end{array}\right) \, .
\end{equation}
The diagonal elements represent the invariant masses of the uncoupled 
$N$- and $(N+1)$-particle systems and 
consist of their kinetic energies. In addition, 
the mass operators $\hat M_N$ and $\hat M_{N+1}$ may also 
contain instantaneous interactions between the particles.
The non-diagonal elements, i.e. $\hat K^\dagger$ and $\hat K$, 
are vertex operators that account 
for the creation and annihilation of the $(N+1)$st particle, respectively. 
They provide the coupling between the two channels.

Applying a Feshbach reduction to the system of equations (\ref{twochannelgeneral}), 
the second channel is eliminated
in favor of the first channel
\begin{equation}\label{opt}
 m|\psi_N\rangle =\left(\hat M_N +\hat K(m-\hat M_{N+1})^{-1}\hat K^\dagger \right)
|\psi_N\rangle
=: \left( \hat M_N +\hat V_{\text{opt}}(m)\right)|\psi_N\rangle.
\end{equation}
This leads to the optical potential 
that describes a one-particle-exchange process. 
In order to compute matrix elements
of the optical potential, which we need to obtain transition amplitudes, 
 from which currents and form factors can be extracted, it is necessary to insert the 
completeness relations (\ref{eq:vcompl}) of the eigenstates of the mass operators for  the 
$N$- and $(N+1)$-particle systems:
\begin{equation}\label{matrixelemetsopt}
 \langle \ul v'; \ul{\vec k}_N',\ul \mu_N';...; \ul{\vec k}'_1;\ul \mu'_1 |
\mathds{1}_{N}\hat K(m-\hat M_{N+1})^{-1}\mathds{1}_{N+1}\hat K^\dagger \mathds{1}_{N}|
\ul v; \ul{\vec k}_N,\ul \mu_N;...; \ul{\vec k}_1;\ul \mu_1\rangle.
\end{equation}

\begin{figure}[h]
  \includegraphics[width=1\textwidth]{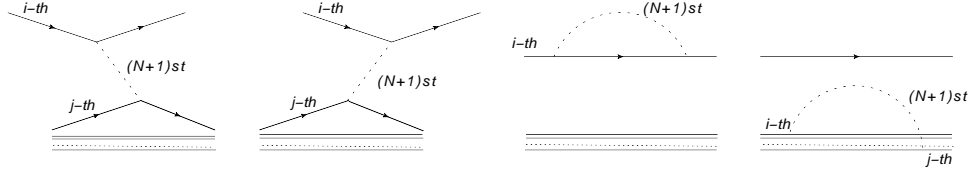}
\caption{All possible time-orderings for the exchange of the $(N+1)$st
particle described by the optical potential (\ref{opt}).}\label{4TimeOrdering2Channels}
\end{figure}

The optical potential $\hat V_\text{opt}(m)$ describes all 
possible time-orderings for the particle exchange
of the $(N+1)$st particle between the $N$ particles, 
including loop contributions, i.e. emission and
absorption by the same particle.  
The propagation of the exchanged particle is represented by
$(m-\hat M_{N+1})^{-1}$, which accounts for retardation effects
 (see Fig.\ref{4TimeOrdering2Channels}). 

Bare vertices of structureless particles are given in quantum field theories 
by interaction Lagrangian densities that couple the fields. 
Such Lagrangian densities are used to fix 
the interaction vertices defined by 
$\hat K$ and $\hat K^\dagger$ as explained in Sec.~\ref{SecVertexLagrangian}.
Unlike quantum-field theoretical vertices, $\hat K$ and $\hat K^\dagger$, however,
may also contain vertex form factors that account for a substructure 
of the interacting particles. If the vertex form factors depend 
on Lorentz invariants only, Poincar\'e invariance will be preserved.

\section{Electron-meson scattering}\label{SecCupledChannelEM}
      
We start with the description of electron-meson scattering with the 
meson being a quark-antiquark bound state. 
We will extract the electromagnetic current from the invariant
one-photon exchange amplitude. 
To describe the process using the coupled-channel approach 
explained above we have to
consider a Hilbert space that is the direct sum of the $eQ\bar q$ and $eQ\bar q\gamma$
Hilbert spaces.\footnote{In view of the fact that we will be interested in heavy-light 
mesons we allow the quark and the anti-quark to have different masses. Without loss of generality
we assume the quark to be the heavier particle which is indicated by a capital ``$Q$'' 
(in contrast to the lower case ``$\bar q$'').} The mass eigenvalue equation has the form
\begin{equation}\label{twochannelEM}
 \left(\begin{array}{cc} \hat M_{e Q\bar q}^{\text{conf}} & \hat K_\gamma \\
 \hat K^\dagger_\gamma & \hat M_{e Q\bar q \gamma}^{\text{conf}}\end{array}\right)
 \left(\begin{array}{l}  |\psi_{e Q\bar q}\rangle \\
 |\psi_{e Q\bar q \gamma}\rangle \end{array}\right)   =
 m \left(\begin{array}{l} |\psi_{e Q\bar q}\rangle \\
 |\psi_{e Q\bar q \gamma}\rangle \end{array}\right) \, .
\end{equation}
$\hat K^\dagger_\gamma$ and $\hat K^\dagger_\gamma$ are the vertex operators that account for the emission 
and absorption of the photon by the (anti)quark or by the electron. 
The confining forces between the mesonic constituents are included in the diagonal of the 
matrix mass operator, i.e.
\begin{eqnarray}
\hat M_{e Q\bar q}^{\text{conf}} &:=& \hat M_{e Q\bar q} +\hat V^{(3)}_{\text{conf}} ,\nonumber \\
\hat M_{e Q\bar q \gamma}^{\text{conf}} &:=& \hat M_{e Q\bar q\gamma} +\hat V^{(4)}_{\text{conf}}, 
\end{eqnarray}
where $\hat V^{(3)}_{\text{conf}}$ and $\hat V^{(4)}_{\text{conf}}$ denote 
the embedding of the confining $Q\bar q$-potential into the 3- and 4-particle Hilbert 
spaces~\cite{Keister:1991sb}.
It is now convenient to introduce (velocity) eigenstates of $\hat M_{e Q\bar q }^{\text{conf}}$ and 
$\hat M_{e Q\bar q \gamma}^{\text{conf}}$ (cf. Sec.~\ref{SecVelocityStates}):
\begin{eqnarray}\label{eq:veigenst}
&&\hat M^{\mathrm{conf}}_{eQ\bar{q}} \vert\,  \underline{v};
\vec{\underline{k}}_e, \underline{\mu}_e;
\vec{\underline{k}}_\alpha,\underline{\mu}_\alpha, \alpha \rangle= (\omega_{\underline{k}_e} +
\omega_{\underline{k}_\alpha} ) \vert\, \underline{v};
\vec{\underline{k}}_e, \underline{\mu}_e;
\vec{\underline{k}}_\alpha,\underline{\mu}_\alpha,  \alpha \rangle, \\
&&\hat M^{\mathrm{conf}}_{eQ\bar{q}\gamma} \vert\,  \underline{v};
\vec{\underline{k}}_e, \underline{\mu}_e;
\vec{\underline{k}}_\gamma, \underline{\mu}_\gamma;
\vec{\underline{k}}_\alpha,\underline{\mu}_\alpha, \alpha \rangle\nonumber\\
&&\qquad\qquad\qquad= (\omega_{\underline{k}_e} +
\omega_{\underline{k}_\gamma}+\omega_{\underline{k}_\alpha} ) \vert\, \underline{v};
\vec{\underline{k}}_e, \underline{\mu}_e;
\vec{\underline{k}}_\gamma, \underline{\mu}_\gamma;
\vec{\underline{k}}_\alpha,\underline{\mu}_\alpha,  \alpha \rangle.
\end{eqnarray}
$\ul \mu_\alpha$ denotes
 the spin projection of the confined $Q\bar q$ bound state, $\alpha$ 
denotes the remaining discrete quantum numbers that specify it uniquely. The 
energy of 
the $Q\bar q$ bound state with quantum numbers $\alpha$ and mass $m_\alpha$ is given by
$\omega_{\ul k_\alpha}:=(m_\alpha^2 +\ul{\vec k}_\alpha^2)^{1/2}$. Underlined velocities, momenta
and spin projections refer to states with a confined $q\bar q$ pair. They have to be distinguished
from eigenstates of the free mass operators $\hat M_{eQ\bar q}$, $\hat M_{eQ\bar q\gamma}$:
\begin{align}
 \hat M_{eQ\bar{q}} \vert\,  v;
\vec{k}_e, &\mu_e;
\vec{k}_Q,\mu_Q; \vec{k}_{\bar q},\mu_{\bar q} \rangle= (\omega_{k_e} +
\omega_{k_Q} +\omega_{k_{\bar q}}) \vert\,  v;
\vec{k}_e, \mu_e;
\vec{k}_Q,\mu_Q; \vec{k}_{\bar q},\mu_{\bar q} \rangle, \\
\hat M_{eQ\bar{q}\gamma} \vert\,  v;
\vec{k}_e, & \mu_e;
\vec{k}_Q,\mu_Q; \vec{k}_{\bar q},\mu_{\bar q}; \vec k_\gamma,\mu_\gamma \rangle\nonumber\\
&= (\omega_{k_e} +
\omega_{k_Q} +\omega_{k_{\bar q}}+\omega_{\gamma}) \vert\,  v;
\vec{k}_e, \mu_e;
\vec{k}_Q,\mu_Q; \vec{k}_{\bar q},\mu_{\bar q}; \vec k_\gamma,\mu_\gamma \rangle.
\end{align}

\begin{figure}[t]
\begin{center}
\includegraphics[width=0.8\textwidth]{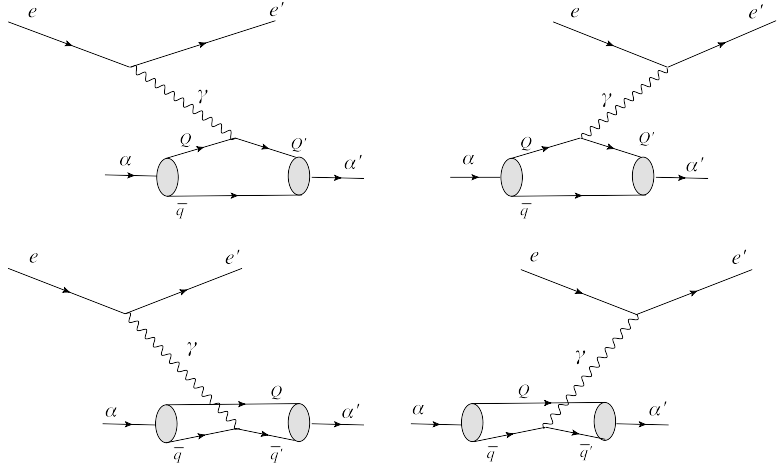}
\caption{The 4 time orderings of photon exchange contributing to 
the optical potential (\ref{eq:voptclust}). The two graphs in the first row contribute 
to $J^\nu_Q$, the other two graphs to $J^\nu_{\bar q}$.}\label{Fig4TemporalOrderingsAlpha}
\end{center}
\end{figure}

The invariant amplitude for the one-photon exchange is obtained  by
calculating appropriate matrix elements of the optical potential. The optical potential 
can be read off from the Feshbach-reduced mass eigenvalue problem:
\begin{equation}\label{optEM}
\left(\hat M_{eQ\bar q}^{\text{conf}} +
\underbrace{\hat K_\gamma(m-\hat M_{eQ\bar q\gamma}^{\text{conf}})^{-1}\hat K_\gamma^\dagger}_{\hat V_{\text{opt}}(m)} \right)
|\psi_{eQ\bar q}\rangle=  m|\psi_{eQ\bar q}\rangle.
\end{equation}
The required matrix elements are obtained by inserting appropriate 
completeness relations between operators (cf. Eq.~(\ref{eq:vcompl}))
\begin{eqnarray}\label{eq:voptclust}
\lefteqn{\langle  \underline{v}^\prime;
\vec{\underline{k}}_e^\prime, \underline{\mu}_e^\prime;
\vec{\underline{k}}_\alpha^\prime,\underline{\mu}_\alpha^\prime,
\alpha \vert\,
\hat{V}_{\mathrm{opt}}(m)
\vert\, \underline{v}; \vec{\underline{k}}_e, \underline{\mu}_e;
\vec{\underline{k}}_\alpha,\underline{\mu}_\alpha,
\alpha \rangle_{\mathrm{os}}} \nonumber \\
&&= \langle  \underline{v}^\prime; \vec{\underline{k}}_e^\prime,
\underline{\mu}_e^\prime;
\vec{\underline{k}}_\alpha^\prime,\underline{\mu}_\alpha^\prime,
\alpha \vert\, \mathds{1}_{e Q \bar{q}}\,\hat{K}_{\gamma}
\mathds{1}_{e Q \bar{q} \gamma}\left(\hat{M}_{e Q\bar q \gamma}^{\text{conf}}
\!-\! m\right)^{-1} \nonumber\\
& &\quad\times \mathds{1}_{e Q \bar{q} \gamma}^{\mathrm{conf}}
\mathds{1}_{e Q \bar{q} \gamma} \, \hat{K}^\dag_{\gamma}
\mathds{1}_{e Q \bar{q}}\, \vert\, \underline{v};
\vec{\underline{k}}_e, \underline{\mu}_e;
\vec{\underline{k}}_\alpha,\underline{\mu}_\alpha,  \alpha
\rangle_{\mathrm{os}}\,
. \nonumber\\
\end{eqnarray}
``os'' means on-shell, this is $m=\omega_{\underline{k}_e}+\omega_{\underline{k}_\alpha} =
\omega_{\underline{k}_e^\prime}+\omega_{\underline{k}_\alpha^\prime}$,
$\omega_{\underline{k}_e}=\omega_{\underline{k}_e^\prime}$ and
$\omega_{\underline{k}_\alpha}=\omega_{\underline{k}_\alpha^\prime}$. The matrix elements 
to be evaluated include wave functions of the confined $Q\bar q$ 
and a free electron (and photon), i. e. 
$\langle v; \vec{k}_e, \mu_e; \vec{k}_Q, \mu_Q; \vec{k}_{\bar{q}},
\mu_{\bar{q}} \vert\, \underline{v}; \vec{\underline{k}}_e,$
$\underline{\mu}_e;
\vec{\underline{k}}_\alpha,\underline{\mu}_\alpha, \alpha
\rangle$,
$\langle v; \vec{k}_e, \mu_e; \vec{k}_Q, \mu_Q; \vec{k}_{\bar{q}},
\mu_{\bar{q}};\vec{k}_{\gamma}, \mu_{\gamma} \vert\,
\underline{v}; \vec{\underline{k}}_e, \underline{\mu}_e;
\vec{\underline{k}}_\alpha, \underline{\mu}_\alpha, \alpha;
\vec{\underline{k}}_\gamma, \underline{\mu}_\gamma\rangle$;
and the transition from a free $Q\bar q e$ state to a free $Q\bar q e\gamma$ state by emission 
(absorption) of a photon,
$ \langle v^\prime; \vec{k}_e^\prime,\! \mu_e^\prime;
\vec{k}_Q^\prime,\! \mu_Q^\prime; \vec{k}_{\bar{q}}^\prime,\!
\mu_{\bar{q}}^\prime; \vec{k}_\gamma^\prime,\! \mu_\gamma^\prime
\vert \,\hat{K}^\dag\, \vert v\,  ; \vec{k}_e,\! \mu_e; \vec{k}_Q,\!
\mu_Q; \vec{k}_{\bar{q}},\! \mu_{\bar{q}} \rangle $.
The latter are related to the interaction Lagrangian density of quantum electrodynamics 
$\mathcal{L}^{\text{em}}_{\text{int}}(x)$~\cite{Klink:2000pp}:
\begin{eqnarray}\label{eq:emvertex}\lefteqn{
\langle v^\prime; \vec{k}_e^\prime,\! \mu_e^\prime;
\vec{k}_Q^\prime,\! \mu_Q^\prime; \vec{k}_{\bar{q}}^\prime,\!
\mu_{\bar{q}}^\prime; \vec{k}_\gamma^\prime,\! \mu_\gamma^\prime
\vert \,\hat{K}^\dag_\gamma \, \vert v\,  ; \vec{k}_e,\! \mu_e; \vec{k}_Q,\!
\mu_Q; \vec{k}_{\bar{q}},\! \mu_{\bar{q}} \rangle} \nonumber\\
&=& N v_0 \delta^3(\vec{v}^\prime - \vec{v})\, \langle
\vec{k}_e^\prime,\! \mu_e^\prime; \vec{k}_Q^\prime,\! \mu_Q^\prime;
\vec{k}_{\bar{q}}^\prime,\! \mu_{\bar{q}}^\prime;
\vec{k}_\gamma^\prime,\! \mu_\gamma^\prime \vert 
\hat{\mathcal{L}}^{\mathrm{em}}_{\mathrm{int}}(0)\, \vert
\vec{k}_e,\! \mu_e; \vec{k}_Q,\! \mu_Q; \vec{k}_{\bar{q}},\!
\mu_{\bar{q}} \rangle\, .\nonumber\\
\end{eqnarray}
$N$ is a uniquely determined normalization factor. 
Explicit analytical expressions for these matrix elements
are given in App.~\ref{AppCoupledChannel}. Inserting all these matrix elements into  
Eq.~(\ref{eq:voptclust}) shows that the on-shell matrix elements of the optical potential
have the structure that one expects from the invariant one-photon-exchange amplitude, i.e. it is 
proportional to the 
contraction of the electron and hadron currents, $j^\mu_e$ and $\tilde J^\nu_{[\alpha]}$ 
times the covariant photon propagator $(-g_{\mu\nu}/\text{Q}^2)$
\begin{eqnarray}\label{eq:voptos}
\lefteqn{\langle  \underline{v}^\prime;
\vec{\underline{k}}_e^\prime, \underline{\mu}_e^\prime;
\vec{\underline{k}}_\alpha^\prime,\underline{\mu}_\alpha^\prime,
\alpha \vert\,
\hat{V}_{\mathrm{opt}}(m)
\vert\, \underline{v}; \vec{\underline{k}}_e, \underline{\mu}_e;
\vec{\underline{k}}_\alpha,\underline{\mu}_\alpha^\prime, \alpha \rangle_{\mathrm{os}}}
\nonumber \\
&&=\underline{v}_0 \delta^3 (\vec{\underline{v}}^{\, \prime} -
\vec{\underline{v}}\, )\, \frac{(2 \pi)^3
}{\sqrt{(\omega_{\underline{k}_e^{\prime}}+
\omega_{\underline{k}_\alpha^{\prime}})^3}
\sqrt{(\omega_{\underline{k}_e^{\phantom{\prime}}}+
\omega_{\underline{k}_\alpha^{\phantom{\prime}}})^3}}\\
&&\quad \times(-e^2)\underbrace{\,\bar{u}_{\underline{\mu}_e^\prime}
(\vec{\underline{k}}_e^\prime)\gamma^\mu u_{\underline{\mu}_e}
(\vec{\underline{k}}_e)}_{j_e^\mu(\vec{\underline{k}}_e^\prime,
\underline{\mu}_e^\prime;\vec{\underline{k}}_e,
\underline{\mu}_e)}\frac{(-g_{\mu \nu})}{\text{Q}^2}\underbrace{ (\mathcal{Q}_Q
J_Q^\nu(\dots) + \mathcal{Q}_{\bar{q}}J_{\bar
q}^\nu(\dots))}_{\tilde{J}_{[\alpha]}^\nu(\vec{\underline{k}}_\alpha^\prime,
\underline{\mu}_\alpha^\prime;\vec{\underline{k}}_\alpha,
\underline{\mu}_\alpha)}\, .\nonumber
\end{eqnarray}
where $\text{Q}^2=-\underline{q}_\mu \underline{q}^\mu$ is the (negative) square of
the space-like 4-momentum-transfer\footnote{It should not be confused with the index $Q$ in italics denoting
the heavy quark.} $\underline{q}^\mu = (\underline{k}_\alpha-
\underline{k}_\alpha^\prime)^\mu = (\underline{k}_e^\prime -
\underline{k}_e)^\mu$, and $\mathcal{Q}_{Q(\bar q)}$ is the charge of the (anti)quark (in terms of multiples
o the electron charge). 
The 4-time-ordered contributions to $\hat V_{\text{opt}}(m)$ are 
sketched in Fig.~\ref{Fig4TemporalOrderingsAlpha}.

In order to identify the hadronic current and to ensure that it has
 the correct normalization, the procedure of 
Refs.~\cite{Biernat:2009my,Biernat:2011mp} is followed, where 
the one-photon-exchange amplitude is compared with the 
analogous amplitude one obtains when the meson is considered as a point-like 
particle with the discrete quantum numbers  $\alpha$. Because the point-like current is known, the 
kinematical factor can be uniquely identified. The hadronic current is a sum
of terms, $J^\nu_Q$ and $J^\nu_{\bar q}$, which correspond to the coupling of the photon to the quark 
and to the antiquark, respectively. For a pseudoscalar meson, $\ul\mu_\alpha=\ul\mu_\alpha'=0$, the 
$Q\bar q$ bound-state has to be such that the current takes on the form
\begin{eqnarray}\label{eq:JQ}
 J^{\nu}_Q
(\vec{\underline{k}}_\alpha^\prime,\vec{\underline{k}}_\alpha)&=&
\frac{\sqrt{\omega_{\underline{k}_\alpha}\omega_{\underline{k}_\alpha^{\prime}}}}{4
\pi} \int\, \frac{d^3\tilde{k}_{\bar{q}}^\prime}{2 \omega_{k_Q}}\,
\sqrt{\frac{\omega_{{k}_Q}+\omega_{{k}_{\bar{q}}}}
{\omega_{{k}^\prime_Q}+\omega_{{k}^\prime_{\bar{q}}}}} \,
\sqrt{\frac{\omega_{\tilde{k}^\prime_Q}+\omega_{\tilde{k}^\prime_{\bar{q}}}}
{\omega_{\tilde{k}_Q}+\omega_{\tilde{k}_{\bar{q}}}}} \,
\sqrt{\frac{\omega_{\tilde{k}_Q} \omega_{\tilde{k}_{\bar{q}}}}
{\omega_{\tilde{k}^\prime_Q} \omega_{\tilde{k}^\prime_{\bar{q}}}}}\nonumber\\
&&  \, \times \bigg\{\!\sum_{\mu_Q,\mu_Q^\prime
=\pm \frac{1}{2}}\!\!\!
\bar{u}_{\mu_Q^\prime}(\vec{k}_Q^\prime)\,\gamma^\nu\,
u_{\mu_Q}(\vec{k}_Q)  \nonumber\\ 
&& \times D^{1/2}_{\mu_Q\mu_Q^\prime}\!
\Big[ \!R_{\mathrm{W}}\!\left(\frac{\tilde{k}_{Q}}{m_Q},
B_c(v_{Q\bar{q}})\right)\,
\!R^{-1}_{\mathrm{W}}\!\left(\frac{\tilde{k}_{\bar{q}}}{m_{\bar{q}}},
B_c(v_{Q\bar{q}})\right)\,\nonumber\\
&&  \times\!R_{\mathrm{W}}\!\left(\frac{\tilde{k}_{\bar q}^\prime}{m_{\bar
q}}, B_c(v_{Q\bar{q}}^\prime)\right)\,
\!R^{-1}_{\mathrm{W}}\!\left(\frac{\tilde{k}_{Q}^\prime}{m_{Q}},
B_c(v_{Q\bar{q}}^\prime)\right)\Big] \bigg\}\, \nonumber\\
&& \times \psi^\ast\,(\vert
\vec{\tilde{k}}_{\bar{q}}^\prime\vert)\,  \psi
 \,(\vert \vec{\tilde{k}}_{\bar{q}}\vert)\,\, .
\end{eqnarray}
The tilde variables refer to the $Q\bar q$ center-of-momentum frame.
The analogous expression for $ J^{\nu}_{\bar q}$ can be obtained by interchanging 
$Q$ and $\bar q$ in Eq. (\ref{eq:JQ}). In the electromagnetic hadron currents one can 
distinguish several parts, which are present in all currents obtained through
this method, namely
\begin{itemize}
 \item 
the overlap of the initial and final meson wave function, which are written  
in terms of  $\vec{\tilde k}_{\bar q}$, $\vec{\tilde k}'_{\bar q}$, i.e. 
the incoming and outgoing spectator momenta in the $Q\bar q$ center-of-momentum frame, 
 \item 
the quark current times a spin rotation factor caused 
by boosting from the incoming to the outgoing meson states,
\item 
kinematical factors that come from the Lorentz transformations and guarantee the 
correct normalization of the current.
\end{itemize}

The (radial) $s$-wave bound state function $\psi(\kappa )$ is normalized according to 
\begin{equation}\label{eq:psinorm}
\int_0^\infty \, d\kappa \kappa^2\, \psi^\ast (\kappa)\,
\psi(\kappa) = 1 \, .
\end{equation}
The angular part resides in the factor $1/4\pi$ in front of the integral. 
The electromagnetic hadron current extracted from Eq.~(\ref{eq:voptos}) 
has the form 
of a spectator current. Here, however, the spectator
condition is not imposed, it comes rather
from the matrix elements of the interaction Lagrangian density through
which the vertex operators are defined. 
\newpage
The spectator momenta $\vec{\tilde k}_{\bar q}$ and $\vec{\tilde k}'_{\bar q}$ 
are related by canonical boosts (cf. App.~\ref{AppCanonicalBoosts}):
\begin{eqnarray}\label{eq:kktildep}
\tilde{k}_{\bar q} &=& B_c^{-1}(v_{Q\bar{q}})\, k_{\bar q}
= B_c^{-1}(v_{Q\bar{q}})\, k_{\bar q}^\prime\nonumber \\
&=& B_c^{-1}(v_{Q\bar{q}})\, B_c(v_{Q\bar{q}}^\prime)\tilde{k}_{\bar
q}^\prime \, .
\end{eqnarray}
Schematically the relationships of the required constituent momenta are given by ($Q$ active,
$\bar q$ spectator):
\begin{center}
\begin{tabular}[c]{ccccccc}
                   &               &           & $\text{INT}$  &                         &                  &  \\
                   &               &           & $ \downarrow $ &                         &                  &   \\
$\vec{\tilde k}_Q$ &$ \rightarrow$ &$\vec k_Q$ & $\longrightarrow$ & $\vec k_Q + \vec q $ & $ \rightarrow $  & $\vec{\tilde k}'_Q$ \\
                  & $B_c(v_{Q\bar q})$         &           &                &                         &     $B_c^{-1}(v'_{Q\bar q})$        &       \\     
$\vec{\tilde k}_{\bar q}$ & $ \rightarrow$    & $\vec k_{\bar q}$& $\longrightarrow$ &   $\vec k_{\bar q}  $  &   $ \rightarrow$ &$\vec{\tilde k}'_{\bar q}$
\end{tabular}    
\end{center}
It is also useful to note that $\vec{\tilde k}_{Q}=-\vec{\tilde k}_{\bar q}$.

   \section{Weak decays}\label{SecCupledChannelWeak}
\begin{figure}[h]
\includegraphics[width=0.9\textwidth]{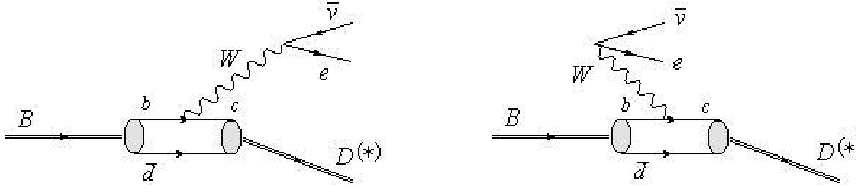}
\caption{The two time-ordered contributions to the semileptonic weak
decay of a $\bar{B}^0$ into a $D^{(\ast)+}$ meson.} \label{fig:decay}\end{figure}

We will use the same procedure to extract weak meson transition currents 
from the invariant 
amplitudes for semileptonic meson decays. We will illustrate it for the particular case of the  
$\bar B^0\to D^{(*)+}e \bar \nu_e$ decay. The coupled-channel mass operator differs from
the electromagnetic case in the number of particles in the initial, final as well
as intermediate states. The matrix mass operator to be considered requires at least four channels
\begin{equation}\label{eq:massopdecay}
 \left(\begin{array}{cccc} \hat M^{\mathrm{conf}}_{b\bar{d}} & 0 &
 \hat{K}_{c\bar{d}W\rightarrow b\bar{d}} &
 \hat{K}_{b\bar{d}We\bar{\nu}_e\rightarrow b\bar{d}}\\
0 & \hat M^{\mathrm{conf}}_{c\bar{d}e\bar{\nu}_e} &
\hat{K}_{c\bar{d}W\rightarrow c\bar{d}e\bar{\nu}_e} &
\hat{K}_{b\bar{d}We\bar{\nu}_e\rightarrow c\bar{d}e\bar{\nu}_e} \\
\hat{K}^\dag_{c\bar{d}W\rightarrow b\bar{d}} &
\hat{K}^\dag_{c\bar{d}W\rightarrow c\bar{d}e\bar{\nu}_e} & \hat
M^{\mathrm{conf}}_{c\bar{d}W} &
0 \\
\hat{K}^\dag_{b\bar{d}We\bar{\nu}_e\rightarrow b\bar{d}} &
\hat{K}^\dag_{b\bar{d}We\bar{\nu}_e\rightarrow c\bar{d}e\bar{\nu}_e}
& 0 & \hat M^{\mathrm{conf}}_{b\bar{d}We\bar{\nu}_e}
\end{array}\right)\, .
\end{equation}
Applying a Feshbach reduction to eliminate the $W$-boson channels one obtains the 
transition potential 
\begin{eqnarray}\label{EqWeakOptPotential}
\lefteqn{\hat V_{\mathrm{opt}}^{b\bar{d}\rightarrow c\bar{d}e\bar{\nu}_e}(m) = \hat{K}_{c\bar{d}W\rightarrow c\bar{d}e\bar{\nu}_e}
(m-M^{\mathrm{conf}}_{c\bar{d}W})^{-1}\hat{K}^\dag_{c\bar{d}W\rightarrow
b\bar{d}}}\nonumber\\
&& + \hat{K}_{b\bar{d}We\bar{\nu}_e\rightarrow
c\bar{d}e\bar{\nu}_e} (m-\hat
M^{\mathrm{conf}}_{b\bar{d}We\bar{\nu}_e})^{-1}
 \hat{K}^\dag_{b\bar{d}We\bar{\nu}_e\rightarrow b\bar{d}}\, .
\end{eqnarray}
Each term accounts for one time-ordered contribution of the $W$ exchange.  
The process is sketched in Fig. \ref{fig:decay}. 

Denoting again the discrete quantum numbers of the confined systems $B$ and 
$D^{(*)}$ by $\alpha$ and $\alpha'$ the invariant decay amplitude becomes
\begin{eqnarray}\label{eq:vopttrans}
&&\hspace{-1cm}\langle \underline{v}^\prime; \vec{\underline{k}}_e^\prime,
\underline{\mu}_e^\prime; \vec{\underline{k}}_{\bar{\nu}_e}^\prime;
\vec{\underline{k}}_{\alpha^\prime}^\prime,\underline{\mu}_{\alpha^\prime}^\prime,
\alpha^\prime \vert \hat V_{\mathrm{opt}}^{b\bar{d}\rightarrow
c\bar{d}e\bar{\nu}_e}(m)\vert \vec{\underline{k}}_{\alpha},
\underline{\mu}_{\alpha}, \alpha \rangle_{\mathrm{os}}\, \nonumber \\
&&\hspace{-1cm}= \langle \underline{v}^\prime; \vec{\underline{k}}_e^\prime,
\underline{\mu}_e^\prime; \vec{\underline{k}}_{\bar{\nu}_e}^\prime;
\vec{\underline{k}}_{\alpha^\prime}^\prime,\underline{\mu}_{\alpha^\prime}^\prime,
\alpha^\prime \vert \mathds{1}_{e\bar\nu_e c \bar d}  \hat{K}_{c\bar{d}W\rightarrow c\bar{d}e\bar{\nu}_e}
(m-M^{\mathrm{conf}}_{c\bar{d}W})^{-1}\mathds{1}_{W c \bar d}\hat{K}^\dag_{c\bar{d}W\rightarrow
b\bar{d}} \mathds{1}_{b\bar d}\nonumber\\
&&\hspace{-1cm} + \hat{K}_{b\bar{d}We\bar{\nu}_e\rightarrow
c\bar{d}e\bar{\nu}_e} (m-\hat
M^{\mathrm{conf}}_{b\bar{d}We\bar{\nu}_e})^{-1}
\mathds{1}_{b\bar d W \bar \nu_e}  \hat{K}^\dag_{b\bar{d}We\bar{\nu}_e\rightarrow b\bar{d}} \mathds{1}_{b\bar d} \vert \vec{\underline{k}}_{\alpha},
\underline{\mu}_{\alpha}, \alpha \rangle_{\mathrm{os}}\,,\nonumber\\
\end{eqnarray}
where ``on-shell'' (``os'') means $m=m_B=\omega_{\ul k_\alpha}=
\omega_{\ul k'_\alpha}+\omega_{\ul k'_e}+\omega_{\ul k'_{\bar \nu_e}}$. It is necessary again to 
introduce completeness relations between the operators that form the optical potential
in order to calculate the matrix elements in Eq.~(\ref{eq:vopttrans}). Thereby we obtain again 
the corresponding expressions for the wave functions and matrix elements of the interaction
vertices. In the case of weak decays the latter are determined by the weak interaction
density $\mathcal L^{\text{wk}}_{\text{int}}(x)$. 

Proceeding in the same way 
in the calculation of matrix elements and wave functions as in the 
electromagnetic case (explicit expressions are given in App.~\ref{AppCoupledChannel}) one obtains for the on-shell 
matrix elements
of $ \hat V_{\mathrm{opt}}^{b\bar{d}\rightarrow c\bar{d}e\bar{\nu}_e}(m)$ the same structure as for the 
invariant $B\to D^{(*)}e\bar \nu_e$ decay amplitude that results from 
leading-order covariant 
perturbation theory\footnote{The covariant structure is a little more difficult to
obtain than in the electromagnetic case. The explicit calculation is given
in App.~\ref{AppWeakPropagator}.}:
\begin{eqnarray}\label{eq:voptcov}
\lefteqn{\langle \underline{v}^\prime; \vec{\underline{k}}_e^\prime,
\underline{\mu}_e^\prime; \vec{\underline{k}}_{\bar{\nu}_e}^\prime;
\vec{\underline{k}}_{\alpha^\prime}^\prime,\underline{\mu}_{\alpha^\prime}^\prime,
\alpha^\prime \vert \hat V_{\mathrm{opt}}^{b\bar{d}\rightarrow
c\bar{d}e\bar{\nu}_e}(m)\vert \vec{\underline{k}}_{\alpha},
\underline{\mu}_{\alpha}, \alpha \rangle_{\mathrm{os}}}\nonumber\\
&=& \underline{v}_0 \delta^3 (\vec{\underline{v}}^{\, \prime} -
\vec{\underline{v}}\, )\, \frac{(2 \pi)^3
}{\sqrt{(\omega_{\underline{k}_e^{\prime}}+\omega_{\underline{k}_{\bar{\nu}_e}^{\prime}}
+\omega_{\underline{k}_{\alpha^\prime}^{\prime}})^3} \sqrt{
\omega_{\underline{k}_\alpha^{\phantom{\prime}}}^3}}\nonumber\\&&\times
\frac{e^2}{2\sin^2\vartheta_{\mathrm{w}}}V_{cb}\,
\frac{1}{2}\underbrace{\bar{u}_{\underline{\mu}_e^\prime}
(\vec{\underline{k}}_e^\prime)\gamma^\mu (1-\gamma^5)
v_{\underline{\mu}_{\bar{\nu}_e}^\prime}
(\vec{\underline{k}}_{\bar{\nu}_e}^\prime) }_{j_{\bar\nu_e\rightarrow
e}^\mu(\vec{\underline{k}}_e^\prime,
\underline{\mu}_e^\prime;\vec{\underline{k}}_{\bar{\nu}_e}^\prime,
\underline{\mu}_{\bar{\nu}_e}^\prime)}\nonumber\\&&\times\frac{(-g_{\mu
\nu})}{(\underline{k}^\prime_e+\underline{k}^\prime_{\bar{\nu}_e})^2-m_W^2}
\frac{1}{2}{J_{\alpha\rightarrow
\alpha^\prime}^\nu(\vec{\underline{k}}_{\alpha^\prime}^\prime,
\underline{\mu}_{\alpha^\prime}^\prime;\vec{\underline{k}}_\alpha,
\underline{\mu}_\alpha)}\, .
\end{eqnarray}
Here $\vartheta_{\mathrm{w}}$ denotes the electroweak mixing angle
and $e$ the usual elementary electric charge and $V_{cb}$ is the 
Cabibbo-Kobayashi-Maskawa
matrix element occurring at the $Wbc$-vertex.

\subsubsection{Pseudoscalar-to-pseudoscalar transitions}
If $\alpha$ and $\alpha'$ are
 the quantum numbers of 
$B$ and $D$ mesons, respectively, the weak transition
current turns out to have the form
\begin{eqnarray}\label{eq:Jwkpsps}
 J^{\nu}_{B\rightarrow D}
(\vec{\underline{k}}_D^\prime;\vec{\underline{k}}_B=\vec{0})&=&
\frac{\sqrt{\omega_{\underline{k}_B}\omega_{\underline{k}_D^{\prime}}}}{4
\pi} \int\, \frac{d^3\tilde{k}_{\bar{q}}^\prime}{2 \omega_{k_b}}\,
\sqrt{\frac{\omega_{\tilde{k}^\prime_c}+\omega_{\tilde{k}^\prime_{\bar{q}}}}
{\omega_{{k}^\prime_c}+\omega_{{k}^\prime_{\bar{q}}}}} \,
\sqrt{\frac{\omega_{\tilde{k}_b} \omega_{\tilde{k}_{\bar{q}}}}
{\omega_{\tilde{k}^\prime_c} \omega_{\tilde{k}^\prime_{\bar{q}}}}}
 \,  \nonumber\\
&& \times \bigg\{\!\sum_{\mu_b,\mu_c^\prime
=\pm \frac{1}{2}}\!\!\!
\bar{u}_{\mu_c^\prime}(\vec{k}_c^\prime)\,\gamma^\nu\,(1-\gamma^5)\,
u_{\mu_b}(\vec{k}_b)\nonumber  \\ 
&&\times D^{1/2}_{\mu_b\mu_c^\prime}\!
\left[\!R_{\mathrm{W}}\!\left(\frac{\tilde{k}_{\bar
q}^\prime}{m_{\bar q}}, B_c(v_{c\bar{q}}^\prime)\right)\,
\!R^{-1}_{\mathrm{W}}\!\left(\frac{\tilde{k}_{c}^\prime}{m_{c}},
B_c(v_{c\bar{q}}^\prime) \right)\right] \bigg\}\,\nonumber\\
&&\times \psi^\ast_{D}\,(\vert \vec{\tilde{k}}_{\bar{q}}^\prime\vert)\, \psi_B
 \,(\vert \vec{\tilde{k}}_{\bar{q}}\vert)\, .
\end{eqnarray}
The structure of the current is, of course, very similar to the electromagnetic case. 
Here the point-like current is the one that comes from the $Wbc$-vertex. $\psi_B$ 
as well as $\psi_D$ (and in the following $\psi_{D^\ast}$)
are normalized like in Eq.~(\ref{eq:psinorm}). The expression is simpler than 
in the electromagnetic case, since the initial state is at rest and therefore 
$\vec{\tilde k}_{\bar q}=\vec{k}_{\bar q}=-\vec{k}_{b}=-\vec{\tilde k}_{b}$.
There are no relativistic spin-rotation effects on the initial state and the Wigner $D$-functions 
refer only to the final $c\bar q$ state.
\subsubsection{Pseudoscalar-to-vector transitions}
In the transition where the spin of the meson also changes, i.e. $B\to D^* e\bar \nu_{e}$, one
has to take into account the corresponding Clebsch-Gordan coefficients that couple 
the quarks to the spin-1 $D^*$ meson. This affects the current such that it differs
from the previous one in the way how the Wigner rotations act on the spin components. In this case
there are independent rotations that act on each of the constituents:
\begin{eqnarray}\label{eq:Jwkpsv}
&&  J^{\nu}_{B\rightarrow D^\ast}
(\vec{\underline{k}}_{D^\ast}^\prime,\underline{\mu}_{D^\ast}^\prime;\vec{\underline{k}}_B=\vec{0})=
\!\!
\frac{\sqrt{\omega_{\underline{k}_B}\omega_{\underline{k}_{D^\ast}^{\prime}}}}{4
\pi} \int\, \frac{d^3\tilde{k}_{\bar{q}}^\prime}{2 \omega_{k_b}}\,
\sqrt{\frac{\omega_{\tilde{k}^\prime_c}+\omega_{\tilde{k}^\prime_{\bar{q}}}}
{\omega_{{k}^\prime_c}+\omega_{{k}^\prime_{\bar{q}}}}} \,
\sqrt{\frac{\omega_{\tilde{k}_b} \omega_{\tilde{k}_{\bar{q}}}}
{\omega_{\tilde{k}^\prime_c} \omega_{\tilde{k}^\prime_{\bar{q}}}}}
 \,\nonumber\\
&& \quad  \times  \bigg\{\!\sum_{\mu_b,\mu_c^\prime,\tilde\mu_c^\prime,\tilde\mu_{\bar
 q}^\prime=\pm \frac{1}{2}}
 \!\!\!\!\!\!
\bar{u}_{\mu_c^\prime}(\vec{k}_c^\prime)\,\gamma^\nu\,
(1-\gamma^5) u_{\mu_b}(\vec{k}_b)  \nonumber \\
&&\quad\quad\quad\quad\quad\quad  \times \sqrt{2} (-1)^{\frac{1}{2}-\mu_b}
C^{1\mu^\prime_{\!D^\ast}}_{\frac{1}{2}\tilde\mu_c^\prime\frac{1}{2}
\tilde{\mu}_{\bar{q}}^\prime}\, D^{1/2}_{\tilde\mu_c^\prime
\mu_c^\prime}\!
\left[\!R^{-1}_{\mathrm{W}}\!\left(\frac{\tilde{k}_{c}^\prime}{m_{c}},
B_c(v_{c\bar{q}}^\prime)\right)\right]\,\nonumber\\
&&\quad\quad\quad\quad\quad\quad \times D^{1/2}_{\tilde\mu_{\bar{q}}^\prime
-\mu_b}\!\left[ \!R^{-1}
_{\mathrm{W}}\!\left(\frac{\tilde{k}^\prime_{\bar q}}{m_{\bar q}},
B^{-1}_c(v_{c\bar{q}}^\prime)\right)\right] \bigg\}\,
\psi^\ast_{D^{\ast}}\,(\vert \vec{\tilde{k}}_{\bar{q}}^\prime\vert)\, \psi_B
 \,(\vert \vec{\tilde{k}}_{\bar{q}}\vert)\, .\nonumber\\
\end{eqnarray}

\section{Dynamical exchange potential}
We have seen that the number of channels to be considered depends on the kind of process one 
is interested in. 
Up till now we have only considered the electroweak structure of $q\bar q$-bound
states that were generated by instantaneous confining forces. In the following we will
also be interested in the electroweak structure of bound states that are caused by dynamical 
particle exchange. 
Treating explicitly the dynamics of the exchange particles that are 
responsible for the 
binding requires 
the introduction of additional channels in the mass operator. 
In Chap.~\ref{ChDeuteronExCurrents} we will investigate the electromagnetic
structure of the deuteron, considered as a neutron-proton bound state caused by dynamical 
$\sigma$-meson exchange. 
The general mass eigenvalue problem for electron-deuteron 
scattering in this case then needs 4-channels,

\begin{equation}  
\left( \begin{array}{cccc} \hat M_{enp}          &    \hat K_\gamma      &     \hat K_\sigma    &     0 \\
                           \hat K_\gamma^\dagger & \hat M_{enp\gamma}   &0    & \hat K_\sigma \\
                           \hat K_\sigma^\dagger  & 0 & \hat M_{enp\sigma}  & \hat K_\gamma   \\
                           0      &     \hat K_\sigma^\dagger    &  \hat K_\gamma^\dagger  &  \hat M_{enp\gamma\sigma} \end{array} \right) \left( \begin{array}{c} |\psi_{enp}\rangle \\ |\psi_{enp\gamma} \rangle  \\ |\psi_{enp\sigma} \rangle \\ | \psi_{enp\gamma\sigma} \rangle \end{array} \right) = m\left( \begin{array}{c} |\psi_{enp}\rangle \\ |\psi_{enp\gamma} \rangle  \\ |\psi_{enp\sigma} \rangle \\ | \psi_{enp\gamma\sigma} \rangle \end{array} \right). 
\end{equation}
Additional relativistic effects become important 
when the retardation of the meson exchange that binds 
the nucleons is comparable 
to the one of the photon exchange. This leads to, so-called, \textit{exchange currents}.

 \newpage
\thispagestyle{empty}

%% file: Ch4.tex
\chapter{Currents and form factors}\label{ch:currents:and:ff}

We will dedicate this chapter to the study of 
 the properties of the current derived by means of the method explained above. 
The current is extracted in each case from the invariant one-boson-exchange 
amplitude for electron-meson scattering and weak semileptonic decays. 
No particular ansatz is made for the current that imposes 
the desired properties that such a current should have. It is therefore necessary to 
examine the properties of our currents, in order to check that the procedure carried out
makes sense. The essential properties that the current should fulfill are:
Lorentz covariance, i.e. the current must 
transform like a 4-vector under Lorentz 
transformations; current conservation for electromagnetic 
scattering, i.e. $\partial_\mu \tilde J^\mu_{[\alpha]}=0$; and cluster 
separability or macrocausality. 
Cluster separability means in this context that 
the hadron currents and the corresponding 
form factors
should depend on the hadron properties only, 
and not on the ones of the particle with which it
interacts. 
Once one is able to understand the properties of the hadron currents, 
it will be possible to provide consistent
analytical expressions for the form factors, as deduced from them. 

  \section{Electromagnetic form factors}
\subsection{Pseudoscalar bound states}\label{SecPseudoscalarBoundStatesCurrents}
\subsubsection{Covariance and current conservation}
We first 
check whether the current transforms like a 4-vector under
Lorentz transformations.
If one looks at the current (\ref{eq:JQ}) and applies a Lorentz transformation $\Lambda$ 
it turns out that $\tilde J^\nu_{[\alpha]}(\ul{\vec{k}}'_\alpha; \ul{\vec{k}}_\alpha)$
 does not transform like a 4-vector.
Instead, the current  transforms by the Wigner
rotation  $R_W(v,\Lambda)$, where $v$ is the overall 4-velocity of the electron-meson
system~\cite{Biernat:2009my}. This is due to the fact that 
$\tilde J^\nu_{[\alpha]}(\ul{\vec{k}}'_\alpha; \ul{\vec{k}}_\alpha)$
is computed using velocity states, which do not transform like 4-vectors 
under Lorentz transformations, 
 they transform by a Wigner rotation instead (cf.~Sec.~\ref{SecVelocityStates}). 
Going back to the physical meson momenta 
$\ul p_\alpha^{(')}=B_c(v)\ul k_\alpha^{(')}$, which do transform like 4-vectors, 
one finds that the current has the desired 
transformation properties:
\begin{equation}
\tilde{J}_{[\alpha]}^\nu(\vec{\underline{p}}_\alpha^\prime;\vec{\underline{p}}_\alpha):=
[B_c(\underline{v})]^\nu_{\phantom{\nu}\rho}
\tilde{J}_{[\alpha]}^\rho(\vec{\underline{k}}_\alpha^\prime;\vec{\underline{k}}_\alpha)\,
.
\end{equation}
$\tilde{J}_{[\alpha]}^\nu(\vec{\underline{p}}_\alpha^\prime;\vec{\underline{p}}_\alpha)$
transforms like a 4-vector and is a conserved current, i.e.
$(\underline{p}_\alpha-\underline{p}_\alpha^\prime)_\nu
\tilde{J}_{[\alpha]}^\nu(\vec{\underline{p}}_\alpha^\prime;\vec{\underline{p}}_\alpha)=0$. 
A more detailed discussion of transformation properties 
and current conservation can be found in Refs.~\cite{Biernat:2009my,Biernat:2011mp}.

\subsubsection{Cluster separability}
The next task is to investigate if the current satisfies the desired cluster separability properties. 
 For a pseudoscalar meson, one expects a current of the form
\begin{equation}\label{em:ff:usual}
J_{[\alpha]}^\nu(\vec{\underline{p}}_\alpha^\prime;\vec{\underline{p}}_\alpha)
=  (\underline{p}_\alpha+\underline{p}^\prime_\alpha)^\nu \,
F(\text{Q}^2).
\end{equation}
It is known, however, that the Bakamjian-Thomas construction leads to problems with 
 cluster separability. 
Basis states which are appropriate to represent Bakamjian-Thomas type mass operators 
(like our velocity states) use variables to represent relative momenta 
 which do not have a physical interpretation in the presence of 
interactions~\cite{Keister:1991sb}. As
a consequence, problems with macroscopic locality appear. A manifestation of such problems
is that our microscopic current contains non-physical contributions.
 It cannot 
be expressed in terms of hadronic covariants only, but one needs an additional
covariant associated with the electron four-momenta~\cite{Biernat:2010tp,Biernat:2011mp}:
\begin{equation}\label{eq:Jemcovdec}
\tilde{J}_{[\alpha]}^\nu(\vec{\underline{p}}_\alpha^\prime;\vec{\underline{p}}_\alpha)
= (\underline{p}_\alpha+\underline{p}^\prime_\alpha)^\nu \,
f(\text{Q}^2,s)+
(\underline{p}_e+\underline{p}^\prime_e)^\nu \, g(\text{Q}^2,s)\, .
\end{equation}
The impossibility of decomposing the current as in Eq.~(\ref{em:ff:usual}) shows up when one tries 
to extract the form factor from the different non-vanishing components of the current,
since it turns out that this cannot be done 
unambiguously.
Furthermore, the form factors associated with the covariants depend not only on the  
4-momentum transfer squared (Mandelstam $t$), but also on the Mandelstam $s=(\ul p_e+\ul p_\alpha)^2$, 
i.e. the square of the invariant mass of the electron-meson system.

The necessity of non-physical covariants and corresponding form factors 
resembles the occurrence of analogous contributions within the covariant light-front 
formulation of Carbonell \textit{et al.}~\cite{Carbonell:1998rj}.
In the covariant light-front approach the unphysical covariants contain 
a 4-vector $\omega^\mu$, which specifies the orientation of the light front and which 
has to be
introduced to render the front-form approach manifestly covariant. 
In the present case, the problem is related to cluster separability violation
caused by the Bakamjian-Thomas construction. 

For a better understanding of these unphysical features we have undertaken a numerical 
study. Since the form factors are functions of Lorentz invariants we are free to choose the frame in
which they are extracted. Without loss of generality we choose a 
center-of-momentum frame  
in which $\vec{v}=\vec{0}$,
i.e.
$\underline{\vec{p}}^{(\prime)}_\alpha=\underline{\vec{k}}^{(\prime)}_\alpha$,
and
\begin{equation}\label{eq:momentumscatt}
\underline{\vec{k}}_\alpha=-\underline{\vec{k}}_e=
\left( \begin{array}{c} -\frac{\text{Q}}{2}\\ 0 \\
\sqrt{\kappa_\alpha^2-\frac{\text{Q}^2}{4}}
\end{array}\right)\,  \quad \hbox{and}\quad \vec{q}=\left(
\begin{array}{c} -\text{Q}\\ 0 \\ 0 \end{array}\right)\, ,
\end{equation}
where $\kappa_\alpha:=|\underline{\vec{k}}_\alpha|=|\underline{\vec{k}}_\alpha^\prime|$.
In this parametrization the modulus of the relative momentum is subject to the
constraint that $\kappa_\alpha^2\geq \text{Q}^2/4$, which
means that $s\geq
m_\alpha^2+m_e^2+\text{Q}^2/2+2\sqrt{m_\alpha^2+\text{Q}^2/4}\sqrt{m_e^2+\text{Q}^2/4}$.     

The only non-vanishing components of the current within this kinematics are 
 $\tilde{J}_{[\alpha]}^0$ and $\tilde{J}_{[\alpha]}^3$ from which the form
factors $f(\text{Q}^2,s)$ and $g(\text{Q}^2,s)$ can be extracted by 
inserting the microscopic
expression (cf.~Eqs.~(\ref{eq:voptos}) and (\ref{eq:JQ})) for
$\tilde{J}_{[\alpha]}^\nu$ on the left-hand side of Eq.~(\ref{eq:Jemcovdec}):
\begin{equation}
 f(\text{Q}^2,s)=\frac{1}{\left( 1+\frac{\sqrt{\kappa_\alpha^2+m_\alpha^2}}{\sqrt{\kappa_\alpha^2+m_e^2}}\right)}
\left(\frac{\tilde J^0_{[\alpha]}(\underline{p}_\alpha,\underline{p}_\alpha')}{2\sqrt{\kappa_\alpha^2+m_e^2}}+
\frac{\tilde J^3_{[\alpha]}(\underline{p}_\alpha,\underline{p}_\alpha')}{2\sqrt{\kappa_\alpha^2-\frac{\text{Q}^2}{4}}}\right),
\end{equation}

\begin{equation} 
 g(\text{Q}^2,s)=\frac{1}{\left(1+\frac{\sqrt{\kappa_\alpha^2+m_e^2}}{\sqrt{\kappa_\alpha^2+m_\alpha^2}}\right)}
\left(\frac{\tilde J^0_{[\alpha]}(\underline{p}_\alpha,\underline{p}_\alpha')}{2\sqrt{\kappa_\alpha^2+m_\alpha^2}}
-\frac{\tilde J^3_{[\alpha]}(\underline{p}_\alpha,\underline{p}_\alpha')}{2\sqrt{\kappa_\alpha^2-\frac{\text{Q}^2}{4}}}\right).
\end{equation}

\begin{figure}[h!]
\begin{center}
 \includegraphics[width=0.8\textwidth]{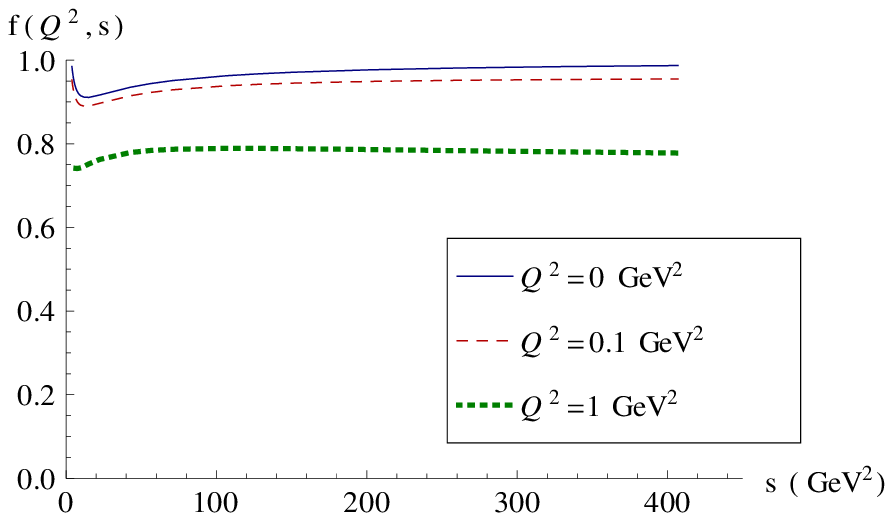}
\includegraphics[width=0.8\textwidth]{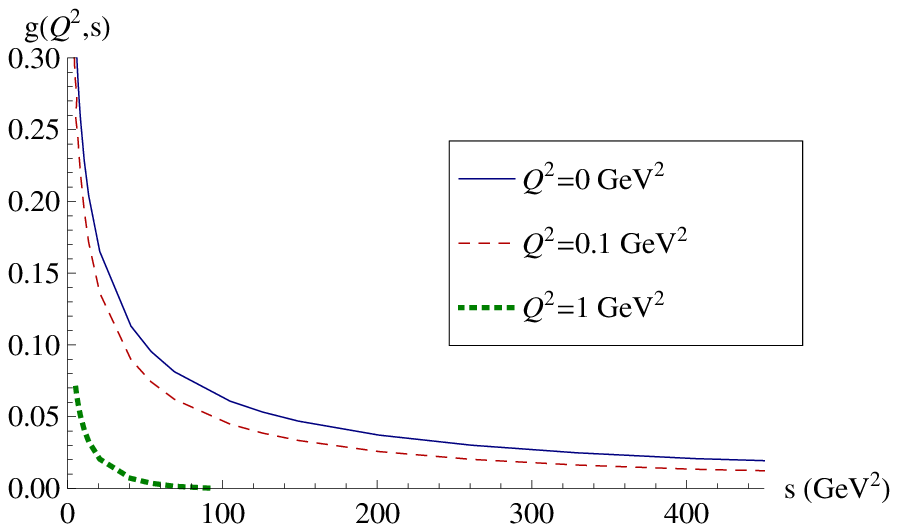}
 \caption{Mandelstam-$s$ dependence of the physical
and spurious $D^+$ 
electromagnetic form factors $f(\text{Q}^2,s)$ and $g(\text{Q}^2, s)$,
respectively, for different values of $\text{Q}^2$ ($0$~GeV$^2$ solid,
$0.1$~GeV$^2$ dashed, $1$~GeV$^2$ dotted) calculated with the
oscillator wave function (\ref{eq:wavefunc}), and (mass)
parameters given in Table~\ref{ParametersHeavyLight}.}\label{fig:sdep}
\end{center}
\end{figure}

\begin{figure}[h!]
\begin{center}
\includegraphics[width=0.8\textwidth]{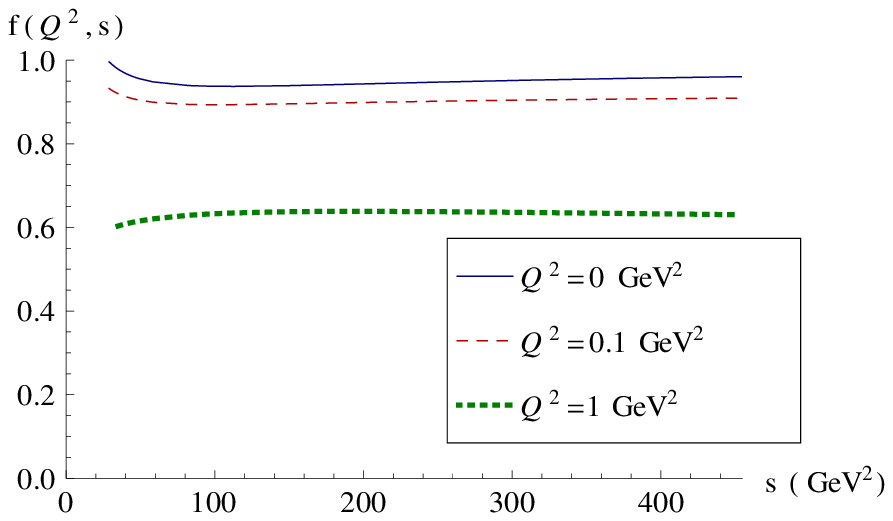}
\includegraphics[width=0.8\textwidth]{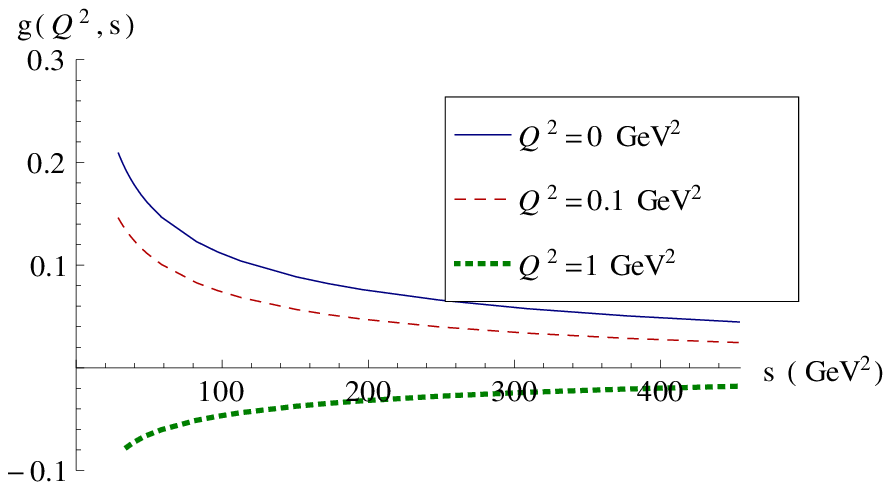}
 \caption{Mandelstam-$s$ dependence of the physical
and spurious $B^-$ 
electromagnetic form factors $f(\text{Q}^2,s)$ and $g(\text{Q}^2, s)$,
respectively, for different values of $\text{Q}^2$ ($0$~GeV$^2$ solid,
$0.1$~GeV$^2$ dashed, $1$~GeV$^2$ dotted) calculated with the
oscillator wave function (\ref{eq:wavefunc}), and (mass)
parameters given in Table~\ref{ParametersHeavyLight}.}\label{fig:sbep}
\end{center}
\end{figure}

For the bound state wave function, we use the  simple harmonic-oscillator form
 (\ref{eq:wavefunc}).
For further comparison we take the oscillator
parameter as well as the constituent-quark masses to be the same as in 
Ref.~\cite{Cheng:1996if} (see also Table~\ref{ParametersHeavyLight}),
 where form factors of heavy-light mesons were calculated 
within the front-form approach. 
The dependence of these form factors on Mandelstam-$s$ is plotted for the
$D^+$ and $B^-$ mesons in 
Figs.~\ref{fig:sdep} and~\ref{fig:sbep} for different values of the momentum transfer $\text{Q}^2$. 
For $s\to \infty$ 
the spurious form factor vanishes and the $s$-dependence of the 
physical form factor disappears with increasing $s$. It is therefore
suggestive to take the $s\to \infty$ limit to get rid of cluster-separability
violating effects and obtain sensible results for the physical form factors.
Taking $s\to \infty$ can be understood as extracting the form factor in the infinite-momentum frame of the
meson. It is equivalent to taking $\kappa_\alpha\to\infty$.

 Similar calculations were done for mesons of equal constituent
masses~\cite{Biernat:2009my,Biernat:2011mp}. 
For light-light systems the resulting analytical expression for the electromagnetic
form factor of a pseudoscalar meson were proved to be equivalent to the usual 
front-form result, obtained for a one-body current in the $q^+=0$ frame~\cite{Biernat:2009my}. 
For heavy-light systems the situation becomes more intricate. One can observe that
the rate of convergence to the $s\to \infty$ limit decreases with increasing the 
heavy-quark mass. In order to extract the Isgur-Wise function one has to be cautious
when taking the heavy-quark limit $m_Q\to \infty$. This matter will be discussed in 
detail in the next
chapter.

\newpage
\newpage
\subsection{Vector bound states}\label{SecVectorCurrent}
Hermiticity, covariance, current conservation, the angular condition and other
properties like cluster separability, were analyzed in detail in Ref.~\cite{Biernat:2011mp} 
for spin-1 bound-state currents of two-body systems with equal constituent masses within 
the point-form approach. We summarize in this section the most important points, since they will 
be necessary to understand further calculations for spin-1 bound states. 
In Chap.~\ref{ChDeuteronExCurrents}
 we will use some elements presented here for the study of 
electromagnetic properties of spin-1 bound states that arise from dynamical 
particle-exchange forces. 
\subsubsection{Cluster separability}
The covariant decomposition of our electromagnetic current becomes more complicated
if one deals with spin-1 bound states. It requires to consider all possible covariants including
those that depend on the electron momenta. 
These are in total 11 covariants, with their associated form factors. The form factors
exhibit also a spurious dependence on Mandelstam-$s$. The most general covariant
decomposition of the current is given by:
\begin{eqnarray}\label{EqVectorCurrent}
&&\tilde J^\mu_{[\alpha]}(\ul{\vec k}_\alpha,\ul\mu_\alpha; 
\ul{\vec k}'_\alpha,\ul \mu'_\alpha; \ul K_e)
=\nonumber \\
&&\;\;\;=\left[  f_1 (\text{Q}^2,s) (\epsilon'^*\cdot \epsilon) + f_2(\text{Q}^2,s)  \frac{(\epsilon'^*\cdot q)(\epsilon^*\cdot q)}{2m_\alpha^2}\right]\ul K_\alpha^\mu \nonumber\\
&&\;\;\;+ g_M(\text{Q}^2,s)   \left[\epsilon'^{*\mu}(\epsilon \cdot q )-\epsilon^\mu (\epsilon'^*\cdot q)\right]\nonumber \\
&&\;\;\;+\frac{m^2_\alpha}{2 \ul K_e\cdot \ul k_\alpha}\Big[ b_1(\text{Q}^2,s)(\epsilon'^*\cdot \epsilon )+b_2(\text{Q}^2,s)
 \frac{(q\cdot \epsilon'^* )(q\cdot \epsilon^*)}{m_\alpha^2}\nonumber\\
&&\;\;\;\qquad\qquad\qquad+ b_3(\text{Q}^2,s)
 m^2_\alpha\frac{(\ul K_e\cdot \epsilon'^*)(\ul K_e\cdot \epsilon^*)}{(\ul K_e\cdot \ul k_\alpha)^2} \nonumber \\
&&\;\;\;\qquad\qquad\qquad+b_4(\text{Q}^2,s)\frac{(q\cdot \epsilon'^*)(\ul K_e\cdot \epsilon)-
(q\cdot \epsilon)(\ul K_e\cdot \epsilon'^*)}{2(\ul K_e\cdot \ul k_\alpha)}\Big]\ul K_e^\mu\nonumber  \\
&&\;\;\;+\Big[ b_5(\text{Q}^2,s)m^2_\alpha \frac{(\ul K_e\cdot \epsilon'^*)(\ul K_e\cdot \epsilon)}{(\ul K_e\cdot \ul k_\alpha)^2} \nonumber\\
&&\;\;\;\qquad+ b_6(\text{Q}^2,s)\frac{(q\cdot \epsilon'^*)(\ul K_e\cdot \epsilon)-(q\cdot \epsilon )(\ul K_e\cdot \epsilon'^*)}{2\ul K_e\cdot \ul k_\alpha}\Big]\ul K_\alpha^\mu \nonumber \\
&&\;\;\;+ b_7 (\text{Q}^2,s)m^2_\alpha \frac{\epsilon'^{*\mu}(\epsilon\cdot \ul K_e)+
\epsilon^\mu(\epsilon'^*\cdot \ul K_e)}{\ul K_e\cdot \ul k_\alpha }\nonumber\\
&&\;\;\;+b_8(\text{Q}^2,s)
 q^\mu \frac{(q\cdot \epsilon'^*)(\ul K_e\cdot \epsilon)+
(q\cdot\epsilon)(\ul K_e\cdot \epsilon'^*)}{2\ul K_e\cdot \ul k_\alpha}.\nonumber\\
&& \qquad
\end{eqnarray}
where the shorthand notations $\ul K_{e(\alpha)}:=\ul k_{e(\alpha)}+\ul k_{e(\alpha)}'$ and 
$\epsilon^{(\prime)}:=\epsilon(\ul{\vec k}_\alpha^{(\prime)},\ul\mu_\alpha^{(\prime)})$, 
have been used, the 
latter being the polarization vectors of the incoming and outgoing 
spin-1 bound state (cf. App.~\ref{PolVectors}). 
Only 3 of the 11 form factors have a
physical meaning, namely $f_1$, $f_2$ and $g_M$. For a detailed discussion about the 
elimination of these spurious contributions 
in the infinite-momentum frame
the reader may consult Ref.~\cite{Biernat:2011mp}.
The numerical analysis carried out in Ref.~\cite{Biernat:2011mp} (that uses the kinematics we have
considered in Eq.~(\ref{eq:momentumscatt})) reveals that 4 of the 8 spurious
contributions cannot be eliminated by simply taking $s\to \infty$ as in the pseudoscalar case.
 The form 
factors $b_5$, $b_6$, $b_7$ and $b_8$ do not vanish in the infinite-momentum 
frame\footnote{Note
that in the 
infinite-momentum frame
the corresponding covariants to these non-vanishing spurious form factors do not depend 
on the strength of the electron momenta, but only on their orientations with respect to 
the polarization vector of the scattered bound state.}. These
spurious contributions in the current are relatively small, but
may have important consequences on some properties of the current 
if they are not treated properly.

\subsubsection{Covariance, current conservation and angular condition}
As in the pseudoscalar case, our microscopic expression for the electromagnetic current of a
spin-1 bound state transforms like a 
4-vector under Lorentz transformations
if one goes back to the physical meson momenta by applying a canonical boost
$\ul p^{(\prime)}_\alpha=B_c(v)\ul k^{(\prime)}_\alpha$~\cite{Biernat:2011mp}. 

Because of the non-vanishing $b_7$ and  $b_8$, the current (\ref{EqVectorCurrent})
is not conserved; and the $b_5$ and $b_7$ violate 
the so-called angular condition. Let us abbreviate the notation by calling 
$B_i(\text{Q}^2):=\lim_{s\to \infty}b_i(\text{Q}^2,s)$ and 
$J^\mu_{\ul\mu_\alpha'\ul\mu_\alpha}:=\tilde J^\mu_{[\alpha]}(\ul{\vec k}_\alpha,\ul\mu_\alpha; 
\ul{\vec k}'_\alpha,\ul \mu'_\alpha; \ul K_e)$. 
Without spurious contributions the physical current matrix elements should satisfy
the angular condition:
\begin{equation}
 (1+2\eta)J^0_{11}+J^0_{1-1}-2\sqrt{2\eta}J^0_{10}-J^0_{00}=0,
\end{equation}
with $\eta=\frac{\text{Q}^2}{4m^2_\alpha}$. The studies carried out in~\cite{Biernat:2011mp}
show that, due to the spurious contributions, one gets:
\begin{equation}
 (1+2\eta)\tilde J^0_{11}+\tilde J^0_{1-1}-2\sqrt{2\eta}\tilde J^0_{10}-\tilde J^0_{00}=-|e|\left( B_5(\text{Q}^2)-B_7(\text{Q}^2)\right).
\end{equation}
Nevertheless, it can be shown (for our kinematics) that there are 3 current matrix elements which do no contain
any spurious contributions in the limit $s\to\infty$. This matrix elements are $J^0_{11}$, 
$J^0_{1-1}$ and $J^2_{11}$. They  can be used to extract the physical form factors 
without ambiguity~\cite{Biernat:2011mp}:
\begin{eqnarray}
 F_1(\text{Q}^2)&:=&\lim_{s\to\infty}f_1(\text{Q}^2,\kappa_\alpha)=-\lim_{s\to\infty}\frac{1}{2\kappa_\alpha}\left(J^0_{11}+J^0_{1-1}\right),\\
 F_2(\text{Q}^2)&:=&\lim_{s\to\infty}f_2(\text{Q}^2,\kappa_\alpha)=-\frac{1}{\eta}\lim_{s\to\infty}\frac{1}{2\kappa_\alpha}J^0_{1-1},\\
 G_M(\text{Q}^2)&:=&\lim_{s\to\infty}g_M(\text{Q}^2,\kappa_\alpha)=-\frac{i}{\text{Q}}J^2_{11}.
\end{eqnarray}

\section{Decay form factors}
An analogous study must be done for the weak current obtained from the transition amplitude 
of radiative decays. 
\subsection{Pseudoscalar-to-pseudoscalar transitions}\label{SubsecPtoP}
\subsubsection{Covariance}
As in the electromagnetic case, the pseudoscalar-to-pseudoscalar transition current
turns out to have the right transformation properties under Lorentz transformations after
 applying the canonical boost $B_c(\ul v)$ that connects the physical
momenta with center-of-mass momenta, 
\begin{equation}\label{eq:Jphyspsps}
J^\nu_{B\rightarrow
D}(\vec{\underline{p}}_D^\prime;\vec{\underline{p}}_B):=
[B_c(\underline{v})]^\nu_{\phantom{\nu}\rho} J_{B\rightarrow
D}^\rho(\vec{\underline{k}}_D^\prime;\vec{\underline{k}}_B)\, .
\end{equation}
The way how to extract the form factors of weak transitions is analogous to the 
electromagnetic case. The covariant decomposition of the weak current for a 
pseudoscalar-to-pseudoscalar transition reads~\cite{Wirbel:1985ji}
\begin{equation}\label{eq:Jpspsphys}
J^\nu_{B\rightarrow
D}(\vec{\underline{p}}_D^\prime;\vec{\underline{p}}_B)
=\left( (\underline{p}_B+\underline{p}_D')^\nu -
\frac{m^2_B-m^2_D}{\underline{q}^2 }\, \underline{q}^\nu
\right)F_1(\underline{q}^2)+
\frac{m_B^2-m_D^2}{\underline{q}^2}\, \underline{q}^\nu
F_0(\underline{q}^2)\, ,
\end{equation}
with the time-like 4-momentum transfer
$\underline{q}=(\underline{p}_B-\underline{p}_D)$.
\subsubsection{Cluster separability}
When one inserts the current (\ref{eq:Jwkpsps}) into Eq.~(\ref{eq:Jpspsphys}), 
and extracts of the form factors $F_1$ and $F_0$ it turns out that the 
solution is unique. The form factors can be determined unambiguously
from the components of the current $J^\nu_{B\to D}$, without the necessity of 
introducing additional spurious
covariants. The form factors do not depend on any other Lorentz invariant quantity 
different from
the 4-momentum transfer $q^2$. 
Unlike in the electromagnetic case, wrong cluster properties 
of the Bakamjian-Thomas
construction do not show up in the structure of the currents and 
the dependence of the form factors on the available Lorentz invariants. 

Since the decay current (\ref{eq:Jphyspsps}) transforms like a 4-vector we can analyze it, without
loss of generality, in the frame in which the decaying $B$-meson is at rest (which
corresponds to $\vec{v}=\vec{0}$). We parametrize the meson momenta by:
\begin{equation}\label{eq:decaykinem}
\underline{k}_B= \left( \begin{array}{c} m_B\\ 0 \\0\\0
\end{array}\right) \quad\hbox{and}\quad
\underline{k}_D'= \left( \begin{array}{c} \sqrt{m_D^2+\kappa_D^{2}}\\
\kappa_D
\\0\\0
\end{array}\right)\end{equation}
with
\begin{equation}\label{Eqkappa}  
\kappa_D^2=\frac{1}{4m_B^2}
(m_B^2+m_D^2-\underline{q}^2)^2-m_D^2\,.
\end{equation}
$\kappa_D:=|\underline{\vec{k}}_D^\prime|$ 
is constrained by the 
condition
$0\leq \kappa_D^2 \leq (m_B^2-m_D^2)^2/(4m_B^2)$, since
\begin{equation}\label{Eqrangeq2}
0\leq \underline{q}^2 \leq (m_B-m_D)^2 \, .
\end{equation}
In order to understand the observation that the decay current  $J^\nu_{B\rightarrow D}$ is not 
spoiled by cluster separability problems whereas the electromagnetic
current $\tilde{J}_{[\alpha]}^\nu$ was, let us note a few points:
\begin{itemize}
 \item  The spurious  
contributions to the electromagnetic current 
had their origin in the fact that the calculation was carried out 
in the center-of-momentum frame of the electron-meson system. 
Cluster problems appear when 
different sets of subsystems cannot be isolated properly.
In a decay, there is no additional participant in the initial state of the process which 
could modify the bound-state wave function. Only the final state (electron-antineutrino-meson)
might be affected by wrong cluster problems.

\item Like in the electromagnetic case, the current (\ref{eq:Jwkpsps}) has only two 
non-vanishing components (for our chosen kinematics). 
But unlike in the electromagnetic case, there are now
 two covariants and associated form factors
in the covariant decomposition (\ref{eq:Jpspsphys}) of the decay current;
these are $(\underline{p}_\alpha+\underline{p}_\alpha')$ and 
$(\underline{p}_\alpha-\underline{p}_\alpha')$. In the case of electromagnetic 
the latter is forbidden by  current conservation. 

\item Form factors are frame independent quantities, therefore one should be able to 
express them as functions of Lorentz invariant quantities only. In the 
electromagnetic case the modulus of the three momentum $|\vec q|$ cannot be expressed
as a function of the squared 4-momentum only, i.e. Mandelstam $t=q^2$, but one  needs
in addition Mandelstam $s$. The modulus of the 3-momentum transfer in the weak decays is,
on the other hand, determined by $q^2$ only. 

\end{itemize}

\subsection{Pseudoscalar-to-vector transitions}
\subsubsection{Covariance}
The current (\ref{eq:Jwkpsv}) transforms also like a 4-vector after applying 
a canonical boost that connects the physical momenta with the center-of-mass 
momenta, as it happens in the pseudoscalar case. In this case, however,
one needs an additional Wigner $D$-function that is associated with the 
rotation of the $D^*$-meson spin:
\begin{align}\label{eq:Jphyspsv}
J^\nu_{B\rightarrow
D^\ast}(\vec{\underline{p}}_{D^\ast}^{\prime},\underline\sigma^\prime_{D^\ast};
\vec{\underline{p}}_B):=&
[B_c(\underline{v})]^\nu_{\phantom{\nu}\rho} J_{B\rightarrow
D^\ast}^\rho(\vec{\underline{k}}_{D^\ast}^{\prime},\underline{\mu}^\prime_{D^\ast};
\vec{\underline{k}}_B)\nonumber\\
& \times
D^{1\ast}_{\underline{\mu}^\prime_{D^\ast}
\underline\sigma^\prime_{D^\ast}}\left[R^{-1}_{\mathrm{W}}\!
\left({\underline{k}}_{D^\ast}^{\prime}/m_{D^\ast},
B_c(v)\right)\right]\, . \nonumber\\
\end{align}
The most general covariant decomposition of the 
current is given by~\cite{Wirbel:1985ji}
\begin{align}\label{eq:Jppsvdec}
J^\nu_{B\rightarrow D^\ast}(\vec{\underline{p}}_{D^\ast}^{\prime},\underline\sigma^\prime_{D^\ast};
\vec{\underline{p}}_B)&= \frac{2 i\epsilon^{\nu\mu\rho\sigma}}{m_B+m_{D^*}}\,\epsilon^*_\mu(\vec{\underline{p}}_{D^\ast}^{\prime}, \underline\sigma^\prime_{D^\ast})\, \underline{p}'_{D^\ast\rho}\, \underline{p}_{B\sigma} \, V(\underline{q}^2)\nonumber\\
& - (m_B+m_{D^*})\, \epsilon^{*\nu}(\vec{\underline{p}}_{D^\ast}^{\prime}, \underline\sigma^\prime_{D^\ast})\, A_1(\underline{q}^2)\nonumber\\
& + \frac{\epsilon^*(\vec{\underline{p}}_{D^\ast}^{\prime}, \underline\sigma^\prime_{D^\ast}) \cdot \underline{q}}{m_B+m_{D^*}}\,(\underline{p}_B+\underline{p}_{D^\ast}')^\nu\, A_2(\underline{q}^2)\nonumber\\
&+2 m_{D^*}\,\frac{\epsilon^*(\vec{\underline{p}}_{D^\ast}^{\prime}, \underline\sigma^\prime_{D^\ast}) \cdot \underline{q}}{\underline{q}^2}\, \underline{q}^\nu \, A_3(\underline{q}^2)\nonumber\\
&- 2m_{D^*}\,\frac{\epsilon^*(\vec{\underline{p}}_{D^\ast}^{\prime}, \underline\sigma^\prime_{D^\ast}) \cdot \underline{q}}{\underline{q}^2}\, \underline{q}^\nu\, A_0(\underline{q}^2)\, ,
\end{align}
with $\epsilon^*(\vec{\underline{p}}_{D^\ast}^{\prime},
\underline\sigma^\prime_{D^\ast})$  being the polarization 4-vector of the $D^\ast$. It appears 
boosted according to the kinematics used in
Eq.~(\ref{eq:decaykinem}), i.e. (cf. App.~\ref{PolVectors})
\begin{eqnarray}\label{eq:Dpol}
\epsilon(\underline{\vec{k}}^\prime_{D^\ast},\pm 1)&=&\frac{1}{\sqrt{2}}(\mp \frac{\kappa_{D^\ast}}{m_{D^\ast}}, \mp\sqrt{1+(\frac{\kappa_{D^\ast}}{m_{D^\ast}})^2},-i,0)\, , \nonumber\\ \epsilon(\underline{\vec{k}}^\prime_{D^\ast},0 )&=&(0,0,0,1)\, .
\end{eqnarray}
$A_3(\underline{q}^2)$ is a linear combination of $A_1(\ul q^2)$ and $A_2(\ul q^2)$, 
namely
$A_3(\underline{q}^2)= \frac{m_B+m_{D^\ast}}{2 m_{D^\ast}}\, A_1(\underline{q}^2) - \frac{m_B-m_{D^\ast}}{2 m_{D^\ast}} A_2(\underline{q}^2) $.
\subsubsection{Cluster separability}
For the same reasons as in the pseudoscalar-to-pseudoscalar case,
the current does not exhibit cluster problems in the form of unphysical contributions 
to the covariant decomposition, and the form factors can be extracted unambiguously
from the independent components of the current.
Let us introduce the shorthand notation 
\begin{equation}
 J^\nu(\underline\mu^\prime_{D^\ast}) := J^\nu_{B\rightarrow
D^\ast}(\vec{\underline{k}}_{D^\ast}^\prime,
\underline\mu^\prime_{D^\ast};\vec{\underline{k}}_B). 
\end{equation}
The non-vanishing 
components of the current for the kinematics (\ref{eq:decaykinem}) are a total of 10, namely
$J^2(0)$, $J^3(0)$, $J^\mu(\pm 1)$, $\mu=0,1,2,3$. Taking into account that $J^\mu(1)=-J^\mu(-1)$, one is 
left with only 6 different matrix elements, 4 of them being independent. As one can see,
$A_0$ and $A_2$ enter only $J^0(1)$ and $J^1(1)$. Thus, the set 
$J^2(0)$, $J^3(0)$, $J^0(1)$ and $J^1(1)$ can be used to extract all the 
$P\to V$ decay form factors. They can be also obtained by means of
appropriate projections:
\begin{eqnarray}\label{eq:V}
V(\underline{q}^2)&=&\frac{i(m_B+m_{D^\ast})}{2 m_B^2 m_{D^\ast}^2}\left[\left(\frac{m_B^2+m_{D^\ast}^2-
\underline{q}^2}{2 m_B m_{D^\ast}} \right)^2-1\right]^{-1}\nonumber\\
&&\times \epsilon_\mu(\vec{\underline{k}}_{D^\ast}^{\prime}, \underline{\mu}^\prime_{D^\ast}=0)\,
\underline{k}'_{D^\ast\rho}\, \underline{k}_{B\sigma} \, \nonumber\\&&\times
\epsilon_{\nu}^{\phantom{\nu}\mu\rho\sigma} J^\nu_{B\rightarrow
D^\ast}(\vec{\underline{k}}_{D^\ast}^{\prime},\underline{\mu}^\prime_{D^\ast}=0; \vec{\underline{k}}_B)\, ,\\
A_0(\underline{q}^2)&=&\frac{1}{\sqrt{2} m_B m_{D^\ast}
}\left[\left(\frac{m_B^2+m_{D^\ast}^2- \underline{q}^2}{2 m_B
m_{D^\ast}} \right)^2-1\right]^{-1/2}\nonumber\\&&\times
\underline{q}_\nu J^\nu_{B\rightarrow
D^\ast}(\vec{\underline{k}}_{D^\ast}^{\prime},\underline{\mu}^\prime_{D^\ast}=1;
\vec{\underline{k}}_B)\, ,\\
A_1(\underline{q}^2)&=&\frac{1}{m_B+m_{D^\ast}}\,
\epsilon_\nu(\vec{\underline{k}}_{D^\ast}^{\prime},
\underline{\mu}^\prime_{D^\ast}=0)\, \nonumber\\ & &\times J^\nu_{B\rightarrow
D^\ast}(\vec{\underline{k}}_{D^\ast}^{\prime},\underline{\mu}^\prime_{D^\ast}=0;
\vec{\underline{k}}_B)\, .
\end{eqnarray}
and
\begin{eqnarray}\label{eq:A2}
A_2(\underline{q}^2)\!&=&\!\!\frac{\underline{q}^2
(m_B+m_{D^\ast})}{4 m_B^2 m_{D^\ast}^2
}\left[\left(\frac{m_B^2+m_{D^\ast}^2- \underline{q}^2}{2 m_B
m_{D^\ast}} \right)^2\!\!\!\!-1\right]^{-1}\nonumber\\
&&\times\Bigg\{
\frac{\sqrt{2}}{m_B}\left[\left(\frac{m_B^2+m_{D^\ast}^2-
\underline{q}^2}{2 m_B m_{D^\ast}}
\right)^2\!\!\!\!-1\right]^{-1/2}\left(
(\underline{p}_B+\underline{p}_{D^\ast}') -
\frac{m^2_B-m^2_{D^\ast}}{\underline{q}^2 }\, \underline{q}
\right)_\nu\nonumber\\
&&\quad\times J_{B\rightarrow
D^\ast}^\nu(\vec{\underline{k}}_{D^\ast}^{\prime},\underline{\mu}^\prime_{D^\ast}=1;
\vec{\underline{k}}_B)\nonumber\\
&& - \left[ 1- \frac{m_B^2 -
m_{D^\ast}^2}{\underline{q}^2}
\right]\epsilon_\nu(\vec{\underline{k}}_{D^\ast}^{\prime},
\underline{\mu}^\prime_{D^\ast}=0 )
 J^\nu_{B\rightarrow
D^\ast}(\vec{\underline{k}}_{D^\ast}^{\prime},\underline{\mu}^\prime_{D^\ast}=0;
\vec{\underline{k}}_B) \Bigg\}\, .\nonumber\\\end{eqnarray}

Having checked the fundamental properties of the electromagnetic and weak currents and 
knowing how to extract the corresponding form factors unambiguously, we are now
in the position
to compute the form factors for many different reactions and compare them with experimental data.
For simplicity,
we will take the same wave function model, namely the harmonic-oscillator wave function used 
for the numerical
studies shown in Fig.~\ref{fig:sdep}. The parameters are those given 
in Table~\ref{ParametersHeavyLight}, which allow
also for comparisons with front-form calculations. 
The discussion of the numerical results will be presented
in Chap.~\ref{ChNumStudiesII}. The method admits, of course, a much wider range of 
binding forces, 
namely all those which are compatible with the Bakamjian-Thomas construction. 

 \newpage
\thispagestyle{empty}

%% file: Ch5.tex
\chapter{Heavy-quark symmetry}\label{HQ:sym}

The formalism presented as far provides a way of calculating electroweak form factors of
two-body bound states and it is general enough to allow for different masses of the 
constituents, such that we are able to study heavy-light mesons. 
A requirement for any approach
 that attempts to describe this kind of systems is to be able to 
 reflect the heavy-quark symmetry predictions in
the limit in which one of the constituent masses goes to infinity.
The aim of this chapter is to examine the features of our formalism that emerge 
in the heavy-quark limit, $m_Q\to \infty$ (a precise definition 
of the limit will be given in the next section).
The heavy-quark limit provides additional symmetries 
beyond QCD~\cite{Neubert:1993mb}. Hadrons containing a single heavy quark share physical 
properties that make them simpler to describe. These
 properties are often used 
to design constituent quark models that describe heavy-light systems.
The work presented here, by contrast, starts from 
the most general case of systems of different constituent masses. It is 
the aim of this section to study if 
the requirements of heavy-quark symmetry emerge if the mass of the 
heavy quark goes to infinity. 

When the mass of the heavy particle of a system is heavy enough 
(in hadrons this means in practice $m_Q\gg\Lambda_{QCD}$) 
the behavior of the light quarks does not depend on the 
flavor of the heavy quark. 
Mathematically, what one obtains is that matrix elements do not depend on 
the heavy quark mass -- flavor symmetry -- 
or on the heavy quark spin -- spin symmetry.
The heavy-quark limit eliminates the 
heavy-quark mass from the description by assuming
$m_Q\simeq m_M$ and  $\frac{m_q}{m_Q}\rightarrow 0$. 
It becomes more convenient to use velocities instead of momenta, 
and the notion of velocity states 
gains thus more relevance. 

The intuitive quantum-field theoretical view of
a meson in he heavy-quark limit is to conceive it  
as a (anti)quark, whose mass is considered infinitely heavy, 
that moves with velocity $v$ and drags along
a cloud of light (anti)quarks and gluons. 
The dynamics of the heavy hadron is thus completely controlled by the heavy 
constituent (anti)quark. 
The main features and consequences of this kind of picture should, of course,
also be reflected by a simplified description of heavy-light mesons via constituent quark models.

\subsubsection{The Isgur-Wise function}

One of the consequences of heavy-quark symmetry is the existence 
of only one  universal form factor which 
is independent on the heavy-constituent mass and on the heavy-constituent spin.
This universal form factor is known as Isgur-Wise function, 
due to N.~Isgur and M.~B.~Wise \cite{Isgur:1989vq,Isgur:1990kf}, and it is 
usually written as function $\xi(v\cdot v')$, where $v$ and $v'$ are the initial
 and final four-velocities of the heavy-light hadron, respectively. 
The scalar product $v\cdot v'$ replaces the momentum 
transfer, which goes to infinity, as will be explained later. The existence of such a 
universal form factor in the heavy-quark limit is an indication of
heavy quark symmetry.

The main task of the present chapter will be to obtain and study this 
universal form factor. 
From the general expression obtained in the previous chapter for form factors
of arbitrary constituent masses we will see analytically as well as numerically
how heavy-quark symmetry arises. By comparison with the result for finite
heavy-quark masses we will be able to study he amount of
 heavy-quark symmetry breaking in the real world.

\subsubsection{Heavy-quark symmetry and the point form of dynamics}
Dirac's point form of dynamics is a framework in which the dependence 
 upon mass is explicit, making it particularly useful for studying 
the heavy-quark limit within the context of  specific models~\cite{Keister:1992wq}.
The model considered here will be the same harmonic-oscillator wave-function model
used in previous chapters. 
The  analytical result, however, allows for  any other bound state solutions.

In the following we will discuss  how the heavy-quark
limit has to be taken, we will examine the 
analytical and numerical consequences in the different processes, 
and provide the physical interpretation. 

\section{Space-like momentum transfer}

Let us start with electron-meson scattering. 
We will examine step by step the consequences 
of taking the heavy-quark mass going to infinity. 

\subsection{Definition of the heavy-quark limit (h.q.l.)}
It is important to keep in mind that the heavy-quark limit  
is not the non-relativistic limit.
The framework is fully relativistic but now one of the constituents 
shares non-relativistic features, while the other one does not. 
This is an important point, since the 4-momentum transfer squared, $q^2$,  
goes to infinite too, when the mass goes to infinity. 
In order to perform the heavy-quark limit the meson momenta 
 are expressed in terms of velocities and the scalar 
product of the initial and final velocities of the meson ($v\cdot v'$) is taken 
as the parameter that replaces the momentum transfer. More precisely,
the heavy-quark limit has to be taken in such a way that the quantity
\begin{equation}
 \ul v_{\alpha}\cdot \ul v'_{\alpha^{(\prime)}} = 
\frac{\ul k_\alpha\cdot \ul k_{\alpha^{(\prime)}}'}{m_\alpha m_{\alpha^{(\prime)}}}
\end{equation}
stays constant. In this limit the binding energy and the light-quark mass become
negligible, which means
\begin{equation}
 m_{Q^{\prime}}= m_{\alpha^{\prime}} \quad \text{and}
\quad \frac{m_q}{m_{Q^{(\prime)}}}= 0 \quad \text{for} 
\quad m_{Q^{(\prime)}}\to\infty.
\end{equation}
This is the precise definition of 
the heavy-quark limit (h.q.l.) that will be used in the following.
\subsection{Meson-electron kinematics in terms of velocities}\label{HQ:limit:def}

The kinematics of the meson in terms of velocities is hence
\begin{equation}
 \ul k_{[\alpha]}=m_\alpha\;\ul v_\alpha,\qquad  
\ul k_{[\alpha]}'=m_\alpha\; \ul v_\alpha',
\end{equation}
with the 4-velocities 
\begin{equation}\label{k:meson:hql}
\ul v_\alpha= \left(\begin{array}{c} \sqrt{1+ |\ul{\vec v}_\alpha|^2} \\ 
        -\sqrt{\frac{\ul v_\alpha\cdot \ul v_\alpha'-1}{2}}
\\ 0 \\ \sqrt{|\vec{\ul v}_\alpha|^2-\frac{1}{2}(\ul v_\alpha\cdot \ul v_\alpha'-1)}
       \end{array}\right);\quad 
\ul v_\alpha'= \left(\begin{array}{c} \sqrt{1+ |\ul{\vec v}_\alpha|^2} \\ 
        \sqrt{\frac{\ul v_\alpha\cdot \ul v_\alpha'-1}{2}}\\ 0 \\ 
\sqrt{|\vec{\ul v}_\alpha|^2-\frac{1}{2}(\ul v_\alpha\cdot \ul v_\alpha'-1)}
       \end{array}\right).
\end{equation}
Analogously, for the electron
\begin{equation}
 \underline{k}_e=\left(\begin{array}{c}\sqrt{m_e^2 +m_\alpha^2 |\vec{\ul v}_\alpha|^2} \\ m_\alpha\sqrt{\frac{\ul v_\alpha\cdot \ul v_\alpha'-1}{2}}\\ 0 \\ -m_\alpha\sqrt{|\vec{\ul v}_\alpha|^2-\frac{1}{2}(\ul v_\alpha\cdot \ul v_\alpha'-1)}\end{array} \right);\quad
 \underline{k}_e'=\left(\begin{array}{c}\sqrt{m_e^2 +m_\alpha^2 |\vec{\ul v}_\alpha|^2}\\-m_\alpha\sqrt{\frac{\ul v_\alpha\cdot \ul v_\alpha'-1}{2}}\\0\\-m_\alpha\sqrt{|\vec{\ul v}_\alpha|^2-\frac{1}{2}(\ul v_\alpha\cdot \ul v_\alpha'-1)}\end{array} \right) .
\end{equation}
The momentum transfer is then parametrized as follows
\begin{equation}\label{eq:Qemvv}
\text{Q}=\sqrt{-(\underline{k}_\alpha-\underline{k}_\alpha^\prime)^2}=
2m_\alpha\sqrt{\frac{\underline{v}_\alpha\cdot \underline{v}_\alpha^{\prime}-1}{2}}=:
2m_\alpha u\, .
\end{equation}
Note that $\ul v_\alpha\cdot \ul v_\alpha'\ge 1$ and that $|\vec v|$ 
is subject to the condition
\begin{equation}\label{minimalv}
 |\vec{\ul v}_\alpha|\ge u.
\end{equation}

\subsection{Currents and form factors in the h.q.l.}\label{HQ:limit:def}

Let us now see in detail how the heavy-quark limit leads to simplifications 
for the current at the hadronic and constituent levels, 
i.e. Eqs.~(\ref{eq:Jemcovdec}) and (\ref{eq:JQ}), 
respectively, leading to a $m_Q$-independent form factor. 
In the h.q.l. the electron momenta can be written 
as\footnote{The meson kinematics in the h.q.l. remains exactly
the same as in (\ref{k:meson:hql}).}
\begin{equation}\label{EqKinematicsmv}
\underline{k}_e\,\rightarrow\,m_\alpha \left(\begin{array}{c}\nu_\alpha \\ 
u\\ 0 \\ -\sqrt{\nu_\alpha^2-u^2}\end{array}\right); \quad
\underline{k}_e'\,\rightarrow\, m_\alpha\left(\begin{array}{c}\nu_\alpha
\\-u\\0\\-\sqrt{\nu_\alpha^2-u^2}\end{array}\right)
\end{equation}
where the notation $\nu_\alpha:=|\ul{\vec v}_\alpha|=|\ul{\vec v}_\alpha'|$
has been introduced.
The covariants that depend on the electron and meson momenta are $(\underline{k}_\alpha+\underline{k}^\prime_\alpha)^\mu$
and $ (\underline{k}_e+\underline{k}^\prime_e)^\mu$
\begin{equation}
 (\underline{k}_\alpha+\underline{k}^\prime_\alpha)=
m_\alpha\left( 2\sqrt{1+\nu_\alpha^2}\;,\;0\;,\;0\;,\;2\sqrt{\nu_\alpha^2-u^2} \right),
\end{equation}
\begin{equation}
 (\underline{k}_e+\underline{k}^\prime_e)=m_\alpha\left( 2 \nu_\alpha\;,\;0\;,\;0\;,\;
-2\sqrt{\nu_\alpha^2-u^2}\right).
\end{equation}

The current at the constituent level, Eq.~(\ref{eq:JQ}), requires more care. 
Using that 
\begin{equation}\label{limits}
  \vec{\ul k}_\alpha^{(\prime)},  \vec{\ul k}_Q^{(\prime)} 
\rightarrow m_\alpha \vec{\ul v}_\alpha^{(\prime)}, 
\quad \frac{|\vec k_{\bar q}^{\prime}|}{m_Q},
\frac{|\vec{ \tilde k}_{\bar q}^{\prime}|}{m_Q},
\frac{|\vec{ \tilde k}_Q^{\prime}|}{m_Q}
\rightarrow 0, \quad
\text{and}
\quad
\vec v_{Q\bar q}^{(\prime)} \rightarrow \vec{\ul v}_\alpha^{(\prime)},
\end{equation}
the pseudoscalar meson current (\ref{eq:JQ}) simplifies considerably. 
The most important effect of the limit is that one of the two 
contributions of the current to the form factor
 found in Eq.~(\ref{eq:voptos}) vanishes, namely 
the term that describes the photon coupling to 
the light antiquark, i.e.
\begin{equation}
\tilde J_{[\alpha]}^\nu = (\mathcal{Q}_Q J_Q^\nu + \mathcal{Q}_{\bar q}J_{\bar q}^\nu ) 
\longrightarrow \mathcal{Q}_Q J_Q^\nu.
\end{equation}

This is easy to understand. In the h.q.l. the momentum transfer goes to infinity.
If the transferred momentum is absorbed by the light quark, the wave-function overlap
vanishes. An infinitely heavy quark, on the other hand, is able to absorb  an
infinite amount of momentum with the wave function overlap staying finite. 

Thus only the contribution where the heavy quark is active survives 
and the meson mass can 
be factored out:

\begin{eqnarray}\label{EM:IW:analytical}
J^\mu_Q (\ul{\vec k}'_\alpha,\ul{\vec k}_\alpha) &\rightarrow & 
m_\alpha \tilde J^\mu_\infty (\ul{\vec v}'_\alpha,\ul{\vec v}_\alpha) = m_\alpha 
\int \frac{d^3 \tilde k'_{\bar q}}{4\pi}
\sqrt{\frac{\omega_{\tilde k_{\bar q}}}{\omega_{\tilde k'_{\bar q}}}} 
\Big\{ \sum_{\mu_Q,\mu'_Q}
\bar u_{\mu'_{\bar q}} (\vec{\ul v}'_{\bar q}) \gamma^\nu u_{\mu_{\bar q}} 
(\vec{ \ul v}_{\bar q} ) \nonumber \\
&& \times\frac{1}{2} D^{1/2}_{\mu_{\bar q}\mu'_{\bar q}}  \left[ 
R^{-1}_W \left(\frac{\tilde k_{\bar q}}{m_{\bar q}}, B_c(v_{\alpha})\right) 
R_W \left( \frac{\tilde k'_{\bar q}}{m_{\bar q}},B_c(v'{_{\alpha}}) \right) 
\right]\Big\} \nonumber \\
&&\times \psi^*(|\vec{\tilde k}'_{\bar q}|) \psi(|\vec{\tilde k}_{\bar q}|).
\end{eqnarray}
The Wigner rotations that act  on the heavy-quark spin
 have turned into the unit matrix and therefore they
disappear. Boost effects however, are still present for  the light degrees of freedom. 
The whole dependence of the integrand on $m_\alpha$ has vanished.
For the kinematics given in Eq.~(\ref{k:meson:hql}) it can be shown that the microscopic
current
$\tilde J^\mu_\infty (\ul{\vec v}'_\alpha,\ul{\vec v}_\alpha)$ has
 only two non-vanishing components (cf. App.~\ref{compspinorcurrent})
\begin{center}
 $\tilde J^\mu_\infty (\ul{\vec v}'_\alpha,\ul{\vec v}_\alpha)=
(\tilde J^0_\infty,0,0,\tilde J^3_\infty).$
\end{center}
 
In the following it will
be seen that $\tilde J^\mu_\infty (\ul{\vec v}'_\alpha,\ul{\vec v}_\alpha)$ 
still contains nonphysical contributions and we will show how we, nevertheless,
can extract the Isgur-Wise function in a sensible way.

\subsubsection{Covariant structure of the current and non-physical
contributions}

As explained in Chap.~\ref{ch:currents:and:ff}, cluster problems inherent in the 
Bakamjian-Thomas construction entail non-physical components in the most general 
covariant decomposition 
of the electromagnetic current of pseudoscalar mesons. 
These unphysical features were seen to vanish for large invariant mass
of the electron-meson system.
It is thus natural to wonder now if they still 
remain in the h.q.l. or if they disappear completely.

The general covariant decomposition of the electromagnetic
 current of pseudoscalar mesons, in the h.q.l. analogous 
 to Eq.~(\ref{eq:Jemcovdec}), but expressed in terms of velocities,
 can be written as
\begin{equation}\label{covestructureinf}
 \tilde J^\mu_\infty(\vec{\ul v'_\alpha},\vec{\ul v_\alpha}) =
(\ul v_\alpha+\ul v_\alpha ')^\mu\tilde f(\ul v_\alpha\cdot \ul v_\alpha',\nu_\alpha) + 
\frac{m_e}{m_\alpha}(\ul v_e+\ul v_e ')^\mu
\tilde g (\ul v_\alpha \cdot \ul v_\alpha', \nu_\alpha),
\end{equation}
where
\begin{equation}
 \frac{m_e}{m_\alpha}(\ul v_e +\ul v'_e)=2(\nu_\alpha,0,0,\sqrt{\nu_\alpha^2-u^2})
\end{equation}
is independent of the heavy-quark mass $m_\alpha$.
The heavy-quark limit
 does obviously not eliminate the second, nonphysical covariant in
Eq.~(\ref{covestructureinf}).
As in the case of finite heavy-quark mass, 
the form factors can, in addition to 
$\ul v_\alpha \cdot \ul v_\alpha'$, depend
 also on the modulus on the meson velocities $\nu_\alpha$. 
The latter replaces
the Mandelstam-$s$ dependence mentioned in the previous chapter, since
\begin{equation}
 \nu_\alpha = \frac{1}{2}\left(\frac{\sqrt{s}}{m_\alpha}- \frac{m_\alpha}{\sqrt{s}}\right),
\quad \text{with}\quad s=m_\alpha^2\left(\ul v_\alpha + \frac{m_e}{m_\alpha}\ul v_e \right)^2.
\end{equation}

\begin{figure}
\begin{center}
  \includegraphics[width=0.8\textwidth]{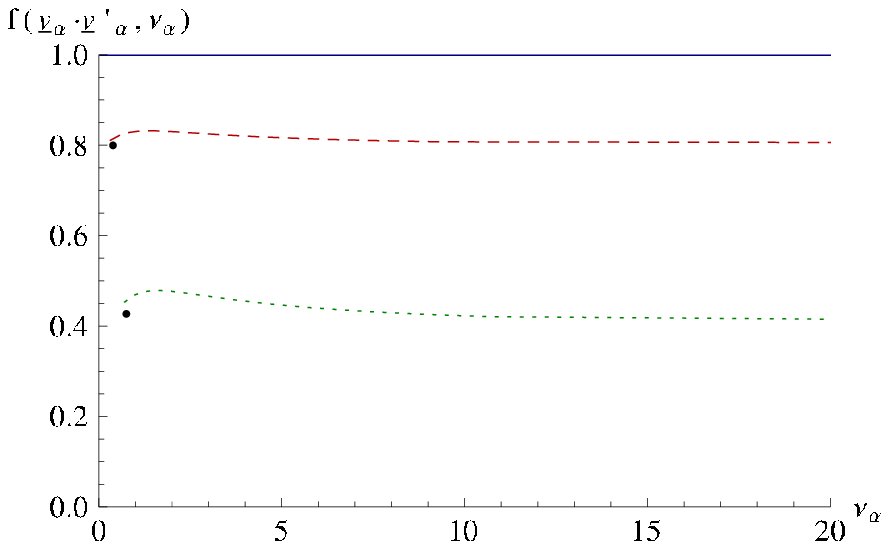}\\
 \includegraphics[width=0.8\textwidth]{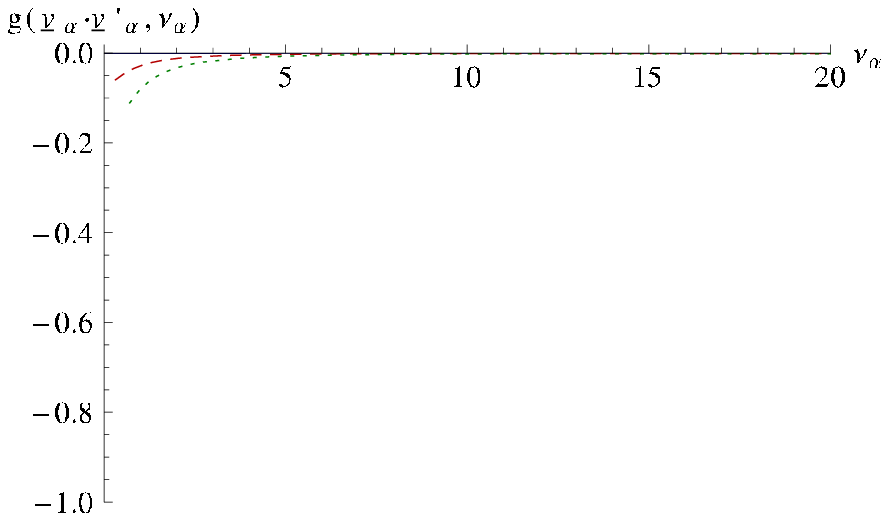}
\caption{Physical and spurious electromagnetic form factors, 
$\tilde f(\ul v_\alpha\cdot\ul v_\alpha',\nu_\alpha)$ and 
$\tilde g(\ul v_\alpha\cdot\ul v_\alpha',\nu_\alpha)$, 
of a heavy-light pseudoscalar
meson as a function of the modulus of the meson velocity 
$\nu_\alpha$ for different fixed values 
of $\ul v_\alpha\cdot\ul v_\alpha'$ (1 solid, 1.2 dashed, 2 dotted). 
The black dots in the upper figure are the values for the 
Isgur-Wise function directly calculated in the Breit 
frame ($\nu_\alpha=u$).}\label{fvandgv}
\end{center}
\end{figure}

Similarly as it was shown in the previous chapter
for finite heavy-quark mass, the dependence of 
$\tilde f(\ul v_\alpha\cdot \ul v_\alpha', \nu_\alpha)$
and $\tilde g(\ul v_\alpha\cdot \ul v_\alpha', \nu_\alpha)$ on $\nu_\alpha$
is displayed in Fig.~\ref{fvandgv} for several fixed values of $\ul v_\alpha\cdot \ul v_\alpha'$.
The $\nu_\alpha$-dependence of the physical form factor $\tilde f(\ul v_\alpha\cdot \ul v_\alpha', \nu_\alpha)$
and the size of the unphysical form factor $\tilde g(\ul v_\alpha\cdot \ul v_\alpha', \nu_\alpha)$ 
 are observed to vanish rather fast with increasing $\nu_\alpha$.

\subsection{The infinite momentum frame and the Breit frame}\label{sec:BreitInf}

In order to get the Isgur-Wise function that depends on 
$\ul v_\alpha\cdot \ul v_\alpha'$ alone we have to fix $\nu_\alpha$.
There are two particular choices that lead to interesting consequences and that
correspond to two particular reference frames.
The first one is the infinite momentum frame,
i.e. $\nu_\alpha\rightarrow \infty$, which has been 
already studied for finite masses in Chap.~\ref{ch:currents:and:ff}. 
The second one corresponds to the minimal possible value of 
$\nu_\alpha$ ($\nu_\alpha=u$), which is characteristic for  
the Breit frame.

\subsubsection{The infinite-momentum frame}

We have already observed that the $\nu_\alpha$-dependence 
of $\tilde f(\ul v_\alpha\cdot \ul v_\alpha', \nu_\alpha)$ 
as well as the spurious form factor
$\tilde g (\ul v_\alpha \cdot \ul v_\alpha', \nu_\alpha)$  vanish 
quickly with increasing $\nu_\alpha$.
It is thus suggestive to identify the Isgur-Wise function 
$\xi(\ul v_\alpha \cdot \ul v_\alpha')$ with the limit $\nu_\alpha\to \infty$ of 
the physical form factor $\tilde f (\ul v_\alpha \cdot \ul v_\alpha',\nu_\alpha)$. 
This corresponds to 
the infinite-momentum frame (IF). In this limit the current acquires the expected 
structure
\begin{equation}
\tilde{J}^{\nu}_\infty(\vec{\underline{v}}_\alpha^\prime,
\vec{\underline{v}}_\alpha)\stackrel{\nu_\alpha\rightarrow\infty}{\longrightarrow}
(\underline{v}_{\alpha}+\underline{v}_{\alpha}^{\prime})^\nu\,
\xi_{\mathrm{IF}}(\underline{v}_\alpha\cdot
\underline{v}_{\alpha}^{\prime})\, .
\end{equation}

The Isgur-Wise function can be extracted from the non-vanishing components of the 
current\footnote{For a detailed explanation about how to extract the form factor 
see App.~\ref{ApInfBreit}.}

\begin{equation}\label{eq:xiif}
\xi_{\mathrm{IF}}(\underline{v}_\alpha\cdot\underline{v}_{\alpha}^{\prime})=
\int\, \frac{d^3\tilde{k}_{\bar{q}}^\prime}{4\pi}\,
\sqrt{\frac{\omega_{\tilde{k}_{\bar{q}}}}
{\omega_{\tilde{k}^\prime_{\bar{q}}}}}\,\mathcal{S}_{\mathrm{IF}}\,
\psi^\ast\,(\vert \vec{\tilde{k}}_{\bar{q}}^\prime\vert)\,  \psi
 \,(\vert \vec{\tilde{k}}_{\bar{q}}\vert)\,\, .
\end{equation}
$\mathcal S_{\text{IF}}$ is the spin rotation factor in this particular frame 
\begin{equation}
\mathcal{S}_{\mathrm{IF}}=\frac{m_{\bar{q}}+\omega_{\tilde{k}^\prime_{\bar{q}}}+\tilde{k}^{
\prime 1}_{\bar{q}}\,
u}{\sqrt{(m_{\bar{q}}+\omega_{\tilde{k}_{\bar{q}}})
(m_{\bar{q}}+\omega_{\tilde{k}^\prime_{\bar{q}}})}}\, .
\end{equation}
$\tilde{k}^\prime_{\bar{q}}$ and 
$\tilde{k}_{\bar{q}}$ are related by Eq.~(\ref{eq:kktildep})
which, in the heavy-quark limit
and for this particular kinematics, leads to the following relation between 
$\omega_{\tilde{k}_{\bar{q}}}$ and
$\omega_{\tilde{k}^\prime_{\bar{q}}}$ (see boosts in App.~\ref{boostLimit}): 

\begin{equation}\label{eq:omegaif}
\omega_{\tilde{k}_{\bar{q}}}=2 \tilde{k}^{\prime 1}_{\bar{q}}\, u +2
\tilde{k}^{\prime 3}_{\bar{q}}\, u^2+
\omega_{\tilde{k}^\prime_{\bar{q}}} (2u^2+1)\, .
\end{equation}

\subsubsection{The Breit frame}

Another widely used
frame to analyze the $\gamma^\ast M_\alpha\rightarrow M_\alpha$
subprocess is the Breit frame (B) in which the energy-transfer between
the meson in the initial and the final states
vanishes~\cite{Lev:1994au,Klink:1998qf}. 
It corresponds to the opposite situation of the infinite-momentum frame, since it
is reached by taking the minimal value of $\nu_\alpha$, this is $\nu_\alpha^2=u^2=(\underline{v}_\alpha\cdot
\underline{v}_{\alpha}^{\prime}-1)/2$ (cf. Eq. (\ref{minimalv})). The structure of the current
 in this case is (see App.~\ref{ApInfBreit}):
\begin{eqnarray}\label{JQBreit}
\tilde{J}^{\nu}_\infty(\vec{\underline{v}}_\alpha^\prime,
\vec{\underline{v}}_\alpha) &\stackrel{\nu_\alpha \rightarrow
u}{\longrightarrow}& (\underline{v}_\alpha+
\underline{v}_{\alpha}^{\prime})^\nu\,
\Big\{\tilde{f}(\underline{v}_\alpha\cdot
\underline{v}_{\alpha}^{\prime},\nu_{\alpha}=u) \nonumber\\
&& + \sqrt{\frac{\underline{v}_\alpha\cdot
\underline{v}_{\alpha}^{\prime}-1}{\underline{v}_\alpha\cdot
\underline{v}_{\alpha}^{\prime}+1}}
\,\tilde{g}(\underline{v}_\alpha\cdot
\underline{v}_{\alpha}^{\prime},\nu_{\alpha}=u)\Big\}\nonumber\\&&=:
(\underline{v}_\alpha+ \underline{v}_{\alpha}^{\prime})^\nu\,
\xi_{\mathrm{B}}(\underline{v}_\alpha\cdot
\underline{v}_{\alpha}^{\prime}).
\end{eqnarray}

Since both  covariants become
proportional to $(\underline{v}_\alpha+\underline{v}_{\alpha}^{\prime})^\nu$, it
is not possible to distinguish the physical and the spurious form factor. 
One is thus led to
identify the Lorentz invariant quantity in Eq.~(\ref{JQBreit}) 
as the Isgur-Wise function obtained in the 
Breit (B) frame. The structure of $\xi_\text{B}(\ul v_\alpha\cdot\ul v_\alpha')$ is the same 
as in Eq.~(\ref{eq:xiif})
\begin{equation}\label{eq:xiB}
\xi_{\mathrm{B}}(\underline{v}_\alpha\cdot\underline{v}_{\alpha}^{\prime})=
\int\, \frac{d^3\tilde{k}_{\bar{q}}^\prime}{4\pi}\,
\sqrt{\frac{\omega_{\tilde{k}_{\bar{q}}}}
{\omega_{\tilde{k}^\prime_{\bar{q}}}}}\,\mathcal{S}_{\mathrm{B}}\,
\psi^\ast\,(\vert \vec{\tilde{k}}_{\bar{q}}^\prime\vert)\,  \psi
 \,(\vert \vec{\tilde{k}}_{\bar{q}}\vert)\,\, ,
\end{equation}
with the spin factor $\mathcal S_\text{B}$ being now
\begin{equation}
\mathcal{S}_{\mathrm{B}}=\frac{m_{\bar{q}}+\omega_{\tilde{k}^\prime_{\bar{q}}}+\tilde{k}^{
\prime 1}_{\bar{q}}\,
\frac{u}{\sqrt{u^2+1}}}{\sqrt{(m_{\bar{q}}+\omega_{\tilde{k}_{\bar{q}}})
(m_{\bar{q}}+\omega_{\tilde{k}^\prime_{\bar{q}}})}}\, .
\end{equation}
The difference between both frames resides in the relation
 between $\tilde{k}_{\bar{q}}^\prime$ and $\tilde{k}_{\bar{q}}$, which are connected
by boosts, that are different in the Breit frame and the 
infinite-momentum frame (cf. App.~\ref{ApInfBreit}). Correspondingly
\begin{equation}\label{eq:omegabreit}
\omega_{\tilde{k}_{\bar{q}}}=2 \tilde{k}^{\prime 1}_{\bar{q}}\, u
\sqrt{u^2+1}+ \omega_{\tilde{k}^\prime_{\bar{q}}} (2u^2+1).
\end{equation}

\subsubsection{Relating both reference frames}

Despite the integrands in (\ref{eq:xiif}) and (\ref{eq:xiB}) are different, 
the numerical results for the  integrals are found to be identical 
 for $\xi_{\mathrm{B}}(\underline{v}_\alpha\cdot\underline{v}_{\alpha}^{\prime})$ and
$\xi_{\mathrm{IF}}(\underline{v}_\alpha\cdot\underline{v}_{\alpha}^{\prime})$. This can be
seen in Fig. \ref{fvandgv}, where the results for 
$\xi_{\mathrm{B}}(\underline{v}_\alpha\cdot\underline{v}_{\alpha}^{\prime})$ are indicated
by black dots. The values for the dots coincide with the values for the curves for large 
$\nu_\alpha$.

It is thus suggestive to look for an analytical  
relation between $\xi_{\text{B}}(\ul v_\alpha\cdot \ul v_\alpha')$
and $\xi_{\text{IF}}(\ul v_\alpha\cdot \ul v_\alpha')$. One can indeed establish an
analytical relation by a simple change of variables. 
The transformation turns out to be the following 
rotation:
\begin{eqnarray}
\begin{pmatrix}
\tilde{k}_{\bar{q}}^{\prime\,1}\\
\tilde{k}_{\bar{q}}^{\prime\,3}
\end{pmatrix}_\mathrm{IF}=\frac{1}{\sqrt{u^2+1}}
\begin{pmatrix}
1 & -u\\ u & 1
\end{pmatrix}
\begin{pmatrix}
\tilde{k}_{\bar{q}}^{\prime\,1}\\
\tilde{k}_{\bar{q}}^{\prime\,3}
\end{pmatrix}_\mathrm{B}\, .
\end{eqnarray}
Applying this change of variables to the integrand in the infinite-momentum
 frame one obtains the same analytical 
 result for the Isgur-Wise function as in the Breit frame. The conclusion 
is then that the Isgur-Wise function obtained by this procedure turns out 
to be the same irrespective of where it is computed, either in the Breit or in the 
infinite-momentum frame. This does not hold, however, for arbitrary
frames (cf. Eq.~(\ref{covestructureinf}) and Fig.~\ref{fvandgv}), and it does not hold 
for finite mass of the heavy quark (cf. Figs.~\ref{fig:sdep} and ~\ref{fig:sbep}).

\subsubsection{The Isgur-Wise function}\label{IW:analitical}
The resulting Isgur-Wise function is thus the same irrespective of
whether it is extracted
in the Breit frame or in the infinite momentum frame.
The subscripts ``IF'' and ``B'' will therefore not
be taken into account any more. 
It will be more convenient for further purposes to use the analytical expression for the 
Isgur-Wise function obtained in the Breit frame. 
So we will take the following expression for the Isgur-Wise function
 in the sequel:

\begin{equation}\label{eq:IWfinal}
\xi(v\cdot v^\prime)= \int\,
\frac{d^3\tilde{k}_{\bar{q}}^\prime}{4\pi}\,
\sqrt{\frac{\omega_{\tilde{k}_{\bar{q}}}}
{\omega_{\tilde{k}^\prime_{\bar{q}}}}}\,\mathcal{S}\,
\psi^\ast\,(\vert \vec{\tilde{k}}_{\bar{q}}^\prime\vert)\,  \psi
 \,(\vert \vec{\tilde{k}}_{\bar{q}}\vert)\,\, ,
\end{equation}
with
\begin{equation}\label{eq:IWomega}
\omega_{\tilde{k}_{\bar{q}}}= \tilde{k}^{\prime 1}_{\bar{q}}\,
\sqrt{(v\cdot v^\prime)^2-1}+ \omega_{\tilde{k}^\prime_{\bar{q}}}\,
(v\cdot v^\prime)
\end{equation}
and
\begin{equation}\label{eq:IWspin}
\mathcal{S}=\frac{m_{\bar{q}}+\omega_{\tilde{k}^\prime_{\bar{q}}}+\tilde{k}^{
\prime 1}_{\bar{q}}\, \sqrt{\frac{{(v\cdot v^\prime)-1}} {{(v\cdot
v^\prime)+1}}}} {\sqrt{(m_{\bar{q}}+\omega_{\tilde{k}_{\bar{q}}})
(m_{\bar{q}}+\omega_{\tilde{k}^\prime_{\bar{q}}})}}\, .
\end{equation}
The underline and the subscript $\alpha$ have been dropped for 
simplicity.
The Isgur-Wise function obtained 
within this procedure depends only on $v\cdot v'$, 
has the correct normalization condition, 
i.e. $\xi (v\cdot v'=1)=1$, and 
is independent of the heavy-quark mass.  
 $\xi (v\cdot v')$ is thus
universal and the same for any meson that contains 
the same light antiquark. This property is called heavy-quark
flavor symmetry and it is the first part of the  proof 
that heavy-quark symmetry is respected by our approach.

\section{Time-like momentum transfer}\label{Electroweak}

Until now we have studied for electron-meson scattering
how heavy-quark symmetry arises sending the heavy-quark mass to infinity.
In this way it disappears from the description, 
leading to a universal form factor, the Isgur-Wise function
$\xi(v\cdot v')$.  
Heavy-quark symmetry goes even further, it has also consequences for processes that 
involve time-like momentum transfers. 

If matrix elements do not depend on the mass of the heavy quark,
transition form factors that involve a change of flavor of the heavy quark
are expected to be identical to those in which
the flavor of the heavy quark is unaltered. One thus may expect  
relations between electromagnetic
and weak form factors in the heavy-quark 
limit. Such relations are indeed given in the literature~\cite{Isgur:1989vq,Neubert:1993mb}.
They will be studied in the present section. 
As in the previous section, starting from the general expression for the 
form factors, the consequences of heavy-quark symmetry will be tested by taking the h.q.l. 
As we will see, both flavor symmetry and spin symmetry will occur in electroweak processes.

\subsection{Kinematics in terms of velocities}

For time-like momentum transfer the parametrization in terms 
of velocities $\underline{v}_B \cdot
\underline{v}^\prime_{D^{(\ast)}}$ leads to the following meson and heavy-quark 
momenta~(cf. Eqs.~(\ref{eq:decaykinem}) and~(\ref{Eqkappa})):
\begin{eqnarray}\label{eq:vdecay}
\underline{k}_B&=&m_B \left( \begin{array}{c} 1\\ 0 \\0\\0
\end{array}\right)=m_B \underline{v}_B\, , \\
\underline{k}_{D^{(\ast)}}^\prime&=& m_{D^{(\ast)}}\left(
\begin{array}{c}
\sqrt{(\underline{v}_B \cdot \underline{v}^\prime_{D^{(\ast)}})^2-1} \\
\underline{v}_B \cdot
\underline{v}^\prime_{D^{(\ast)}}
\\0\\ 0
\end{array}\right)=m_{D^{(\ast)}}\underline{v}_{D^{(\ast)}}\,
.
\end{eqnarray}
The (time-like) momentum transfer is given here by
\begin{eqnarray}\label{eq:qdecay}
0\leq \underline{q}^2&=&(\underline{k}_B-\underline{k}_{D^{(\ast)}})^2 \\
&&\hspace{-1.5cm}=m_B^2+m_{D^{(\ast)}}^2 -2 m_B
m_{D^{(\ast)}}\underline{v}_B \cdot
\underline{v}^\prime_{D^{(\ast)}}\leq (m_B-m_{D^{(\ast)}})^2\,
.\nonumber
\end{eqnarray}
From Eqs.~(\ref{eq:vdecay}) and (\ref{eq:qdecay}) one can deduce that
$\underline{v}_B \cdot
\underline{v}^\prime_{D^{(\ast)}}$ is restricted by the condition
\begin{equation}
1\leq \underline{v}_B \cdot \underline{v}^\prime_{D^{(\ast)}}\leq
1+\frac{(m_B-m_{D^{(\ast)}})^2}{2 m_B m_{D^{(\ast)}}}\,.
\end{equation}
Direct comparisons of form factors for space-like and time-like processes can 
be done within this interval. 

\subsection{Flavor symmetry}

First we will study heavy-quark flavor symmetry by comparing the Isgur-Wise 
function~(\ref{eq:IWfinal}) from electromagnetic scattering
with the one from the $B\to D e\bar \nu_e $
transition.                                                            
Flavor symmetry predicts that in a $Q\bar q$-system 
the behavior of the light quark 
appears blind to the flavor of the heavy one. 
This implies that the form factors
obtained for a system like $\bar u b$ should be identical to 
the ones obtained for $\bar u c$. 
Comparing the covariant structure 
of the electromagnetic and weak currents, (\ref{eq:Jpspsphys})
and (\ref{em:ff:usual}) respectively, it can be demonstrated that
the following relations should be fulfilled when the mass of the heavy quark
goes to infinity \cite{Isgur:1989vq,Neubert:1993mb}
\begin{equation}\label{eq:f0xi}  
R\,\left[1-\frac{q^2}{(m_B+m_D)^2}\right]^{-1}\,
F_0(q^2)\stackrel{\mathrm{h.q.l.}}{\longrightarrow}\,
\xi(\underline{v}_B\cdot\underline{v}^\prime_{D}),
\end{equation}
\begin{equation}\label{eq:f1xi}
R\,\, F_1(q^2)\stackrel{\mathrm{h.q.l.}}{\longrightarrow}\,
\xi(\underline{v}_B\cdot\underline{v}^\prime_{D})\, ,
\end{equation}
with
\begin{equation}\label{eq:R}
R=\frac{2\sqrt{m_B m_D}}{m_B+m_D}\, .
\end{equation}
and $\xi(\underline{v}_B\cdot\underline{v}^\prime_{D})$ being the h.q.l. of the electromagnetic
form factor $F(Q^2)$ (considered as function of 
$\ul v_\alpha\cdot \ul v_\alpha'$).

\subsection{Currents and form factors \\ in pseudoscalar-to-pseudoscalar meson transitions}

The quark current (\ref{eq:Jwkpsps}) in the h.q.l.
 takes on the form
\begin{eqnarray}\label{eq:hqljBD}
J^{\nu}_{B\rightarrow D}
(\vec{\underline{k}}_D^\prime,\vec{\underline{k}}_B)
&\stackrel{\mathrm{h.q.l.}}{\longrightarrow}&  \sqrt{m_B
m_D}\, \tilde{J}^{\nu}_{B\rightarrow D}
(\vec{\underline{v}}_D^\prime,\vec{\underline{v}}_B)\nonumber\\
&=& \sqrt{m_B
m_D}\, \int\, \frac{d^3\tilde{k}_{\bar{q}}^\prime}{4\pi}\,
\sqrt{\frac{\omega_{\tilde{k}_{\bar{q}}}}
{\omega_{\tilde{k}^\prime_{\bar{q}}}}}
 \,  \bigg\{\!\sum_{\mu_b,\mu_c^\prime
=\pm \frac{1}{2}}\!\!\!
\bar{u}_{\mu_c^\prime}(\vec{\underline{v}}_D^{\,\prime})\,\gamma^\nu\,
u_{\mu_b}(\vec{\underline{v}}_B) \nonumber \\ && \times\frac{1}{2}\,
D^{1/2}_{\mu_b\mu_c^\prime}\!
\left[\!R_{\mathrm{W}}\!\left(\frac{\tilde{k}^\prime_{\bar{q}}}{m_{\bar{q}}},
B_c(\underline{v}_{D}^\prime)\right)\, \, \right] \bigg\}\,
\psi^\ast\,(\vert \vec{\tilde{k}}_{\bar{q}}^\prime\vert)\,  \psi
 \,(\vert \vec{\tilde{k}}_{\bar{q}}\vert)\,\, .\nonumber \\
\end{eqnarray}

As in the electromagnetic case, the Wigner $D$-function
that acts on the heavy-quark degrees of freedom becomes the 
unit matrix. The expression~(\ref{eq:hqljBD}) is simpler 
than in the electromagnetic case due to the kinematics 
of the weak decay processes, where the initial 
state is at rest (cf. Sec.~\ref{SecCupledChannelWeak}). 
This is the reason why the second Wigner rotation that depends on the initial velocity
is absent here.
When one imposes the condition $\ul{\vec v}_\alpha=0$ 
in the electromagnetic case (\ref{EM:IW:analytical}) one recovers exactly Eq.~(\ref{eq:hqljBD}).
Note also that, due to the condition $\ul{\vec{v}}_B=0$, 
the kinematics resembles the one in the Breit frame, where the whole
process also takes place only along one direction. 
 
Using the properties of the Wigner $D$-functions
one can write the point-like quark current as
(cf. App.~\ref{WeakPointLikeCurrents}):
\begin{equation}
 \bar{u}_{\mu_b}(\vec{\underline{v}}_D^{\,\prime})\,\gamma^\nu\,
 u_{\mu_b}(\vec{\underline{v}}_B)=\sqrt{\frac{2}{\underline{v}_B
 \cdot \underline{v}^\prime_{D}+1}}\,
 (\underline{v}_B+\underline{v}^\prime_{D})^\nu \,.
\end{equation}
The meson transition current can therefore be expressed in terms of the  covariant 
$(\underline{v}_B+\underline{v}^\prime_{D})^\nu$ alone:
\begin{equation}\label{eq:Jpspshql}
\tilde{J}^{\nu}_{B\rightarrow D}
(\vec{\underline{v}}_D^\prime,\vec{\underline{v}}_B)=
(\underline{v}_B+\underline{v}^\prime_{D})^\nu \,
\xi_W(\underline{v}_B\cdot\underline{v}^\prime_{D})\, .
\end{equation}
The resulting analytical expression for $\xi_W(\underline{v}_B\cdot\underline{v}^\prime_{D})$ 
extracted in this manner from the semileptonic weak (`$W$') process is

\begin{equation}\label{eq:IWweak}
\xi_W(\ul v_B\cdot \ul v_D^\prime)= \int\,
\frac{d^3\tilde{k}_{\bar{q}}^\prime}{4\pi}\,
\sqrt{\frac{\omega_{\tilde{k}_{\bar{q}}}}
{\omega_{\tilde{k}^\prime_{\bar{q}}}}}\,\mathcal{S}_W\,
\psi^\ast\,(\vert \vec{\tilde{k}}_{\bar{q}}^\prime\vert)\,  \psi
 \,(\vert \vec{\tilde{k}}_{\bar{q}}\vert)\,\, .
\end{equation}
with
\begin{equation}
\omega_{\tilde{k}_{\bar{q}}}= \tilde{k}^{\prime 1}_{\bar{q}}\,
\sqrt{(\ul v_B\cdot \ul v_D^\prime)^2-1}+ \omega_{\tilde{k}^\prime_{\bar{q}}}\,
(\ul v_B\cdot \ul v_D^\prime)
\end{equation}
and
\begin{equation}\label{eq:IWspinWeak}
\mathcal{S}_W=\frac{m_{\bar{q}}+\omega_{\tilde{k}^\prime_{\bar{q}}}+\tilde{k}^{
\prime 1}_{\bar{q}}\, \sqrt{\frac{{(\ul v_B\cdot \ul v_D^\prime)-1}} {{(\ul v_B\cdot
\ul v_D^\prime)+1}}}} {\sqrt{(m_{\bar{q}}+\omega_{\tilde{k}_{\bar{q}}})
(m_{\bar{q}}+\omega_{\tilde{k}^\prime_{\bar{q}}})}}\, .
\end{equation}

The Isgur-Wise function for this weak heavy-to-heavy decay is thus 
identical with the one extracted from electron-meson scattering  
(cf. Eqs.~(\ref{eq:IWfinal})-(\ref{eq:IWspin})). This is an important result
showing that the description 
of the electroweak structure of mesons is properly done within our approach.
It guarantees heavy-quark flavor symmetry and provides the correct relations between 
space- and time-like form factors in the h.q.l.

\subsection{Spin symmetry}

Heavy-quark symmetry allows also to relate form factors involving pseudoscalar mesons with
corresponding ones involving vector mesons in the h.q.l. This symmetry emerges from the decoupling 
of the heavy-quark spin. 
For weak pseudoscalar-to-vector transitions the form factors 
are related by~\cite{Neubert:1991xw}:
\begin{equation}\label{eq:a1xi}
R^\ast\,\left[1-\frac{q^2}{(m_B+m_D^\ast)^2}\right]^{-1}\,
A_1(q^2)\stackrel{\mathrm{h.q.l.}}{\longrightarrow}\,
\xi(\underline{v}_B\cdot\underline{v}^\prime_{D^\ast})\, ,
\end{equation}
\begin{equation}\label{eq:vxi}
R^\ast\,\, V(q^2)\stackrel{\mathrm{h.q.l.}}{\longrightarrow}\,
\xi(\underline{v}_B\cdot\underline{v}^\prime_{D^\ast})\, ,
\end{equation}
and
\begin{equation}\label{eq:a02xi}
R^\ast\,\, A_i(q^2)\stackrel{\mathrm{h.q.l.}}{\longrightarrow}\,
\xi(\underline{v}_B\cdot\underline{v}^\prime_{D^\ast})\, ,\quad i=0,2\, ,
\end{equation}
with
\begin{equation}\label{eq:Rs}
R^\ast=\frac{2\sqrt{m_B m_{D^\ast}}}{m_B+m_{D^\ast}}\, ,
\end{equation}
where the form factors $A_i(q^2)$, $A_1(q^2)$ and $V(q^2)$ are those 
introduced in Eq.~(\ref{eq:Jppsvdec}).
In the following we will take as a representative example the
$B\to D^* e\bar \nu_e$ transition. 

\subsection{Currents and form factors \\ in pseudoscalar-to-vector meson transitions}

The weak transition current~(\ref{eq:Jwkpsv}) becomes 
in the heavy quark limit 

\begin{align}\label{eq:Jhqlpsv}
J^{\nu}_{B \rightarrow D^\ast}&
(\vec{\underline{k}}_{D^\ast}^\prime,\underline{\mu}_{D^\ast}^\prime;\vec{\underline{k}}_B)
\stackrel{\text{h.q.l.}}{\longrightarrow}  \sqrt{m_B m_D}\, \tilde{J}^{\nu}_{B\rightarrow D^\ast}
(\vec{\underline{v}}_{D^\ast}^\prime,\underline\mu_{D^\ast}^\prime;\vec{\underline{v}}_B)\nonumber\\
& =\sqrt{m_B
m_D}\,  \int\, \frac{d^3\tilde{k}_{\bar{q}}^\prime}{4 \pi}\,
\sqrt{\frac{\omega_{\tilde{k}_{\bar{q}}}}
{\omega_{\tilde{k}^\prime_{\bar{q}}}}} \,
 \bigg\{\!\sum_{\mu_b,\mu_c^\prime,\tilde\mu_{\bar
 q}^\prime=\pm \frac{1}{2}}
 \!\!\!\!\!\!
\bar{q}_{\mu_c^\prime}(\underline{\vec{v}}_{D^\ast}^\prime)\,\gamma^\nu\,
(1-\gamma^5) u_{\mu_b}(\underline{\vec{v}}_B) \nonumber \\ 
&  \times
\sqrt{2} (-1)^{\frac{1}{2}-\mu_b}
C^{1\underline\mu^\prime_{\!D^\ast}}_{\frac{1}{2}\mu_c^\prime\frac{1}{2}
\tilde{\mu}_{\bar{q}}^\prime}\, \nonumber\\
&\times D^{1/2}_{\tilde\mu_{\bar{q}}^\prime
-\mu_b}\!\left[ \!R^{-1}
_{\mathrm{W}}\!\left(\frac{\tilde{k}^\prime_{\bar q}}{m_{\bar q}},
B^{-1}_c(\underline{v}_{D^\ast}^\prime)\right)\right] \bigg\}\,
\psi^\ast_{D^{\ast}}\,(\vert \vec{\tilde{k}}_{\bar{q}}^\prime\vert)\, \psi_B
 \,(\vert \vec{\tilde{k}}_{\bar{q}}\vert)\, .
\end{align}

In the h.q.l. the covariant structure of (\ref{eq:Jppsvdec}) goes over into
\begin{eqnarray}\label{eq:Jpsvhqlcov}
\tilde{J}^{\nu}_{B\rightarrow D^\ast}
(\vec{\underline{v}}_{D^\ast}^\prime,\underline{\mu}_{D^\ast}^\prime;\vec{\underline{v}}_B)&=&
i\,\epsilon^{\nu\alpha\beta\gamma}\, \epsilon_\alpha(m_{D^\ast} \vec{\underline{v}}_{D^\ast}^\prime,\underline{\mu}_{D^\ast}^\prime)
\,\underline{v}^{\prime}_{D^\ast\beta}\, \underline{v}_{B\gamma} \,
\xi(\underline{v}_B\cdot\underline{v}^\prime_{D^\ast})\, \nonumber\\
&&-\Big[\epsilon^\nu(m_{D^\ast} \vec{\underline{v}}_{D^\ast}^\prime,\underline{\mu}_{D^\ast}^\prime)
\,(\underline{v}_B\cdot\underline{v}^\prime_{D^\ast}+1) \\
&& \qquad -\underline{v}^{\prime\nu}_{D^\ast}\,\epsilon(m_{D^\ast} 
\vec{\underline{v}}_{D^\ast}^\prime,\underline{\mu}_{D^\ast}^\prime)\cdot\underline{v}_B
\Big]\, \xi(\underline{v}_B\cdot\underline{v}^\prime_{D^\ast})\,. \nonumber
\end{eqnarray}
By comparison of Eqs.~(\ref{eq:Jhqlpsv}) and~(\ref{eq:Jpsvhqlcov}) 
one can see that the Isgur-Wise function 
$\xi(\underline{v}_B\cdot\underline{v}^\prime_{D^\ast})$ is the same as in 
Eqs.~(\ref{eq:IWweak})-(\ref{eq:IWspinWeak}).
This shows how heavy-quark spin symmetry, which has its origin in the decoupling of 
the heavy-quark spin from the spin of the light degrees of freedom, arises when the 
mass goes to infinity.
This proves that heavy-quark spin symmetry is also respected by our
approach. 

In the next chapter numerical results for the Isgur-Wise function 
will be presented and compared with results for the case of finite
heavy-quark masses. Heavy-quark symmetry breaking due to finite masses will be
discussed.

 \newpage
\thispagestyle{empty}

%% file: Ch6.tex
\chapter{Numerical studies I}\label{ChNumStudiesI}

In this chapter we will study the electroweak (transition) form factors of heavy-light mesons 
numerically. By comparing the numerical results for these form factors, obtained with physical masses for the heavy
quarks, with the outcome in the heavy-quark limit, we will estimate the amount of heavy-quark-symmetry
breaking for the physical masses.

\section{Meson wave function}
The form factors, and thus the Isgur-Wise function, are solely determined by the $Q\bar q$ bound-state wave
function and the constituent masses. We take a harmonic-oscillator wave function which is defined
as follows:
\begin{equation}\label{eq:wavefunc}
 \psi(\kappa)=
 \frac{2}{\pi^{\frac{1}{4}}a^{\frac{3}{2}}}
 \exp\left(-\frac{\kappa^2}{2a^2}\right)\,.
\end{equation}
There are mainly two reasons for choosing such a simple wave function.
On the one hand it is the main goal of this work to demonstrate 
that the kind of relativistic coupled-channel approach we are using is
 general enough to provide sensible results for the description of the electroweak 
structure of heavy-light systems. We do not want to give 
quantitative predictions for electroweak form factors based on
sophisticated constituent-quark models. On the other hand this wave function will allow 
to do a direct comparison with analogous calculations carried out 
within a front-form approach~\cite{Cheng:1996if}.
The numerical calculations could, of course, be carried out 
using any other model wave function obtained from a
particular bound-state problem. 
The numerical results presented in this chapter 
have been computed using the model parameters quoted in 
Table~\ref{ParametersHeavyLight}, which have been 
taken from Ref.~\cite{Cheng:1996if}.

 \begin{table}[h!]\label{ParametersHeavyLight}
\begin{center}
    \begin{tabular}{c c c c}
        $m_{u,d}$ & $m_b$   & $m_c$   & $a$       \\ \hline\hline
        0.25 GeV  & 4.8 GeV & 1.6 GeV &   0.55 GeV  \\ \hline
    \end{tabular}\caption{Model parameters used for the numerical calculations 
presented in this chapter. 
Physical masses as well as the harmonic-oscillator parameter $a$ 
are taken from Ref.~\cite{Cheng:1996if}.
The Cabibbo-Kobayashi-Maskawa matrix element $|V_{cb}|$ as well as the physical meson masses 
are the actual values quoted by the 
Particle Data Group~\cite{Beringer:2012}.}
\end{center}
\end{table}
\vspace{-0.2cm}
\section{The Isgur-Wise function}

\begin{figure}[h!]
\begin{center}
\includegraphics[width=0.8\textwidth]{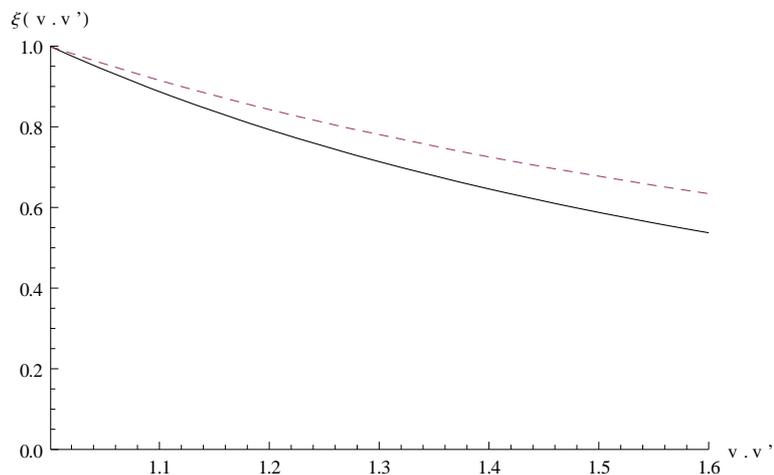}
\caption{The Isgur-Wise function (solid line) as given in 
Eqs.~(\ref{eq:IWfinal})-(\ref{eq:IWspin}) with the model parameters of
Table~\ref{ParametersHeavyLight}. 
The dashed line corresponds to spin-rotation factor $\mathcal S=1$.}\label{IW}
\end{center}
\end{figure}

The solid line in Fig.~\ref{IW} shows the numerical result for the 
Isgur-Wise function as derived in Sec.~\ref{IW:analitical}. 
The result is the same for the electromagnetic case
computed either in the Breit frame or in the infinite-momentum frame 
and agrees also with the one for weak decays, 
despite those processes involve
space- and time-like momentum transfers, respectively. 
Our numerical result coincides with the Isgur-Wise function obtained within
the light-front quark model of Ref.~\cite{Cheng:1996if}.
The dashed line corresponds to spin-rotation factor $\mathcal S=1$; 
this is the result one would have for spinless quarks. 
The difference between both lines 
indicates the importance of the appropriate treatment of relativistic spin  
rotations when boosting the initial to the final $Q\bar q$-bound-state wave function. 

A comparison with experimental data will be done later when we present form-factor results
for finite heavy-quark masses. The Isgur-Wise function 
will then be used as a reference quantity to estimate 
the amount of heavy-quark symmetry breaking.
\newpage

 \section[Heavy-quark symmetry breaking in e.m. processes~~]{Heavy-quark symmetry breaking \\ in electromagnetic processes}
Heavy-quark symmetry is broken for finite heavy-quark masses. 
Form factors of heavy-light mesons differ from the Isgur-Wise function 
since the physical masses are finite. 
Once the Isgur-Wise function has been obtained, it can be used to estimate 
quantitatively to which extend the realistic case, corresponding to physical
masses, deviates from the heavy-quark limit. 
This gives us an estimate of the amount of heavy-quark symmetry breaking
and provides conditions under which the heavy-quark 
limit turns out to be a good approximation.

\begin{figure}[h!]
\begin{center}
\includegraphics[width=0.8\textwidth]{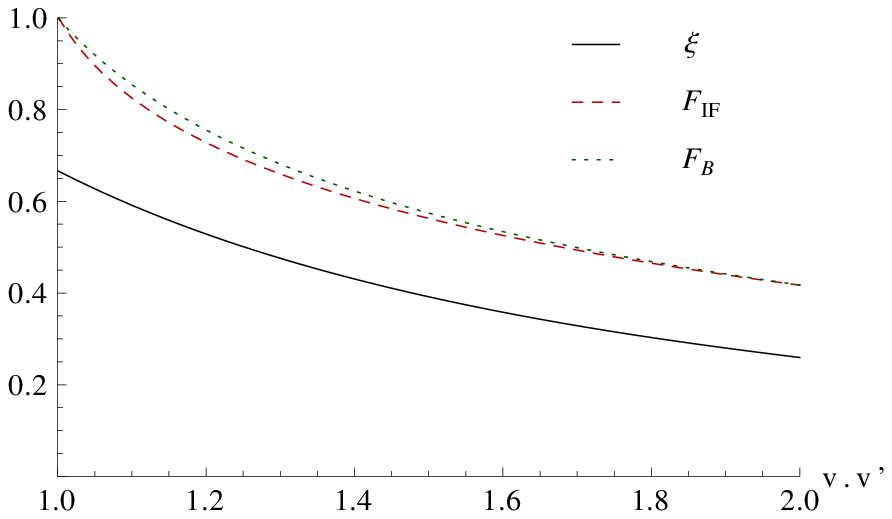}\\
\includegraphics[width=0.8\textwidth]{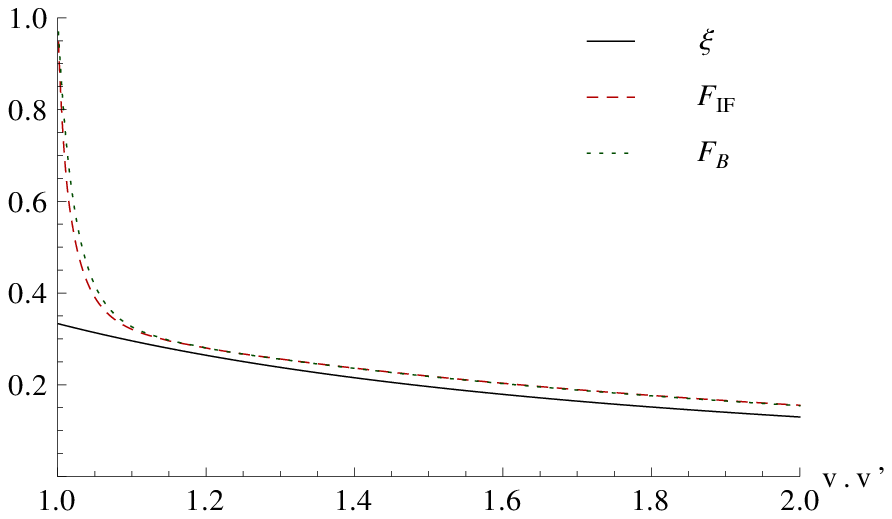}
\caption{Electromagnetic form factors of the $D^+$
(top) and $B^-$ (down) mesons calculated in the Breit frame
(dotted line) and infinite-momentum frame (dashed line) in
comparison with the Isgur-Wise function (solid line). For direct
comparison the Isgur-Wise function is multiplied by $|\mathcal{Q}_Q|$, i.e.
the charge of the heavy quark. Model parameters are taken from
Table~\ref{ParametersHeavyLight}.} \label{fig:scattpsps}
\end{center}
\end{figure}

Fig.~\ref{fig:scattpsps} shows the electromagnetic form factors for the 
$D^+$ and $B^-$ mesons, 
as measured in the space-like momentum transfer region, as functions of
$v\cdot v'$.  
They have to be compared with the Isgur-Wise function (solid line) which
is normalized to the charge of the heavy quark in each case. 
The electromagnetic form factor is the result of the sum of
two contributions that correspond to the photon coupling to the heavy and 
to the light quarks (cf. Eq.~\ref{eq:voptos}). 
Each of them is weighted with the charge of the 
corresponding quark. 
In the heavy-quark limit the contribution in which the light quark is active 
vanishes.  
For finite heavy-quark masses it causes a peak at $v\cdot v'\to 1$, which 
becomes more pronounced with increasing the mass of the heavy quark and disappears 
completely in the heavy-quark limit, since the light-quark contribution 
decreases faster with increasing $v\cdot v'$ than the heavy-quark contribution. 
In case of the $B⁻$ meson (bottom) the light-quark
contribution dies out rather fast, such that nearly 
the whole form factor is dominated by the heavy-quark 
contribution. It lies above the Isgur-Wise function, being 
about 20\% larger.
The heavy-quark contribution starts to dominate at $v\cdot v'\gtrsim 1.1$  in this case
(which corresponds to $\text{Q}^2\gtrsim 5$~GeV$^2$).

In the case of the $D^+$-meson the dominance of the heavy-quark 
contribution sets in at about the
same momentum transfer ($\text{Q}^2\gtrsim 5$~GeV$^2$), which corresponds 
to (cf. Eq.~(\ref{eq:qdecay}))
$v\cdot v^\prime \gtrsim 1.7$. The absolute magnitude of the form factor at $v\cdot v'\sim 2$
deviates from the Isgur-Wise function by abot 60\%. This is due to the smallness of the 
charm-quark mass.
For the $B^-$ meson the heavy-quark limit thus seems to be a reasonable approximation whereas the 
charm quark mass is obviously too small to be well approximated by the heavy-quark limit. 
\newpage

\newpage
\section[Heavy-quark symmetry breaking in weak processes~~]{Heavy-quark symmetry breaking \\in weak processes}

In the following, numerical results for the weak form factors as measured in
the $B^-\to D^0e^-\bar \nu_e$ and $B^-\to D^{0(*)}e^-\bar \nu_e$ decay processes 
will be discussed.

\vspace{-0.3cm}
\subsubsection{Heavy-quark flavor symmetry breaking}
\begin{figure}[h!]
\includegraphics[width=0.8\textwidth]{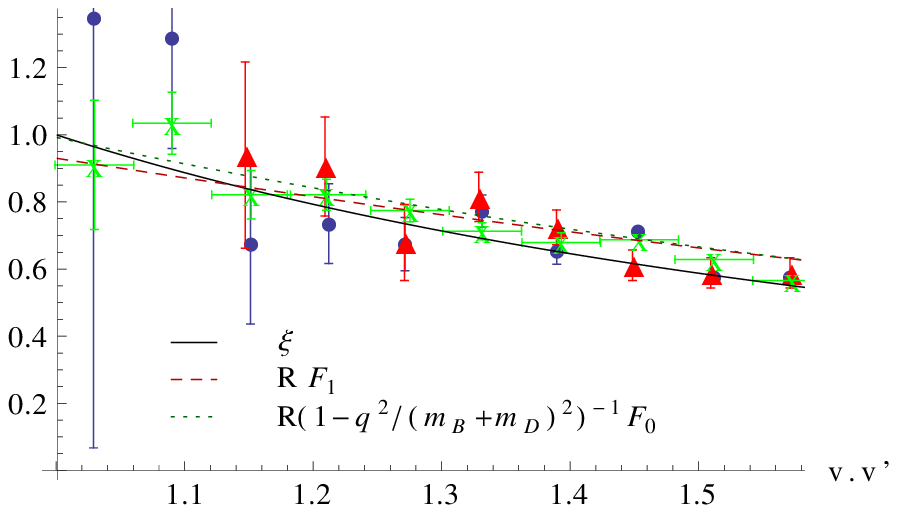}\\
\includegraphics[width=0.8\textwidth]{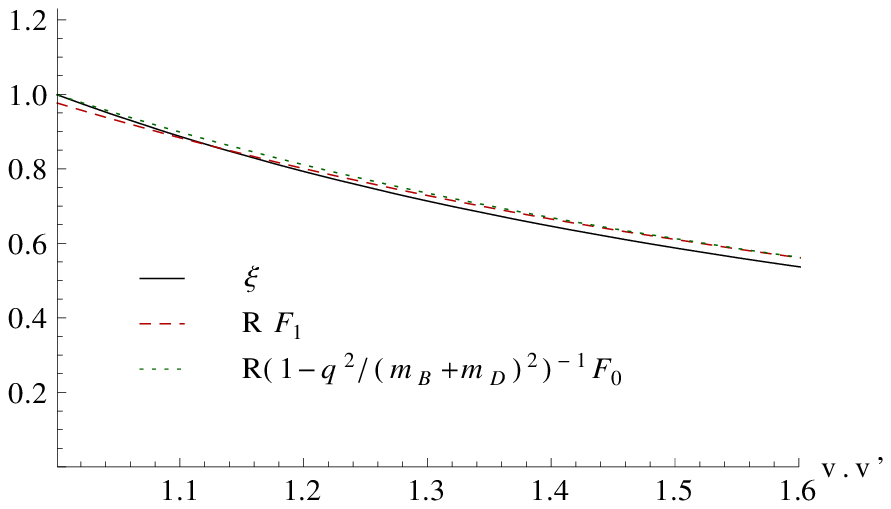}\\
\caption{Top: weak $B^- \rightarrow D^0$
decay form factors  (multiplied by appropriate kinematical
factors, cf.~(\ref{eq:f0xi}) - (\ref{eq:R})) for physical
heavy-quark masses in comparison with the Isgur-Wise function. 
 Experimental data have been taken from 
Belle~\cite{Abe:2001yf} (dots), CLEO~\cite{Bartelt:1998dq}
(triangles) and BABAR~\cite{Aubert:2009ac} (crosses) assuming that
$|V_{cb}|=0.0409$, i.e. the central value given by the Particle Data
Group~\cite{Beringer:2012}. Model parameters are taken from Table~\ref{ParametersHeavyLight}. 
Bottom: $c$ and $b$-quark masses are multiplied by a factor
$6.25$ such that $m_c=10$~GeV and meson masses are taken equal to
the corresponding quark masses.} \label{fig:decpsps}
\end{figure}

\vspace{-0.1cm}
In order to test heavy-quark flavor symmetry numerically and to see
how its breaking takes place,
we calculate the decay form factors for finite masses of the heavy quarks
(cf. Table~\ref{ParametersHeavyLight}).
Starting from the physical values for the quark masses,
it can be tested numerically how heavy-quark symmetry arises by
scaling up $m_{b}$ and $m_{c}$. Fig.~\ref{fig:decpsps} 
shows the 
$F_1$ and $F_0$ form factors that can be measured in the 
$B^-\to D^0 e^-\bar\nu_e$ decay,
 for physical 
values of $m_b$ and $m_c$ and values that are about 6 times larger. 
The form factors are multiplied by the 
corresponding kinematical factors that relate
them to the Isgur-Wise function (cf.~(\ref{eq:f0xi})-(\ref{eq:R})). 
The deviation from the Isgur-Wise function indicates the amount of heavy-quark 
symmetry breaking.
As it was shown analytically, heavy-quark symmetry predicts that 
the quantities $RF_1$ and $R(1-q^2/(m_B+m_D)^2)^{-1}F_0$ agree in the heavy-quark limit and they go 
over into the 
Isgur-Wise function. Taking the  
physical masses of the heavy quarks the differences between the resulting 
quantities turn out to be less than 7\% of their 
absolute values and they become
smaller with increasing $v\cdot v'$ (cf. Fig \ref{fig:decpsps} (top)).
the deviation from the Isgur-wise function, however, is about 15\%.

Taking $b$- and $c$-masses 6.25 times larger than the physical masses (which means
$m_c=10$ GeV), $RF_1$ and $R(1-q^2/(m_B+m_D)^2)^{-1}F_0$ are already much closer 
to the Isgur-Wise function (cf. Fig \ref{fig:decpsps}, bottom). The discrepancy between
 $RF_1$, $R(1-q^2/(m_B+m_D)^2)^{-1}F_0$ and $\xi(v\cdot v')$ is in this case
 less than 10\%. 

\vspace{-0.3cm}
\subsubsection{Heavy-quark spin symmetry breaking}
\begin{figure}[h!]
\begin{center}
 \includegraphics[width=0.8\textwidth]{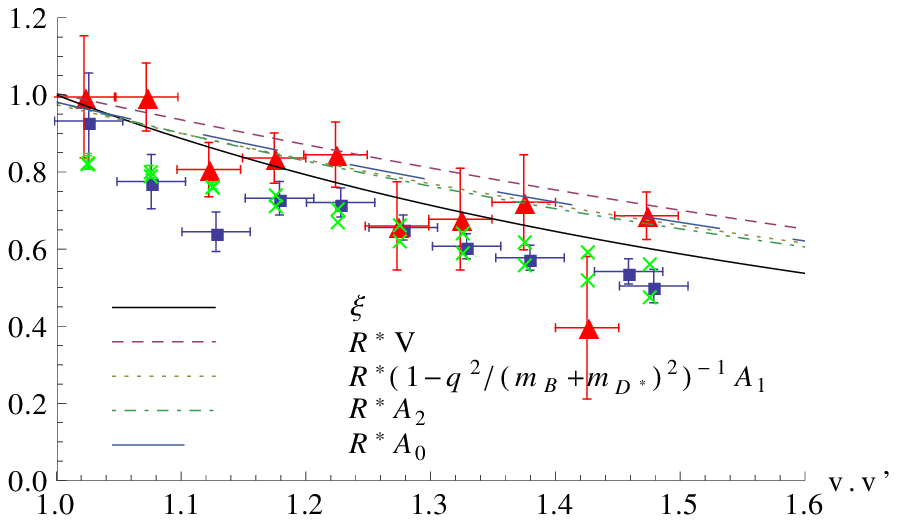}\\
\includegraphics[width=0.8\textwidth]{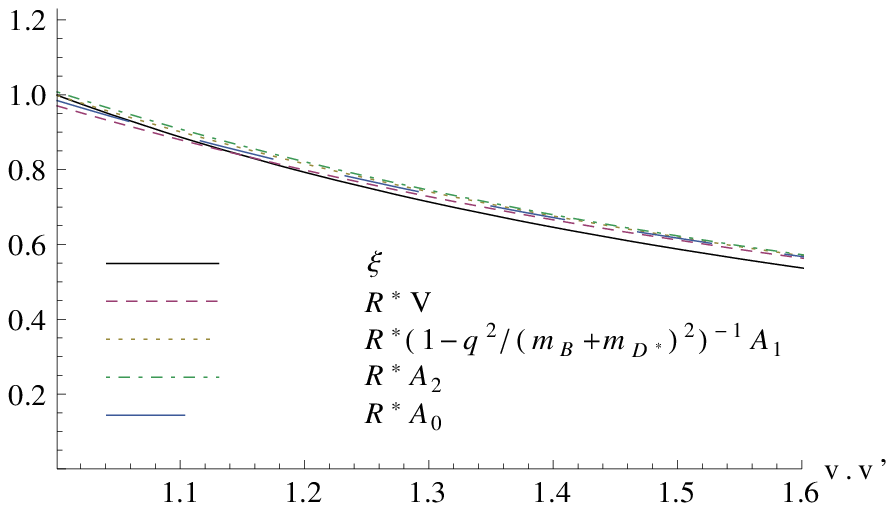}
\caption{Top: weak $B^- \rightarrow D^{0\ast}$ decay form factors  
(multiplied by appropriate
kinematical factors, cf.~(\ref{eq:a1xi}) - (\ref{eq:Rs})) for
physical heavy-quark masses in comparison with the Isgur-Wise
function. Model parameters are taken from 
Table~\ref{ParametersHeavyLight}. Experimental data have been taken from 
Belle~\cite{Abe:2001cs} (dots), CLEO~\cite{Adam:2002uw}
(triangles) and BABAR~\cite{Aubert:2004bw} (crosses) assuming that
$|V_{cb}|=0.0409$, i.e. the central value given by the Particle Data
Group~\cite{Beringer:2012}. 
In the lower figure $c$ and $b$-quark masses
are multiplied by a factor $6.25$ such that $m_c=10$~GeV and meson
masses are taken equal to the corresponding quark masses.}
\label{fig:decpsv}
\end{center}
\end{figure}

\vspace{-0.1cm}
Analogously to heavy-quark flavor symmetry, heavy-quark spin symmetry can be also
 tested numerically.  Fig.~\ref{fig:decpsv} shows how the form factors, multiplied by
appropriate kinematical factors given by the relations (\ref{eq:a1xi})-(\ref{eq:Rs}) approach 
the Isgur-Wise function by increasing the heavy-quark mass. 

The prediction is that $R^*V$, $R^*A_0$, $R^*A_2$ and
$R^*(1-q^/(m_B+m_{D^*})^2)^{-1}A_1$ coincide with the Isgur-Wise function 
in the heavy-quark limit. These quantities are shown in 
Fig.~\ref{fig:decpsv} (top) for  physical masses (cf. Table~\ref{ParametersHeavyLight}). 
The maximum 
difference between them are about 5\% and they differ from the Isgur-Wise
function by about 20\%. In order to see numerically the restoration 
of heavy-quark symmetry for large masses the form factors (multiplied by the 
corresponding kinematical factors) are plotted in Fig.~\ref{fig:decpsv} (down)
for $b$- and $c$- masses of about one order of magnitude larger, so that $m_c=10$ GeV.
\newpage
\newpage

\newpage

\subsubsection{Other observables}

The quantity that can be directly extracted experimentally from the (unpolarized) 
semileptonic decay rate,
$d\Gamma_{B\to De\bar\nu}/d\omega\propto (\omega^2-1)^{3/2}|V_{cb}|^2|F_D(\omega)|^2$, 
is  
$V_{cb}F_D(\omega):=V_{cb}RF_1(q^2(\omega))$, with $\omega:=v\cdot v'$. 
Results measured 
in Refs.~\cite{Abe:2001yf,Bartelt:1998dq,Aubert:2009ac} are plotted in
Fig.~\ref{fig:decpsps} divided by the current value of $V_{cb}$.
They can be compared with this model predictions (dashed line).
The simple harmonic-oscillator model reproduces the experiments reasonably well 
and they are comparable with other constituent quark models \cite{Choi:1999nu,Ebert:2006nz}.
The resulting branching ratio is in good agreement with experiment:
\begin{eqnarray}
 BR(B^0\to D^+l^-\nu_l)&=&2.3\% , \\
 BR^{\text{exp}}(B^0\to D^+l^- \nu_l)&=&(2.18\pm0.12)\% .
\end{eqnarray}

Other quantities that characterize the process are the value of  
$F_D$ at zero recoil, i.e. $F_D(\omega=1)=RF_1(\omega=1)$ and the slope at zero recoil
$\rho^2_D:=-F'_D(\omega=1)/F_D(\omega=1)$. The results obtained within the present
model are
\begin{eqnarray}
 F_D(\omega=1)&=&0.93, \\
 \rho^2_D&=&0.59.
\end{eqnarray}
It is interesting to look at these values in the heavy-quark limit.  
$F_D(\omega)$ goes over into $\xi(\omega)$ in the h.q.l., so its value at
zero recoil is obviously 1. The slope becomes then 
$\rho^2:=-\xi'_D(\omega=1)=1.24$.
The result differs considerably from the 
corresponding one for physical masses of the heavy quarks. The 
experimental value, as quoted by the Heavy Flavor Averaging Group~\cite{Asner:2010qj}
is
\begin{equation} 
\rho^2_D=1.18\pm 0.06,
\end{equation}
which is closer to the result of the model in the h.q.l.

\subsubsection{Remarks}
For the discussion presented above we have used the same harmonic-oscillator
wave function with oscillator parameter $a=0.55$ GeV, for both the 
 $B^-$ and $D^{*}$ mesons. Any flavor dependence in the wave function has been 
ignored. It has been deliberately done in this way
 for the purpose of studying those effects that are exclusively 
consequences of heavy quark symmetry breaking due to taking the masses of 
the heavy quarks finite. 
Any kind of flavor dependence in the wave function, as it could be for example, a different 
oscillator parameter for each meson, would have influenced
the numerical results. 
For instance, taking $a_D=0.465$ GeV 
as it is suggested in a front-form analysis of heavy-meson decay 
constants~\cite{Hwang:2010hw}, 
while keeping $a_B=0.55$ GeV, the result for the slope at zero recoil would have been
$\rho^2_D=0.65$, which is about 10\% larger than the one obtained for 
$a_B=a_D=0.55$ GeV.

\subsection{Cluster properties}

Let us discuss now the spurious dependencies that appear in the 
form factors due to wrong cluster properties inherent 
in the Bakamjian-Thomas construction. The Mandelstam-$s$ dependence
does not spoil the Poincar\'e invariance of the 1-photon-exchange amplitude, 
it is rather a consequence of the non-local character of the photon-meson 
vertex. If one considered the $\gamma^*M\to M$ subprocess only, the 
$s$-dependence could be interpreted as a frame dependence in the description 
of the subprocess.
We have seen that there are two special reference frames (for $\gamma^*M\to M$), 
corresponding to
two particular values of $s$. Namely,
the  minimum $s$ to reach a particular momentum transfer $\text{Q}^2$ and 
$s\to\infty$ (with $\text{Q}^2$ fixed). 
They have been called the 
Breit frame and the infinite-momentum frame, respectively. 
We have shown (cf. Sec.~\ref{sec:BreitInf}) that in both cases the covariant structure of the 
current can be expressed in terms of physical meson covariants only.
In Fig.~\ref{fig:scattpsps} the electromagnetic form factors
for the $D^+$ and $B^-$ mesons computed in these two particular reference frames
are shown.
The difference between the Breit-frame result and the infinite-momentum-frame result is determined by the ratio
$m_Q^2/\text{Q}^2$. In both frames the form factors at $\text{Q}^2=0$ are normalized to 1. Therefore the Breit-frame result tends to 
approach the infinite-momentum-frame result faster for the light $D^+$-meson than for the $B^-$-meson.

\section{Comparison with front-form results}\label{SecComparisonff}

There are several calculations that have been done in a front-form 
approach to which the analytical and numerical results presented here can be compared. 
Ref.~\cite{Cheng:1996if}, where the model parameters have been taken
from, provides two different analytical expressions
corresponding to two different wave function models, namely a Gaussian-type one
and the flavor dependent 
Wirbel-Stech-Bauer wave function~\cite{Wirbel:1985ji}. The Isgur-Wise 
functions in~\cite{Cheng:1996if} are
obtained by taking the heavy-quark limit of the form factors of
the $B\to D$ and $B\to D^*$ decays. The authors obtain the same numerical
result for the Isgur-Wise function in the $B\to D$ and $B\to D^*$ processes 
when using a Gaussian wave function. 
However, their result 
turns out to be different when using  a flavor dependent 
Wirbel-Stech-Bauer wave function~\cite{Wirbel:1985ji}. Their conclusion is thus that 
the Wirbel-Stech-Bauer wave function does not preserve heavy-quark symmetry. 
In this work we have used the Gaussian wave function type and the 
numerical result for the Isgur-Wise function turns out to be identical to
the one of Ref.~\cite{Cheng:1996if}. The value of the slope at
zero recoil is also found to be the same, $\rho^2= -\xi'(1)=1.24$.

Unlike in the case of systems of equal constituent masses 
\cite{Biernat:2009my,Biernat:2011mp}, where the equivalence between
point- and front-form results was already 
shown analytically
 for electromagnetic form factors, 
such an equivalence is not easy to establish analytically for heavy-light systems 
in the heavy-quark limit.
There are, however, several hints suggesting that the equivalence may hold also
in this case. 
In the case of the pion, the analytical result for 
 the form factor for space-like momentum transfers 
was shown to be identical to the front-form expression that results from 
the +-component of a one-body current in the 
$q^+=0$ frame~\cite{Biernat:2009my}. 
If such an equivalence extended to the case of unequal masses 
and generalized  at least to those electroweak transition form factors 
that are not affected by $Z$-graphs contributions, the heavy-quark limit of electroweak
heavy-light meson (transition) form factors in the point form should lead to the 
same Isgur-Wise function as the front-form approach. We see this numerically, 
but we have not attempted to establish the relation analytically so far. 

Derivations of the Isgur-Wise function carried out within a front-form approach in,
e.g.~\cite{Cheng:1996if,Cheng:1997au}, used as a starting point the weak transition form factor, which
is extracted in the time-like momentum transfer region. In order to do calculations
that involve time-like momentum transfers it is necessary to give up the $q^+=0$
condition and consequently the absence of $Z$-graphs is not guaranteed any more. 
This is confirmed by an analysis of the triangle diagram for $B\to D^{(*)}$ decays within 
a simple covariant model~\cite{Bakker:2003up}.
In Ref.~\cite{Bakker:2003up} it is demonstrated that computing the 
weak form factors for $\bar \nu_e B\to eD^{(*)}$ scattering in a $q^+=0$ frame and
 applying an analytical continuation ($q_\perp \to i q_\perp$) in order to go to
the time-like momentum transfer region is equivalent to computing the decay form factors
in the time-like momentum transfer region (where $q^+\neq 0$), provided that 
the $Z$-graph contributions
are appropriately taken into account. The importance of the $Z$-graph contributions 
decreases when increasing the mass of the heavy quark and vanishes completely in 
the heavy-quark limit. This is due to the impossibility of creating an
infinitely heavy quark-antiquark pair out of the vacuum.

\begin{figure}[h]
\begin{center}
 \includegraphics[width=0.8\textwidth]{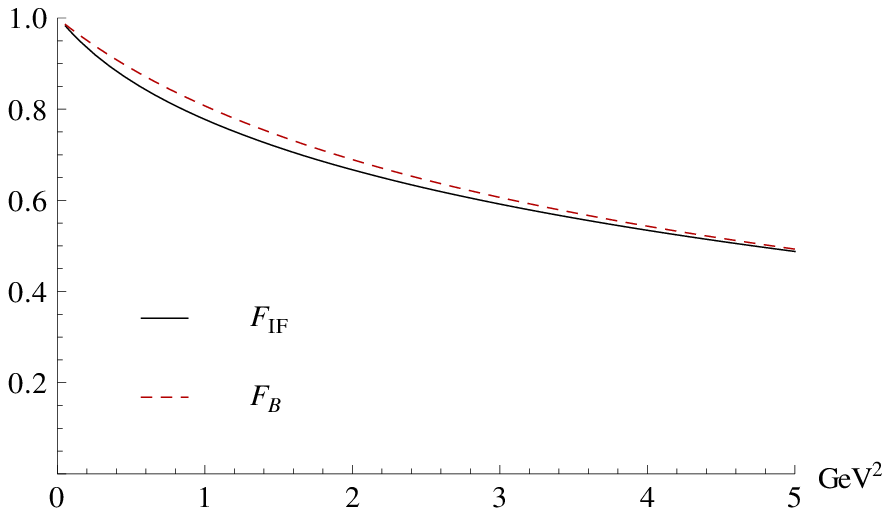}\\
\includegraphics[width=0.8\textwidth]{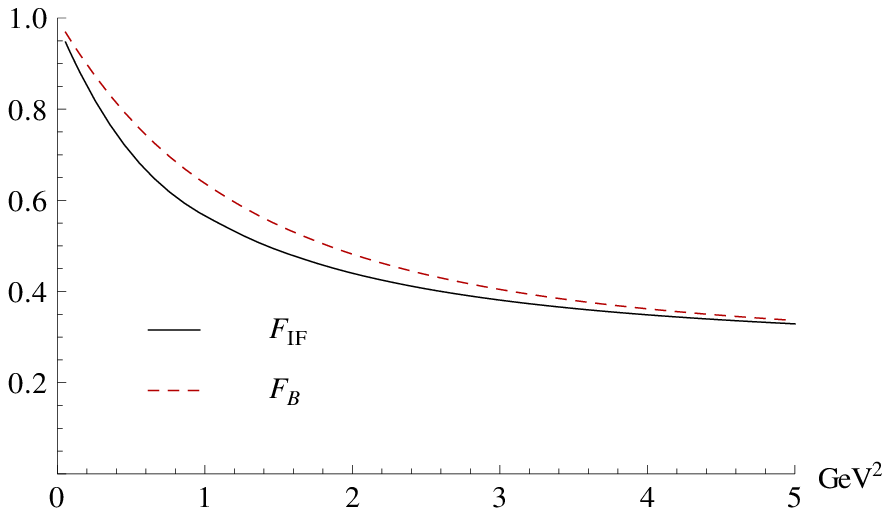}
\caption{Electromagnetic form factor of the $D^+$ (top) and $B^-$ (bottom) 
mesons calculated
in the Breit (B) and infinite-momentum (IF)
frames as a function of the (space-like) 4-momentum transfer squared $\text{Q}^2=-q^2$. 
Model parameters are taken from Table~\ref{ParametersHeavyLight}.}
\label{BreitInfinityQ2}
\end{center}
\end{figure}

For finite masses we obtain results for the 
weak $M\to M'$ decay form factors which differ from those obtained within 
the front-form approach.
Fig.~\ref{BreitInfinityQ2} may also provide a hint at the importance of 
the $Z$-graph contributions 
for finite masses when computing 
in $q^+\neq 0$ frames. Part of the discrepancies between the form factors 
extracted in the infinite-momentum frame and in the Breit frame may be 
attributed to missing $Z$-graph contributions in the Breit frame~\cite{Simula:2002vm}.

\newpage

 \newpage
\thispagestyle{empty}

%% file: Ch7.tex
\chapter{Numerical studies II}\label{ChNumStudiesII}

\section{Introduction}
In the previous chapter the numerical calculations where focused on the study of 
 heavy-quark symmetry 
 as well as on cluster-separability properties of our point-form approach. 
For that purpose the same 
harmonic-oscillator wave function with parameter 
$a=0.55$ GeV was used for all numerical studies. In this chapter 
we introduce a flavor dependence in the wave function, by assuming a different harmonic
oscillator parameter for each meson. We will
extract the form factors for several semileptonic decays, both
for heavy-to-heavy and
 for heavy-to-light transitions, as a function of 
the 4-momentum transfer squared, i.e. $F_0(q^2)$ and $F_1(q^2)$ 
in the pseudoscalar-to-pseudoscalar case,
 and $V(q^2)$, $A_0(q^2)$, $A_1(q^2)$ and $A_2(q^2)$ 
in the pseudoscalar-to-vector 
case. 

Even taking into account this flavor dependence, 
the model remains very simple. It is certainly not sophisticated enough
to establish quantitative predictions which could be compared with experiments. 
Nonetheless, it is worth carrying out such calculations for
several decays in order to see how our approach
compares with other approaches and to learn at least qualitatively
how the transition form factors depend on the kind of transition considered.
We are particularly interested in comparisons with front-form results
and the role of non-valence contributions in the description of currents and 
form factors. In front-form such non-valence contributions turn out 
to become important when one goes from 
space-like to time-like momentum transfers and they may play a role in our approach as well. 
As mentioned already in Sec.~\ref{SecComparisonff}, for time-like momentum transfer it is
not possible to use the $q^+=0$ frame in front form. As a consequence non-valence
configurations leading to $Z$-graph contributions (quark-antiquark pairs created 
from the vacuum) can occur. Such $Z$-graph contributions have been 
analyzed in Ref.~\cite{Bakker:2003up}. 
Applying analytic continuation ($q_\perp\to iq_\perp$) from the space-like to 
the time-like momentum transfer region to transition form factors calculated 
in a $q^+=0$ frame for space-like momentum transfers it is shown that the outcome is 
the same as the result from a direct
calculation of the decay form factors in the time-like region 
(where $q^+\neq 0$), provided that the 
 $Z$-graph contributions are appropriately taken into account. 
The importance of the $Z$-graph contributions decreases with increasing the mass 
of the heavy quark
and it vanishes in the heavy-quark limit, since an infinitely heavy quark-antiquark pair cannot 
be produced out of the vacuum.  
Our numerical values for the Isgur-Wise function agree with those obtained within a front form quark 
model~\cite{Cheng:1996if}. As soon as the decay form factors are calculated for finite physical masses
of the heavy quarks, differences between the point- and front-form approach appear. 
These differences may be attributed to the different roles played by $Z$-graphs in either approach.

The major aim of this chapter is to provide the numerical results 
for time-like form factors using identical parameters and wave functions as
in Ref.~\cite{Cheng:1996if} (they are quoted in
Table~\ref{ParametersFlavorDependent}) and perform a numerical
comparison of our point-form approach with the front-form one. 

A second issue we would like to address with these comparisons concerns the 
frame dependence that appears in the calculation of form factors 
of pseudoscalar-to-vector transitions
in the front-form approach. 
In the light-front quark model of Ref.~\cite{Cheng:1996if}, for instance,
the authors choose a frame in which the momentum transfer is 
purely longitudinal, i.e. $q_\perp=0$, $q^2=q^+q^-$.
Working in this way, form factors of processes that involve vector
mesons, cannot be extracted unambiguously, 
and the form factors exhibit a dependence on whether the daughter meson goes in the positive
 or negative
$z$-direction. 
We, on the other hand, showed in Chap.~\ref{ch:currents:and:ff} 
that in our case 
there is no frame dependence of the form factors and
they can be determined without any ambiguity 
from the different components
of the current.  
This will be discussed Sec.~\ref{SecPtoV}.

\begin{table}[h!]\label{ParametersFlavorDependent}
\begin{center}
    \begin{tabular}{cccccccc}
     $a_\pi$  &  $a_\rho$  & $a_K$  & $a_{K^*}$  & $a_D$ & $a_{D^*}$  & $a_B$  & $a_{B^*}$  \\     \hline \hline
	0.33  & 0.30 & 0.38 & 0.31  & 0.46  & 0.47  & 0.55  & 0.55 \\ \hline
    \end{tabular}\caption{Harmonic-oscillator parameters (in GeV) for the meson wave functions 
used for the calculation of transition form factor in this chapter. They have been taken from  
Ref.~\cite{Cheng:1996if} where they were determined by fitting the wave functions to 
the experimental values for the decay constants.}
\end{center}
\end{table}

\section{Pseudoscalar-to-pseudoscalar transitions}  
For pseudoscalar-to-pseudoscalar transitions, in order to allow for comparison with other 
work, besides $F_0(q^2)(=f_+(q^2))$ 
and $F_1(q^2)$, also $f_-(q^2)$ 
is depicted for all computed decays. $f_-(q^2)$ and $f_+(q^2)$ are defined by 
\begin{equation}
 J^\mu(p_1,p_2) = f_+(q^2) (p_1+p_2)^\mu +f_-(q^2) (p_1-p_2)^\mu,
\end{equation}
where $p_1$ and $p_2$ are the initial and final meson 4-momenta. 
Their relation with $F_0(q^2)$  and $F_1(q^2)$ is given by (cf. Sec.~\ref{SubsecPtoP}):
\begin{equation}
 F_1(q^2)=f_+(q^2), \qquad F_0(q^2)=f_+(q^2)+\frac{q^2}{M_1^2-M_2^2}f_-(q^2).
\end{equation} 
The values at $q^2=0$ for $F_1(0)$, or equivalently
for $f_+(0)$, are shown in Table~\ref{fplus} together with the results obtained within 
the light-front quark model~\cite{Cheng:1996if}. For heavy-to-heavy 
transitions, i.e. $B\to D$, as well as for $B\to\pi(K)$ transitions both, front-form and point-form
results, seem to agree quite well, whereas 
they differ slightly for $D\to\pi(K)$. 

We do not have a definitive explanation for this fact,
but we suspect that these differences 
are due to the different way in which $Z$-graphs 
enter the form factors 
in either approach. There is a particular
frame, namely the $q^+=0$ frame, in the front form, 
where $Z$-graphs disappear. In point form a particular $q^+=0$ frame
can be realized for lepton-hadron scattering by taking the 
limit of infinite large Mandelstam $s$, which corresponds to the 
infinite-momentum frame of the hadron. This explains, e.g., the equality of our point-form
results for electromagnetic meson form factors (for $q^2<0$) with corresponding front-form results.
In the $q^+=0$ frame however, weak decays cannot take place, 
since the process is necessarily time-like ($q^2=q^+q^- -q_\perp > 0$) or light-like
at the point for maximal recoil ($q^2=0$). In
the light-front quark model of Ref.~\cite{Cheng:1996if}, 
the calculations are done in a 
frame where the momentum transfer is purely
longitudinal, this is $q_\perp=0$, $q^2=q^+q^-$. 
At $q^2=0$ either $q^+$ or $q^-$ must vanish which corresponds to the 
daughter meson going either in + or in $-$ $z$-direction, respectively.
Since the pseudoscalar decay form factors do not depend on whether the daughter
meson goes into + or $-$ $z$-direction one can assume $q^+=0$. 
This implies, however, that $Z$-contributions
vanish at the maximum recoil point. 
For $q^2>0$ there is, however, no argument to exclude $Z$-graph
contributions in the decay form factors. 
In point form one does  not even have an argument at $q^+=0$
(apart of the mass of the produced $Q\bar Q$-pair) that $Z$-graphs should vanish. 

As in Ref.~\cite{Cheng:1996if}, a quantitative estimate of the $Z$-graph contribution
is not within the scope of this work. We have seen, however, in the previous section
that the point-form results reproduce the front-form ones exactly in the heavy-quark limit. 
The reason is that in these $Z$-graph contributions vanish, 
 since it is
not possible to create a infinitely heavy $Q\bar Q$-pair out of the vacuum. One can therefore expect 
that for heavy-to-heavy transitions point-form and front-form results show a greater resemblance than for heavy-to-light
transitions. For heavy-to-light
processes non-valence contributions are expected to be more important.
It is thus not surprising that
the results differ in both approaches.  In the $D\to K$ and
$D\to \pi$ cases, point- and front-form results differ considerably, the front-form results being somewhat
closer to the experimental data~\cite{Beringer:2012}. 

\begin{figure}[h!]
\centering
\begin{center}
  \includegraphics[width=0.8\textwidth]{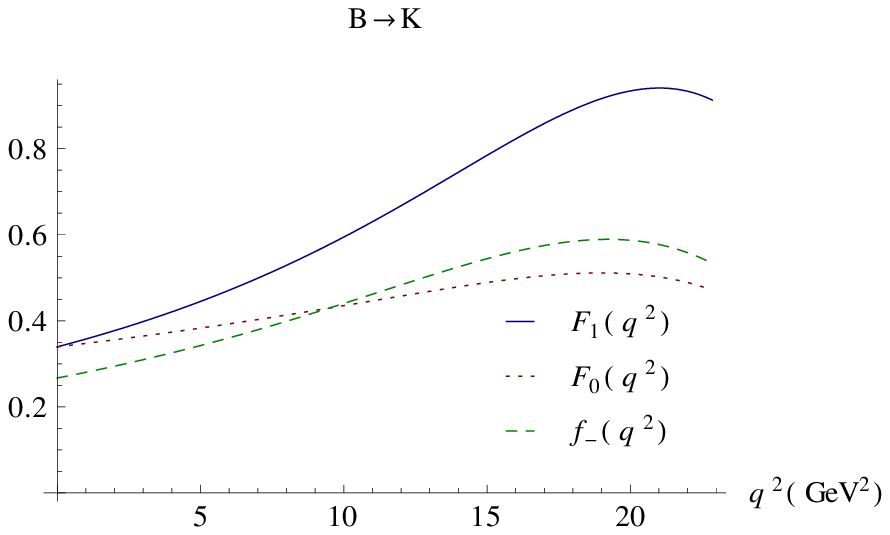}
  \includegraphics[width=0.8\textwidth]{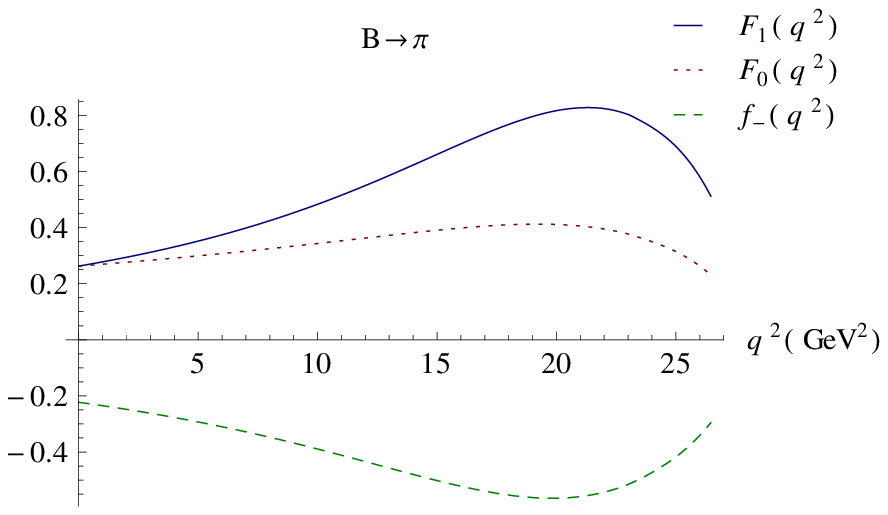}
\caption{$B\to K$ and $B\to \pi$ transition form factors in the whole range
 $0\leq q^2\leq(M_B-M_{K(\pi)})^2$. Parameters for the quark masses and harmonic-oscillator
wave functions are taken from 
Tables~\ref{ParametersHeavyLight} and ~\ref{ParametersFlavorDependent}, respectively. 
For the meson masses the current values given by the Particle Data Group are 
taken~\cite{Beringer:2012}.}\label{fig:BtoPiK}
\end{center}
  \end{figure}

\begin{figure}[h!]
\begin{center}
  \includegraphics[width=0.8\textwidth]{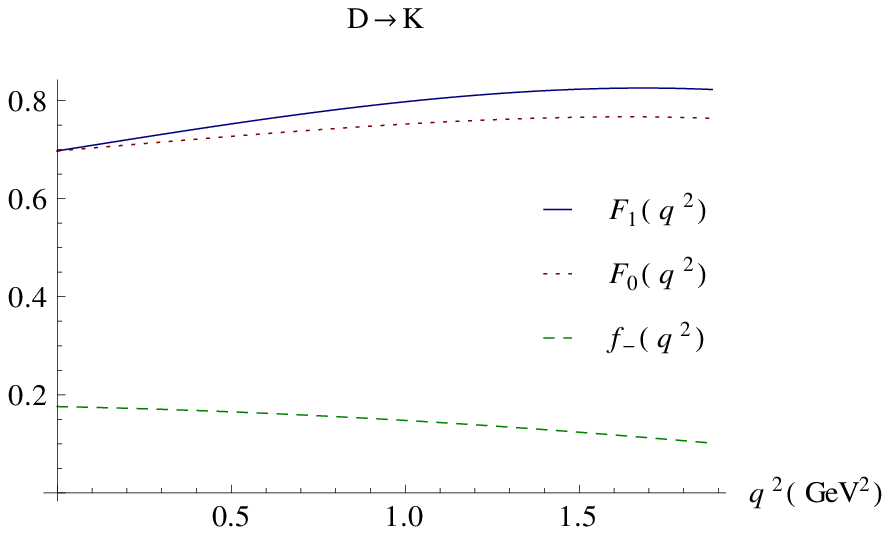}
  \includegraphics[width=0.8\textwidth]{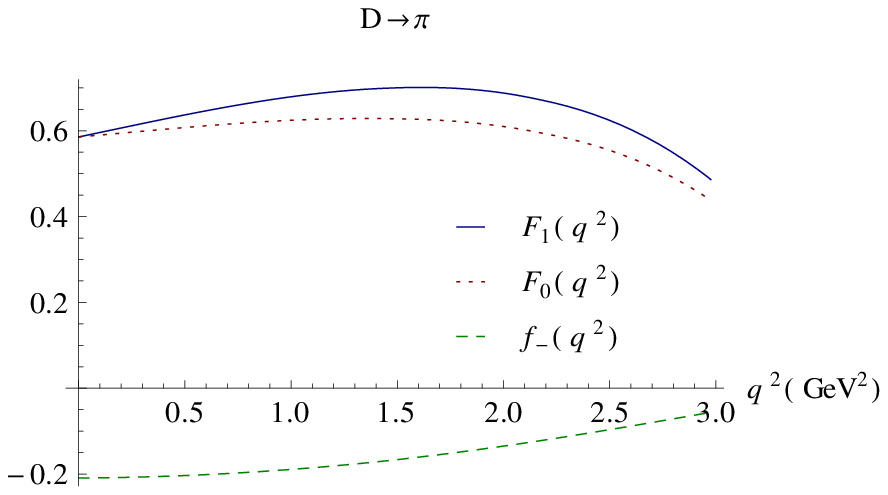}
\caption{$D\to K$ and $D\to \pi$ transition form factors in the whole range
 $0\leq q^2\leq(M_D-M_{K(\pi)})^2$. 
Parameters for the quark masses and harmonic-oscillator
wave functions are taken from 
Tables~\ref{ParametersHeavyLight} and ~\ref{ParametersFlavorDependent}, respectively. 
For the meson masses the current values given by the Particle Data Group are 
taken~\cite{Beringer:2012}.}\label{fig:DtoPiK}
\end{center}
  \end{figure}

Another resemblance with the front-form results is that $f_-(q^2)\sim - f_+(q^2)$ for $B\to \pi$ 
 and to less extend for $D\to \pi$ (cf. Figs.~\ref{fig:BtoPiK} and~\ref{fig:DtoPiK}).
Near zero recoil (where $q^2$ is maximum) heavy-quark
symmetry predicts $(f_+ +f_-)^{B(D)\pi}\sim \frac{1}{\sqrt{m_{B(D)}}}$. In our case we rather have
\begin{equation}
(f_+ +f_-)^{B\pi}_{q_{\text{max}}}\sim 0.22,\qquad (f_+ +f_-)^{D\pi}_{q_{\text{max}}}\sim 0.43,
\end{equation}
whereas $1/\sqrt{m_B}\sim 0.43$ and $1/\sqrt{m_D}\sim 0.73$. Like in front form these kind of 
relations are badly violated for $B(D)\to K$.
\begin{table}[h!]\label{fplus}
\begin{center}
    \begin{tabular}{c|ccc}
		   &    Front form~\cite{Cheng:1996if} &  Point form (this work)  & Experiment~\cite{Beringer:2012}  \\            \hline \hline
        $B\to D$   &  0.70      &  0.68  &  \\
	$B\to\pi$  &  0.26      &  0.26	 &  \\
	$B\to K$   &  0.34      &  0.34  &  \\
	$D\to\pi$  &  0.64      &  0.58  & 0.661$\pm$0.022 \\
	$D\to K$   &  0.75      &  0.70  & 0.727$\pm$0.011 \\
	 \hline
    \end{tabular}\caption{$F_1(0)$ (or equivalently $f_+(0)$) form factor for pseudoscalar-to-pseudoscalar
transitions, corresponding to Figs.~\ref{fig:BtoD} - \ref{fig:DtoPiK}.}
\end{center}
\end{table}

\vspace{-1cm}
\begin{figure}[h!]
\begin{center}
  \includegraphics[width=0.8\textwidth]{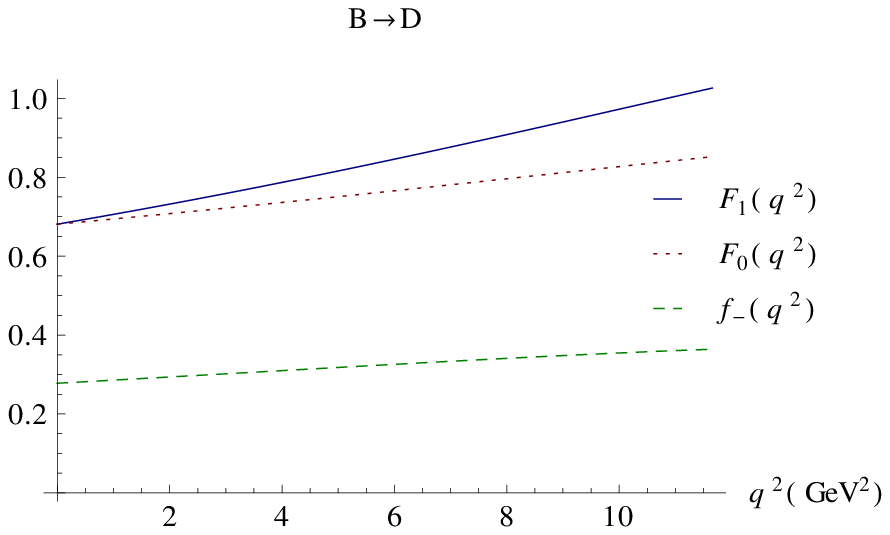}
\caption{$B\to D$ transition form factors in the whole range $0\leq q^2\leq(M_B-M_D)^2$. Parameters for the quark masses and harmonic-oscillator
wave functions are taken from 
Tables~\ref{ParametersHeavyLight} and~\ref{ParametersFlavorDependent}, respectively. 
For the meson masses the current values given by the Particle Data Group have been  
taken~\cite{Beringer:2012}.}\label{fig:BtoD}
\end{center}
  \end{figure}

\newpage

\newpage

\newpage
\section{Pseudoscalar-to-vector transitions}\label{SecPtoV}
More interesting is the comparison
for transitions that involve mesons with spin. 
In the light-front quark model~\cite{Cheng:1996if}, 
the form factors for pseudoscalar-to-vector meson transitions  extracted in the $q_\perp=0$ frame,
exhibit a certain
frame dependence. For a given $q^2$, the form factors depend on whether the recoiling daughter
moves in the positive ``+'' or negative ``$-$'' $z$-direction relative to the parent 
meson. In the light-front quark model the results for the 
form factors are larger in the ``+'' frame than in the ``$-$'' one. 
The exact vanishing of $Z$-graphs at $q^2=0$ in the ``+'' frame is taken as an argument in 
Ref.~\cite{Cheng:1996if} to conclude that $Z$-graphs are less important in the ``+'' frame
than in the ``$-$'' frame.
In Table~\ref{TableBtoDstar} results for both frames together with the point-form
results obtained in this work are given at $q^2=0$. 
The authors of~\cite{Cheng:1996if} interpret the difference between the results at $q^2=0$ 
in the ``+'' and ``$-$'' frames as a measure for the $Z$-graph contribution present in the ``$-$'' frame. 
In the point form, as explained in Chap.~\ref{ChCoupledChannel} 
and \ref{ch:currents:and:ff}, all form factors can be extracted without ambiguity 
and no frame dependence appears in our description of weak decays. 
The 
current level of this
work does not 
allow to give a serious estimate of $Z$-graph contributions. One could perhaps guess that they are of the 
same order of magnitude as the difference between ``+'' and ``$-$'' frames in front from.

In Tabs.~\ref{TableBtoDstar}-\ref{TableDtoRho} our from-factor results at $q^2=0$
are compared with those of Ref.~\cite{Cheng:1996if} for several decays.
One observes that the results obtained in the point form for $A_0(0)$, $A_1(0)$ and $A_2(0)$ are very similar 
in all the computed transitions, whereas they differ notably in the front form. There seems to be a good agreement
between both approaches for $V(0)$ and $A_0(0)$. For these two form factors one sees 
that for the heavy-to-heavy
transition the point-form result lies between the obtained ones in the front form in the ``+'' and
``$-$'' frames, being closer to the ``+'' one. $A_1(0)$ and $A_2(0)$ turn out to be larger in the point
form in all cases. 

\begin{figure}[h!]
\begin{center}
  \includegraphics[width=0.8\textwidth]{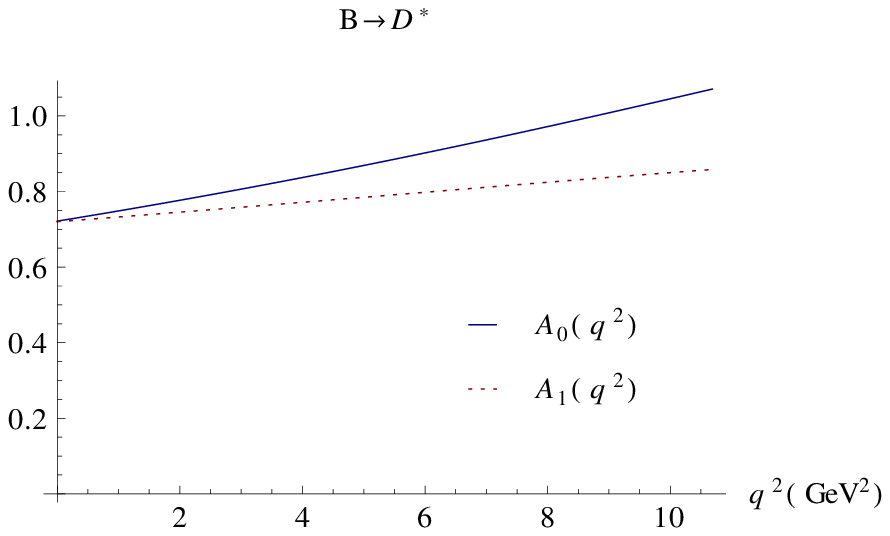}
  \includegraphics[width=0.8\textwidth]{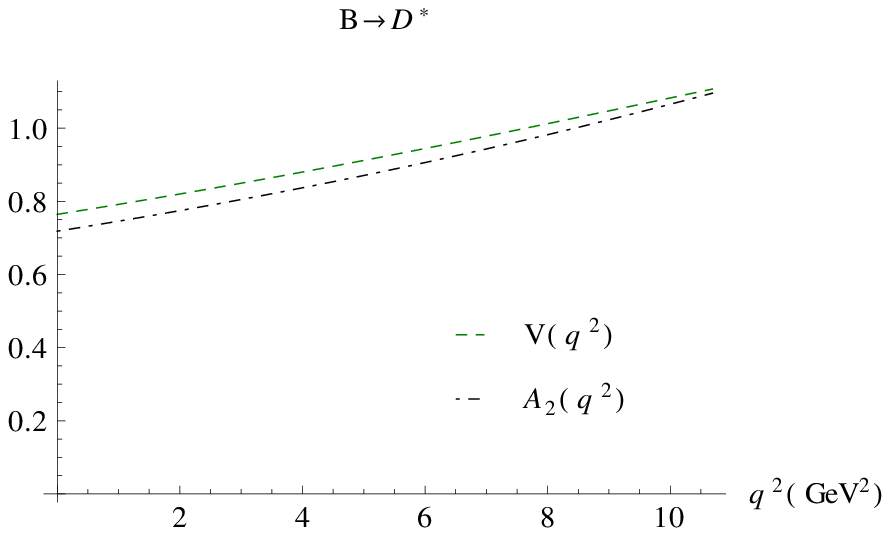}
\caption{$B\to D^*$ transition form factors in the whole range
 $0\leq q^2\leq(M_{B}-M_{D^*})^2$. 
Parameters for the quark masses and harmonic-oscillator
wave functions are taken from 
Tables~\ref{ParametersHeavyLight} and~\ref{ParametersFlavorDependent} respectively. 
For the meson masses the current values given by the Particle Data Group are 
taken~\cite{Beringer:2012}.}\label{FigBtoD}
\end{center}
  \end{figure}

\begin{figure}[h!]
\begin{center}
  \includegraphics[width=0.8\textwidth]{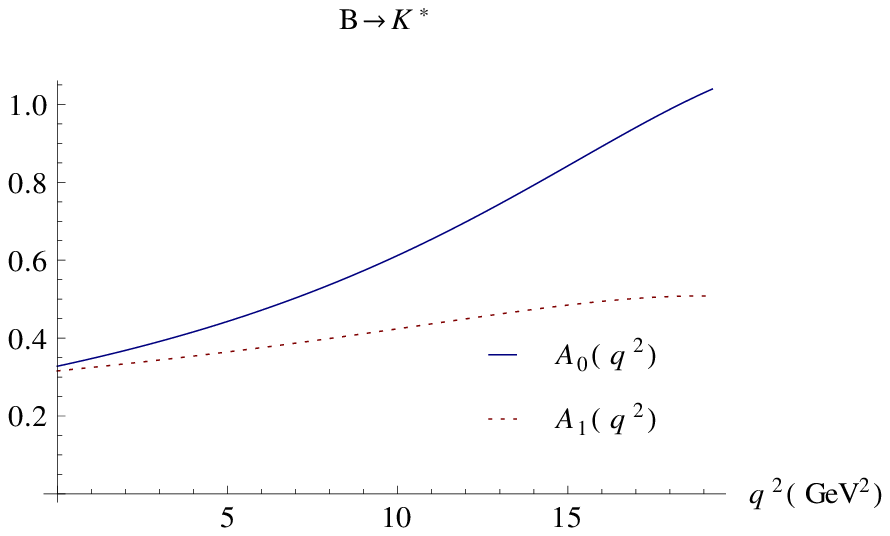}
  \includegraphics[width=0.8\textwidth]{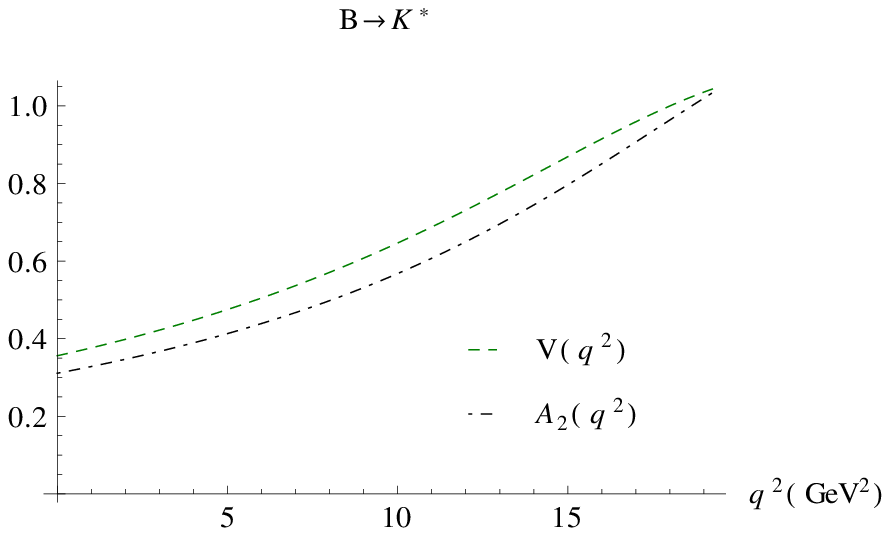}
\caption{$B\to K^*$ transition form factors in the whole range
 $0\leq q^2\leq(M_{B}-M_{K^*})^2$. 
Parameters for the quark masses and harmonic-oscillator
wave functions are taken from 
Tables~\ref{ParametersHeavyLight} and~\ref{ParametersFlavorDependent} respectively. 
For the meson masses the current values given by the Particle Data Group are 
taken~\cite{Beringer:2012}.}\label{FigBtoKstar}
\end{center}
  \end{figure}

\begin{figure}[h!]
\begin{center}
  \includegraphics[width=0.8\textwidth]{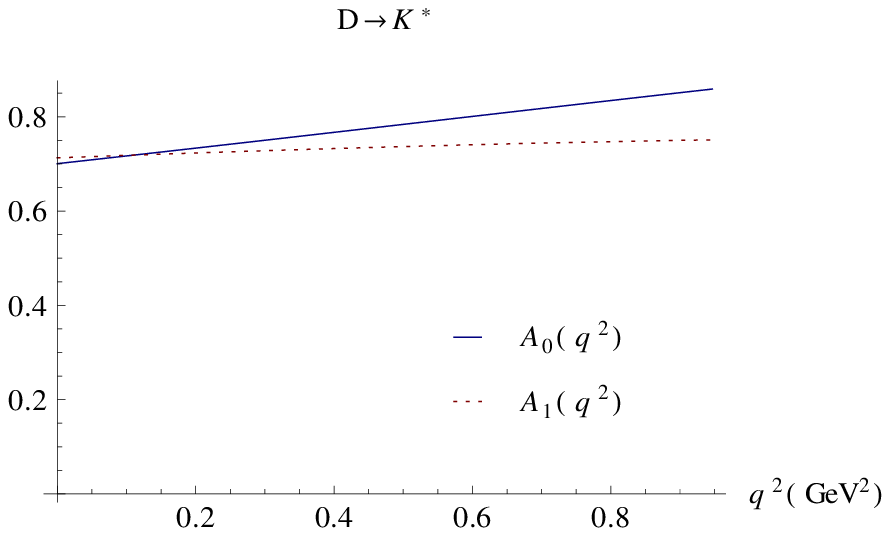}
  \includegraphics[width=0.8\textwidth]{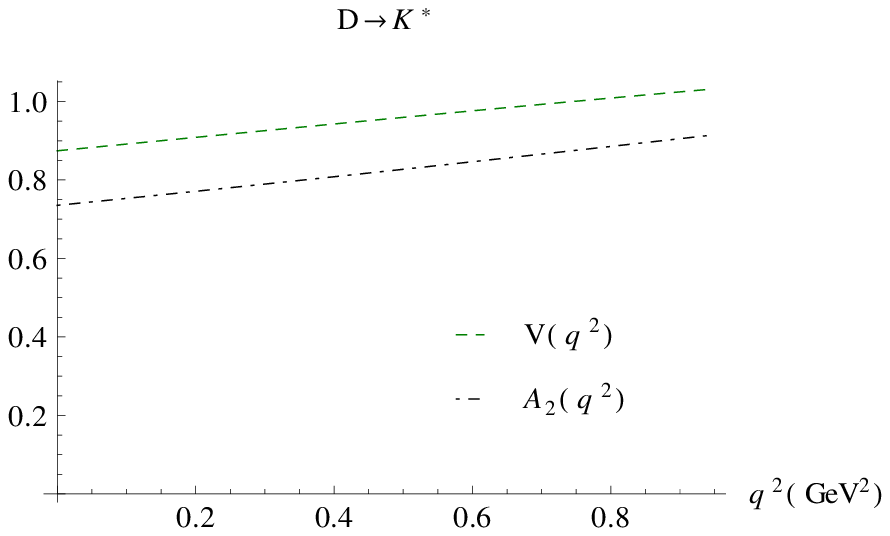}
\caption{$D\to K^*$ transition form factors in the whole range
 $0\leq q^2\leq(M_{D}-M_{K^*})^2$. 
Parameters for the quark masses and harmonic-oscillator
wave functions are taken from 
Tables~\ref{ParametersHeavyLight} and~\ref{ParametersFlavorDependent} respectively. 
For the meson masses the current values given by the Particle Data Group are 
taken~\cite{Beringer:2012}.}\label{FigDtoKstar}
\end{center}
  \end{figure}

\begin{figure}[h!]
\begin{center}
  \includegraphics[width=0.8\textwidth]{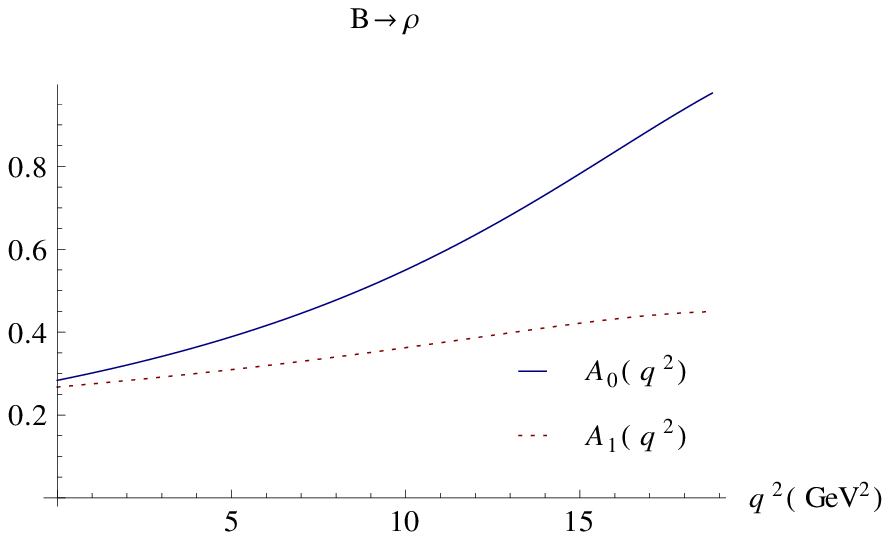}
  \includegraphics[width=0.8\textwidth]{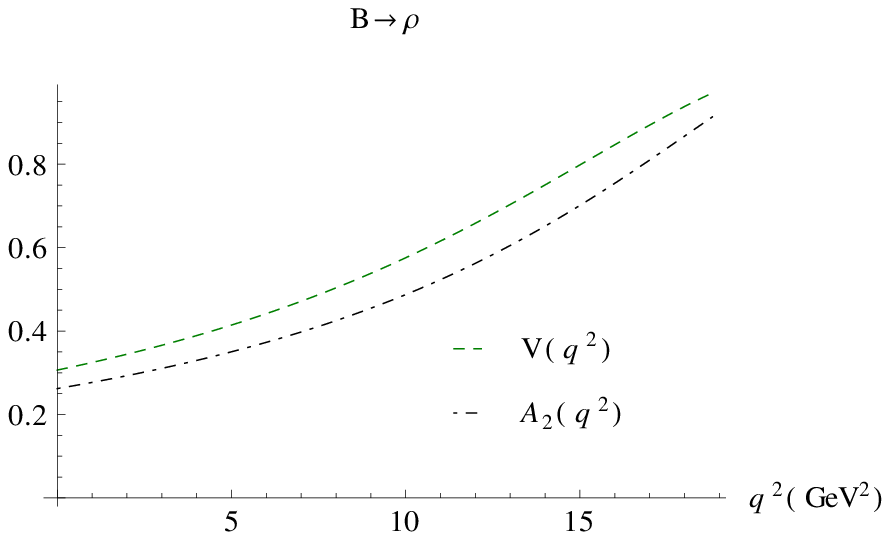}
\caption{$B\to \rho$ transition form factors in the whole range
 $0\leq q^2\leq(M_{B}-M_{\rho})^2$. 
Parameters for the quark masses and harmonic-oscillator
wave functions are taken from 
Tables~\ref{ParametersHeavyLight} and~\ref{ParametersFlavorDependent} respectively. 
For the meson masses the current values given by the Particle Data Group are 
taken~\cite{Beringer:2012}.}\label{FigBtorho}
\end{center}
  \end{figure}

\begin{figure}[h!]
\begin{center}
  \includegraphics[width=0.8\textwidth]{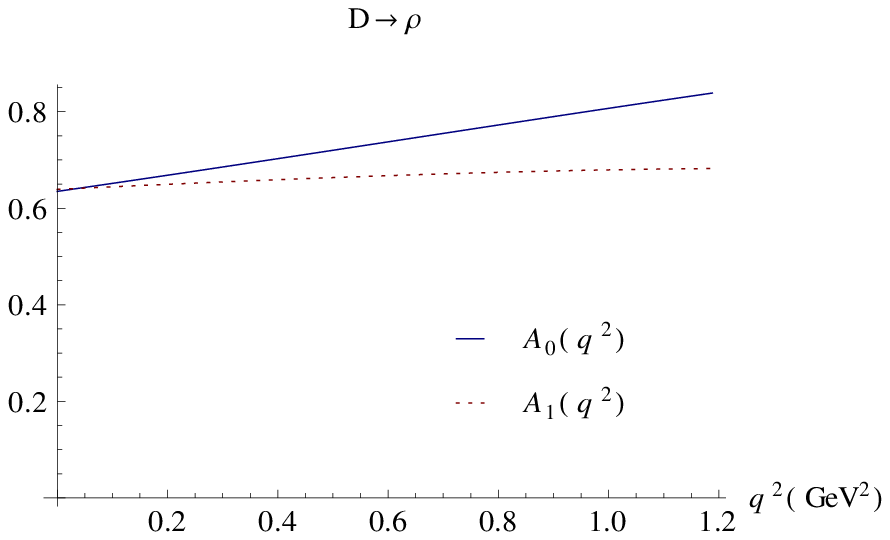}
  \includegraphics[width=0.8\textwidth]{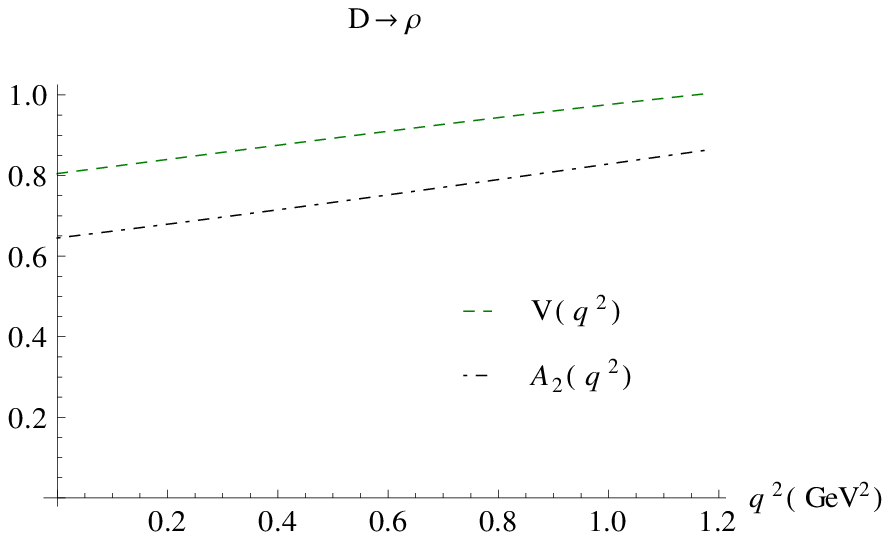}
\caption{$D\to \rho$ transition form factors in the whole range
 $0\leq q^2\leq(M_{B}-M_{\rho})^2$. 
Parameters for the quark masses and harmonic-oscillator
wave functions are taken from 
Tables~\ref{ParametersHeavyLight} and~\ref{ParametersFlavorDependent} respectively. 
For the meson masses the current values given by the Particle Data Group are 
taken~\cite{Beringer:2012}.}\label{FigDtorho}
\end{center}
  \end{figure}

For the whole $q^2$ range, i.e. $0\leq q^2\leq (M_1-M_2)^2$, the
form factors $V(q^2)$, $A_{0}(q^2)$, $A_1(q^2)$ and $A_2(q^2)$  are depicted in 
Figs.~\ref{FigBtoD}-\ref{FigDtorho}. If one compares with 
the corresponding plots in Ref.~\cite{Cheng:1996if} the observations made already for $q^2=0$ are 
confirmed. For the $B$ decays our form factors resemble very much those of Ref.~\cite{Cheng:1996if} (in
the ``+'' frame) with $A_2(q^2)$ showing the biggest deviations. For $D$-decays larger differences
can be observed, in particular for $A_1(q^2)$ and $A_2(q^2)$, but the qualitative features of the form factors are 
still quite similar.
This discrepancy is, of course, foreseeable since the point- and front-form approaches are not equivalent 
as long as one does not include non-valence contributions.
The equivalence is only reached in the heavy-quark limit, where the same Isgur-Wise function is obtained.

\newpage
\begin{table}[h!]
\begin{center}
    \begin{tabular}{c|cccc}\label{TableBtoDstar}
	 $B\to D^*$  &   $V(0)$ & $A_0(0)$ & $A_1(0)$ & $A_2(0)$    \\            \hline \hline
	Front form~\cite{Cheng:1996if} in the ``+'' frame &    0.78     & 0.73 & 0.68 &   0.61  	   \\
Front form~\cite{Cheng:1996if} in the ``$-$'' frame &  0.62 & 0.58 & 0.59 & 0.61\\
	Point form (this work)  &     0.76   & 0.72 & 0.72 & 0.72           \\
	 \hline
    \end{tabular}\caption{Form factors at $q^2=0$ for the $B\to D^*$ transition obtained 
within the light-front quark model in Ref.~\cite{Cheng:1996if} in the frames where the recoiling daughter
moves in the positive  $z$-direction (``+'' frame) and negative  $z$-direction
(``$-$'' frame)
in comparison with the results obtained in the point form.}
\end{center}

\begin{center}
    \begin{tabular}{c|cccc}
	 $B\to K^*$  &   $V(0)$ & $A_0(0)$ & $A_1(0)$ & $A_2(0)$    \\            \hline \hline
	Front form~\cite{Cheng:1996if}     & 0.35 &  0.32     & 0.26  & 0.23  	   \\
	Point form [this work]            & 0.36 &  0.33     & 0.32  &  0.31        \\
	 \hline
    \end{tabular}\caption{Form factors at $q^2=0$ for the $B\to K^*$ transition obtained 
within the front-form quark model in the frame where  where the recoiling daughter
moves in the positive  $z$-direction , i.e. ``+'' frame, and in the point form
of relativistic quantum mechanics.}\label{TableBtoKstar}
\end{center}

\begin{center}
    \begin{tabular}{c|cccc}
	 $D\to K^*$  &   $V(0)$ & $A_0(0)$ & $A_1(0)$ & $A_2(0)$    \\            \hline \hline
	Front form~\cite{Cheng:1996if}     & 0.87 &  0.71     & 0.62  &  0.46 	   \\
	Point form [this work]            & 0.87   &  0.70     & 0.71  &  0.73     \\
	 \hline
    \end{tabular}\caption{Same comparison as in Table~\ref{TableBtoKstar} but for 
the $B\to K^*$ transition.}\label{TableDtoKstar}
\end{center}

\begin{center}
    \begin{tabular}{c|cccc}
	 $B\to \rho$  &   $V(0)$ & $A_0(0)$ & $A_1(0)$ & $A_2(0)$    \\            \hline \hline
	Front form~\cite{Cheng:1996if}     & 0.30  &  0.28     & 0.20  &  0.18 \\
	Point form [this work]            & 0.31   &  0.28     & 0.27  &  0.26     \\
	 \hline
    \end{tabular}\caption{Same comparison as in Table~\ref{TableBtoKstar} but for 
the $B\to\rho$ transition.}\label{TableBtoRho}
\end{center}

\begin{center}
    \begin{tabular}{c|cccc}
	 $D\to \rho$  &   $V(0)$ & $A_0(0)$ & $A_1(0)$ & $A_2(0)$    \\            \hline \hline
	Front form~\cite{Cheng:1996if}  &  0.78     & 0.63  &  0.51   & 0.34  	   \\
	Point form [this work]          &  0.80     & 0.63  &  0.64   & 0.64       \\
	 \hline
    \end{tabular}\caption{Same comparison as in Table~\ref{TableBtoKstar} but for 
the $D\to\rho$ transition.}\label{TableDtoRho}
\end{center}
\end{table}

 \newpage
\newpage

\newpage
\newpage
\newpage

\newpage
\section{Conclusions and outlook}

In this chapter numerical results for a simple harmonic-oscillator wave 
function with different 
harmonic-oscillator parameters for each meson have been 
presented in the range $0\leq q^2\leq(M_1-M_2)^2$. Form factors are extracted without ambiguity
for both, pseudoscalar-to-pseudoscalar and pseudoscalar-to-vector meson transitions. 

The harmonic-oscillator wave function seems still to be too simple to do 
quantitative predictions that can be compared with
experiments, but qualitative features of the point form approach can be studied numerically. 
To compare with, analogous calculation in the front-form approach have been considered.
While in the front form the obtained results for pseudoscalar-to-vector transitions 
exhibits a certain dependence on the reference frame, i.e.
on whether the recoiling daughter moves in the positive or negative $z$-direction relative to the parent
meson, in the point form all form factors are determined unambiguously. 

In the heavy-quark limit, as was shown in Chap.~\ref{ChNumStudiesI}, point-form and front-form calculations
yield the same numerical result for the Isgur-Wise function. 
This equivalence is possible
because in the heavy-quark limit nonvalence contributions in the form of $Z$-graph
vanish. 
For finite heavy-quark masses, the point form and the front form are not equivalent, since 
the role that the $Z$-graph contributions play in either approach differs. This is shown by
the numerical results presented in this chapter. 
The non-valence contributions enter in a different way in the point and in the front forms.  

As long as $Z$-graph contributions are not introduced, 
it will be not possible to give a full answer in the whole $q^2$-range.
It will be the subject of future work to introduce non-valence contributions in the coupled channel approach
and to investigate how they affect the form factors. 
Similar studies on this subject have been carried out 
already in the front form~\cite{Bakker:2003up}. Like
in Ref.~\cite{Bakker:2003up} an estimate on $Z$-graph contributions within our approach could be obtained by 
calculating the transition form factors in the space-like region, where one can go into the infinite-momentum
frame and continue those results analytically to the time-like momentum-transfer region. The difference 
with the present calculation should then give an estimate of the size of $Z$-graph contributions.

%% file: Ch8.tex
\chapter{Dynamical binding forces}\label{ChDeuteronExCurrents}

\section{Introduction}

So far we have considered two-body states that are bound by instantaneous forces. 
In this chapter we will have a look at binding forces that are caused by dynamical particle exchange.
 Taking 
into account the retardation effects of the exchange particle
 requires the introduction of more
channels in the mass eigenvalue problem.
We will illustrate this with a simple example. We consider the deuteron as a system
of two nucleons which interact via $\sigma$- and $\omega$-exchange, i.e. a Walecka 
type model~\cite{Walecka:1974qa}. 
This is part of a benchmark calculation 
initiated during a workshop at ECT* Trento in 2009 with the goal of 
comparing relativistic effects as resulting from different approaches using
common parameters. 
The starting point was nonrelativistic quantum mechanics~\cite{Biernat:2011mp,Bakker:2010zz} 
with the static approximation of $\sigma$- and $\omega$-exchange to compute the common parameters.
This provided the first step for the investigation of relativistic effects in different approaches. 
Here we make a next step towards the understanding of the role that relativity plays
in the description of electromagnetic properties of strongly bound states.

Dynamical meson exchange gives rise to relativistic corrections.
These corrections will not only change the binding energy of the deuteron, they
also modify the bound-state wave function and hence the electromagnetic properties of
the deuteron as tested in electron-deuteron scattering.
The photon exchanged between the electron and the deuteron can resolve the internal structure
of the bound state.
Its coupling to the nucleons may take place during the meson exchange and 
furthermore, if the exchanged mesons are charged, the 
photon can also couple to them.
Such meson-exchange currents become important at  momentum transfer squared of a few GeV$^2$.
In order to account for these processes a consistent relativistic description for both, the 
bound-state wave function and for the electron-deuteron 
scattering amplitude, is necessary. In the following we will present
an appropriate theoretical framework and we will give the numerical solution for the relativistic wave function as resulting from
the dynamical boson-exchange responsible for the binding in the Walecka model.

\section{The Walecka model}
The interaction term of the Lagrangian density in the Walecka model is given by
\begin{equation}
\hat{\mathcal{L}}_{\text{int}}= g_\sigma\hat{\bar \psi}_N\hat{ \psi}_N\hat \sigma + 
ig_\omega\hat{\bar \psi}_N \gamma_\mu \hat\psi_N\hat \omega^\mu
+ \frac{f_\omega}{4m_N}\hat{\bar \psi}_N \sigma_{\mu\nu}\hat\psi_N
(\partial^\mu \hat\omega^\nu-\partial^\nu \hat\omega^\mu),
\end{equation}
where $\hat\psi_N(x)$ is the nucleon field of mass $m_N$, $\hat\sigma(x)$ is a neutral scalar
meson field of mass $m_\sigma$ and $\hat\omega(x)$ a neutral vector meson field of mass 
$m_\omega$. 
The $\sigma$-exchange is responsible for the binding of the nucleons, while the $\omega$ is associated
to repulsive effects.
The Walecka model requires regularization.
We introduce Pauli-Villars particles~\cite{Pauli:1949zm} which we call $\sigma_{PV}$ and
$\omega_{PV}$, with masses $\Lambda_\sigma$ and $\Lambda_\omega$, respectively. 
The Lagrangian density terms of this fields have
identical form to the ones corresponding to the physical particles but are multiplied by a factor $i$.

The notation used in this chapter will be: capital $N$ for ``nucleon'' 
and lower-case $n$ and $p$ for ``neutron'' and
``proton'', respectively.
The parameters calculated in the static approximation~\cite{Biernat:2011mp,Bakker:2010zz}
will be used 
in the following. They are shown in Table~\ref{ParametersWaleckaStatic}.
\begin{center}
\begin{table}[h]
\begin{center}
    \begin{tabular}{c | c c}
	Parameters	  &    Biernat \textit{et al.}~\cite{Biernat:2011mp,Bakker:2010zz}     & \cite{Biernat:2011mp,Bakker:2010zz} no PV  \\            \hline \hline
	$E_B$ (MeV)	  &  -2.224575   &  -3.36772	   \\
	$a_t$ (fm)	  &  5.4151      &  4.58658        \\
	$g^2_\sigma/4\pi$ &  6.31        &  6.31           \\
	$g^2_\omega/4\pi$ &  18.617      & 18.617          \\
	$m_\sigma$ (MeV/$c^2$)&  400          & 400     \\
	$m_\omega$ (MeV/$c^2$)&  782.7        & 782.7    \\
	$\Lambda_\sigma$ (MeV/$c^2$)&1000  &$\infty$       \\
	$\Lambda_\omega$ (MeV/$c^2$)& 1500 & $\infty$       \\
         \hline
    \end{tabular}\caption{Walecka model parameters considered for the numerical calculations
of this chapter.
Masses and coupling constants were chosen such that the deuteron binding energy
$E_B$ and the scattering length $a_t$ are 
reproduced~\cite{Biernat:2011mp,Bakker:2010zz}. In accordance with realistic models the tensor
coupling of the $\omega$ is neglected, i.e. $f_\omega=0$.}\label{ParametersWaleckaStatic}
\end{center}
\end{table}
 
\end{center}

The next step to understand the role of relativity in the description
of two-body bound states requires to abandon the static approximation and to take the 
dynamics of the exchanged particles fully into account. 
For the sake of simplicity we will still consider the $\omega$-meson exchange, which 
provides the short-range repulsion, in the static limit; 
only the retardation of the $\sigma$-exchange, which is responsible for the binding,
will be taken into account. 
The fact that in this
model we consider only neutral mesons makes the problem much simpler
since the photon cannot couple to
them. 
We will study the relativistic effects that are
due to the retardation of the $\sigma$-meson.

\section{The deuteron bound-state problem}
\begin{figure}[h!]
 \begin{center}
  \includegraphics[width=0.4\textwidth]{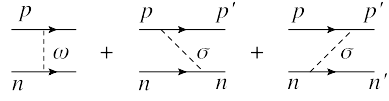}\caption{Graphical
representation of 
the one-boson exchange kernel occurring in the deuteron bound-state 
problem (cf. Eq.~(\ref{EqEigenvalueDeuteronCoupledChannel})). 
There is a repulsive core described by an instantaneous $\omega$-exchange.
The attraction is provided by $\sigma$-exchange. The $\sigma$-dynamics is fully
taken into account.}\label{FigDeuteronBoundState}
\end{center}
\end{figure}

We present in this section the bound-state problem from 
which the deuteron wave function will be 
obtained. 
The deuteron wave function for our Walecka-type model 
was already computed numerically by Biernat~\cite{Biernat:2011mp,Bakker:2010zz} using the point-form 
approach
in the approximation where the $\sigma$- and $\omega$-exchanges were 
instantaneous~\cite{Biernat:2011mp,Bakker:2010zz}. 
Let us first consider the generalization in which both, the $\sigma$- and the 
$\omega$-exchange, are treated dynamically.
The mass-eigenvalue equation for the 2-nucleon system in this case can be 
formulated as the following coupled-channel problem:
\begin{equation}\label{coupledsigmaomega}
\left( \begin{array}{ccc}
  \hat M_{np} & \hat K_\sigma & \hat K_\omega  \\
\hat K^\dagger_\sigma  &  \hat M_{np\sigma} & 0\\
\hat K^\dagger_\omega & 0 & \hat M_{np\omega}
 \end{array}\right) 
\left( \begin{array}{c}
 | \psi_{np}\rangle \\ 
|\psi_{np\sigma}\rangle\\
|\psi_{np\omega}\rangle
 \end{array}\right) = m
\left( \begin{array}{c}
  |\psi_{np}\rangle \\ 
|\psi_{np\sigma}\rangle\\
|\psi_{np\omega}\rangle
 \end{array}\right).
\end{equation}
The vanishing matrix elements impose the condition that no more than one meson can
be exchanged simultaneously. Applying
a Feshbach reduction one obtains the reduced eigenvalue equation for 
the 2-nucleon component we want to solve:
\begin{equation}\label{EqEigenvalueBothDynamical}
 \hat M_{np}|\psi_{np}\rangle + \hat K_\sigma (m-\hat M_{np\sigma})^{-1} \hat K_\sigma^\dagger |\psi_{np}\rangle 
+ \hat K_\omega (m-\hat M_{np\omega})^{-1} \hat K_\omega^\dagger |\psi_{np}\rangle =m |\psi_{np}\rangle. 
\end{equation}
 For simplicity, we will allow retardation only for the $\sigma$-exchange
and we will keep the static limit in the $\omega$-exchange. 
In the approximation where the $\omega$-exchange is instantaneous, 
Eq.~(\ref{EqEigenvalueBothDynamical}) can be written 
as (see also Fig.~\ref{FigDeuteronBoundState})\footnote{For simplicity
we have here neglected the Pauli-Villar particles. In the numerical calculations they are taken into account.}:
\begin{align}\label{EqEigenvalueDeuteronCoupledChannel}
\left(\hat M_{np} +\hat V_{\omega}^{\text{inst}}+\hat K_\sigma(m-\hat M_{np\sigma})^{-1}\hat K^\dagger_\sigma \right)|\psi_{np}\rangle 
=m |\psi_{np}\rangle.
\end{align}
The mass operators $\hat M_{np}$ and $\hat M_{np\sigma}$ account for the relativistic 
kinetic energies of the free particles, i.e.
\begin{align}
 \hat M_{np}|v; \vec k_p,\mu_p;\vec k_n,\mu_n\rangle&= 
\left( \omega_{k_p}+\omega_{k_n}\right)|v; \vec k_p,\mu_p;\vec k_n,\mu_n\rangle,\\
 \hat M_{np\sigma}|v; \vec k_p,\mu_p;\vec k_n,\mu_n;\vec k_\sigma\rangle&=
\left(  \omega_{k_p}+\omega_{k_n}+\omega_{k_\sigma} \right)|v; \vec k_p,\mu_p;\vec k_n,\mu_n;\vec k_\sigma\rangle . 
\end{align} 
We are now interested in bound-state solutions of Eq.~(\ref{EqEigenvalueDeuteronCoupledChannel}),
which have the quantum numbers of the deuteron. 
$|\psi_{np}\rangle$ is thus a 1-particle velocity state with the 
(discrete) quantum numbers $\alpha_D$ of the deuteron, i.e. $|\psi_{np}\rangle= |\ul v;\alpha_D\rangle$.
Since neither the instantaneous $\omega$-exchange nor the scalar $\sigma$-exchange can couple $s$-
and $d$-waves, our deuteron is still a pure $s$-wave bound state. 
This means that the matrix element 
$\langle \ul v; \vec{k}_p,\mu_p;\vec{k}_n,\mu_n|\psi_{np}\rangle$ can be written in the form:
\begin{align}
 \langle v; \vec{k}_p,\mu_p;\vec{k}_n,\mu_n|\psi_{np}\rangle=&(2\pi)^{9/2}\ul v_0 
\delta^3(\vec{\ul v}-\vec v)\sqrt{\frac{2}{m_D}}
\sqrt{\frac{2\omega_{k_n}2\omega_{k_p}}{(\omega_{k_n}+\omega_{k_p})^3}}\nonumber\\
& \times C^{1\mu_D}_{\frac{1}{2}\mu_n\frac{1}{2}\mu_p} 
\text{u}_{D}(|\vec k_p|)\text{Y}_{00}(\hat k_p).
\end{align}
Multiplying Eq.~(\ref{EqEigenvalueDeuteronCoupledChannel}) by $\langle \ul v; \vec{k}_p,\mu_p;\vec{k}_n,\mu_n|$
and using this relation one obtains
the integral equation for the deuteron wave function:
\begin{equation}\label{EqEigenvalueDeuteronIntegral}
\left( \omega_{k_p}+\omega_{k_n}\right) \tilde{\text{u}}_{D}(|\vec k|) + \int \frac{d^3k'}{(2\pi)^3}\left( \tilde v_{\text{int}}^{N\omega}(\vec k,\vec k')+\tilde v_{\text{int}}^{N\sigma}(\vec k,\vec k')\right)
\tilde{\text{u}}_{D}(|\vec k'|)=m_D \tilde{\text{u}}_{D}(|\vec k|)
\end{equation}
with 
\begin{align}
&\tilde v_{\text{int}}^{N\omega}(\vec k,\vec k')=\; \frac{g_\omega^2}{(\vec k-\vec k')^2+m^2_\omega}-
\frac{g_\omega^2}{(\vec k-\vec k')^2+\Lambda^2_\omega},\\
& \tilde v_{\text{int}}^{N\sigma}(\vec k,\vec k')=
\left(\frac{\omega_{k'_p}}{\omega_{k_p}}\right)^{3/2}\sum_{\mu_p'\mu_n'}\frac{\bar u_{\mu_n}(\vec k_n)u_{\mu'_n}(\vec k_n')}{2\omega_{k'_n}}
\frac{\bar u_{\mu_p}(\vec k_p)u_{\mu'_p}(\vec k_p')}{2\omega_{k'_p}}\nonumber\\
&\;\;\times\left(\frac{g_\sigma^2}{\omega_{k_\sigma}(m_D-\omega_{k_p}-\omega_{k'_p}-\omega_{k_\sigma})}-\frac{g_\sigma^2}{\omega_{k_{\sigma PV}}(m_D-\omega_{k_p}-\omega_{k'_p}-\omega_{k_{\sigma PV}})}\right).
\end{align}
 where the notation 
$\tilde{\text{u}}_{D}(|\vec k^{(\prime)}|):=(\omega_{k_p^{(\prime)}})^{-1/2}\text{u}_{D}(|\vec k^{(\prime)}|)$ and 
$\vec k:=\vec k_p=-\vec k_n$, has
been used. Note that now the kernel depends on the eigenvalue $m_D$.

The problem can be solved numerically using Gaussian quadrature, 
following analogous steps as in
Refs.~\cite{Biernat:2011mp,vanIersel:2000sa,Bakker:2010zz}.
The wave-function solution of the integral equation
will be used for the calculation of
transition amplitudes and currents analogously to  Chaps.~\ref{ChCoupledChannel}
and~\ref{ch:currents:and:ff}.
 $\text{u}_{D}(|\vec k|)$ has to be appropriately normalized such that 
\begin{equation}\label{EqPsiNorm}
 \langle \psi_{np}|\psi_{np}\rangle+\langle\psi_{np\sigma}|\psi_{np\sigma}\rangle=
(2\pi)^3\frac{2}{m^2_D}v^0\delta^3(\vec v'-\vec v),
\end{equation}
where  $\vec v'$ is the velocity of the bra and $\vec v$ the one of the ket.
To compute $|\psi_{np\sigma}\rangle$, one can use:
\begin{equation}
|\psi_{np\sigma}\rangle =\left( m_D -M_{np\sigma} \right)^{-1} \hat K^\dagger | \psi_{np}\rangle,
\end{equation} 
which follows from Eq.~(\ref{coupledsigmaomega}).
Hence the matrix element$\langle v; \vec{k}_p,\mu_p;\vec{k}_n,\mu_n;$ $ \vec k_\sigma|\psi_{np\sigma}\rangle$
can be expressed in terms of the deuteron wave function $\text{u}_{D}(|\vec k_p|)$.
The analytical expression is given in App.~\ref{AppWaveFunction}.

\section{The electron-deuteron scattering problem}
The bound-state wave function obtained in this manner 
will be used in the coupled-channel problem
that describes electron-deuteron scattering.
The corresponding mass-eigenvalue problem needs 4-channels:
 \begin{equation}  
\hspace{-1cm} \left( \begin{array}{cccc} \hat M_{enp} +\hat V_{\omega}^{\text{inst}}          &    \hat K_\gamma      &     \hat K_\sigma    &     0 \\
                           \hat K_\gamma^\dagger & \hat M_{enp\gamma} +\hat V_{\omega}^{\text{inst}}  &0    & \hat K_\sigma \\
                           \hat K_\sigma^\dagger  & 0 & \hat M_{enp\sigma}  & \hat K_\gamma   \\
                           0      &     \hat K_\sigma^\dagger    &  \hat K_\gamma^\dagger  &  \hat M_{enp\gamma\sigma} \end{array} \right) \left( \begin{array}{c} |\psi_{enp}\rangle \\ |\psi_{enp\gamma} \rangle  \\ |\psi_{enp\sigma} \rangle \\ | \psi_{enp\gamma\sigma} \rangle \end{array} \right) 
= m\left( \begin{array}{c} |\psi_{enp}\rangle \\ |\psi_{enp\gamma} \rangle  \\ |\psi_{enp\sigma} \rangle \\ | \psi_{enp\gamma\sigma} \rangle \end{array} \right).
\end{equation}
The diagonal part represents the kinetic energy of the free particles plus instantaneous
forces. Here the eigenvalue is
$m=\omega_e +\omega_D$.
In analogy to the electron-meson scattering problem, the system of equations can be reduced to a single 
equation for the $|\psi_{enp}\rangle$ state
 \begin{eqnarray}\label{Feshbach3}
&&\{ \hat M_{eD}  + \nonumber\\
&+& \hat K_\gamma (m-\hat M_{eD\gamma})^{-1}\hat K^\dagger_\gamma \nonumber\\
&+& \hat K_\gamma (m-\hat M_{eD\gamma})^{-1} \hat K_\sigma (m-\hat M_{enp\sigma \gamma})^{-1}\hat K^\dagger_\gamma (m-\hat M_{enp\sigma})^{-1}\hat K_\sigma^\dagger \nonumber\\
&+&\hat K_\sigma (m-\hat M_{enp\sigma})^{-1}\hat K_\gamma (m-\hat M_{enp\sigma\gamma})^{-1}\hat K_\gamma^\dagger (m-\hat M_{enp\sigma})^{-1}\hat K_\sigma^\dagger \nonumber\\
&+& \hat K_\sigma (m-\hat M_{enp\sigma})^{-1}\hat K_\gamma(m-\hat M_{enp\sigma\gamma})^{-1}\hat K_\sigma^\dagger  (m-\hat M_{eD\gamma})^{-1} \hat K_\gamma^\dagger \nonumber\\
&+& \hat K_\sigma (m-\hat M_{enp\sigma})^{-1}\hat K_\gamma (m-\hat M_{enp\sigma\gamma})^{-1}\hat K_\sigma^\dagger (m-\hat M_{eD\gamma})^{-1}  \nonumber \\
&&\times \hat K_\sigma (m-\hat M_{enp\sigma\gamma})^{-1}\hat K_\gamma^\dagger (m-\hat M_{enp\sigma})^{-1}\hat K_\sigma^\dagger \}|\Psi_{enp}\rangle = m|\Psi_{enp}\rangle, \nonumber\\
&&
\end{eqnarray}

where the following notation has been used:
 \begin{equation}\label{EqDefMD}
 \hat M_{eD} :=\hat M_{enp} + \hat V^{\text{inst}}_{\omega} +\hat K_\sigma(m-\hat M_{enp\sigma})^{-1}\hat K^\dagger_\sigma.
 \end{equation}
Now  $\hat M_{eD}$ and $\hat M_{eD\gamma}$ have eigenstates
 \begin{align}
  \hat M_{eD}\,|\ul v; \ul{\vec k}_e, \ul\mu_e  ;\ul{\vec k}_D, \ul\mu_D \rangle 
&= (\omega_{k_e} +\omega_{k_D} )\,|\ul v; \ul{\vec k}_e, \ul\mu_e  ;\ul{\vec k}_D, \ul\mu_D \rangle,  \\
  \hat M_{eD\gamma}\,|\ul v; \ul{\vec k}_e, \ul\mu_e  ;\ul{\vec k}_D, \ul\mu_D ;\ul{\vec k}_\gamma, \ul\mu_\gamma\rangle 
&= (\omega_{k_e} +\omega_{k_D} +\omega_{k_\gamma})\,|\ul v; \ul{\vec k}_e, \ul\mu_e  ;\ul{\vec k}_D, \ul\mu_D ;\ul{\vec k}_\gamma, \ul\mu_\gamma\rangle . 
 \end{align}
with $\omega_{k_D}=\sqrt{m_D^2+\vec k_D^2}$,   $\omega_{k_e}=\sqrt{m_e^2+\vec k_e^2}$, and
 $\omega_{k_\gamma}=|\vec k_\gamma|$.

Eq.~(\ref{Feshbach3}) describes the electron-deuteron scattering process 
in analogy
 to what was done for two-body systems that are bound by instantaneous confining forces.
Now, however, the binding is generated by the dynamical $\sigma$-exchange, which leads to additional terms in 
the 1-photon-exchange optical potential. In the following we will discuss those terms. 
By the replacement (\ref{EqDefMD}) one absorbs the sigma exchange 
into the deuteron wave function. 

\subsection{Graphical representation}\label{SecGraphRep}

\subsubsection{One-body currents}
The first term in Eq.~(\ref{Feshbach3})  represents the kinetic energy of the electron-deuteron
system, with the deuteron binding being determined by dynamical $\sigma$-exchange and an instantaneous 
$\omega$-exchange. The other terms in Eq.~(\ref{Feshbach3})  represent the 1-photon-exchange potential and thus 
determine the electromagnetic deuteron current.
For instance, the second term
\begin{equation}\label{TermNoSigma}
\hat K_\gamma (m-\hat M_{eD\gamma})^{-1}\hat K^\dagger_\gamma
\end{equation}
has the same form as the optical
potential in Eq.~(\ref{optEM}) for 
the case of systems which are bound by an instantaneous potential. It gives rise to the 
four time-ordered contributions that are sketched 
in Fig.~\ref{OneBody4timeorderings}.
The gray ovals represent the deuteron.
Already these graphs contain ralativistic corrections which go beyond instantaneous binding forces,
namely the retardation effect of the $\sigma$ in the deuteron wave function.
As in Sec.~\ref{ChCoupledChannel} the electromagnetic deuteron current extracted from these
graphs is a one-body current.

\subsubsection{Exchange currents}
\begin{figure}[h!]
 \begin{center}
\includegraphics[width=0.5\textwidth]{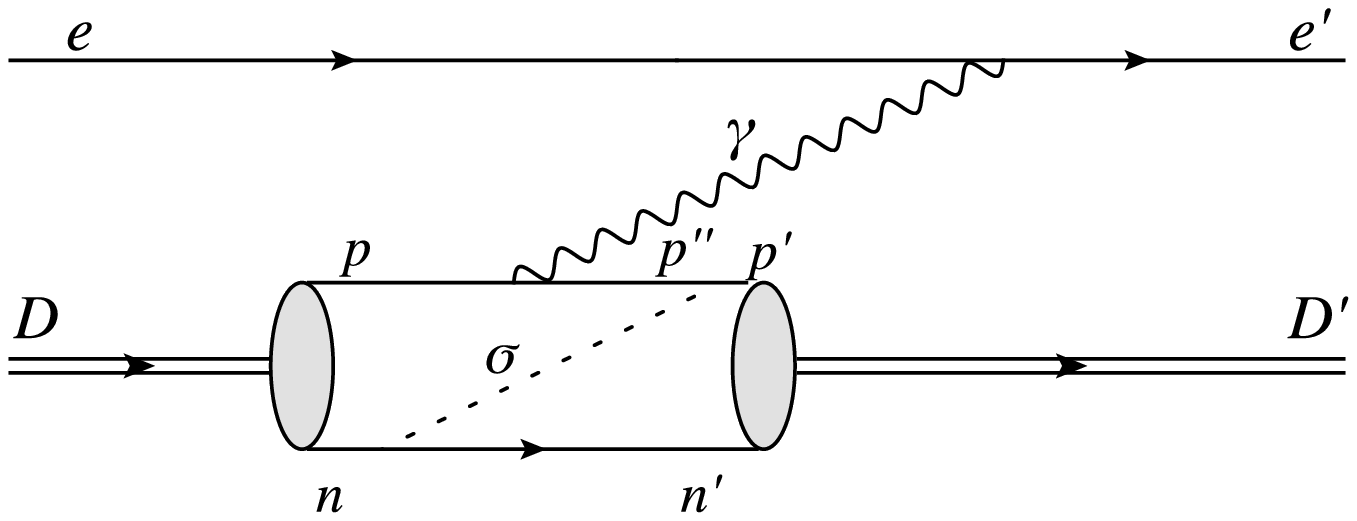}\includegraphics[width=0.5\textwidth]{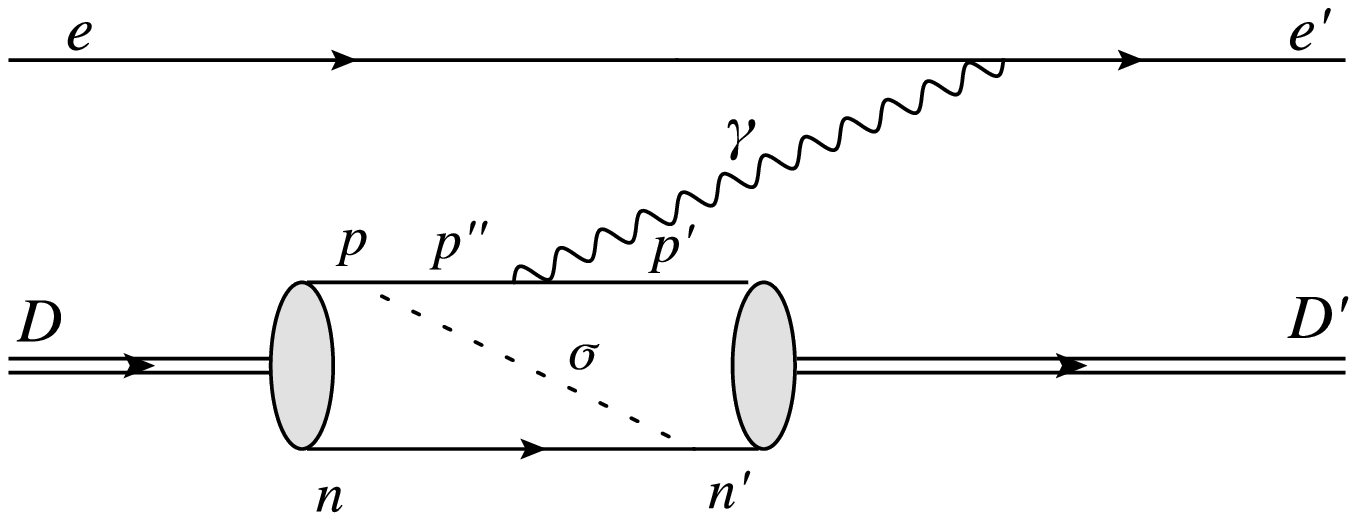}
\end{center}\caption{Graphical representation of the one-photon exchange 
in electron-deuteron scattering corresponding to the term 
$\hat K_\gamma (m-\hat M_{eD\gamma})^{-1} \hat K_\sigma (m-\hat M_{enp\sigma \gamma})^{-1}\hat K^\dagger_\gamma (m-\hat M_{enp\sigma})^{-1}\hat K_\sigma^\dagger $
in the optical potential (\ref{Feshbach3}). Graphs in which the photon couples to the neutron are not shown.
}\label{FigDouble2}
\end{figure}

\begin{figure}[h!]
\begin{center}
\includegraphics[width=0.5\textwidth]{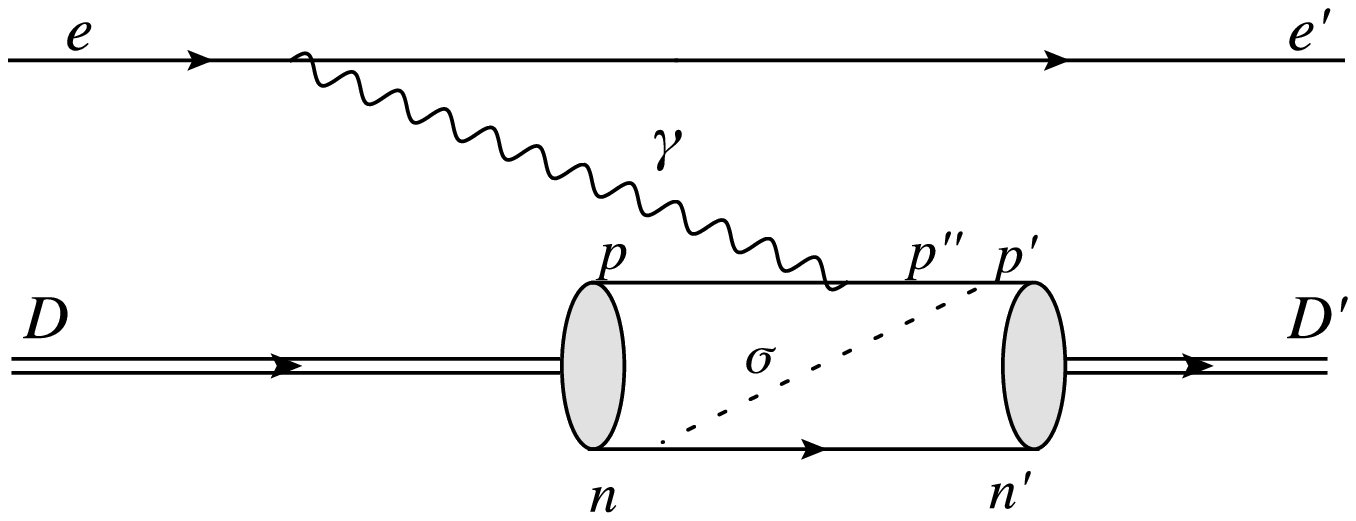}\includegraphics[width=0.5\textwidth]{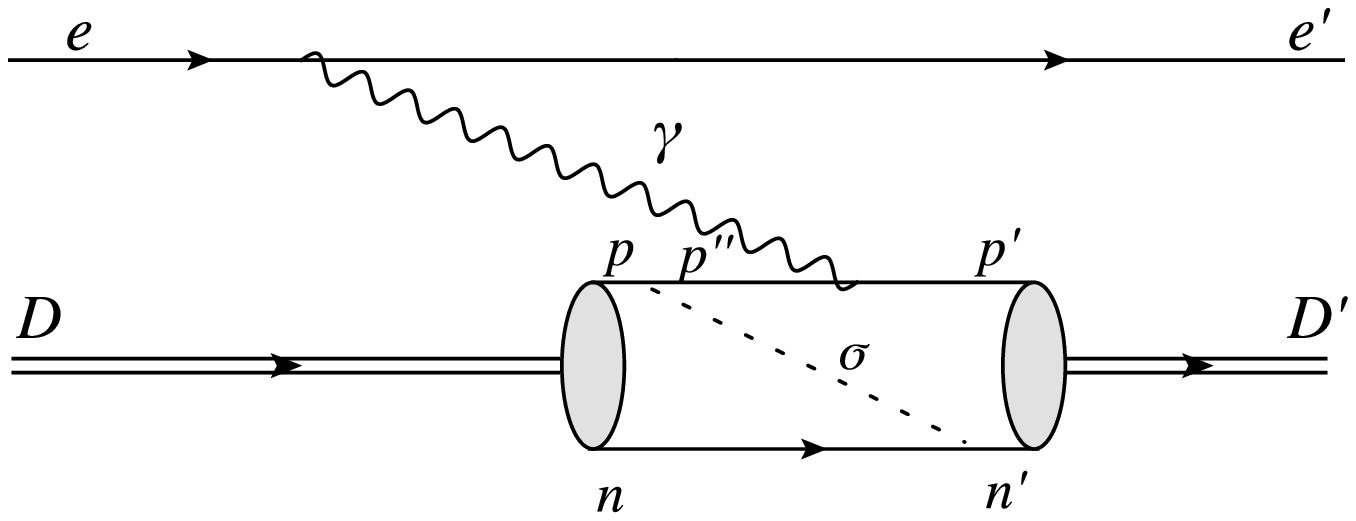}
\end{center}\vspace{-0.4cm}\caption{Graphical representation of the one-photon exchange 
in electron-deuteron scattering corresponding to the term 
$\hat K_\sigma (m-\hat M_{enp\sigma})^{-1}\hat K_\gamma(m-\hat M_{enp\sigma\gamma})^{-1}\hat K_\sigma^\dagger  (m-\hat M_{eD\gamma})^{-1} \hat K_\gamma^\dagger$
in the optical potential (\ref{Feshbach3}). 
Graphs in which the photon couples to the neutron are not shown.}\label{FigDouble4}
\end{figure}

\begin{figure}[h!]
 \begin{center}
  \includegraphics[width=1\textwidth]{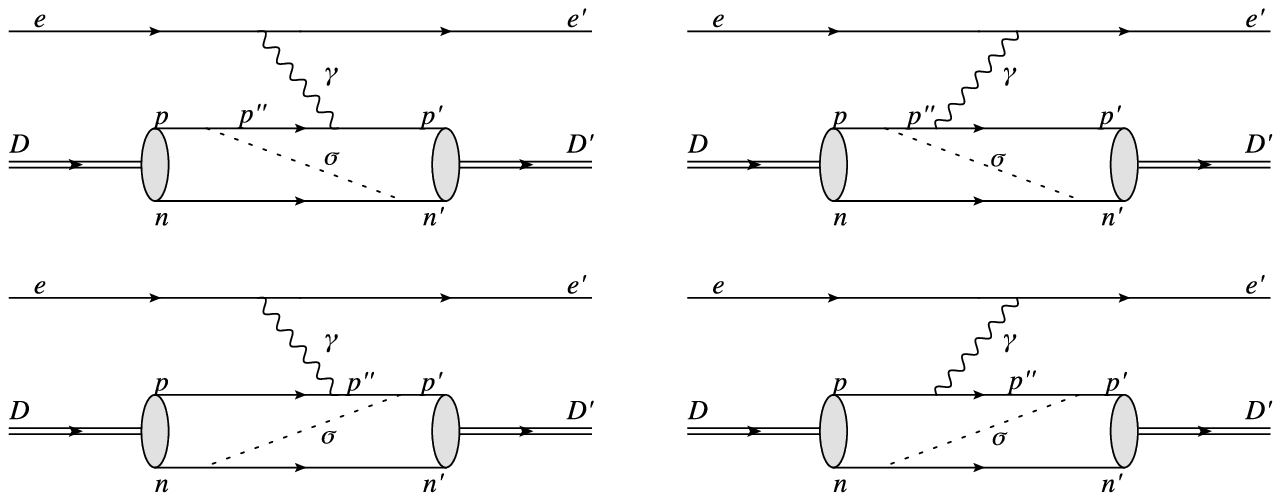}
\caption{Graphical representation of the one-photon exchange 
in electron-deuteron scattering corresponding to the term 
$\hat K_\sigma (m-\hat M_{enp\sigma})^{-1}\hat K_\gamma (m-\hat M_{enp\sigma\gamma})^{-1}\hat K_\gamma^\dagger (m-\hat M_{enp\sigma})^{-1}\hat K_\sigma^\dagger $
in the optical potential (\ref{Feshbach3}). Graphs in which the photon couples to the neutron are not shown.
}\label{FigDouble2112p}
 \end{center}
\end{figure}

In the following we discuss those contributions to 
electron-deuteron scattering which give rise to  
the, so-called, \textit{exchange currents}. 
These are the terms in the optical potential that describe 
the coupling of the photon  
to the nucleons during the $\sigma$-meson exchange.  
Figures~\ref{FigDouble2} and~\ref{FigDouble4} are graphical representations of (part of)
the photon and $\sigma$-exchanges described by the third and fifth terms in Eq.~(\ref{Feshbach3}).
Initially, a  $\sigma$-meson is emitted by one of
the nucleons, the $enp\sigma$-system propagates freely until a photon is emitted by
one of the nucleons; then the $enp\sigma\gamma$-system propagates freely until the 
$\sigma$-meson is absorbed by the second nucleon. Finally, the deuteron-bound state
propagates together with the photon and the electron, until the photon is absorbed by the electron.
The fourth term (see Fig.~\ref{FigDouble2112p}) 
represents the case in which the whole 
$\gamma$-exchange (emission and absorption) occurs during the $\sigma$-exchange process.

\begin{figure}[h!]
 \begin{center}
  \includegraphics[width=1\textwidth]{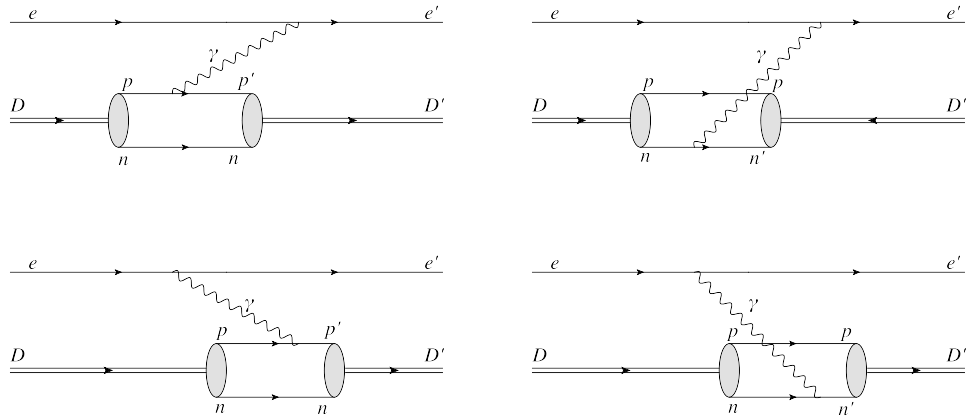}\caption{Graphical representation of the one photon exchange 
in electron-deuteron scattering corresponding to the term (\ref{TermNoSigma}) of the optical potential.
The sketch represents the four possible time orderings of this particular contribution, where the 
photon-nucleon coupling occurs during the time in which no $\sigma$-meson exchange takes place. The bound
state, however, represented by the gray oval, accounts for the dynamical exchange given by the 
$\sigma$-potential, as well as for the instantaneous one corresponding to the $\omega$.}\label{OneBody4timeorderings}
\end{center}
\end{figure}

\begin{figure}[h!]
 \begin{center}
  \includegraphics[width=0.6\textwidth]{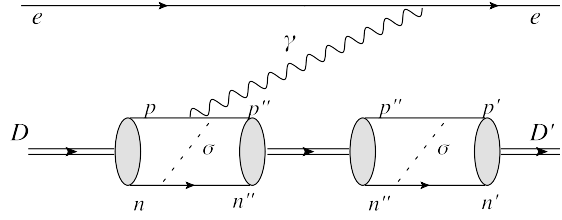}
\caption{Graphical representation of the one-photon exchange 
in electron-deuteron scattering corresponding to the term $\hat K_\sigma (m-\hat M_{enp\sigma})^{-1}\hat K_\gamma (m-\hat M_{enp\sigma\gamma})^{-1}\hat K_\sigma^\dagger (m-\hat M_{eD\gamma})^{-1} 
\hat K_\sigma (m-\hat M_{enp\sigma\gamma})^{-1}\hat K_\gamma^\dagger (m-\hat M_{enp\sigma})^{-1}\hat K_\sigma^\dagger$
in the optical potential (\ref{Feshbach3}). Graphs in which the photon couples to the neutron and in which the 
time orderings of the $\gamma$- and $\sigma$-exchanges are reversed are not shown.
}\label{FigTwoSigmas}
 \end{center}
\end{figure}

A representative example of a one-photon-exchange graph corresponding to the sixth 
term is given in  Fig.~\ref{FigTwoSigmas}.
The significance of this type of graphs is still not clear to us. 
They could be necessary to get the correct normalization  of the deuteron form factors at the end. 
Therefore we will concentrate on the 
graphs shown in Figs.~\ref{OneBody4timeorderings}-\ref{FigDouble4}.

Electromagnetic self-energy graphs in which the photon is emitted and absorbed by the same particle 
(electron or nucleon) are also contained in Eq.~(\ref{Feshbach3}). But they are not of interest here, because they do 
not contribute to photon-exchange between the electron and the deuteron. More interesting, however, are
those graphs in which the $\sigma$ is emitted and reabsorbed by the same nucleon while the photon couples
to the nucleon. There are also contained in Eq.~(\ref{Feshbach3}) and represent strong vertex correction to the
photon-nucleon vertex. Such corrections, however, should, at least partly, be contained in the electromagnetic nucleon 
form factors. We therefore omit them here too. 

\newpage
\vspace{-3cm}

\newpage
\subsection{Currents and form factors}
In this section we present the analytical expression for the electromagnetic currents
as extracted from the one-photon-exchange optical potential in Eq.~(\ref{Feshbach3}) 
(see also Figs.~\ref{OneBody4timeorderings}-\ref{FigTwoSigmas}).
Two kinds of contributions to the current can be distinguished 
from the discussion
presented above, namely the one corresponding to one-body currents, i.e. those
contributions where the exchanged $\sigma$ is in the deuteron wave function, and those 
in which $\sigma$-exchange happens at the same time as the photon exchange; 
these are called \textit{exchange currents}. 
Some remarks on the covariance and cluster separability properties of these
 currents and form factors will be discussed.
\subsubsection{One-body currents}
The transition amplitude is computed exactly in the same way as it was done for 
the case of mesons (cf. Chap.~\ref{ChCoupledChannel}):
\begin{eqnarray}\label{EqOptPotOneBody}
&&\langle \ul{v'}; \vec{\ul k}'_e, \ul \mu'_e; \vec{\ul k}'_D, \ul \mu'_D \vert  \hat K_\gamma (m-\hat M_{eD\gamma})^{-1}\hat K^\dagger_\gamma    \vert \ul{v}; \vec{\ul k}_e, \ul \mu_e; \vec{\ul k}_D ,\ul \mu_D\rangle \nonumber\\
&& \qquad= \ul v_0 \delta(\vec{\ul v}'-\vec{\ul v}) \frac{(2\pi)^3}{\sqrt{(\omega_{k'_e}+\omega_{\ul k'_D})^3(\omega_{k_e} + \omega_{\ul k_D})^3}}\nonumber\\
&&\qquad\qquad\times e\;\underbrace{\bar{u}_{\mu_e'}(\vec k_e')\gamma^\mu u_{\mu_e}(\vec k_e)}_{j_e^\mu}\frac{(-g_{\mu\nu})}{\text{Q}^2}|e|\underbrace{ (J_p^\nu +J_{n}^\nu )}_{J_D^\nu}, 
\end{eqnarray}
with the proton current
\begin{eqnarray}\label{EqCurrentOneBody}
&& J^{\nu}_p(\vec k'_D,\mu'_D;\vec k_D,\mu_D)= 
\sqrt{\omega_{\ul{k}'_D}\omega_{\ul{k}_D}}\int \frac{d^3 \tilde k_p'}{\omega_{k_p}}
\sqrt{\frac{\omega_{k_p}}{\omega_{k_p'}}}
\sqrt{\frac{m_{np}}{ m_{np}'}}  \nonumber \\
&&\quad\times \sum_{\mu_p'\mu_n\mu_p}\sum_{\tilde \mu_n\tilde \mu_p \tilde \mu_n'\tilde \mu_p'}
\bar u_{\mu_p'}(\vec k_p')\Gamma^\nu_p u_{\mu_p}(\vec k_p)
\; \text{u}^*_{D}(|\vec{\tilde k}_p'|)Y^*_{00}(\hat{\vec{\tilde{k}}}_p') \text{u}_{D}(|\vec{\tilde k}_p|)Y_{00}(\hat{\vec{\tilde{k}}}_p)
\nonumber\\
&&\quad\times  C^{1\mu_D'}_{\frac{1}{2}\tilde \mu_p'\frac{1}{2}\tilde \mu_n'}\; D^{1/2}_{\tilde\mu_{p}'\mu_{p}'}\left[ R_W^{-1}\left( \frac{\tilde k_p'}{m_p},B_c\left( v_{np}\right)\right)\right] D^{1/2}_{\mu_{p}\tilde\mu_{p}}
\left[ R_W\left( \frac{\tilde k_p}{m_p},B_c\left( v_{np}\right)\right)\right] \nonumber\\
&&\quad\times C^{1\mu_D}_{\frac{1}{2}\tilde \mu_p\frac{1}{2}\tilde \mu_n}
D^{1/2}_{\tilde\mu_{n}'\tilde\mu_{n}}\left[ R_W^{-1}\left( \frac{\tilde k_n'}{m_n},B_c\left( v_{np}'\right)\right)
 R_W\left( \frac{\tilde k_n}{m_n},B_c\left( v_{np}\right)\right)\right]. \nonumber\\
\end{eqnarray}
An analogous expression is obtained for the neutron current.
The $\Gamma^\mu_N$ in the nucleon current contains
the nucleon structure:
\begin{equation}
\Gamma^\mu_N :=\left( F_{1N}(\text{Q}^2)\gamma^\mu + F_{2N}(\text{Q}^2) 
\frac{iq_\nu \sigma^{\nu\mu}}{2m_N}\right),\quad
N=n,p.
\end{equation}
where $F_{1N}(\text{Q}^2)$ and $F_{2N}(\text{Q}^2)$ are the electromagnetic
 form factors of the active nucleon, either the proton
or the neutron, and $q^\mu=(k_N'^\mu-k_N^\mu)$, $\sigma^{\mu\nu}=\frac{i}{2}(\gamma^\mu\gamma^\nu-\gamma^\nu\gamma^\mu)$.
For the calculations presented here we will later on consider the nucleons as point-like 
particles, using $F_{1p}(\text{Q}^2)=1$ and $F_{2p}(\text{Q}^2)=0$ for the proton, and 
$F_{1n}(\text{Q}^2)=F_{2n}(\text{Q}^2)=0$ for the neutron. This is $\Gamma^\mu\to\gamma^\mu$
and the photon coupling to the neutron vanishes.

\subsubsection{Exchange currents}\label{SubSecDeutExch}
There are several terms in the potential (\ref{Feshbach3}) that contribute to the 
exchange currents we want to analyze in this chapter. 
They correspond to the case in which the photon couples to one of the nucleons
 while the process of $\sigma$-exchange, that keeps the system bound, takes place. 
The interaction between the nucleons should show up in the
structure of the system, affecting therefore the form factors. 
The photon \textit{feels} not only the interaction of the particle to which it couples 
since it is transfered to a system of three particles, one of them providing the  binding 
interaction. 
We will present here first
the analytical result for the exchange current corresponding  to the third term in  
Eq.~(\ref{Feshbach3}). It yields four graphs, two of them are illustrated in Fig.~\ref{FigDouble2}. 
We will explain the way how to extract  the rest of them, 
which is straightforward with the correct interpretation of the diagrams shown in
the previous section.
Matrix elements of this term of the optical potential have the same structure as 
Eq.~(\ref{EqOptPotOneBody}), but now the constituent current is much
more complicated. One contribution that comes out 
from the third term in 
Eq.~(\ref{Feshbach3}), namely the one corresponding to the left panel in Fig.~\ref{FigDouble2} reads
\begin{align}\label{EqExchangeCurrentsqJp}
& J_{p}^{\mu,\text{ex}}(\vec k_D',\vec k_D; \mu'_D,\mu_D)  = \sqrt{2\omega_{k_D}2\omega_{k'_D}} \int d^3\tilde k'_p  \int \frac{d^3\tilde q}{2\omega_{k_\sigma}} 
\sqrt{\frac{m_{np}}{m'_{np}}}\sqrt{\frac{\omega_{k_n}}{\omega_{k'_n}}}\nonumber\\
&\quad\times \left(\frac{1}{m-\omega_{k_e}-\omega_{k''_p}-\omega_{k'_n}-\omega_{k_\sigma}-\omega_{k_\gamma}}\right)\left( \frac{1}{m-\omega_{k_e}-\omega_{k_p}-\omega_{k'_n}-\omega_{k_\sigma}}\right) \nonumber\\
&\quad\times \sum_{\mu_p'\mu_n\mu_p}\sum_{\tilde \mu_n\tilde \mu_p \tilde \mu_n'\tilde \mu_p'} \frac{\bar u_{\mu'_p}(\vec k''_p)\Gamma^\nu u_{\mu_p}(\vec k_p)}{2\omega_{k_p}} 
\;g^2_\sigma  \; 
\frac{\bar u_{\mu'_p}(\vec k'_p) u_{\mu'_p}(\vec k''_p)}{2\omega_{k''_p}} \;
\frac{ \bar u_{\mu_n}(\vec k'_n) u_{\mu_n}(\vec k_n)}{2\omega_{k_n}} \nonumber\\
&\quad\times  C^{1\mu_D'}_{\frac{1}{2}\tilde \mu_p'\frac{1}{2}\tilde \mu_n'}\; D^{1/2}_{\tilde\mu_{p}'\mu_{p}'}\left[ R_W^{-1}\left( \frac{\tilde k_p'}{m_p},B_c\left( v_{np}\right)\right)\right] D^{1/2}_{\mu_{p}\tilde\mu_{p}}
\left[ R_W\left( \frac{\tilde k_p}{m_p},B_c\left( v_{np}\right)\right)\right] \nonumber\\
&\quad\times C^{1\mu_D}_{\frac{1}{2}\tilde \mu_p\frac{1}{2}\tilde \mu_n}
D^{1/2}_{\tilde\mu_{n}'\tilde\mu_{n}}\left[ R_W^{-1}\left( \frac{\tilde k_n'}{m_n},B_c\left( v_{np}'\right)\right)
 R_W\left( \frac{\tilde k_n}{m_n},B_c\left( v_{np}\right)\right)\right] \nonumber\\
&\quad\times  \text{u}^*_{D}(|\vec{\tilde k}_p'|)Y^*_{00}(\hat{\vec{\tilde{k}}}_p') \text{u}_{D}(|\vec{\tilde k}_p|)Y_{00}(\hat{\vec{\tilde{k}}}_p),
\end{align}
where $\vec{\tilde q}$ is defined as $\vec{\tilde q}:=\vec k_p-\vec k_p'$ (see App.~\ref{AppMatrixElementsandCurrents}).
 Interchanging primed and
unprimed variables one obtains the terms corresponding to the right panel of Fig.~\ref{FigDouble4}. 
The right graph in Fig.~\ref{FigDouble2} is  
is identical to the left one except for the propagators. It is directly obtained by applying the changes
\begin{align}
&(m-\omega_{k_e}-\omega_{k''_p}-\omega_{k'_n}-\omega_{k_\sigma}-\omega_{k_\gamma})^{-1}
(m-\omega_{k_e}-\omega_{k_p}-\omega_{k'_n}-\omega_{k_\sigma})^{-1} \nonumber\\
& \to (m-\omega_{k_e}-\omega_{k'_p}-\omega_{k_n}-\omega_{k_\sigma}-\omega_{k_\gamma})^{-1}
(m-\omega_{k_e}-\omega_{k_p''}-\omega_{k_n}-\omega_{k_\sigma})^{-1} \nonumber\\
\end{align}
and 
\begin{eqnarray}
 \bar u_{\mu'_p}(\vec k''_p)\Gamma^\nu u_{\mu_p}(\vec k_p) &\to&  \bar u_{\mu_p'}(\vec k'_p)\Gamma^\nu u_{\mu_p}(\vec k_p''),\\
\bar u_{\mu'_p}(\vec k'_p) u_{\mu'_p}(\vec k''_p)  &\to& \bar u_{\mu_p}(\vec k''_p) u_{\mu_p}(\vec k_p) .
\end{eqnarray}
From the obtained result, by interchanging primed and unprimed variables one obtains the left
graph in Fig.~\ref{FigDouble4}.

In addition to the integration over $\vec{\tilde k}'_p$,
 a second integral appears, which 
runs over the intermediate state $\vec{\tilde k}''_p$, and accounts for the fact 
that momentum is transfered
by the $\sigma$-meson from one nucleon to the other. 
By a change of variables one can 
go over to an integration over the momentum transfer, which we call $\vec{\tilde q}$.
 The change requires some work, because $\vec{\tilde k}''_p$ and $\vec{\tilde q}$ are defined in 
different reference frames. 
One has to transform $\vec{\tilde k}_p''$ to $\vec k_p''$ 
(see Eq.~\ref{EqChangeTildeNoTilde}), which
is related to $\vec{\tilde q}$  by a simple translation.

It is not possible to combine all contributions in one term in a simple way
 as it was done
for one-body currents. However, we will see that this will not be a problem
for our purpose, since in the infinite-momentum frame,
where the extraction of the form factors will be carried 
out (cf. Chap.~\ref{ch:currents:and:ff}), the binding energy becomes negligible and
every propagator reduces to the same form, allowing to 
write the exchange currents as a one-term expression.  We will see this in the following section.

\subsection{Properties of the currents}
The description of our electromagnetic spin-1 current requires 
 11 covariants and form factors (cf. Sec.~\ref{SecVectorCurrent}). 
It is convenient to take the infinite momentum frame, i.e. 
 $s\to\infty$, in order to get rid of most of the spurious contributions and to be
 able to determine the form factors from those matrix 
elements of the current that 
contain only physical contributions. 
The introduction of the $\sigma$-exchange does not alter the structure of the covariant decomposition. 
It modifies
instead the microscopic structure of the current. 
The inclusion of these additional degrees of freedom might reduce, however,
the strength of the non-physical quantities. In order to check this, a numerical study
of the exchange currents and form factors is needed. In this work we will provide 
the analytical prerequisites. 
In the following, we present analytical results for the exchange current as resulting in the 
infinite-momentum frame. To do this study the kinematics of Eq.~(\ref{eq:momentumscatt}) (with $\alpha=D$)
is chosen as a starting point. The infinite-momentum frame is then reached by taking $\kappa_D\to\infty$.

\subsubsection{The infinite-momentum frame}
In the infinite-momentum frame the physical form factors can be extracted from three independent
matrix elements of the current, $J^0_{11}$, $J^0_{1-1}$ and $J^2_{11}$ 
(see Sec.~\ref{SecVectorCurrent}).
The infinite-momentum frame
reduces all and eliminates some of the spurious contributions 
that appear due to the cluster problems inherent in the 
Bakamjian-Thomas construction~\cite{Keister:1991sb,Biernat:2010tp,Biernat:2011mp}. 
Here we give the resulting
current in the limit. 
It can be compared with the one-body current, also obtained previously 
in Ref.~\cite{Biernat:2011mp}. For 
details of taking the limit see App.~\ref{LimitsInfMomentumFrame}.

The momenta in the initial and final states 
are related by canonical boosts (cf.~Eq.~(\ref{eq:kktildep}) and App.~\ref{AppCanonicalBoosts}).
 This relation
in the infinite momentum frame is:
\begin{equation}
\vec{\tilde k}_p\rightarrow \left( \begin{array}{c} \tilde k'^{[1]}_p-\left( \frac{1}{2}-\frac{\tilde k'^{[3]}_p}{m'_{np}}\right) \text{Q}-\tilde q^{[1]}\\
\tilde k'^{[2]}_p-\tilde q^{[2]}\\
\tilde k'^{[3]}_p \frac{m_{np}}{m'_{np}} \end{array} \right).
\end{equation}
Similarly, the free invariant $np$ mass in the initial state can be  written in terms of variables 
given in the final state:
\begin{align}
m^2_{np}\rightarrow \quad &  \text{Q}^2\frac{m'_{np}-2\tilde k'^{[3]}_p}{m'_{np}+2\tilde k'^{[3]}_p} +\text{Q} \frac{4m'_{np}}{m'_{np}+2\tilde k'^{[3]}_p}(\tilde q^{[1]}-\tilde k'^{[1]}_p)+m_{np}'^{2}+\nonumber \\
&+\frac{m_{np}'^{2}}{m_{np}'^{2}-4\tilde k'^{[3]2}_p}4\left( \tilde q^{[1]} (\tilde q^{[1]} -2\tilde k'^{[1]}_p)+\tilde q^{[2]} (\tilde q^{[2]}-2\tilde k'^{[2]}_p)\right),
\end{align}
where the indices within brackets in $\tilde k_p^{(\prime)}$ and $\vec{\tilde q}$ indicate the coordinate.
It is remarkable that for both expressions, $\vec{\tilde k}_p$ and $m^2_{np}$, 
one recovers the corresponding
expressions in the case of one-body currents \cite{Biernat:2011mp} 
by setting $\vec{\tilde q}=0$.
The momentum transfer $\vec{\tilde q}$ is completely arbitrary, i.e. all its components can be
different from zero. 
If one considers the photon-nucleon vertex, which has four components, 
each of 
them depending
 on the initial and final spin projections, one sees that the $\vec{\tilde q}$ enters only  the 
first and second spatial components.

Another relevant simplification concerns the propagators in 
Eq.~(\ref{EqExchangeCurrentsqJp}). The combinations of the propagators 
are different depending on 
the time ordering. To obtain the transition amplitude that results from leading-order 
covariant perturbation theory is not as simple as in the case of the one-body current. 
However, in the infinite-momentum frame one sees that all 
terms coming from the different time-orderings acquire the same form and 
the propagators contain only the momentum transfers of the photon and $\sigma$-meson.
This can be understood if one keeps in mind that in the infinite-momentum frame the binding energy 
becomes negligible in comparison with the kinetic energies and thus,
$\omega_{k_D}\sim \omega_{k_p} +\omega_{k_n}$. 

To compute the physical deuteron form factors it will be only necessary to consider 
the 0- and 2-components of the current (see Sec.~\ref{SecVectorCurrent} and 
Refs.~\cite{Biernat:2011mp,Carbonell:1998rj}). 
For point-like nucleons the 0-component of the exchange current $J^{0,\text{ex}}_{\mu_D'\mu_D}$ 
simplifies in the  $\kappa_D\to\infty$ limit, to
(cf.~App.~\ref{LimitsInfMomentumFrame} to see the general expression for all components of the current):

\begin{align}
\lim_{\kappa_D\to \infty}J^{0,\text{ex}}_{\mu_D'\mu_D} =& 2\kappa_D\int \frac{d^3 \tilde k'_n}{4\pi} \int d^3\tilde q \sqrt{\frac{m_{np}}{m'_{np}}}
g^2_\sigma \left(\frac{1}{\omega_\sigma}\right)^2 \frac{1}{\text{Q}+\omega_\sigma} \nonumber\\
&\times \;\mathcal{S}_{\mu_D'\mu_D}\; \text{u}^*_{D}(|\vec{\tilde k}_p'|)
\text{u}_{D}(|\vec{\tilde k}_p|), 
\end{align}
with the spin factor 
\begin{align}
 \mathcal{S}_{\mu_D'\mu_D}&:=\sum_{\tilde \mu_n\tilde \mu_p \tilde \mu_n'\tilde \mu_p'} 
C^{1\mu_D'}_{\frac{1}{2}\tilde \mu_p'\frac{1}{2}\tilde \mu_n'}\; D^{1/2}_{\tilde\mu_{p}'\tilde\mu_{p}}
\left[ R_W^{-1}\left( \frac{\tilde k_p'}{m_p},B_c\left( v'_{np}\right)\right) R_W\left( \frac{\tilde k_p}{m_p},B_c\left( v_{np}\right)\right)\right] \nonumber\\
&\quad\times C^{1\mu_D}_{\frac{1}{2}\tilde \mu_p\frac{1}{2}\tilde \mu_n}
D^{1/2}_{\tilde\mu_{n}'\tilde\mu_{n}}\left[ R_W^{-1}\left( \frac{\tilde k_n'}{m_n},B_c\left( v_{np}'\right)\right)
 R_W\left( \frac{\tilde k_n}{m_n},B_c\left( v_{np}\right)\right)\right].\nonumber\\
\end{align}

\subsubsection{Form factors}
The extraction of the form factors has to be done from the most general
covariant decomposition of the current 
(cf.~Sec.~\ref{SecVectorCurrent} and Ref.\cite{Biernat:2011mp}). 
The most general covariant decomposition requires 11 form factors, 4 of them 
disappear completely in the infinite-momentum frame. The exchange-current contributions to the
three physical form 
factors $F_1^{\text{ex}}$, $F_2^{\text{ex}}$ and $G_M^{\text{ex}}$ can 
be uniquely extracted from the matrix elements
$J^{0,\text{ex}}_{11}$, $J^{0,\text{ex}}_{1-1}$ and $J^{2,\text{ex}}_{11}$ (cf. Section~\ref{SecVectorCurrent}
and Ref.~\cite{Biernat:2011mp}). 
The expressions for the form factors obtained in this way are then:
\begin{eqnarray}
 F_1^{\text{ex}}(\text{Q}^2)&:=&\lim_{\kappa_D\to\infty}f_1^{\text{ex}}(\text{Q}^2,\kappa_D)
=-\lim_{\kappa_D\to\infty}\frac{1}{2\kappa_D}\left(J^{0,\text{ex}}_{11}+J^{0,\text{ex}}_{1-1}\right)\nonumber\\
         &=& \int \frac{d^3 \tilde k'_n}{4\pi} \int d^3\tilde q \sqrt{\frac{m_{np}}{m'_{np}}}
g^2_\sigma \left(\frac{1}{\omega_\sigma}\right)^2 \frac{1}{\text{Q}+\omega_\sigma} \nonumber\\
&&\times \;(\mathcal{S}_{11}+\mathcal{S}_{1-1})\; \text{u}^*_{D}(|\vec{\tilde k}_p'|)
\text{u}_{D}(|\vec{\tilde k}_p|),
\end{eqnarray}

\begin{eqnarray}
 F_2^{\text{ex}}(\text{Q}^2)&:=&\lim_{\kappa_D\to\infty}f_2^{\text{ex}}(\text{Q}^2,\kappa_D)
=-\frac{1}{\eta}\lim_{\kappa_D\to\infty}\frac{1}{2\kappa_D}J^{0,\text{ex}}_{1-1}\nonumber\\
&=&\int \frac{d^3 \tilde k'_n}{4\pi} \int d^3\tilde q \sqrt{\frac{m_{np}}{m'_{np}}}
g^2_\sigma \left(\frac{1}{\omega_\sigma}\right)^2 \frac{1}{\text{Q}+\omega_\sigma} \nonumber\\
&&\times \;\mathcal{S}_{1-1}\; \text{u}^*_{D}(|\vec{\tilde k}_p'|)
\text{u}_{D}(|\vec{\tilde k}_p|).
\end{eqnarray}
The form factor $G_M$ is 
\begin{eqnarray}
  G_M^{\text{ex}}(\text{Q}^2)=-\frac{i}{\text{Q}}\lim_{\kappa_\alpha\to \infty}J^{2,\text{ex}}_{11}.
\end{eqnarray}
with $\eta=\frac{\text{Q}^2}{4m_D^2}$.
Now we are in the position to start numerical studies.
 We will provide  
numerical results for the relativistic wave function, deuteron mass and binding energy and 
leave the detailed numerical analysis of form factors  and contributions of
the exchange currents to future work. 
\section{Numerical results: the bound-state problem}

\begin{figure}[h]
\begin{center}
 \includegraphics[width=0.9\textwidth]{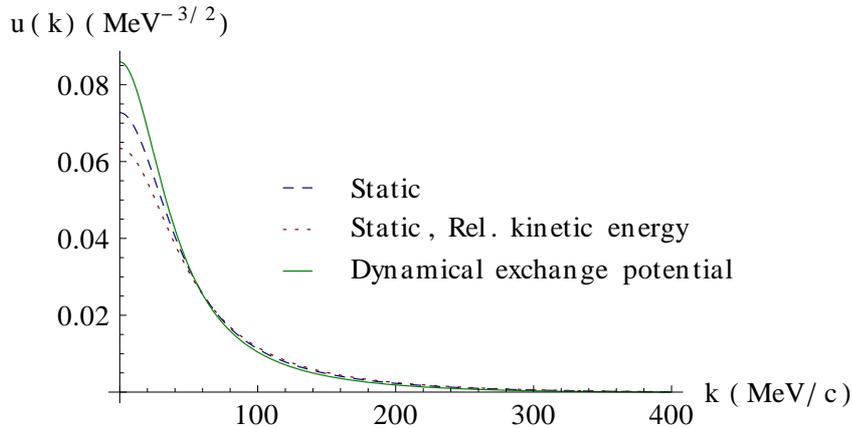}
\caption{Solutions of the mass eigenvalue equation for the Walecka-type model 
in the non-relativistic approximation, where the kinetic energy is nonrelativistic and 
$\sigma$- and $\omega$-exchange are taken in the 
static limit~\cite{Biernat:2011mp,Bakker:2010zz} (dashed line). It is to be compared
 with the same static 
approximation but relativistic kinetic energy~\cite{Biernat:2011mp,Bakker:2010zz} 
(dotted line) and with the final result 
obtained within this work, were the kinetic energy is relativistic, the $\omega$-exchange
is still static, but the 
 $\sigma$-meson exchange is dynamical (solid line).}\label{WFComparisons}
\end{center}
\end{figure}

The numerical solution of the integral equation (\ref{EqEigenvalueDeuteronIntegral})  
for the deuteron wave function $\text{u}_D(k)$ is
 shown in 
Fig.~\ref{WFComparisons}. 
The binding energy and the deuteron mass, together with results from previous work using  
the nonrelativistic or static approximations
are quoted in Table~\ref{BindingEnergyMassDeuteron}.

\begin{table}[h]
\hspace{-1cm}\begin{tabular}{r|c c}
    & $E_B$ & $m_D$ \\
\hline\hline
Nonrelativistic approx.~\cite{Bakker:2010zz,Biernat:2011mp}  & -2.224575 MeV &  1875.61 MeV \\
Relativistic kinetic energy + static potential~\cite{Bakker:2010zz,Biernat:2011mp}  & -2.73414 MeV &  1875.44 MeV \\
  Dynamical $\sigma$-exchange potential (this work)        &  - 1.73192 MeV & 1876.1 MeV\\ 
\hline
\end{tabular}\caption{Binding energies and deuteron masses in the Walecka-type model
for the static approximation of $\sigma$- and $\omega$-exchange~\cite{Biernat:2011mp} (first and second rows)
with non-relativistic and relativistic 
kinetic energies, respectively. The third row corresponds to the 
dynamical treatment of the binding $\sigma$-exchange.}\label{BindingEnergyMassDeuteron}
\end{table}
The first row in Table~\ref{BindingEnergyMassDeuteron} corresponds 
to the nonrelativistic approximation, where the parameters of the 
Walecka model were calculated in such a way that the experimental values for the binding 
energy and the scattering length were reproduced. 
Replacing the kinetic energies of the nucleons by the relativistic ones, and keeping the static
approximation for the interactions, one obtains the 
strongest binding energy, shown in the second row of Table~\ref{BindingEnergyMassDeuteron}.
The calculation carried out in this work shows the relativistic effects that are due to 
the relativistic kinetic energies as well as the 
retardation of the $\sigma$-exchange (cf.~Eq.~(\ref{EqEigenvalueDeuteronIntegral})). 
The nucleons are less bound than in the nonrelativistic case, with the absolute 
value of the binding energy being
approximately 22\% smaller.  
The difference between our result and the one in the second row of 
Table~\ref{BindingEnergyMassDeuteron} is the relativistic 
effect that is exclusively due
to the retardation of the meson exchange, which reduces the binding energy by about 37\%.

The $\omega$-exchange has been considered in the static approximation, 
as was done in previous calculations. One can guess from the previous results
that the dynamical $\omega$-exchange is less repulsive than the
static approximation, since in the nonrelativistic limit both, $\sigma$- and $\omega$-terms, have the
same structure with opposite signs. 
It is therefore expectable
that treating also the $\omega$-exchange dynamically leads again to a stronger binding. 
Nevertheless, we are not yet in the position to make a definitive conclusion since 
$\omega$-exchange includes also important spin effects.
Regarding the wave function, as one can see,
 the strongest binding leads to the broadest wave function. This is 
reasonable, since the average constituent momentum is larger for strong binding than for weak binding.

%% file: Ch9.tex
\chapter{Front-form chiral multiplets}\label{ChFFChiralMultiplets}

In this chapter we will present the angular-momentum decomposition of chiral multiplets 
in front form. This is an interesting application of the 
Clebsch-Gordan coefficients of the Poincar\'e group. We provide the unitary 
transformation that relates  the $q\bar q$ chiral basis $\{R;IJ^{PC}\}$ with the 
nonrelativistic-inspired $\{I;^{2S+1}L_J\}$ classification scheme in 
a front-front form framework. 
The following discussion can also be found in Ref.~\cite{GomezRocha:2012yk}.

In relativistic
composite systems the 
internal degrees of freedom transform among themselves nontrivially
under rotations~\cite{Keister:1991sb}.
Dirac's forms of 
dynamics~\cite{Dirac:1949cp}, which are defined according to the way how the interactions
enter, lead to the definition of different kinds of spin bases. In relativistic quantum mechanics, 
any kind of spin is fully defined by a certain 
type of boost (cf. App.~\ref{Spins}). In this sense, one can speak about ``forms'' of dynamics even 
in the context of free systems or where the interactions can be neglected,
refering to a particular basis. 
In particular, we refer to \textit{spin bases}, on which spin states can be defined.  
We present an  example in which the three most 
important kinds of spin, i.e. \text{canonical spins}, \textit{helicity spins} and \textit{front-form spins},
come into play in the angular momentum decomposition of chiral states.
This example shows 
the nontriviality of the description of the angular momentum 
in relativistic composite systems, even in absence of interactions.

\section{Motivation}
It has been shown in Ref.~\cite{Glozman:2007at} that there is a unitary transformation that relates 
the $q\bar q$ chiral basis, usually represented as  $\{R; I J^{PC} \}$, 
and the $\{I; $ $^{2S+1}L_J \}$ basis, which regards the spin-orbit angular momentum coupling used  
in nonrelativistic quantum mechanics.
Here $R$ is the index of the chiral representation 
($R=(0,0)$, $(1/2,1/2)_a$, $(1/2,1/2)_b$ or $(0,1) + (1,0)$), $I$ the quantum number of isospin, and
$J^{PC}$ indicates the total angular momentum of the state with definite parity and charge.
This allows one to write a particular state belonging to 
a chiral multiplet with quantum numbers $J^{PC}$ as a superposition
of states of the nonrelativistically-inspired
$\{I; ^{2S+1}L_J\}$ classification scheme.

A chiral state with definite parity $|R; I J^{PC}\rangle $ 
can be decomposed  as a superposition of helicity states without definite parity 
$|J\lambda_1\lambda_2\rangle$  
 through  \cite{Glozman:2007at,Glozman:2007ek}
\begin{equation}\label{Leonid1}
|R;IJ^{PC}\rangle =\sum_{\lambda_1 \lambda_2}\sum_{i_1 i_2} 
\chi^{RPI}_{\lambda_1 \lambda_2} C^{Ii}_{(1/2)i_1 (1/2)i_2} 
|i_1\rangle |i_2\rangle |J\lambda_1\lambda_2\rangle,
\end{equation}
where $i_{1(2)}$ and $\lambda_{1(2)}$ are individual isospins and  helicities respectively. 
The coefficients $\chi^{RPI}_{\lambda_1 \lambda_2}$ relate the helicity basis to 
the chiral  basis with definite parity in the state. They can be found in Refs.
 \cite{Glozman:2007at,Glozman:2007ek}.
$C^{JM}_{s_1\sigma_1 s_2\sigma_2}$ are the 
usual $SU(2)$ Clebsch-Gordan coefficients.

Two-particle helicity states $ |J\lambda_1\lambda_2\rangle $ can be written in terms of vectors 
in the $\{I; ^{2S+1}L_J\}$ basis once one knows the expression for the matrix elements \cite{Landau}
\begin{equation}\label{Leonid3}
\langle J\lambda_1 \lambda_2| ^{2S+1}L_J\rangle = 
\sqrt{\frac{2L+1}{2J+1}}C^{S\Lambda}_{(1/2)\lambda_1(1/2)-
\lambda_2}C^{J\Lambda}_{L0S\Lambda}.
\end{equation}
It represents the angular momentum coupling of a two-particle state with individual
helicities $\lambda_1$,  $\lambda_2$ (with $\Lambda=\lambda_1-\lambda_2$; this should
not be confused with the Lorentz-transformation matrix used in previous chapters) to 
a system of total spin $S$ and orbital 
angular momentum $L$. 

Combining Eqs.~(\ref{Leonid1}) and (\ref{Leonid3}) one finds
\begin{eqnarray}\label{chiral:to:LS}
|R;IJ^{PC}\rangle &=& \sum_{LS}\sum_{\lambda_1 \lambda_2}
\sum_{i_1 i_2}\chi^{RPI}_{\lambda_1 \lambda_2}C^{Ii}_{(1/2)i_1(1/2)i_2} 
|i_1\rangle |i_2\rangle \nonumber  \\
&&\times \sqrt{\frac{2L+1}{2J+1}}C^{S\Lambda}_{(1/2)\lambda_1 (1/2)-\lambda_2}
C^{J\Lambda}_{L0S\Lambda}|^{2S+1}L_J\rangle .
\end{eqnarray}
As an example, the $\rho$-like state 
which belongs to the chiral multiplets $|(0,1)+(1,0);11^{--}\rangle$ and 
$|(1/2,1/2)_b ;11^{--}\rangle$ can be represented as \cite{Glozman:2007at}
\begin{eqnarray}
|(0,1)+(1,0);11^{--}\rangle =\sqrt{\frac{2}{3}}|1;^3S_1\rangle +\sqrt{\frac{1}{3}}|1;^3D_1\rangle,\label{rho1} \\
|(1/2,1/2)_b;11^{--}\rangle =\sqrt{\frac{1}{3}}|1;^3S_1\rangle -\sqrt{\frac{2}{3}}|1;^3D_1\rangle.\label{rho2}
\end{eqnarray}

Since both, the chiral and $^{2S+1}L_J$ representations are complete for two-particle 
systems  with
the quantum numbers $I, J^{PC}$, the angular momentum expansion is uniquely determined 
for each
chiral state.

 Chiral symmetry imposes strong restrictions on the spin and angular momentum 
 distribution of a system. 
Such a decomposition has been used in Ref.~\cite{Glozman:2009rn} 
to test the chiral-symmetry breaking of the
 $\rho$ meson in the infrared, and at the same time, to 
reconstruct its spin and orbital angular momentum content in terms of partial waves.
This was achieved by using interpolators that transform according to 
$|(0,1)+(1,0);11^{--}\rangle$ and $|(1/2,1/2)_b;11^{--}\rangle$.
If chiral symmetry were not broken there would be only two possible chiral states in the meson; while chiral symmetry
breaking would imply a superposition of both. 
The obtained result in Ref.~\cite{Glozman:2009rn}
indicates that the $q\bar q$ component of the $\rho$-meson in the infrared is indeed a superposition of the 
$|(0,1)+(1,0);11^{--}\rangle$ and $|(1/2,1/2)_b;11^{--}\rangle$ chiral states, and therefore chiral 
symmetry turns out to be broken. By using the transformations
(\ref{rho1}) and (\ref{rho2}) the partial wave content can be extracted, leading for the 
particular case of
 the $\rho$ meson to a nearly pure $^3S_1$ state~\cite{Glozman:2009rn}. 
This is an example for a physical application of the angular-momentum decomposition given in
Eq.~(\ref{chiral:to:LS})
(see also Refs.~\cite{Glozman:2007at,Glozman:2009bt}). 

It is however, not the aim of this work to
 discuss  problems in which the chiral basis or its transformation
can play a role as was done in Refs.~\cite{Glozman:2007at,Glozman:2009rn} or~\cite{Glozman:2009bt}, for instance. 
The problem we want to address here
 is more technical and related to the transformation (\ref{chiral:to:LS})
itself.
The unitary transformation (\ref{chiral:to:LS}) was obtained in the 
instant 
form of relativistic quantum mechanics. 
In this work we
investigate the corresponding expression one should use in the context of 
 approaches that use light-front 
quantization~\cite{Brodsky:1997de} or front-form relativistic quantum mechanics~\cite{Keister:1991sb}.
We pose the question whether the transformation  (\ref{chiral:to:LS}) is identical
 in any other form~\cite{Dirac:1949cp,Keister:1991sb}
or if it is a special feature of those that use canonical spin, 
such as the instant- or the point-forms. 
 The problem is not trivial since
in relativistic composite systems the internal degrees of freedom transform among themselves 
nontrivially under rotations~\cite{Keister:1991sb}. Relativity mixes spatial and temporal
components and, as a consequence, one is not allowed to treat boosts and 
angular momentum separately in general. The election of a particular representation matters and in some cases
some of the symmetry properties of the Poincar\'e group  
might not be manifest. 
The front form is of special interest, since rotations do not form a subgroup of the kinematical 
group
and rotational invariance is not manifest. 
On the other hand, front-form boosts form a subgroup 
of the Poincar\'e group and, as a result, the front-form Wigner rotation becomes the identity~\cite{Keister:1991sb}. 

In this work we will  show that the unitary transformation derived in Ref.~\cite{Glozman:2007at}
in instant form
is indeed identical in the front form of relativistic quantum mechanics. 
The argument resides in the fact that the \textit{generalized
Melosh rotation} that transforms front-form spins to helicity ones, becomes the identity 
when the mass goes to zero \cite{Diehl:2003ny,Soper:1972xc,Lepage:1980fj}. 

\section{Instant-form decomposition}

Due to rotational and translational invariance in 
nonrelativistic quantum mechanics,
the angular momentum coupling of two particles with individual spin and 
orbital angular momentum $(\vec s_1,\vec l_1)$ and
$(\vec s_2,\vec l_2)$ to a composite system of total spin and 
orbital angular momentum $(\vec S,\vec L)$ is easily realized by using 
the $SU(2)$ Clebsch-Gordan coefficients.
Relativity involves, however, a change of representation 
in which the single-particle momenta and spins are replaced by an overall
system momentum and internal angular momentum \cite{Keister:1991sb}. It is customary to 
use the Clebsch-Gordan coefficients of the Poincar\'e group \cite{Keister:1991sb}.

The kind  of spin vector can be fully determined through 
the choice of a certain type of boost (cf.~App.~\ref{Spins}).
Canonical boosts are rotationless. 
Spin vectors defined through canonical boosts have the advantage 
that in the center-of-momentum frame they transform under rotations
 in the same way as in nonrelativistic quantum mechanics and therefore for a composite system 
one can find a direct decomposition in terms of $SU(2)$ Clebsch-Gordan coefficients. 
The reason is that 
in the canonical case
the Wigner rotation associated with a pure rotation turns out
to be the rotation itself~\cite{Keister:1991sb}.
This does not hold in general. In the front form 
an angular momentum decomposition in terms of
Clebsch-Gordan coefficients requires additional transformations.

Expression (\ref{Leonid3}) can be achieved in a straightforward manner in 
the instant-form of dynamics, as well as 
in any other form that uses canonical spin. 
The derivation of (\ref{Leonid3}) can be found in Ref. \cite{Landau}.
 We will reproduce it here in a basis for the two-particle representation space of 
 the Poincar\'e group
 in order to be able to refer the most important steps when we go to the analogous decomposition in 
front form in the next
section. 
We decompose the spin part of a two-particle state with total 
canonical angular momentum $J$ and $\hat z$-component $M$, orbital angular momentum 
$L$ and total spin $S$, in terms of quantum numbers of the constituents in the 
center-of-momentum frame ($ \vec P = \vec 0$), where the relative momentum is expressed as
$ \vec k= \vec k_1=-\vec k_2$,

\begin{equation}\label{LandauFormel}
|[LS]|\vec k| J; \vec 0\, M\rangle =
\sum_{M_LM_S}\sum_{\sigma_1\sigma_2}\int d\hat{k}\,|\vec k \sigma_1 -\vec k\sigma_2\rangle 
 C^{SM_S}_{s_1\sigma_1 s_2\sigma_2}
 Y_{LM_L}(\hat{ k})\; C^{JM}_{LM_LSM_S},
\end{equation}
where $|\vec k \sigma_1 -\vec k\sigma_2\rangle  := |\vec k \sigma_1\rangle |-\vec k\sigma_2\rangle$, 
$s_{1(2)}$ and $\sigma_{1(2)}$ are the individual 
canonical spins and their $\hat z$-projections, respectively, and $\hat k=\vec k/|\vec k|$. 

Given a particular direction $\hat{ n}$, the tensor product state can be written as
\begin{equation}
 \langle \hat{ n}|\vec k \sigma_1 -\vec k\sigma_2\rangle :=\psi_{s_1\sigma_1}(\vec k) 
\psi_{s_2\sigma_2}(-\vec k)\delta^2(\hat{  k} - \hat{ n} ),
\end{equation}
and one can introduce the wave function
\begin{eqnarray}
 \psi_{JLSM}(\vec k)&:=&\langle \hat{ n}|[LS]|\vec k| J; \vec 0\, M\rangle  \\
&=&\sum_{M_LM_S}\sum_{\sigma_1\sigma_2}\psi_{s_1\sigma_1}(\vec k) 
\psi_{s_2\sigma_2}(-\vec k)
 C^{SM_S}_{s_1\sigma_1 s_2\sigma_2} Y_{LM_L}(\hat{k})\; C^{JM}_{LM_LSM_S}.\nonumber
\end{eqnarray} 
In order to express $ \psi_{JLSM}(\vec k)$ in terms of 
helicities one needs to transform states with canonical spin to a basis of states with 
helicity spin. 
The unitary transformation that provides this is a Wigner rotation 
whose argument corresponds to the angle between the $z$-axis and the direction of motion 
$\hat{ k}:=\vec k/|\vec k|$
 \begin{eqnarray}\label{c:to:h1}
 &&\psi_{s_1\sigma_1}(\vec k)=\sum_{\lambda_1}D^{(s_1)}_{\lambda_1\sigma_1}(\hat{ k})\psi_{s_1 \lambda_1}(\vec k),\\
 &&\psi_{s_2\sigma_2}(-\vec k)=\sum_{\lambda_2}D^{(s_2)}_{-\lambda_2\sigma_2}(\hat{ k})\psi_{s_2 -\lambda_2}(\vec k).
 \end{eqnarray}
Inserting these relations into Eq.~(\ref{LandauFormel}) one gets
  \begin{eqnarray}\label{helicity:decomp}
  \psi_{JLSM}(\vec k)
  &=&\sum_{M_SM_L} \sum_{\sigma_1\sigma_2}\sum_{\lambda_1\lambda_2}D^{(s_1)}_{\lambda_1\sigma_1}(\hat{ k})\psi_{s_1 \lambda_1} (\vec k)
  D^{(s_2)}_{-\lambda_2\sigma_2}(\hat{ k})\psi_{s_2 -\lambda_2}(\vec k)\; \nonumber\\
  &&\times Y_{LM_L}(\hat{ k})\;  
  C^{SM_S}_{s1\sigma_1 s2\sigma_2}\; C^{JM}_{LM_LSM_S}.
 \end{eqnarray}

It is now convenient to write the spherical harmonics in 
terms of Wigner $D$-functions\footnote{Our notation differs from Ref.~\cite{LandauQM} 
by a factor $i^L$ in the definition of the 
phase of the spherical harmonics.}
\begin{equation}
Y_{LM_L}(\hat{ k})= \sqrt{\frac{2L+1}{4\pi}}D^{L}_{0M_L}(\hat{ k})
\end{equation}
in such a way that one can make use of the relation for the product of Wigner $D$-functions 
with the same argument for axially symmetric systems~\cite{LandauQM}, 
\begin{equation}\label{D:same:arguments}
 D^{(j_1)}_{\text{m}_1'\text{m}_1}(\hat{w})D^{(j_2)}_{\text{m}_2'\text{m}_2}(\hat{w}) 
=\sum_{j} C^{j\text{m}'}_{j_1\text{m}_1'j_2 \text{m}_2'}D^{(j)}_{\text{m}\text{m}'}(\hat{w}) C^{j\text{m}}_{j_1\text{m}_1j_2 \text{m}_2}, 
\end{equation}
with $\text{m}=\text{m}_1+\text{m}_2$,  $\text{m}'=\text{m}_1'+\text{m}_2'$, and $\hat{w}$ 
accounting for the Euler angles.
This leads to 
\begin{eqnarray}\label{total:expansion}
\psi_{JLSM}(\vec k)&=&\sum_{\lambda_1 \lambda_2}  \sqrt{\frac{2J+1}{4\pi}}D^{J}_{\Lambda M_J}
 (\hat{k})\psi_{s_1\lambda_1}(\vec k) \psi_{s_2-\lambda_2}(\vec k)\nonumber \\
&&\times \sqrt{\frac{2L+1}{2J+1}}C^{S\Lambda}_{s_1 \lambda_1 s_2 -\lambda_2} 
C^{J\Lambda}_{L0S\Lambda}.
\end{eqnarray}

The fact that the Wigner $D$-functions in Eq.~(\ref{helicity:decomp}) have the same argument 
is a particular feature of the instant form and it
is restricted to the rest frame~\cite{Keister:1991sb}.

It is now easy to identify the needed matrix elements as
  \begin{equation}\label{sinCG}
\psi_{JLSM}(\vec k)=\sum_{\lambda_1 \lambda_2}\psi_{JM\lambda_1\lambda_2}(\vec k)
\langle JM\lambda_1\lambda_2 
  |^{2S+1}L_JM\rangle,
  \end{equation}
with
 \begin{equation}
  \psi_{JM\lambda_1\lambda_2} (\vec k):=  \sqrt{\frac{2J+1}{4\pi}}D^{J}_{\Lambda M_J} (\hat{ k})\psi_{s_1 \lambda_1}(\vec k)\psi_{s_2 -\lambda_2}(\vec k)
 \end{equation}
and 
 \begin{equation}\label{final}
 \langle JM\lambda_1\lambda_2 |^{2S+1}L_JM\rangle =
 \sqrt{\frac{2L+1}{2J+1}}C^{S\Lambda}_{s_1 \lambda_1 s_2 -\lambda_2} C^{J\Lambda}_{L0S\Lambda}.
 \end{equation}
This permits the translation from two-particle helicity states with total angular momentum $J$, 
to a 
state of overall orbital angular momentum $L$ and intrinsic spin $S$. The connection with chirality
is immediately given by Eq.~(\ref{Leonid1}).

\section{Front-form decomposition}
Equation (\ref{LandauFormel}) describes the angular momentum decomposition of a representation of 
canonical spin into a superposition of representations with canonical spin. 
Because in the front form rotations do not form a subgroup of the kinematical group of
the Poincar\'e group, 
 the decomposition (\ref{LandauFormel}) is not feasible \textit{a priori}.
In order to analyze the coupling of two representations with individual spin to a 
superposition of representations with total spin for an arbitrary case in relativistic 
quantum mechanics, it is necessary to use a consistent expression of the Clebsch-Gordan 
coefficients of the Poincar\'e group~\cite{Keister:1991sb}.
Front-form angular momentum coupling is well known and it has been widely applied 
to hadron and nuclear problems in front-form relativistic quantum mechanics. A relation of the type  
(\ref{chiral:to:LS}), however, has not been established yet in the front form. 
This is the aim of the present 
section.

In the following we will use the normalization and notation of  Ref.~\cite{Keister:1991sb}.
The light-front 
components of the four-momentum are defined by 
$\vec{\tilde p}:=(p^+=p^0+p^3,\vec p_\perp=(p^1,p^2))$, $p^-=p^0-p^3$.
$|  \vec{\tilde p} \mu \rangle_f$ represents a single particle state belonging to the front-form 
basis (labeled by $f$), with $\hat z$-spin projection $\mu$. 
The expression 
for the Clebsch-Gordan coefficients of the Poincar\'e group in the front form 
for an arbitrary frame is given by~\cite{Keister:1991sb}:
\begin{eqnarray}\label{ff:ClebschGordan}
&&  _f\langle \vec{\tilde p}_1 \mu_1 \vec {\tilde p}_2 \mu_2 | \,[LS]\,|\vec k| J; 
\vec{\tilde P} \,M\rangle_f 
 \nonumber\\
&&\quad=\delta(\vec {\tilde P} - \vec {\tilde p}_1- \vec {\tilde p}_2 )\; 
\frac{1}{|\vec k|^2}
\delta(|\vec k (\vec{\tilde p}_1,\vec{\tilde p}_2)|-|\vec k|)
\left|\frac{\partial (\vec{\tilde P},\vec{k})}{\partial(\vec{\tilde p}_1,\vec{\tilde p}_2)}\right|^{1/2}\nonumber \\
&&\quad\times \sum_{\sigma_1\sigma_2} D^{(s_1)}_{\mu_1\sigma_1} 
[\text R_{fc}(\vec k, m_1)]D^{(s_2)}_{\mu_2\sigma_2} [\text R_{fc}(-\vec k,m_2)]\nonumber \\
&&\quad\times Y^L_{M_L}(\hat{\vec k})C^{SM_S}_{s_1\sigma_1 s_2 \sigma_2} C^{JM}_{LM_L SM_S},
\end{eqnarray}
where
$_f\langle \vec{\tilde p}_1 \mu_1 \vec {\tilde p}_2 \mu_2|$ 
represents a tensor-product state of two particles with individual momenta 
$ \vec{\tilde p}_1$ and $ \vec{\tilde p}_2$ and spin $\hat z$-projections 
$\mu_1$ and $\mu_2$, respectively. 
The system of two particles moves with a total front-form  momentum $ \vec{\tilde P}$ 
and the individual spins couple to give a total angular momentum $J$ with orbital and 
spin contributions $[LS]$ in the rest frame in the canonical form, and total angular momentum projection 
on the $\hat z$-direction, $M$. Finally, 
 $ \vec k=\vec k_1=-\vec k_2$
is used to denote the individual momenta in the rest frame in the canonical form, 
and $m_1$ and $m_2$ denote the individual constituent masses (they should not 
be confused with the spin projections, which appear in Roman in equation (\ref{D:same:arguments})). 
The arguments of the Wigner $D$-functions are Melosh rotations
which transform states with canonical spin to states with front-form spin and vice versa. Note that
 the rotation depends on the mass in general, producing a different effect on each constituent. Unless
we are dealing with a system of identical constituent masses (e.g. the chiral case),
we will not be able to use the properties of the $D$-function with the same
argument (cf. Eq.~\ref{D:same:arguments}) as was done in the instant form. 

The Clebsch-Gordan coefficient (\ref{ff:ClebschGordan}) is consistent with the normalization 
condition for single states
\begin{equation}
 _f \langle \vec{\tilde p}' \mu' |  \vec{\tilde p} \mu \rangle_f =
\delta(\vec{\tilde p}-\vec{\tilde p}')\,\delta_{\mu\mu'}
\end{equation}
and for state vectors of overall momentum $\vec{\tilde P}$
\begin{eqnarray}
&& _f \langle [L'S']\,|\vec k'|J'; \vec {\tilde P}' \,M' |\, [LS]\,|\vec k|J; \vec {\tilde P} \,M\rangle_f \\
 &&\quad \quad= \delta_{M'M}\delta_{J'J}\delta_{L'L}\delta_{S'S}\delta(P'^+-P^+) \delta^2(\vec P'_\perp-\vec P_\perp)\frac{1}{|\vec k|^2}\delta(|\vec k|-|\vec k'|).\nonumber
\end{eqnarray}

The problem now is to couple a state of total front-form angular momentum $J$ 
and spin projection $M$, $| [LS]|\vec k|, J; \vec {\tilde P} \,M\rangle_f $, 
to a tensor-product state of two particles with individual spins described in 
terms of helicities $ _h\langle \vec{ p}_1 \lambda_1 \vec { p}_2 \lambda_2 |$.

Irreducible representations with different types of spin are related 
to each other through a unitary transformation~\cite{Keister:1991sb}.
The unitary transformation that relates helicity spin to front-form spin  becomes the identity
for massless particles~\cite{Diehl:2003ny,Soper:1972xc,Lepage:1980fj}. This means:
\begin{equation}\label{h:to:f}
  |\vec{\tilde p}_1 \mu_1 \vec{\tilde p}_2\mu_2\rangle_f\stackrel{m\to 0} {\longrightarrow}\sum_{\lambda_1\lambda_2} 
 |\vec{\tilde p}_1 \lambda_1 \vec{\tilde p}_2\lambda_2\rangle_h \delta_{\lambda_1\mu_1}\delta_{\lambda_2\mu_2},
\end{equation}
where the subindex $h$  labels helicity states. Light-cone spins and 
helicity spins coincide in the chiral limit and
one is allowed to make use of them without distinction. 
Using (\ref{h:to:f}) in (\ref{ff:ClebschGordan}), one obtains the Clebsch-Gordan coefficient
that couples two-particle helicity states to an overall state of the front-form basis, 
\begin{eqnarray}\label{ff:helicity:final}
&&  _h\langle \vec{\tilde{ p}}_1 \lambda_1 \vec{\tilde { p}}_2 \lambda_2 |\, [LS]\,|\vec k |J;
 \vec{\tilde P} \,M\rangle_f \nonumber\\
&&\quad =\delta(\vec{\tilde P} - \vec{\tilde p}_1- \vec{\tilde p}_2 )\; 
\frac{1}{|\vec k|^2}\;
\delta(|\vec k (\vec{\tilde p}_1,\vec {\tilde p}_2)|-|\vec k|) \left|\frac{\partial (\vec{\tilde P},\vec{k})}{\partial(\vec{\tilde p}_1,\vec{\tilde p}_2)}\right|^{1/2}\nonumber \\
&&\quad\times \sum_{\sigma_1\sigma_2} D^{(s_1)}_{\lambda_1\sigma_1} [\text R_{hc}(\hat{k})]D^{(s_2)}_{\lambda_2\sigma_2} [\text R_{hc}(-\hat{k})]\nonumber \\
&&\quad\times Y^L_{M_L}(\hat{\vec k})C^{SM_S}_{s_1\sigma_1 s_2\sigma_2}C^{JM}_{LM_LS M_S}. 
\end{eqnarray}
Now the Melosh rotations  
$D^{(s_1)}_{\lambda_1\sigma_1} [\text R_{hc}(\hat{ k})]$ and $D^{(s_2)}_{\lambda_2\sigma_2} 
[\text R_{hc}(-\hat{ k})]$ are equivalent to the Wigner rotations and they only depend on the 
direction of $ \vec k$.
They have exactly the same significance as 
in Eq.~(\ref{helicity:decomp}): they transform canonical spins into helicity spins.
We are now in the position to write the expression for the state in which we are interested:
\begin{eqnarray}\label{state:ff}
 | \,[LS]\,|\vec k| J;
 \vec {\tilde P} \,M\rangle_f 
&=&\sum_{\lambda_1\lambda_2} \int d^3 \tilde p_1 d^3 \tilde p_2 
| \vec{\tilde p}_1 \lambda_1 \vec {\tilde p}_2 \lambda_2 \rangle_{h} \nonumber\\
&&\times _h\langle \vec{\tilde p}_1 \lambda_1 \vec{\tilde p}_2 \lambda_2 |\, [LS]\,|\vec k|,J;
 \vec{\tilde P} \,M\rangle_f ,
\end{eqnarray}
where 
$\mathds{1}=\sum\int d^3\tilde p_1 d^3\tilde p_2| \vec{\tilde p}_1 \lambda_1 
\vec{\tilde p}_2 \lambda_2 \rangle_{h} \, _h\langle \vec{\tilde p}_1 \lambda_1 \vec{\tilde p}_2 \lambda_2 |$
has been introduced.

Going now to the center-of-momentum frame, 
$ \vec{\tilde P}= \vec{\tilde 0}:=
(2p^0,0,0,0)$, $\vec{\tilde p}_1=-\vec{\tilde p}_2=\vec{\tilde k}$, we have
\begin{eqnarray}
| \,[LS]\,|\vec k|J;
 \vec {\tilde 0} \,M\rangle_f 
&=&\sum_{\lambda_1\lambda_2} \sum_{\sigma_1\sigma_2} \int d\hat{\vec k} |\vec k\lambda_1 -\vec k \lambda_2\rangle  \nonumber\\
&&\times D^{(s_1)}_{\lambda_1\sigma_1} [\text R_{hc}(\hat{\vec k})]D^{(s_2)}_{\lambda_2\sigma_2} [\text R_{hc}(-\hat{\vec k})]\nonumber \\
&&\times Y^L_{M_L}(\hat{\vec k})C^{SM_S}_{s_1\sigma_1 s_2\sigma_2}C^{JM}_{LM_LS M_S}. 
\end{eqnarray}
Choosing a particular direction of relative motion $ \hat n$, the integral over 
$d\hat{ k}$ can be carried out by means of 
\begin{equation}
 \langle \hat n |\vec k\lambda_1 -\vec k \lambda_2\rangle:= 
\psi_{s_1\lambda_1} (\vec k)\psi_{s_2\lambda_2} (-\vec k) 
\delta(\hat{ k}-\hat{ n})
\end{equation}
and we define
\begin{eqnarray}\label{ff:JLMS}
 \psi_{JLSM}(\vec k)&:=& \langle \hat n|\,[LS]\,|\vec k|J;
 \vec{\tilde 0} \,M\rangle_f \nonumber\\
&=&\sum_{\lambda_1\lambda_2} \sum_{\sigma_1\sigma_2} \psi_{s_1\lambda_1} (\vec k)\psi_{s_2-\lambda_2} (\vec k)  D^{(s_1)}_{\lambda_1\sigma_1} [\text R_{hc}(\hat{ k})]D^{(s_2)}_{-\lambda_2\sigma_2} [\text R_{hc}(\hat{ k})]\nonumber \\
&&\times Y^L_{M_L}(\hat{ k})C^{SM_S}_{s_1\sigma_1 s_2\sigma_2}C^{JM}_{LM_LS M_S}. 
\end{eqnarray}

Treating  
the spherical harmonics and the Wigner 
$D$-functions in the same way as in the previous section one obtains

\begin{equation}
\psi_{JLSM}(\vec{k})=\sum_{\lambda_1 \lambda_2}\psi_{JM\lambda_1\lambda_2}(\vec{k})\langle JM\lambda_1\lambda_2 
|^{2S+1}L_JM\rangle
\end{equation}
with
\begin{equation}\label{ff:sinCG}
 \psi_{JM\lambda_1\lambda_2} (\vec{k}):=  \sqrt{\frac{2J+1}{4\pi}}D^{J}_{\Lambda M_J} 
(\hat{k})\psi_{s_1\lambda_1} (\vec k)\psi_{s_2-\lambda_2} (\vec k),
\end{equation}
and 
\begin{equation}\label{repeated:final}
\langle JM\lambda_1\lambda_2 |^{2S+1}L_JM\rangle =
\sqrt{\frac{2L+1}{2J+1}}C^{S\Lambda}_{s_1 \lambda_1 s_2 -\lambda_2} C^{J\Lambda}_{L0S\Lambda}.
\end{equation}
Having found (\ref{repeated:final}), the validity of decomposition
(\ref{chiral:to:LS}) is demonstrated. 

Unlike in instant form, the combination 
of the Wigner $D$-functions 
would not have been possible if we had 
considered particles 
of different masses. Only in the chiral limit, 
or for equal masses,
the eigenstates in the rest frame transform in the same way as 
in nonrelativistic quantum mechanics. 
Note that in general the coupling (\ref{ff:ClebschGordan}) 
involves rotations that depend on the masses, namely 
$D^{(s_1)}_{\mu_1\sigma_1} [\text R_{fc}(\vec k, m_1)]$ and 
$D^{(s_2)}_{\mu_2\sigma_2}[\text R_{fc}(-\vec k,m_2)]$. 
This would have prevented the application of Eq.~(\ref{D:same:arguments}), 
since the $D$-functions would not have the same arguments, 
and the dependence on the masses would have entered the decomposition, making it
impossible to write Eq.~(\ref{ff:JLMS}) in the form of a product of (\ref{ff:sinCG}) 
and (\ref{repeated:final}). Moreover, a further rotation would have been necessary in order to 
transform front-from spins to helicity spins, which in the chiral limit 
turns out to be trivial by means of (\ref{h:to:f}).

The result is that the decomposition (\ref{chiral:to:LS}) can also be 
used to expand chiral states as a
superposition of vectors of the $\{I;^{2S+1}L_J\}$-basis within a front-form framework. 
They can be expressed as
\begin{eqnarray}\label{ffchiral:to:LS}
 |R; IJ^{PC}\rangle_f 
&=& \sum_{LS}\sum_{\lambda_1 \lambda_2}
\sum_{i_1 i_2}\chi^{RPI}_{\lambda_1 \lambda_2}C^{Ii}_{(1/2)i_1(1/2)i_2} 
|i_1\rangle |i_2\rangle \nonumber  \\
&&\times \sqrt{\frac{2L+1}{2J+1}}C^{S\Lambda}_{(1/2)\lambda_1 (1/2)-\lambda_2}
C^{J\Lambda}_{L0S\Lambda}|^{2S+1}L_J\rangle_f.
\end{eqnarray}

To summarize, we have derived the unitary transformation 
that relates the $q\bar q$ chiral basis to the 
$\{I; ^{2S+1}L_J\}$-basis in a front-form framework. 
The result turns out to be the same as
in instant form~\cite{Glozman:2007at}.

Spin vectors belonging to different representations can be related through a unitary 
transformation~\cite{Keister:1991sb}. 
We have used the feature of the generalized Melosh rotation 
that relates helicity and 
front-form spins, which becomes the identity for massless particles. 
The limit $m\to 0$ eliminates the mass-dependence in the Wigner $D$-functions 
making it possible to
express the product of $D$-functions with the same argument through a Clebsch-Gordan series for 
axially symmetric systems.
 This simplifies the 
the Clebsch-Gordan coefficient of the Poincar\'e group to an
easier expression, in terms of $SU(2)$ Clebsch-Gordan coefficients.

As a last remark, let us also mention that it would have been possible to develop such a 
decomposition for any type of spin. The Clebsch-Gordan
coefficients of  the Poincar\'e group for an arbitrary form are given in Ref.~\cite{Keister:1991sb}.
Proceeding in an analogous way as before, it is possible to see 
that again the Wigner $D$-functions do not have the same argument, and it 
is not possible to bring them together to an overall rotation by means
of $SU(2)$ Clebsch-Gordan coefficients. 
Only in the chiral limit the rotations are again the same. 
In general, a further transformation on such arbitrary spins  
into helicity spins is necessary to establish the relation to chirality.

%% file: Ch10.tex
\chapter{Summary and conclusions}\label{ChConclusions}

\subsubsection{Heavy-light systems with instantaneous confining forces}
The first goal of this project was to extend and generalize the previous work on the
electromagnetic structure of spin-0 and spin-1 two-body bound states
consisting of equal-mass particles~\cite{Biernat:2010tp,Biernat:2009my,Biernat:2011mp}
to unequal-mass constituents and to
weak decay form factors in the time-like momentum transfer region.  
Working within the point form of relativistic quantum mechanics and
using a constituent-quark model with instantaneous confining forces
we have derived electroweak current matrix elements and (transition)
form factors for heavy-light mesons in the space- and time-like
momentum-transfer regions. 
Starting point of this derivation was a
multichannel formulation of the physical processes in which these
form factors are measured, i.e electron-meson scattering and
semileptonic weak decays. This formulation accounts fully for the
dynamics of the exchanged gauge boson ($\gamma$ or $W$). Poincar\'e
invariance is guaranteed by adopting the Bakamjian-Thomas
construction with gauge-boson-fermion vertices taken from quantum
field theory. Vector and axial-vector currents of the mesons can
then be uniquely identified from the one-boson-exchange ($\gamma$ or
$W$) amplitudes. These currents have already the right
Lorentz-covariance properties and the electromagnetic current of any
pseudoscalar meson is conserved. But wrong cluster properties,
inherent in the Bakamjian-Thomas construction~\cite{Keister:1991sb},
give rise to spurious dependencies of the electromagnetic current on
the electron momenta. For pseudoscalar mesons these unwanted
dependencies are eliminated by taking the invariant mass of the
electron-meson system large
enough~\cite{Biernat:2010tp,Biernat:2009my,Biernat:2011mp}. The
resulting electromagnetic form factor of a pseudoscalar meson is
then equivalent to the one obtained in front form from the
$+$-component of a one-body current in a $q^+=0$ frame. The weak
pseudoscalar-to-pseudoscalar and 
pseudoscalar-to-vector transition currents are not plagued by such
spurious contributions. They can be expressed in terms of physical
covariants and form factors with the form factors depending on the
(time-like) momentum transfer squared, as it should be. In front
form one observes some frame dependence of the $B\rightarrow D^\ast$
decay form factors if they are extracted from the $+$-component of a
simple one-body current~\cite{Cheng:1996if}. This is attributed to a
missing non-valence ($Z$-graph) contribution, which makes the triangle
diagram, from which the form factors are calculated,
covariant~\cite{Cheng:1996if,Bakker:2003up}. In the case of the
point form it is, of course, also not excluded that $Z$-graphs may
play a role, but they are not necessary to ensure covariance of the
current, since Lorentz boosts are purely kinematical and thus do not
mix in higher Fock states.

Having derived comparably simple analytical expressions for the
electromagnetic  form factor of a pseudoscalar heavy-light meson and
the $B\rightarrow D^{(\ast)}$ decay form factors we discussed the
heavy-quark limit. We found that the decay form factors (multiplied
with appropriate kinematical factors) go over into one universal
function, the Isgur-Wise function, as demanded by heavy-quark
symmetry. For the electromagnetic form factors we observed that the
heavy-quark limit does not completely remove the spurious dependence
on the electron momentum. One still has a spurious covariant and the
$s$-dependence of the form factors goes over into a dependence on
the (common) modulus of the incoming and outgoing 3-velocities of
the heavy meson. This dependence on the modulus of the meson
velocities vanishes by taking it large enough. In the limit of
infinitely large meson velocities we found a rather simple
analytical expression for the Isgur-Wise function which turned out
to be (apart from a change of integration variables) the same as the
expression which we got from the decay form factors. Interestingly,
we have also got the same result for the Isgur-Wise function for the
minimum value of the meson velocities that is necessary to reach a
particular value of $v\cdot v^\prime$ (the
argument of the Isgur-Wise function). For minimum velocities it is
not possible to separate physical and spurious contributions since
the respective covariants become proportional. The dependence of the
electromagnetic pseudoscalar meson form factor on Mandelstam-$s$ and
the dependence of the resulting Isgur-Wise function on the modulus
of the meson velocities may be interpreted as a frame dependence of
the $\gamma^\ast M\rightarrow M$ subprocess. The $s\rightarrow
\infty$ (velocities $\rightarrow\infty$) limit corresponds to the
infinite-momentum frame, whereas minimum $s$ (minimum velocities)
corresponds to the Breit frame. Our finding thus means that it does
not matter whether we calculate the Isgur-Wise function in the
infinite-momentum frame or the Breit frame. In the heavy-quark limit
the results are the same and agree with the heavy-quark limit of the
decay form factors. Numerical agreement was also found with the
front-form calculation of Ref.~\cite{Cheng:1996if}.

As a first application and numerical check of our approach we have
calculated electromagnetic $D^+$- and $B^-$ form factors, the
$B\rightarrow D^{(\ast)}$ decay form factors and the Isgur-Wise
function with a simple (flavor independent) Gaussian wave function.
For the electromagnetic $B^-$ form factor and for the $B\rightarrow
D^{(\ast)}$ decay form factors the effect of heavy-quark symmetry
breaking due to finite physical masses of the heavy quarks turned
out be $15-20\%$. For the electromagnetic $D^+$ form factor it
rather amounted to about $60\%$.

Discrepancies between the point and front-form approach show up 
as soon as the decay form factors are calculated for finite, 
physical masses of the heavy-quarks. We have also applied our 
formalism to several heavy-to-heavy and heavy-to-light transition  
form factors using a flavor-dependent Gaussian wave function 
and compared our results with the front-form calculation of 
Ref.~\cite{Cheng:1996if}, with identical parameters. 
For finite quark masses differences between the front-form and 
the point-form approach are observed.
These differences increase with decreasing quark masses.
Most likely, they can be attributed to the different roles played 
by $Z$-graphs, i.e. non-valence contributions, in either approach. 
In the heavy-quark limit $Z$-graphs do not contribute to the current, 
neither in the front form nor in the point form, which explains why the
results agree for the Isgur-Wise function. For finite quark masses, 
however, the inclusion of $Z$-graphs seems to be crucial for 
the frame independence of the decay form factors in front form, 
whereas it is not the case in the point form (as discussed above).

Our approach is general enough to deal with additional dynamical 
degrees of freedom, such that it is possible to consider 
non-valence Fock-state contributions in the meson~\cite{Kleinhappel:2011is}. 
It will be the topic of future work to investigate the role
of $Z$-graph contributions, which indeed can be easily 
accommodated within our multichannel approach. It could help to 
reduce the spurious dependencies of the electromagnetic current on 
the electron momenta. It might explain, e.g., the discrepancy
between the electromagnetic form factors calculated in the 
infinite-momentum frame and in the Breit frame for the $B^-$ and $D^+$ mesons 
(see Fig.~\ref{BreitInfinityQ2}). 

To conclude this part, we have presented a relativistic point-form formalism
for the calculation of the electroweak structure of heavy-light
mesons within constituent quark models with instantaneous confining
forces. This formalism provides the electromagnetic form factor of
pseudoscalar heavy-light systems for space-like momentum transfers
and weak pseudoscalar-to-pseudoscalar as well as pseudoscalar-to-vector 
decay form factors for time-like momentum transfers. It exhibits the 
correct heavy-quark-symmetry properties in the heavy-quark limit. 

\subsubsection{Dynamical binding forces}
Another goal of this thesis was to generalize the point-form approach 
to systems that are bound by dynamical particle-exchange.
We have considered relativistic effects coming from the retardation of a dynamical 
particle-exchange interaction. We have investigated such retardation 
effects for electron-deuteron scattering taking a Walecka-type model 
for the deuteron, where the binding is caused by $\sigma$- and $\omega$-exchanges. 
With the approximation that only the $\sigma$-exchange is considered dynamically, 
whereas the $\omega$-exchange is still taken in the static limit, 
we have obtained the relativistic wave function for the deuteron and we have 
studied the relativistic effects that modify the binding energy as compared with  
the static approximation of the $\sigma$-exchange ~\cite{Biernat:2011mp,Bakker:2010zz}. 
The retardation of the $\sigma$ reduces the binding energy and enhances 
the small-$k$ part of the wave function. The next goal would be to examine 
how large the effects of exchange currents to the form factors are and whether 
they restore (partly) the cluster properties and reduce the spurious form factors 
as sometimes suspected.

\subsubsection{Chiral multiplets in relativistic quantum mechanics}
Because in relativistic composite systems the internal degrees of freedom transform among
themselves nontrivially under rotations, we have emphasized the importance of considering the 
Clebsch-Gordan coefficients of the Poincar\'e group in the spin coupling of relativistic 
composite systems~\cite{Keister:1991sb}. The Clebsch-Gordan
coefficients of the Poincar\'e group convert, through an intermediate step, 
any kind of spin to canonical spin in the rest frame, in such a way that they 
can be added using $SU(2)$ Clebsch-Gordan coefficients. 

Considering the Clebsch-Gordan coefficients of the Poincar\'e group 
and making use of the known property of the front-form spin, 
that becomes equivalent to helicity in the chiral limit, we have presented the 
angular momentum decomposition of chiral multiplets in an instant-form and 
in a front-form basis. We have shown that such decompositions are identical for massless particles. 
With such a unitary transformation it is possible to relate the $q\bar q$ chiral basis 
$\{R;IJ^{PC}\}$ with the nonrelativistically-inspired $\{I;^{2S+1}L_J\}$ classification scheme
as it was done in Ref.~\cite{Glozman:2009rn} for the purpose of studying the angular 
momentum content of the rho-meson in lattice QCD. Although such decompositions are possible 
in any other spin basis, they are not identical to the both presented here in general,
since an additional rotation is necessary to establish the connection with chirality.

%% file: App1.tex
\chapter{Notation and conventions}

We present here some more details about the conventions and notations used in this work. 
Some basic facts about the Poincar\'e group and the covering group of the Poincar\'e group 
which we need repeatedly are summarized. 
The appendix is not intended as 
detailed explanation on how to construct operators, etc., within relativistic quantum mechanics. 
For a more comprehensive 
presentation we refer to Ref.~\cite{Keister:1991sb}, which we are following.

\section{Notation}

We use the Einstein convention on summation. Thereby repeated indices are implicitly summed 
over ($a^i b^i:=\sum_{i} a^i b^i$ ).  
Repeated Lorentz (Greek) indices are used to express the scalar product in Minkowski space:
\begin{equation}
 a\cdot b := a^\mu b_\mu  = g^{\mu\nu}a_\mu b_\nu =a^0 b^0 -a^i b^i,
\end{equation}
with $\mu,\nu=0,1,2,3$. Latin indices are reserved for the spatial vector components.
The metric tensor $g^{\mu\nu}$ of the Minkowski space is:
\begin{equation}
 g^{00}=-g^{11}=-g^{22}=-g^{33}=1.
\end{equation}
In general, the kinetic energy of a particle with mass $m$ and 3-momentum $\vec k$ is represented as:
\begin{equation}
 \omega_k:=\sqrt{m^2 +\vec k^2}.
\end{equation}

The units we use are such that $\hbar=c=1$.

\subsection{The Poincar\'e group}
A Poincar\'e transformation on a 4-vector $x^\mu$ belonging to the Minkowski space
is represented by
\begin{equation}\label{LorentzTransformation}
 x^\mu\to x'^\mu=\Lambda^\mu_{\phantom{\nu}\nu}x^\nu+ a^\mu,
\end{equation}
where $a^\mu$ is a constant 4-vector that represents a space-time translation and 
$\Lambda^\mu_{\phantom{\nu}\nu}$ is a constant $(4\times 4)$-matrix that 
represents a Lorentz transformation. $\Lambda^\mu_{\phantom{\nu}\nu}$ leaves the metric invariant:
\begin{equation}
 g^{\mu\nu}=\Lambda^\mu_{\phantom{\nu}\rho}\Lambda^\nu_{\phantom{\nu}\sigma} g^{\rho\sigma}.
\end{equation}
The composition of two Poincar\'e transformations is given by
\begin{equation}
 (\Lambda_2,a_2)\circ (\Lambda_1,a_1)=(\Lambda_2 \Lambda_1,\Lambda a_1 +a_2).
\end{equation}
and the inverse and identity are
\begin{equation}
 (\Lambda,a)^{-1}=(\Lambda^{-1},-\Lambda^{-1}a), \qquad I=(\mathds 1,0).
\end{equation}
\subsubsection{Commuting self-adjoint operators}
The only two commuting self-adjoint operators 
that can be constructed as independent
 polynomial functions of the generators of the 
Poincar\'e group, and then be used to label irreducible 
representations of the Poincar\'e group are the 
square of the \textit{mass operator} $ \hat M^2$
 and the square of the \textit{Pauli-Lubansky operator} 
$\hat W^2:=-\hat M^2\hat j^2$, with $\hat j^2$ being the total
spin operator of the system. They are the Casimir operators of the Poincar\'e group and 
are defined as follows:
\begin{eqnarray}
 \hat M:=\sqrt{\hat P^\mu \hat P_\mu}, \qquad 
\hat W^\mu := -\frac{1}{2}\varepsilon^{\mu\alpha\beta\gamma}\hat P_\alpha \hat J_{\beta\gamma},
\end{eqnarray}
with $\hat P^\mu$ being the generators of space-time translations
and $\hat J^{\mu\nu}$ the antisymmetric tensor operator that contains 
the rotations and boost generators, $\hat J^i$ ($=\frac{1}{2}\epsilon^{ijk}\hat J^{jk}$) and
$\hat K^i$ ($=\hat J^{0i}$), respectively.

\subsubsection{Boosts}\label{AppCanonicalBoosts}
There is an infinite number of vector valued functions of the generators that satisfy the 
angular-momentum algebra and thus could serve as spin vectors. 
It is thus necessary to specify the type of spin one refers to. 
The type of spin can be distinguished by how they transform under Lorentz boosts. 
Any kind of spin can be defined by a certain type of boost.

The boost operators $B_g^{-1}(\hat V)^\nu_{\phantom{\nu}\mu}$ (where $\hat V:=\hat P/\hat M$ 
is the four-velocity operator) represent
 Lorentz transformations that map $\hat P^\mu$ to $(\hat M,0,0,0)$ 
and have the properties~\cite{Keister:1991sb}:
\begin{eqnarray}\label{BgDef}
&& B_g(\hat V)^\mu_{\phantom{\nu}\nu}(1,0,0,0)^\nu=\hat M^{-1}\hat P^\mu\\
&&\hat B_g(\hat 1,0,0,0)^\nu_{\phantom{\nu}\mu}=g^\nu_{\phantom{\nu}\mu}
\end{eqnarray}
These relations have to be understood as operator relations. 
The subscript $g$ denotes the type of boost used to define the spin. The most relevant types of
boost are \textit{canonical} boosts, $B_c$, \textit{helicity} boosts, $B_h$, and 
\textit{front-form} boosts, $B_f$. 

Canonical boosts are rotationless and are used in the instant form and in the point form. 
The action of a canonical boost with 4-velocity $v^\mu$ (rapidity $\omega=\text{arcsinh}|\vec v|$, 
$v_\mu v^\mu$=1)
on 4-vectors in Minkowski space is described by the matrix:
\begin{equation}\label{EqBc}
 B_c(v):=\left(\begin{array}{cc} 
                v^0 & \vec v^T\\
	      \vec v & \mathds{1}+\frac{v^0-1}{\vec v^2}\vec v \vec v^T 
               \end{array}\right)
\end{equation}
Helicity boosts are defined as a canonical boost in the $\hat z$-direction to the 
velocity of desired magnitude $|\vec v|$, followed by a rotation to the axis that defines the direction of 
the velocity ($\hat{\vec v}=\vec v/|\vec v|$):
\begin{equation}
 B_h(v)=R(\hat z\to \hat v)B_c(|\vec v|\hat z).
\end{equation}
(the hat here denotes unitary vectors).
Finally, front-form boosts are Lorentz transformations that leave the light-front $x^+=0$ invariant.
They form a subgroup of the Poincar\'e group. 
Introducing light-cone coordinates $a^{\pm}=(a^+\pm a^-)$ and $a_\perp=(a^1,a^2)$, 
the action of a front-form boost on a 4-vector $\tilde a=(a^+,a_\perp,a^-)$, represented in these coordinates, is 
given by:
\begin{equation}
 B_f(v)=\frac{1}{\sqrt{v^+}}\left( \begin{array}{ccc}
v^+& v_\perp^T& 0\\
v_\perp & \mathds 1 & 0\\
v_\perp^2/v^+ & 2v_\perp^T/v^+ & 1/v^+ \end{array}\right)
\end{equation}

\subsubsection{Spin vectors}\label{Spins}
A spin vector of the type $g$ is fully determined by the corresponding kind of boost $g$, i.e. 
\begin{equation}
 (0,\hat{\vec j}_g):=\frac{1}{\hat M}B_g^{-1}(\hat{V})^\mu_{\phantom{\nu}\nu} \hat{W}^\nu.
\end{equation}
Using the inverse of the boost operator it is possible to express $\hat{\vec j}_g$ in terms of 
$\hat W^\nu$. Applying thus another type of boost that maps $\hat P^\mu$ to the rest frame 
(cf. Eq.~\ref{BgDef}) one ends up with another type of spin. 
In this way one finds a relation between
two kinds of spin, which turns out to be:
\begin{equation}
 \hat j^j_a=B^{-1}_a(\hat V)^j_{\phantom{j}\nu}B_b(\hat V)^\nu_{\phantom{j}k} \hat{j}^k_b = 
R_{ab}(\hat V)^j_{\phantom{j}k}\hat j^k_b.
\end{equation}
The rotation $R_{cf}$ that transforms canonical spins $c$ into front-form spins $f$, or
vice versa,  is called \textit{Melosh rotation}~\cite{Melosh:1974cu}. The 
general transformation is customary called \textit{generalized Melosh rotation}.
\newpage
In summary,  the notation and properties of the different types of boost and their associated spins
are given by
\begin{center}
 \begin{tabular}[c]{c| c  c}
Type of boost & Subscript  & Properties \\\hline\hline
canonical & $c$ & rotationless\\
front form & $f$ & they form a subgroup\\
helicity  & $h$ & $  B_h(v)=R(\hat z\to \hat v)B_c(|\vec v|\hat z)$\\
any other & $g$  & -\\ \hline
\end{tabular}\\
\end{center}

\subsection{The covering group of the Poincar\'e group}
Elements of the covering group of the Poincar\'e group 
$ISL(2,\mathds C)$ (or inhomogeneous $SL(2,\mathds C)$) 
are $2\times 2$ matrices ($\ul \Lambda, \ul a$) with 
$\det{(\ul \Lambda)}=1$ and $\ul a$ Hermitian~\cite{Bargmann:1954gh}. 
Whereas the Poincar\'e group acts on 4-vectors belonging 
to the Minkowski space, the covering group of the Poincar\'e group acts on spin vectors 
belonging to the spin space. The relation between both groups can be understood through the 
relation between the space-time coordinate $x^\mu$ and the 
corresponding $2\times 2$ Hermitian matrix~\cite{Keister:1991sb}:
 
 \begin{equation}
  \ul X:=x^\mu \sigma_\mu = \left( 
 \begin{array}{cc}
  x^0+x^3 & x^1 -i x^2\\
x^1 + i x^2 & x^0 -x^3 
 \end{array}
 \right),
 \qquad x^\mu = \frac{1}{2}\text{Tr}(\sigma_\mu \ul X),
 \end{equation}
 where $\sigma_\mu$ are the Pauli matrices (including $\sigma_0=\mathds 1_{2\times 2}$). 
A Poincar\'e transformation is given by
\begin{equation}
 \ul X \to \ul X'=\ul\Lambda\,\ul X\,\ul\Lambda^\dagger + \ul a.
\end{equation}
A composition of two $(\ul \Lambda ,\ul a)$ transformations is then:
\begin{equation}
 (\ul \Lambda_2 ,\ul a_2)\circ(\ul \Lambda_1 ,\ul a_1)
=(\ul \Lambda_2\ul\Lambda_1,\ul\Lambda_2 \ul a_1 \ul\Lambda_2^\dagger +\ul a_2),
\end{equation}
and the inverse:
\begin{equation}
 (\ul\Lambda,\ul a )^{-1}=(\ul \Lambda^{-1},-\ul \Lambda^{-1}\ul a (\ul \Lambda^\dagger)^{-1}).
\end{equation}
The relation between $(\Lambda^\mu_{\phantom{\mu}\nu},a)$ and $(\ul \Lambda,\ul a)$ is given
by:
\begin{equation}
 \Lambda^\mu_{\phantom{\mu}\nu}:= 
\frac{1}{2}\text{Tr}(\sigma_\mu \ul \Lambda\sigma_\nu\ul \Lambda^\dagger), \quad \text{and}
\quad a^\mu=\frac{1}{2}\text{Tr}(\sigma_\mu \ul a).
\end{equation}

Hence,  the canonical boost matrix belonging to the covering group of the Poincar\'e group
is given by:
\begin{equation}
 \ul B_c(v):=\sqrt{\frac{v^0+1}{2}}\sigma^0 +\frac{\vec \sigma\cdot \vec v}{\sqrt{2(v^0+1)}}.
\end{equation}
The Wigner rotation defined by means of canonical boosts is in this case
\begin{equation}
 \ul R_W (v,\ul \Lambda)=\ul B_c^{-1}( \Lambda v)\,\ul \Lambda \,\ul B_c( v).
\end{equation}
In order to simplify the notation, 
since elements of the $ISL(2,\mathds{C})$ can be
expressed as a function of elements of the Poincar\'e group, we have avoided the underline in all
the expressions.
\subsubsection{Polarization vectors}\label{PolVectors}
The polarization vectors $\epsilon^\mu(\vec 0,\sigma)$ of a spin-1 
particle with spin projection $\sigma$ are defined at rest by
\begin{eqnarray}
 \epsilon(\vec 0,0)&:=& (0,0,0,1);\\
 \epsilon(\vec 0,\pm 1)&:=& \mp \frac{1}{\sqrt 2}(0,1,\pm i,0).
\end{eqnarray}
Considering polarization vectors in an arbitrary frame requires to perform a boost, that in our case
is canonical (cf. Eq.~\ref{EqBc}):
\begin{equation}
 \epsilon^\mu(\vec k,\sigma)=B_c(v)^\mu_{\phantom{\nu}\nu}\epsilon^\nu(\vec 0,\sigma).
\end{equation}
\subsection{Field operators}
For the calculation of the vertex matrix elements, e.g. Eq.~(\ref{EqEMVertex}), from the corresponding
interaction Lagrangian densities we need the plane-wave expansions of spin-$\frac{1}{2}$ and spin-1 field
operators.
\subsubsection{Dirac field}
\begin{equation}
 \hat \psi (x)=\sum_{\sigma=\pm \frac{1}{2}}\int \frac{d^3 p}{(2\pi)^3 2\omega_{p}}
\left( e^{ip\cdot x}\; v_\sigma(\vec p)\;\hat d^\dagger_\sigma (\vec p) +
e^{-i p\cdot x}\; u_\sigma(\vec p) \; \hat c_\sigma (\vec p)\right), 
\end{equation}
\begin{equation}
 \hat{\bar \psi} (x)=\sum_{\sigma=\pm \frac{1}{2}}\int \frac{d^3 p}{(2\pi)^3 2\omega_{p}}
\left( e^{-ip\cdot x}\; \bar v_\sigma(\vec p)\;\hat d_\sigma (\vec p) +
e^{i p\cdot x}\;\bar u_\sigma(\vec p) \; \hat c_\sigma^\dagger (\vec p)\right). 
\end{equation}
\subsubsection{Maxwell field}
\begin{equation}
 \hat A^\mu (x)=\sum_{\sigma=0}^3\int \frac{d^3 p}{(2\pi)^3 2\omega_{p}} (-g^{\sigma\sigma})
\left( e^{ip\cdot x}\;\epsilon^\mu(\vec p,\sigma) \hat a^\dagger_\sigma (\vec p) +
e^{-i p\cdot x}\;\epsilon^{*\mu}(\vec p,\sigma) \; \hat a_\sigma (\vec p)\right).
\end{equation}
\subsubsection{Dirac spinors}
Spinors and the Dirac matrices are taken in the Dirac representation:
\begin{equation}
 u_\rho (\vec k) = \sqrt{\omega_k+m}\left( \begin{array}{c} \varsigma_{\rho} \\ \frac{\vec \sigma \cdot ¸\vec k}{\omega_k + m} \varsigma_{\rho}\end{array}\right), \; v_{\rho}=\-\sqrt{\omega_k +m}\left( \begin{array}{c}\frac{\vec \sigma \cdot \vec p}{\omega_k + m} \varepsilon \varsigma_\rho \\
\varepsilon \varsigma_\rho                                                                                                \end{array}\right),
\end{equation}
\begin{equation}
 \varepsilon=i\sigma_2, \quad \varsigma_{\rho}=\left( \begin{array}{c} \frac{1}{2}+\rho \\ \frac{1}{2}-\rho \end{array} \right).
\end{equation}
\subsubsection{Dirac matrices}
\begin{equation}
 \gamma^0 = \left( \begin{array}{cc} \mathds{1} & 0 \\ 0 &-\mathds{1}\end{array}\right), \quad \vec{\gamma}=\gamma^0 \vec \alpha=\left( \begin{array}{cc} 0 & \vec{\sigma} \\ -\vec{\sigma} & 0 \end{array} \right),\quad   
\end{equation}

\begin{equation} 
\gamma^5=\left( \begin{array}{cc} 0 & \mathds{1}  \\ \mathds{1} & 0 \end{array}\right),\quad \vec \alpha = \left( \begin{array}{cc} 0 & \vec \sigma  \\ \vec \sigma & 0 \end{array}\right).
\end{equation}

%% file: App2.tex
\chapter{Matrix elements}\label{AppCoupledChannel}

The following matrix elements are needed for the computation of matrix elements of the optical
potential that describes the one-photon-exchange scattering process given in Eq.~(\ref{eq:voptclust}). 
They are inner products of free and cluster velocity states, consistent with the normalization
given by Eq.~(\ref{eq:vnorm}). $\alpha$ represents the discrete quantum numbers of the $q\bar q$ cluster, i.e. 
$(n,j,\tilde m_j,[\tilde l,\tilde s])$, see also Refs.~\cite{Krassnigg:2003gh,Biernat:2011mp}. 
 For example,
for a pseudoscalar meson we consider $(n,0,0,[0,0])$, and for a vector meson we will have
  $(n,1,\tilde m_j,[0,1])$. This affects the Clebsch-Gordan coefficients, and the subindices in the 
Wigner $D$-functions. For weak decays we will write these quantum numbers explicitly, 
to distinguish pseudoscalar-to-pseudoscalar from pseudoscalar-to-vector transitions.
\vspace{-0.3cm}
\section{Electromagnetic scattering}

For electron scattering off a meson with spin $j$ we have used for the computations 
of Sec.~\ref{SecCupledChannelEM}\footnote{We are not specifying here which of the 
quarks is heavy; we give the general expression for a $q\bar q$ bound state, 
and where we use the notation
$\omega_{k_{q\bar q}}:=\omega_{k_{q}}+\omega_{k_{\bar q}}$.}:
\vspace{-0.5cm}

\begin{eqnarray}
&& \langle v; \vec{k}_e, \mu_e; \vec{k}_q, \mu_q; \vec{k}_{\bar{q}},
\mu_{\bar{q}} \vert\, \underline{v}; \vec{\underline{k}}_e,
\underline{\mu}_e;
\vec{\underline{k}}_\alpha,\underline{\mu}_\alpha, \alpha
\rangle\nonumber\\
&&\quad=(2\pi)^{15/2}\ul v_0\delta^3(\vec v-\ul{\vec{v}})\delta^3(\vec{k}_e-\ul{\vec{k}}_e)\delta_{\mu_e\ul \mu_e}\nonumber\\
&&\qquad \times\sqrt{\frac{2\omega_{\ul k_e}2\omega_{\ul k_\alpha}}{(\omega_{\ul k_e}+\omega_{\ul k_\alpha})^3}}
\sqrt{\frac{2\omega_{k_e}2\omega_{\ul k_{q\bar q}}}{(\omega_{k_e}+\omega_{\ul k_{q\bar q}})^3}}
\sqrt{\frac{2\omega_{\tilde k_q}2\omega_{\tilde k_{\bar q}}}{(\omega_{\tilde k_q}+\omega_{\tilde k_{\bar q}})}}\nonumber\\
&&\qquad \times\sum_{\tilde m_l=-\tilde l}^{\tilde l}\sum_{\tilde m_s=-\tilde s}^{\tilde s}
\sum_{\tilde \mu_q\tilde \mu_{\bar q}=\pm1/2} C^{j\tilde m_j}_{\tilde l\tilde m_l\tilde s \tilde m_s}
C^{\tilde s\tilde m_s}_{\frac{1}{2}\tilde \mu_q\frac{1}{2} \tilde \mu_{\bar q}}\; u_{n\tilde l}(|\vec{ \tilde k}_q|)Y_{\tilde l\tilde m_l}(\hat{\tilde k}_q)\nonumber\\
&&\qquad \times D^{1/2}_{\mu_q\tilde \mu_{q}}\left[R_W\left(\frac{\tilde k_q}{m_q},B_c(v_{q\bar q})\right)\right]
D^{1/2}_{\mu_{\bar q}\tilde \mu_{\bar q}}\left[R_W\left(\frac{\tilde k_{\bar q}}{m_{\bar q}},B_c(v_{q\bar q})\right)\right],\nonumber\\
\end{eqnarray}
 \vspace{-0.95cm}
\begin{eqnarray}
&&\langle v; \vec{k}_e, \mu_e; \vec{k}_q, \mu_q; \vec{k}_{\bar{q}},
\mu_{\bar{q}};\vec{k}_{\gamma}, \mu_{\gamma} \vert\,
\underline{v}; \vec{\underline{k}}_e, \underline{\mu}_e;
\vec{\underline{k}}_\alpha, \underline{\mu}_\alpha;
\vec{\underline{k}}_\gamma, \underline{\mu}_\gamma; \alpha\rangle\nonumber\\
&&\quad=(2\pi)^{21/2}\ul v_0\delta^3(\vec v-\ul{\vec{v}})\delta^3(\vec{k}_e-\ul{\vec{k}}_e)\delta_{\mu_e\ul \mu_e}\delta^3(\vec{k}_\gamma-\ul{\vec{k}}_\gamma)(-g_{\mu_\gamma\ul\mu_\gamma})\nonumber\\
&&\qquad \times\sqrt{\frac{2\omega_{\ul k_e}2\omega_{\ul k_\alpha}2\omega_{\ul k_\gamma}}{(\omega_{\ul k_e}+\omega_{\ul k_\alpha}+\omega_{\ul k_\gamma})^3}}
\sqrt{\frac{2\omega_{k_e}2\omega_{\ul k_{q\bar q}}2\omega_{\ul k_\gamma}}{(\omega_{k_e}+\omega_{\ul k_{q\bar q}}+\omega_{\ul k_\gamma})^3}}
\sqrt{\frac{2\omega_{\tilde k_q}2\omega_{\tilde k_{\bar q}}}{(\omega_{\tilde k_q}+\omega_{\tilde k_{\bar q}})}}\nonumber\\
&&\qquad \times\sum_{\tilde m_l=-\tilde l}^{\tilde l}\sum_{\tilde m_s=-\tilde s}^{\tilde s}
\sum_{\tilde \mu_q\tilde \mu_{\bar q}=\pm1/2} C^{j\tilde m_j}_{\tilde l\tilde m_l\tilde s \tilde m_s}
C^{\tilde s\tilde m_s}_{\frac{1}{2}\tilde \mu_q\frac{1}{2} \tilde \mu_{\bar q}}\; u_{n\tilde l}(|\vec{ \tilde k}_q|)Y_{\tilde l\tilde m_l}(\hat{\tilde k}_q)\nonumber\\
&&\qquad \times D^{1/2}_{\mu_q\tilde \mu_{q}}\left[R_W\left(\frac{\tilde k_q}{m_q},B_c(v_{q\bar q})\right)\right]
D^{1/2}_{\mu_{\bar q}\tilde \mu_{\bar q}}\left[R_W\left(\frac{\tilde k_{\bar q}}{m_{\bar q}},B_c(v_{q\bar q})\right)\right],\nonumber\\
\end{eqnarray}
\vspace{-0.90cm}
{\allowdisplaybreaks
\begin{eqnarray}\label{EqEMVertex} \lefteqn{
\langle v^\prime; \vec{k}_e^\prime, \mu_e^\prime; \vec{k}_q^\prime,
\mu_q^\prime; \vec{k}_{\bar{q}}^\prime, \mu_{\bar{q}}^\prime;
\vec{k}_\gamma^\prime, \mu_\gamma^\prime \vert \,\hat{K}_\gamma^\dagger\, \vert v;
\vec{k}_e, \mu_e; \vec{k}_q, \mu_q; \vec{k}_{\bar{q}}, \mu_{\bar{q}}
\rangle}
\nonumber\\
&=& \langle v;
\vec{k}_e, \mu_e; \vec{k}_q, \mu_q; \vec{k}_{\bar{q}}, \mu_{\bar{q}}  \vert \,\hat{K}_\gamma \, \vert v^\prime; \vec{k}_e^\prime, \mu_e^\prime; \vec{k}_q^\prime,
\mu_q^\prime; \vec{k}_{\bar{q}}^\prime, \mu_{\bar{q}}^\prime;
\vec{k}_\gamma^\prime, \mu_\gamma^\prime
\rangle^* \nonumber\\
&=& v_0 \, \delta^3(\vec{v}^\prime-\vec{v})\,
\frac{(2\pi)^3}{\sqrt{(\omega_{k_e^\prime}+\omega_{k_q^\prime}+
\omega_{k_{\bar{q}}^\prime}+ \omega_{k_{\gamma}^\prime})^3}
\sqrt{(\omega_{k_e}+\omega_{k_q^{\phantom{\prime}}}
+\omega_{k_{\bar{q}}})^3}} \nonumber
\\ & &\times  \langle  \vec{k}_e^\prime, \mu_e^\prime; \vec{k}_q^\prime,
\mu_q^\prime; \vec{k}_{\bar{q}}^\prime, \mu_{\bar{q}}^\prime;
\vec{k}_\gamma^\prime, \mu_\gamma^\prime \vert \,\left(
\hat{\mathcal{L}}_{\mathrm{int}}^{e \gamma}(0) +
\hat{\mathcal{L}}_{\mathrm{int}}^{q \gamma}(0) \right) \vert 
\vec{k}_e, \mu_e; \vec{k}_q, \mu_q; \vec{k}_{\bar{q}}, \mu_{\bar{q}}
\rangle \nonumber\\
&=& v_0 \, \delta^3(\vec{v}^\prime-\vec{v})\,
\frac{(2\pi)^3}{\sqrt{(\omega_{k_e^\prime}+\omega_{k_M^\prime}+
\omega_{k_{\gamma}^\prime})^3}
\sqrt{(\omega_{k_e}+\omega_{k_M^{\phantom{\prime}}})^3}}(-1)\nonumber\\
&&\times \left[ \, Q_e\,
\bar{u}_{\mu_e^\prime}(\vec{k}_e^\prime)\gamma_\nu
u_{\mu_e}(\vec{k}_e)\,
\epsilon^\nu(\vec{k}_{\gamma}^\prime,\mu_{\gamma}^\prime)\, (2
\pi)^3 2 \omega_{k_q} \delta^3(\vec{k}_q^\prime - \vec{k}_q)\delta_{\mu_q\mu'_q} \, 
\right. \nonumber \\
& & \left. \;\;\qquad \times 
(2\pi)^3 2 \omega_{k_{\bar{q}}} \delta^3(\vec{k}_{\bar{q}}^\prime -
\vec{k}_{\bar{q}})\delta_{\mu_{\bar q}\mu'_{\bar q}}\right. \nonumber \\
& & \left. \;\; + \, Q_q\,
\bar{u}_{\mu_q^\prime}(\vec{k}_q^\prime)\gamma_\nu
u_{\mu_q}(\vec{k}_q)\,
\epsilon^\nu(\vec{k}_{\gamma}^\prime,\mu_{\gamma}^\prime)\, 
(2\pi)^3 2 \omega_{k_e} \delta^3(\vec{k}_e^\prime - \vec{k}_e)\delta_{\mu_e\mu'_e} \,
\right. \nonumber \\
& & \left. \;\;\qquad \times (2
\pi)^3 2 \omega_{k_{\bar{q}}} \delta^3(\vec{k}_{\bar{q}}^\prime -
\vec{k}_{\bar{q}}) \delta_{\mu_{\bar q}\mu'_{\bar q}}\right. \nonumber \\
& & \left. \;\; + \, Q_{\bar{q}}\,
\bar{v}_{\mu_{\bar{q}}}(\vec{k}_{\bar{q}})\gamma_\nu
v_{\mu_{\bar{q}}^\prime}(\vec{k}_{\bar{q}})\,
\epsilon^\nu(\vec{k}_{\gamma}^\prime,\mu_{\gamma}^\prime)\, (2
\pi)^3 2 \omega_{k_e} \delta^3(\vec{k}_e^\prime - \vec{k}_e)\delta_{\mu_e\mu'_e} \,
\right. \nonumber \\
& & \left. \;\;\qquad \times (2
\pi)^3 2 \omega_{k_q} \delta^3(\vec{k}_q^\prime - \vec{k}_q)\delta_{\mu_{ q}\mu'_{ q}}
\right]\, ,
\end{eqnarray}}
It is also useful to take into account the relation
\begin{equation}\label{EqChangeTildeNoTilde}
 \frac{d^3k_1}{2\omega_{k_1} 2\omega_{k_2}}=  
\frac{d^3\tilde k_1}{2\omega_{\tilde k_1} 2\omega_{\tilde k_2}}
\frac{(\omega_{\tilde k_1}+\omega_{\tilde k_2})}{(\omega_{ k_1}+\omega_{ k_2})}.
\end{equation}
\vspace{-0.9cm}
\section{Weak decays}

The matrix elements needed for the calculation of matrix 
elements of the optical potential (\ref{EqWeakOptPotential}) are\footnote{Here we
specify the $q\bar q$ bound state wave functions to be pure $s$-wave and either that of 
a pseudoscalar meson as that of a vector meson. }
\begin{eqnarray} 
& &  \langle  \vec{v}'; \vec{k}_c',\mu_c'; \vec{k}_{\bar d}',\mu_{\bar d}'; \vec{k}_W', \mu_W'|
\hat{K}_{c\bar dW\to b\bar d}^\dagger|\vec{v};  \vec{k}_b,\mu_b; \vec{k}_{\bar d},\mu_{\bar d}\rangle  \nonumber \\
&&\quad=\langle \vec{v};  \vec{k}_b,\mu_b; \vec{k}_{\bar d},\mu_{\bar d}|
\hat{K}_{c\bar dW\to b\bar d} | \vec{v}'; \vec{k}_c',\mu_c'; \vec{k}_{\bar d}',\mu_{\bar d}'; \vec{  k}_W', \mu_W'\rangle^*\nonumber\\
& & \quad=v_0'\delta^3(\vec{v}'-\vec{v})\frac{(2\pi)^3}{(\omega_{k_c'}+\omega_{k_{\bar d}'}+ \omega_{k'_W})^{3/2}(\omega_{k_b}+\omega_{k_{\bar d}})^{3/2}}
\delta^3(\vec{k}_{\bar d}'-\vec{k}_{\bar d})(2\pi)^3 2 \omega_{k_{\bar d}} \nonumber \\
& &\qquad \times \epsilon^*_\mu(\vec{k}_W',\mu'_W)\delta_{\mu'_{\bar d} \mu_{\bar d}}\frac{-i e V_{cb}}{\sqrt{2}\sin \vartheta_W}\bar{u}_{\mu'_c}(\vec{k}'_c)\gamma^\mu \frac{(1-\gamma^5)}{2}u_{\mu_b}(k_b),
\end{eqnarray}

\begin{eqnarray}
 & & \langle \vec{v}'; \vec{k}_c',\mu_c'; \vec{k}_{\bar d}',\mu_{\bar d}' ;\vec{  k}_e', \mu_e';\vec{ k}_{\bar \nu_e}', \mu_{\bar \nu_e}'|
\hat{K}_{c\bar d W\to c\bar d e\bar \nu_e}^\dagger|  \vec{v}; \vec{k}_c,\mu_c; \vec{k}_{\bar d},\mu_{\bar d}; \vec{  k}_W ,\mu_W\rangle \nonumber \\
&&  \quad=\langle  \vec{v}; \vec{k}_c,\mu_c; \vec{k}_{\bar d},\mu_{\bar d}; \vec{  k}_W ,\mu_W|
\hat{K}_{c\bar d W\to c\bar d e\bar \nu_e}| \vec{v}'; \vec{k}_c',\mu_c'; \vec{k}_{\bar d}',\mu_{\bar d}' ;\vec{  k}_e', \mu_e';\vec{ k}_{\bar \nu_e}', \mu_{\bar \nu_e}'\rangle^* \nonumber \\
& &\quad =v_0'\delta^3(\vec{v}'-\vec{v})\frac{(2\pi)^3}{(\omega_{k'_c}+\omega_{k_{\bar d}'}+ \omega_{k'_{\bar\nu}}+\omega_{k'_e})^{3/2}(\omega_{k_c}+\omega_{k_{\bar d}}+\omega_{k_W})^{3/2}}  \nonumber \\
& & \qquad\times \delta^3(\vec{k}_{\bar d }'-\vec{k}_{\bar d})(2\pi)^3 2 \omega _{k_{\bar d}}\delta^3(\vec{k}_c'-\vec{k}_c)(2\pi)^3 2 \omega_{k_c}  \epsilon_\mu(\vec{k}_W,\mu_W)\delta_{\mu'_{\bar d}\mu_{\bar d}}\delta_{\mu'_c\mu_c}\nonumber \\
& & \qquad\times \frac{-i e V_{cb}}{\sqrt{2}\sin \vartheta_W}\bar u_{\mu_e'} (\vec k_e')\gamma^\mu \frac{(1-\gamma^5 )}{2}v_{\mu_{\bar\nu}'}(\vec k'_{\bar\nu}).
\end{eqnarray}

\subsubsection{Pseudoscalar meson transitions}

\begin{eqnarray}
& &  \langle\vec{v}; \vec{k}_c, \mu_c; \vec{k}_{\bar d}, \mu_{\bar d}; \vec{  k}_e,\mu_e;\vec{k}_{\bar \nu_e}, \mu_{\bar \nu_e}| \vec{ \ul v};\vec{ \ul k}_e,\ul \mu_e;\vec{\ul k}_{\bar \nu_e},\ul \mu_{\bar \nu_e};\vec{\ul k}_D, \ul n,0,0,[0,0] \rangle  \nonumber \\
& & \quad= (2\pi)^{21/2}\ul v_0\delta^3(\vec{\ul v}-\vec{v})\delta^3(\vec{\ul k}_e-\vec{k}_e)\delta_{\ul \mu_e\mu_e}\delta^3(\vec{\ul k}_{\bar\nu_e}-\vec{k}_{\bar\nu_e})\delta_{\ul \mu_{\bar\nu_e}\mu_{\bar\nu_e}}  \nonumber \\
& & \qquad\times  \sqrt{\frac{2\ul \omega _{k_e} 2 \ul \omega_{k_{\bar\nu}} 2 \ul \omega_{k_D}}{(\ul \omega_{k_e} + \ul \omega_{k_{\bar\nu}} + \ul \omega_{k_D})^3}}\sqrt{\frac{2 \omega_{k_e} 2  \omega_{k_{\bar\nu}} 2 ( \omega_{k_c}+\omega_{k_{\bar d}})}{( \omega_{k_e} +  \omega_{k_{\bar\nu}} +  \omega_{k_c} + \omega_{k_{\bar d}})^3}}\sqrt{\frac{2 \omega_{\tilde k_c} 2  \omega_{\tilde k_{\bar d}}}{2( \omega_{\tilde k_c} +  \omega_{\tilde k_{\bar d}})}}  \nonumber \\
& & \qquad\times \sum_{\tilde \mu_c, \tilde \mu_{\bar d}=\pm \frac{1}{2}} C^{00}_{0000}C^{00}_{\frac{1}{2}\tilde \mu_c\frac{1}{2}\tilde \mu_{\bar u}}\;  u_{n0}( |\vec{ \tilde k}_c|) Y_{00}(\hat {\tilde k}_c) \nonumber\\
&& \qquad\times D^{1/2}_{\mu_c\tilde \mu_c}\left[R_W \left(\frac{\tilde{\vec k}_c}{m_c},B(v'_{c \bar d})\right)\right]D^{1/2}_{\mu_{\bar d}\tilde \mu_{\bar d}}\left[R_W \left(\frac{\tilde{\vec k}_{\bar d}}{m_{\bar d}},B(v'_{c \bar d})\right)\right],  \nonumber \\
\end{eqnarray}
\vspace{-0.5cm}
\begin{eqnarray} 
& &\langle \vec{v}; \vec{k}_c,\mu_c; \vec{k}_{\bar d},\mu_{\bar d}; \vec{  k}_W ,\mu_W|   \vec{\ul v}; \vec{\ul k}_W,\ul\mu_W;\vec{\ul k}_D,\ul n,0,0,[0,0]\rangle  = \nonumber \\
& & \quad= (2\pi)^{15/2}\ul v_0\delta^3(\vec{\ul v}-\vec{ v}) \delta^3(\vec{ k}_W-\vec{\ul k}_W)\left( -g^{ \mu_W \ul \mu_W}\right)  \nonumber \\
& & \qquad\times \sqrt{\frac{2 \ul \omega_{k_W} 2 \ul \omega_{k_D}}{(\ul \omega_{k_W}+\ul \omega_{k_D})^3}}
\sqrt{\frac{2  \omega_{k_W} 2 (\omega_{k_c}+\omega_{k_{\bar d}})}{( \omega_{k_W}+ \omega_{k_c}+\omega_{k_{\bar d}})^3}}\sqrt{\frac{2\omega_{\tilde k_c} 2  \omega_{ \tilde k_{\bar d}}}{2( \omega_{\tilde k_c}+  \omega_{\tilde k_{\bar d}})}} \nonumber \\
& & \qquad\times  \sum_{\tilde \mu_c, \tilde \mu_{\bar d }=\pm \frac{1}{2}} C^{00}_{0000}C^{00}_{\frac{1}{2}\tilde \mu_c\frac{1}{2}\tilde \mu_{\bar d}}\; u_{n0}( |\tilde{\vec k}_c|) Y_{00}(\tilde{\vec k}_c) \nonumber\\
&& \qquad\times D^{1/2}_{\mu_c\tilde \mu_c}\left[R_W \left(\frac{\tilde{\vec k}_c}{m_c},B(v_{c \bar d})\right)\right]D^{1/2}_{\mu_{\bar d}\tilde \mu_{\bar d}}\left[R_W \left(\frac{\tilde{\vec k}_{\bar d}}{m_{\bar d}},B(v_{c \bar d})\right)\right], \nonumber \\
\end{eqnarray} 
\vspace{-0.5cm}
\begin{eqnarray}
& & \langle \vec{v};  \vec{k}_b,\mu_b; \vec{k}_{\bar d},\mu_{\bar d} |\vec{\ul v}; \vec{\ul k}_B,n,0,0,[0,0]\rangle  = \nonumber \\
& & \quad =(2\pi)^{9/2} \ul v_0\delta^3(\vec{ v}-\vec{\ul v})
\sqrt{\frac{2 \ul \omega_{k_B}}{\ul \omega_{k_B}^3}} \sqrt{\frac{ 2 (\omega_{k_b}+\omega_{k_{\bar d}})}{( \omega_{k_b}+\omega_{k_{\bar d}})^3}}\sqrt{\frac{2 \omega_{\tilde k_b} 2  \omega_{\tilde k_{\bar d}}}{2( \omega_{\tilde k_b} + \omega_{\tilde k_{\bar d}})}}  \nonumber \\
& & \qquad\times  \sum_{\tilde \mu_b, \tilde \mu_{\bar u }=\pm \frac{1}{2}} C^{00}_{0000}C^{00}_{\frac{1}{2}\tilde \mu_b\frac{1}{2}\tilde \mu_{\bar d}} \; u_{n0}( |\tilde{\vec k}_b|) Y_{00}(\tilde{\vec k}_b) \nonumber\\
&& \qquad\times D^{1/2}_{\mu_b\tilde \mu_b}\left[R_W \left(\frac{\tilde{\vec k}_b}{m_b},B(v_{b \bar d})\right)\right]D^{1/2}_{\mu_{\bar d}\tilde \mu_{\bar d}}\left[R_W \left(\frac{\tilde{\vec k}_{\bar d}}{m_{\bar d}},B(v_{b \bar d})\right)\right]. \nonumber \\
 \end{eqnarray}

\subsubsection{Vector meson transitions}
This transition concerns the  quantum numbers of the final state $\alpha':=(j'_j, m'_j, [l',s'])$ 
which now has spin-1.
This affects to the Clebsch-Gordan coefficients and the matrix elements are

 \begin{eqnarray}
& & \langle \vec{ \ul v};\vec{ \ul k}_e,\ul \mu_e;\vec{\ul k}_\nu,\ul \mu_\nu,\vec{\ul k}_D , \ul n,1,\ul\mu_D,[0,1]
|\vec{v}; \vec{k}_c,\mu_c; \vec{k}_{\bar d},\mu_{\bar d}; \vec{  k}_e, \mu_e;\vec{ k}_{\bar \nu_e}, \mu_{\bar \nu_e}\rangle  = \nonumber \\
& & \quad= (2\pi)^{21/2}\ul v_0\delta^3(\vec{\ul v}-\vec{v})\delta^3(\vec{\ul k}_e-\vec{k}_e)\delta_{\ul \mu_e\mu_e}\delta^3(\vec{\ul k}_{\bar\nu_e}-\vec{k}_{\bar\nu_e})\delta_{\ul \mu_{\bar\nu_e}\mu_{\bar\nu_e}}  \nonumber \\
& &  \qquad\times  \sqrt{\frac{2\ul \omega _{k_e} 2 \ul \omega_{k_{\bar\nu}} 2 \ul \omega_{k_D}}{(\ul \omega_{k_e} + \ul \omega_{k_{\bar\nu}} + \ul \omega_{k_D})^3}}\sqrt{\frac{2 \omega_{k_e} 2  \omega_{k_{\bar\nu}} 2 ( \omega_{k_c}+\omega_{k_{\bar d}})}{( \omega_{k_e} +  \omega_{k_{\bar\nu}} +  \omega_{k_c} + \omega_{k_{\bar d}})^3}}\sqrt{\frac{2  \omega_{\tilde k_c} 2 \omega_{\tilde k_{\bar d}}}{2( \omega_{\tilde k_c} + \omega_{\tilde k_{\bar d}})}} \nonumber \\
& & \qquad \times  \sum_{\tilde \mu_c, \tilde \mu_{\bar d}=\pm \frac{1}{2}}   C^{1\ul \mu_D}_{\frac{1}{2}\tilde \mu_c\frac{1}{2}\tilde \mu_{\bar d}} \; u_{n0}^*( |\tilde{ \vec k}_c|) Y_{00}^*(\hat {\tilde k}_c)   \nonumber\\
&&  \qquad\times D^{*1/2}_{\mu_c\tilde \mu_c}\left[R_W \left(\frac{\tilde{\vec k}_c}{m_c},B(v_{c \bar d})\right)\right]D^{*1/2}_{\mu_{\bar d}\tilde \mu_{\bar d}}\left[R_W \left(\frac{\tilde{\vec k}_{\bar d}}{m_{\bar d}},B(v_{c \bar d})\right)\right],  \nonumber \\
\end{eqnarray}

\begin{eqnarray}
 & &\langle \vec{v}; \vec{k}_c,\mu_c; \vec{k}_{\bar d},\mu_{\bar d}; \vec{  k}_W, \mu_W
|\vec{\ul v}; \vec{\ul k}_W,\ul\mu_W;\vec{\ul k}_D, \ul n,1,\ul\mu_D,[0,1]\rangle  = \nonumber \\
& &\quad = 
(2\pi)^{15/2}\ul v_0\delta^3(\vec{ v}-\vec{\ul v})\delta^3(\vec{ k}_W-\vec{\ul k}_W)\left( -g^{\ul \mu_W \mu_W}\right)  \nonumber \\
& & \qquad\times  \sqrt{\frac{2 \ul \omega_{k_W} 2 \ul \omega_{k_D}}{(\ul \omega_{k_W}+\ul \omega_{k_D})^3}}\sqrt{\frac{2  \omega_{k_W} 2 (\omega_{k_c}+\omega_{k_{\bar d}})}{( \omega_{k_W} + \omega_{k_c}+\omega_{k_{\bar d}})^3}}\sqrt{\frac{2 \omega_{\tilde k_c} 2 \omega_{\tilde k_{\bar u}}}{2(\omega_{\tilde k_v}+ \omega_{\tilde k_{\bar u}})}} \nonumber \\
& & \qquad\times \sum_{\tilde \mu_c, \tilde \mu_{\bar d }=\pm \frac{1}{2}}  C^{1\ul \mu_D}_{\frac{1}{2}\tilde \mu_c\frac{1}{2}\tilde \mu_{\bar d}}   u_{n0}( |\tilde{\vec k}_c|) Y_{00}(\tilde{\vec k}_c)\nonumber\\
&&\qquad \times D^{1/2}_{\mu_c\tilde \mu_c}\left[R_W \left(\frac{\tilde{\vec k}_c}{m_c},B(v_{c \bar d})\right)\right]D^{1/2}_{\mu_{\bar d}\tilde \mu_{\bar d}}\left[R_W \left(\frac{\tilde{\vec k}_{\bar d}}{m_{\bar d}},B(v_{c \bar d})\right)\right]  \nonumber\\
\end{eqnarray}

\subsubsection{Derivation of the covariant $W$ propagator}\label{AppWeakPropagator}
The two time ordered contributions to the transition amplitude given by the matrix elements of 
the potential (\ref{EqWeakOptPotential}) in the weak decay (sketched in Fig.~\ref{fig:decay}) differ only in the propagators. Their
sum leads to the covariant $W$ propagator. 
Taking into account that $m=\omega_{k_B}=\omega_{k_D'}+\omega_{k_e'}+\omega_{k_{\bar \nu_e}'}$
and that $(\omega_{k_e'}+\omega_{k_{\bar \nu_e}'})^2-\omega_{k_W'}^2=(k'_e+k'_{\bar\nu_e})^2+(\vec k'_e+\vec k'_{\bar\nu_e})^2-k_W'^2-\vec k_W'^2$, 
one obtains:
\begin{eqnarray}
&&\frac{1}{2\omega_{k_W'}}\left( \frac{1}{m-\omega_{k_{D^{(*)}}'}-\omega_{k_W'}}
+\frac{1}{m-\omega_{k_B}-\omega_{k_W'}-\omega_{k_{\bar \nu_e}'}-\omega_{k_e'}}\right)\nonumber\\
&&=\frac{1}{(\omega_{k_e'}+\omega_{k_{\bar \nu_e}'})^2-\omega_{k_W'}^2}=\frac{1}{(k'_e+k'_{\bar\nu_e})^2-m_W^2}.
\end{eqnarray}

 \newpage
\thispagestyle{empty}

%% file: App3.tex
\chapter{Limits and frames}

\section{The heavy-quark limit}\label{limit:details}
\subsection{Boosts}\label{boostLimit}
In the h.q.l. and for the kinematics specified in Eqs.~(\ref{EqKinematicsmv}) and (\ref{limits}) 
the matrices for the canonical boosts $B_c(v^{(\prime)}_\alpha)$ ocurring in Eq.~(\ref{EM:IW:analytical}) are:
\begin{eqnarray}
B_c (v'_{\alpha})&=& \left(\begin{array}{cccc} \sqrt{1+\nu_\alpha^2} & u &0& \sqrt{\nu^2_\alpha-u^2}\\
                                u & 1+\frac{\left( \sqrt{1+\nu_\alpha^2}-1\right)u^2}{\nu_\alpha^2} & 0& \frac{\left(\sqrt{1+\nu_\alpha^2}-1\right)\sqrt{(1+2\nu_\alpha^2-v\cdot v')u}}{\sqrt 2\nu_\alpha^2} \\
0&0&1&0\\
\sqrt{\nu^2_\alpha-u^2} &\frac{\left(\sqrt{1+\nu_\alpha^2}-1\right)\sqrt{(1+2\nu_\alpha^2-v\cdot v')(v\cdot v'-1)}}{2\nu_\alpha^2}&0&\sqrt{1+\nu_\alpha^2}-\frac{\left( \sqrt{1+\nu_\alpha^2}-1\right)u^2}{\nu_\alpha^2}                                \end{array}\right),\nonumber\\
\end{eqnarray}

\begin{eqnarray}
B_c^{-1} (v_{\alpha})= \left(\begin{array}{cccc} \sqrt{1+\nu_\alpha^2} & u & 0 & -\sqrt{\nu^2_\alpha-u^2}\\
                                u & 1+\frac{\left( \sqrt{1+\nu_\alpha^2}-1\right)u^2
 }{\nu_\alpha^2} & 0& -\frac{\left(\sqrt{1+\nu_\alpha^2}-1\right)\sqrt{(1+2\nu_\alpha^2-v\cdot v')u}}{\sqrt 2\nu_\alpha^2} \\
0&0&1&0\\
-\sqrt{\nu^2_\alpha-u^2} &-\frac{\left(\sqrt{1+\nu_\alpha^2}-1\right)\sqrt{(1+2\nu_\alpha^2-v\cdot v')u}}{\sqrt 2\nu_\alpha^2}&0&\sqrt{1+\nu_\alpha^2}-\frac{\left( \sqrt{1+\nu_\alpha^2}-1\right)u^2}{\nu_\alpha^2}                                \end{array}\right).\nonumber\\
\end{eqnarray}
In the infinite-momentum frame, $\nu_\alpha\to\infty$, the product $B_c^{-1}(v_\alpha)B_c(v_\alpha')$,
which we need to relate $\tilde k'_Q$ and $\tilde k_Q$ (cf. Eq.~(\ref{eq:kktildep})) is still well defined:
\begin{equation}
\left(B_c^{-1}(v_\alpha)B_c(v_\alpha')\right)_{\text{IF}}= \left(\begin{array}{cccc} v\cdot v' & 2u &0& 2u^2\\
                                 2u & 1 & 0&2u\\
0&0&1&0\\
-2u^2 & -2u & 0 & 2-v\cdot v'                               \end{array}\right).\nonumber\\
\end{equation}
In the Breit frame, $\nu_\alpha=u$, this product becomes:
\begin{equation}
\left(B_c^{-1}(v_\alpha)B_c(v_\alpha')\right)_{\text{B}}
= \left(\begin{array}{cccc} v\cdot v' & \sqrt{v\cdot v'^2-1} &0& 0\\
                                 \sqrt{v\cdot v'^2-1} & v\cdot v' & 0 & 0\\
0&0&1&0\\
0& 0& 0 & 1                               \end{array}\right).\nonumber\\
\end{equation}

\subsection{Currents and form factors}
\subsubsection{Components of the point-like quark currents in the h.q.l.}\label{compspinorcurrent}

The electromagnetic vertex 
$j^\mu_{\mu_Q'\mu_Q}=\bar{u}_{\mu_Q^\prime}(\vec{k}_Q^\prime)\gamma^\mu u_{\mu_Q}(\vec{k}_Q)$
in the h.q.l. for every possible combination of initial and final spin projection is\footnote{To
simplify the notation the subindex $\alpha$ and the underline has been dropped in $v$ and 
$v'$ in this whole section.}:
\begin{equation}
 j_{\frac{1}{2}\frac{1}{2}} \rightarrow m_Q\left(\begin{array}{c} 
\frac{3+2\nu_\alpha^2 +2 \sqrt{1+\nu_\alpha^2}-v\cdot v'}{1+\sqrt{1+\nu_\alpha^2}} 
\\ 0 \\ i\sqrt{2(v\cdot v' -1)} \\ 2\sqrt{\frac{1}{2}+\nu_\alpha^2-\frac{v\cdot v'}{2}} \end{array} \right),
\quad
 j_{\frac{1}{2}-\frac{1}{2}} \rightarrow m_Q\left(\begin{array}{c} -
\frac{\sqrt{1+2\nu_\alpha^2-v\cdot v'}\sqrt{v\cdot v' -1}}{1+\sqrt{1+\nu_\alpha^2}} 
\\ 0 \\ 0 \\ -\sqrt{2(v\cdot v'-1)} \end{array} \right),
\end{equation}
\begin{equation}
 j_{-\frac{1}{2}\frac{1}{2}} \rightarrow m_Q  \left(\begin{array}{c} 
\frac{\sqrt{1+2\nu_\alpha^2-v\cdot v'}\sqrt{v\cdot v' -1}}{1+\sqrt{1+\nu_\alpha^2}}  
\\ 0 \\ 0 \\ \sqrt{2(v\cdot v'-1)} \end{array} \right),
\quad
 j_{-\frac{1}{2}-\frac{1}{2}} \rightarrow m_Q \left(\begin{array}{c}\frac{3+2\nu_\alpha^2 +
2 \sqrt{1+\nu_\alpha^2}-v\cdot v'}{1+\sqrt{1+\nu_\alpha^2}}\\ 0 \\ -i\sqrt{2(v\cdot v'-1)} 
\\2\sqrt{\frac{1}{2}+\nu_\alpha^2-\frac{v\cdot v'}{2}}\end{array} \right).
\end{equation}

Let us define
\begin{eqnarray}
\mathcal{S}&:=&\sum_{\mu_Q,\mu'_Q} \frac{1}{2}
\frac{\bar u_{\mu'_Q} (\vec k'_Q) \gamma^0 u_{\mu_Q} (\vec k_Q )}{(k_Q+k_Q')^0}
D^{1/2}_{\mu_Q\mu'_Q}  \left[ 
R^{-1}_W \left(\frac{\tilde k_{\bar q}}{m_{\bar q}}, B_c(v_{\alpha})\right) 
R_W \left( \frac{\tilde{k}'_{\bar q}}{m_{\bar q}},B_c(v'_{\alpha}) \right) \right]\nonumber \\
&& =  \frac{1}{2} \bar u_{\frac{1}{2}} (\vec k'_{\bar q}) \gamma^0 u_{\frac{1}{2}} \; 2 \text{Re}\left\{D^{1/2}_{\frac{1}{2}\frac{1}{2}}  \left[ 
R^{-1}_W \left(\frac{\tilde{k}_{\bar q}}{m_{\bar q}}, B_c(v_{\alpha})\right) 
R_W \left( \frac{\tilde{k}'_{\bar q}}{m_{\bar q}},B_c(v'_{\alpha}) \right) \right]\right\},\nonumber\\
\end{eqnarray}
where the property of the $D$-functions,
\begin{equation}
 D^j_{\sigma\sigma'}(R_W)=(-1)^{\sigma'-\sigma}D^{j*}_{-\sigma-\sigma'}(R_W)
\end{equation}
and the observation that $j^0_{\frac{1}{2}\frac{1}{2}}=j^0_{-\frac{1}{2}-\frac{1}{2}}$ and 
$j^0_{\frac{1}{2}-\frac{1}{2}}=-j^0_{-\frac{1}{2}\frac{1}{2}}$ has been used.

\subsubsection{The infinite-momentum frame}
In the infinite momentum frame, i.e. $\nu_\alpha\to \infty$, one obtains 
\begin{equation}
 \mathcal{S}_{IF}= \frac{2\omega_{\tilde k_{\bar q}'} + 2 m_{\bar q} +\sqrt{2}  \tilde{k}'^{1}_{\bar q} \sqrt{v\cdot v'-1}}
{2\sqrt{(\omega_{\tilde k_{\bar q}'}+ m_{\bar q})(m_{\bar q} + \sqrt{2}\tilde{k}'^{1}_{\bar q} \sqrt{v\cdot v'-1} +
\tilde{k}'^{3}_{\bar q}(v\cdot v'-1)+\omega_{\tilde k_{\bar q}'} v\cdot v')}}.
\end{equation}
In the limit $\nu_\alpha\to\infty$ one gets
\begin{equation}
\omega_{\tilde{k}_{\bar{q}}}= 2 \tilde{k}^{\prime 1}_{\bar{q}}\, u +2
\tilde{k}^{\prime 3}_{\bar{q}}\, u^2+
\omega_{\tilde{k}^\prime_{\bar{q}}} (2u^2+1)\, , 
\end{equation}
so that one can write the spin factor in the simple form:
\begin{equation}
\mathcal{S}_{\mathrm{IF}}=\frac{m_{\bar{q}}+\omega_{\tilde{k}^\prime_{\bar{q}}}+\tilde{k}^{
\prime 1}_{\bar{q}}\,
u}{\sqrt{(m_{\bar{q}}+\omega_{\tilde{k}_{\bar{q}}})
(m_{\bar{q}}+\omega_{\tilde{k}^\prime_{\bar{q}}})}}\, .
\end{equation}

\subsubsection{The Breit frame}
Similarly, for the Breit frame, i.e. $\nu_\alpha=u:=\sqrt{(v\cdot v'-1)}$, we have
\begin{equation}
 j_{\frac{1}{2}\frac{1}{2}} \rightarrow m_Q\left(\begin{array}{c} 2 \\ 0 \\ i\sqrt{2(v\cdot v'-1)} \\ 0 \end{array} \right), \qquad
 j_{\frac{1}{2}-\frac{1}{2}} \rightarrow m_Q\left(\begin{array}{c} 0 \\ 0 \\ 0 \\ -\sqrt{2(v\cdot v'-1)} \end{array} \right),
\end{equation}
\begin{equation}
 j_{-\frac{1}{2}\frac{1}{2}} \rightarrow m_Q\left(\begin{array}{c} 0 \\ 0 \\ 0 \\ \sqrt{2(v\cdot v'-1)} \end{array} \right), \qquad
 j_{-\frac{1}{2}-\frac{1}{2}} \rightarrow m_Q\left(\begin{array}{c} 2 \\ 0 \\ -i\sqrt{2(v\cdot v'-1)} \\ 0 \end{array} \right).
\end{equation}
In this case one gets
\begin{equation}
\frac{ \bar{u}_{\mu_Q^\prime}(\vec{k}_Q^\prime)\gamma^0 u_{\mu_Q}(\vec{k}_Q)}{(k_Q+k_Q')^0}=
\delta_{\mu'_Q\mu_Q}\sqrt{\frac{2}{1+v\cdot v'}},
\end{equation}
and the spin factor becomes:
\begin{equation}\label{EqSBraitlong}
\mathcal{S}_{\text{B}}=\frac{\left( \tilde{k}'^{1}_{\bar q} \sqrt{v\cdot v'-1}+
(\omega_{\tilde k_{\bar q}'}+m_{\bar q})
\sqrt{1+v\cdot v'}\right)}{\sqrt{(\omega_{\tilde k_{\bar q}'}
+m_{\bar q})(1+v\cdot v')(m_{\bar q}+
\omega_{\tilde k_{\bar q}'} v\cdot v' + \tilde{k}'^{1}_{\bar q}\sqrt{(v\cdot v')^2 -1})}}.
\end{equation}
Furthermore, in the Breit frame one has
\begin{equation}
\omega_{\tilde{k}_{\bar{q}}}=2 \tilde{k}^{\prime 1}_{\bar{q}}\, u
\sqrt{u^2+1}+ \omega_{\tilde{k}^\prime_{\bar{q}}} (2u^2+1),
\end{equation}
and the spin factor simplifies finally to:
\begin{equation}
\mathcal{S}_{\mathrm{B}}=\frac{m_{\bar{q}}+\omega_{\tilde{k}^\prime_{\bar{q}}}+\tilde{k}^{
\prime 1}_{\bar{q}}\,
\frac{u}{\sqrt{u^2+1}}}{\sqrt{(m_{\bar{q}}+\omega_{\tilde{k}_{\bar{q}}})
(m_{\bar{q}}+\omega_{\tilde{k}^\prime_{\bar{q}}})}}\, .
\end{equation}

\subsubsection{Weak point-like currents}\label{WeakPointLikeCurrents}
For weak pseudoscalar-to-pseudoscalar transitions only the vector part, $\,$
$ \bar{u}_{\mu_b}(\vec{\underline{v}}_D^{\,\prime}) \,\gamma^\nu\,
 u_{\mu_b}(\vec{\underline{v}}_B)$, contributes and the axial-vector term vanishes in the h.q.l.
For every possible combination of initial and final spin projection, and for the particular
kinematics in weak decays the h.q.l. of the current matrix elements becomes:
\begin{eqnarray}
&& j_{\frac{1}{2}\frac{1}{2}}=
m\left( \begin{array}{c} \sqrt{2}\sqrt{1+v\cdot v'} \\ 0\\ 0\\ \sqrt{2}\sqrt{-1+v\cdot v'}\end{array}\right),\quad 
 j_{\frac{1}{2}-\frac{1}{2}}=
m\left( \begin{array}{c} 0 \\ \sqrt{2}\sqrt{-1+v\cdot v'} \\ -i\sqrt{2}\sqrt{-1+v\cdot v'}\\ 0\end{array}\right),\nonumber\\
\end{eqnarray}
\begin{eqnarray}
&& j_{-\frac{1}{2}\frac{1}{2}}=
m\left( \begin{array}{c} 0\\ -\sqrt{2}\sqrt{-1+v\cdot v'}\\ -i\sqrt{2}\sqrt{-1+v\cdot v'}\\ 0\end{array}\right),\quad 
 j_{-\frac{1}{2}-\frac{1}{2}}=
m\left( \begin{array}{c} \sqrt{2}\sqrt{1+v\cdot v'} \\ 0 \\ 0\\ \sqrt{2}\sqrt{-1+v\cdot v'}\end{array}\right).\nonumber\\
\end{eqnarray}
After combining 
with the Wigner $D$-functions one gets for the total spin factor:
\begin{equation}
 \mathcal S_{\text{W}}
\frac{\left( \tilde{k}'^{1}_{\bar q} \sqrt{v\cdot v'-1}+(\omega_{\tilde k_{\bar q}'}+m_{\bar q})
\sqrt{1+v\cdot v'}\right)}{\sqrt{(\omega_{\tilde k_{\bar q}'}+m_{\bar q})(1+v\cdot v')(m_{\bar q}+
\omega_{\tilde k_{\bar q}'} v\cdot v' + \tilde{k}'^{1}_{\bar q}\sqrt{(v\cdot v')^2 -1})}},
\end{equation}
which turns out to be identical to (\ref{EqSBraitlong}), and thus
\begin{equation}
\mathcal{S}_{\mathrm{W}}=\frac{m_{\bar{q}}+\omega_{\tilde{k}^\prime_{\bar{q}}}+\tilde{k}^{
\prime 1}_{\bar{q}}\,
\frac{u}{\sqrt{u^2+1}}}{\sqrt{(m_{\bar{q}}+\omega_{\tilde{k}_{\bar{q}}})
(m_{\bar{q}}+\omega_{\tilde{k}^\prime_{\bar{q}}})}}\, ,
\end{equation}
with
\begin{equation}
\omega_{\tilde{k}_{\bar{q}}}=2 \tilde{k}^{\prime 1}_{\bar{q}}\, u
\sqrt{u^2+1}+ \omega_{\tilde{k}^\prime_{\bar{q}}} (2u^2+1).
\end{equation}

\section{Extraction of form factors}\label{ApInfBreit}
Here we elucidate why the infinite-momentum frame and the Breit frame are particularly
convenient for determining the electromagnetic form factor of a pseudoscalar meson.
\subsection{The infinite-momentum frame}
In the infinite-momentum frame the kinematics for electron-meson scattering 
(cf. Eq.~(\ref{eq:momentumscatt})) is such that
\begin{equation}
 (\underline{p}_\alpha+\underline{p}^\prime_\alpha)=
2\kappa_\alpha\left( 1\;,\;0\;,\;0\;,\;1\right),
\end{equation}
and
\begin{equation}
 (\underline{p}_e+\underline{p}^\prime_e)=2\kappa_\alpha\left(1\;,\;0\;,\;0\;,\;-1\right).
\end{equation}

Since $(\underline{p}_\alpha+\underline{p}^\prime_\alpha)^0= 
(\underline{p}_e+\underline{p}^\prime_e)^0=
(\underline{p}_\alpha+\underline{p}^\prime_\alpha)^3= 
-(\underline{p}_e+\underline{p}^\prime_e)^3$,
it is not possible to distinguish two different form factors in the 
decomposition (\ref{covestructureinf}). One has instead:
\begin{eqnarray}
\tilde{J}_{\infty}^0(\vec{\underline{p}}_\alpha^\prime;\vec{\underline{p}}_\alpha)
&=& (\underline{p}_\alpha+\underline{p}^\prime_\alpha)^0 \,
\left(f(\text{Q}^2,s)+g(\text{Q}^2,s) \right)\nonumber \\
&=:& (\underline{p}_\alpha+\underline{p}^\prime_\alpha)^0 \,
\tilde{\text{f}}(\text{Q}^2,s),
\end{eqnarray}
\begin{eqnarray}
\tilde{J}_{\infty}^3(\vec{\underline{p}}_\alpha^\prime;\vec{\underline{p}}_\alpha)
&=& (\underline{p}_\alpha+\underline{p}^\prime_\alpha)^3 \,
\left(f(\text{Q}^2,s)-g(\text{Q}^2,s) \right)\nonumber \\
&=:& (\underline{p}_\alpha+\underline{p}^\prime_\alpha)^0 \,
\tilde{\tilde{\text{f}}}(\text{Q}^2,s),
\end{eqnarray}
where we have introduced $\tilde{\text{f}}$ and $\tilde{\tilde{\text{f}}}$ for the sum and the difference of the 
physical and spurious form factors, respectively. 
Adding both equations one gets
\begin{eqnarray}
 \tilde{J}_{[\alpha]}^0(\vec{\underline{p}}_\alpha^\prime;\vec{\underline{p}}_\alpha)+
\tilde{J}_{[\alpha]}^3(\vec{\underline{p}}_\alpha^\prime;\vec{\underline{p}}_\alpha)&=&
(\underline{p}_\alpha+\underline{p}^\prime_\alpha)^0 \, (2 f(\text{Q}^2,s))= :
\tilde{J}_{[\alpha]}^+(\vec{\underline{p}}_\alpha^\prime;\vec{\underline{p}}_\alpha) \nonumber\\
 &=& (\underline{p}_\alpha+\underline{p}^\prime_\alpha)^0 (\tilde{\text{f}} (\text{Q}^2,s) + \tilde{\tilde{\text{f}}}(\text{Q}^2,s))
\end{eqnarray}
subtraction gives
\begin{eqnarray}
 \tilde{J}_{[\alpha]}^0(\vec{\underline{p}}_\alpha^\prime;\vec{\underline{p}}_\alpha)-
\tilde{J}_{[\alpha]}^3(\vec{\underline{p}}_\alpha^\prime;\vec{\underline{p}}_\alpha)&=&
(\underline{p}_\alpha+\underline{p}^\prime_\alpha)^0 \, (-2 g(\text{Q}^2,s))= :
\tilde{J}_{[\alpha]}^-(\vec{\underline{p}}_\alpha^\prime;\vec{\underline{p}}_\alpha) \nonumber\\
 &=& (\underline{p}_\alpha+\underline{p}^\prime_\alpha)^0 (\tilde{\text{f}}(\text{Q}^2,s) -
\tilde{\tilde{\text{f}}}(\text{Q}^2,s))
\end{eqnarray}
If one introduces now $f^+$ and $f^-$ for the form factors that are associated with the light-cone
components of the currents 
$\tilde{J}_{[\alpha]}^+(\vec{\underline{p}}_\alpha^\prime;\vec{\underline{p}}_\alpha) $
and $\tilde{J}_{[\alpha]}^-(\vec{\underline{p}}_\alpha^\prime;\vec{\underline{p}}_\alpha) $
respectively, one finds the following relations:
\begin{eqnarray}
 && \tilde{\text{f}}(\text{Q}^2,s) +\tilde{\tilde{\text{f}}}(\text{Q}^2,s) = 2 f^+(\text{Q}^2,s) = 2 f (\text{Q}^2,s),\\
 && \tilde{\text{f}}(\text{Q}^2,s) -\tilde{\tilde{\text{f}}}(\text{Q}^2,s) = 2 f^-(\text{Q}^2,s) = -2 g (\text{Q}^2,s).
\end{eqnarray}
The physical form factor $f (\text{Q}^2,s)$ can thus be identified with $ f^+(\text{Q}^2,s)$, 
while the spurious one
$g (\text{Q}^2,s)$, may be associated with $f^-(\text{Q}^2,s)$.
In the infinite momentum frame one sees numerically, as well as analytically, 
that $\tilde{J}_{\infty}^0(\vec{\underline{p}}_\alpha^\prime;\vec{\underline{p}}_\alpha)=
\tilde{J}_{\infty}^3(\vec{\underline{p}}_\alpha^\prime;\vec{\underline{p}}_\alpha)$. Therefore
 both, $g (\text{Q}^2,s)$ and  $f^-(\text{Q}^2,s)$, are zero in the infinite momentum frame. 

\subsubsection{Remarks}
If $\tilde{J}_{\infty}^0(\vec{\underline{p}}_\alpha^\prime;\vec{\underline{p}}_\alpha)
=\tilde{J}_{\infty}^3(\vec{\underline{p}}_\alpha^\prime;\vec{\underline{p}}_\alpha)$
and $(\ul p_{[\alpha]}+\ul p_{[\alpha]}')^0=(\ul p_{[\alpha]}+\ul p_{[\alpha]}')^3$ one 
finds the relation with the +-component of the current customary used in the front form:
\begin{equation}
 \frac{\tilde{J}_{\infty}^0(\vec{\underline{p}}_\alpha^\prime;\vec{\underline{p}}_\alpha)}{(\ul p_{[\alpha]}+\ul p_{[\alpha]}')^0}= 
\frac{\tilde{J}_{\infty}^3(\vec{\underline{p}}_\alpha^\prime;\vec{\underline{p}}_\alpha)}{(\ul p_{[\alpha]}+\ul p_{[\alpha]}')^3}= 
\frac{\tilde{J}_{\infty}^0(\vec{\underline{p}}_\alpha^\prime;\vec{\underline{p}}_\alpha)+\tilde{J}_{\infty}^3(\vec{\underline{p}}_\alpha^\prime;\vec{\underline{p}}_\alpha)}{(\ul p_{[\alpha]}+\ul p_{[\alpha]}')^0+(\ul p_{[\alpha]}+\ul p_{[\alpha]}')^3}=
\frac{\tilde{J}_{\infty}^+(\vec{\underline{p}}_\alpha^\prime;\vec{\underline{p}}_\alpha)}{(\ul p_{[\alpha]}+\ul p_{[\alpha]}')^+}
\end{equation}

\subsection{The Breit frame}
In the Breit frame,  $\nu_\alpha=u$, the kinematics is given by:
\begin{eqnarray}
 &&\underline{p}_\alpha=\left(m_\alpha\sqrt{\frac{v\cdot v'+1}{2}}\;,\;-m_\alpha u\;,\;0\;,\;0\right),\\
 &&\underline{p}_\alpha'=\left(m_\alpha\sqrt{\frac{v\cdot v'+1}{2}}\;,\;m_\alpha u\;,\;0\;,\;0\right),
\end{eqnarray}
\begin{equation}
 (\underline{p}_\alpha+\underline{p}^\prime_\alpha)=\left(2 m_\alpha\sqrt{\frac{v\cdot v'+1}{2}}\;,\;0\;,\;0\;,\;0\right);
\end{equation}
and for the electron
\begin{eqnarray}
 &&\underline{p}_e=\left(m_\alpha u\;,\;m_\alpha u\;,\;0\;,\;0\right),\\
 &&\underline{p}_e'=\left(m_\alpha u\;,\;-m_\alpha u\;,\;0\;,\;0\right),
\end{eqnarray}
\begin{equation}
 (\underline{p}_e+\underline{p}^\prime_e)=\left( 2 m_\alpha u\;,\;0\;,\;0\;,\;0\right).
\end{equation}
since $(\underline{p}_\alpha+\underline{p}^\prime_\alpha)^\nu\propto
 (\underline{p}_e+\underline{p}^\prime_e)^\nu$ it is not possible to separate the 
physical and the spurious form factors. From Eq.~(\ref{eq:Jemcovdec}) one
rather obtains
\begin{align}
 \tilde{J}_{[\infty]}^\mu(\vec{\underline{p}}_\alpha^\prime;\vec{\underline{p}}_\alpha)
&= (\ul p_\alpha +\ul p'_\alpha)^\mu\left\{ f(\text{Q}^2,s) + u\sqrt{\frac{2}{v\cdot v'+1}}\, g(\text{Q}^2,s)\right\}\nonumber\\
&=: (\ul p_\alpha +\ul p'_\alpha)^\mu f_\text{B}(\text{Q}^2,s).
\end{align}
Note, however that $f_\text{B}(\text{Q}^2,s)$ differs from $\lim_{\nu_\alpha\to u}f(\text{Q}^2,s)$. 
This means that one obtains different results, depending on whether the physical form factor is directly
extracted in the Breit frame or physical and spurious form factors are extracted in a frame
different from the Breit frame and then the Breit-frame limit is taken.

%% file: App4.tex
\chapter{Exchange currents}

\section{The deuteron bound-state problem}

\subsection{The $np\sigma$ wave function}\label{AppWaveFunction}
\begin{figure}[h!]
 \begin{center}
  \includegraphics[width=0.8\textwidth]{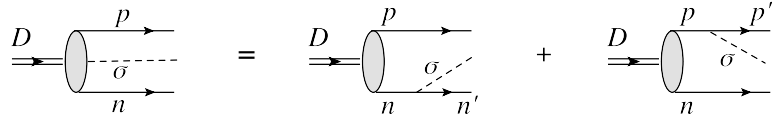}\caption{Graphical 
representation of Eq.~(\ref{npsigmawf}).}\label{FigPsisigmapn}
\end{center}
\end{figure}

In order to obtain the normalization of the full deuteron wave function it is necessary to 
consider also the $|\psi_{np\sigma}\rangle$ state. 
It follows from Eq.~(\ref{EqEigenvalueDeuteronCoupledChannel}) that
\begin{equation}
 \hat K_\sigma^\dagger | \psi_{np}\rangle =\left( m_D-\hat M_{np\sigma}\right) | \psi_{np\sigma}\rangle 
\Rightarrow |\psi_{np\sigma}\rangle =\left( m_D -\hat M_{np\sigma} \right)^{-1} \hat K^\dagger_\sigma | \psi_{np}\rangle.
\end{equation} 
$|\psi_{np\sigma}\rangle$ represents the 3-particle component of the deuteron
where the $\sigma$-meson is in flight (see also Fig.~\ref{FigPsisigmapn}).
Taking the inner product of the equation with 
$\langle v';\vec k_n', \mu_n';\vec k_p', \mu_p'; \vec k_\sigma'|$, 
introducing the appropriate completeness relations and 
the corresponding expressions for the $\sigma$-vertices one obtains an expression in terms of
the matrix elements $\langle v;\vec k_p, \mu_p;\vec k_n ,\mu_n| \psi_{np}\rangle$
\begin{eqnarray}\label{npsigmawf}
 &&\langle v';\vec k_n', \mu_n';\vec k_p', \mu_p'; \vec k_\sigma'| \psi_{np\sigma}\rangle
  = \langle v´;\vec k_n', \mu_n';\vec k_p', \mu_p'; \vec k_\sigma'
|\left( m_D - \hat M_{np\sigma} \right)^{-1} \hat K^\dagger_\sigma | \psi_{np}\rangle\nonumber \\
&&=\sum_{\mu_n\mu_p}\int \frac{d^3 v}{(2\pi)^3 v_0}\frac{d^3 k_p}{(2\pi)^3 2\omega_{k_p}}
\frac{(\omega_{k_n}+\omega_{k_p})^3}{2\omega_{k_n}} (m_D -\omega_{k_n'}-\omega_{k_p'}-
\omega_{k_\sigma'})^{-1}\nonumber\\
&&\quad \langle v';\vec k_n'', \mu_n';\vec k_p', \mu_p'; \vec k_\sigma'|\hat K^\dagger_\sigma 
|v;\vec k_p, \mu_p;\vec k_n ,\mu_n\rangle \langle v;\vec k_p, \mu_p;\vec k_n ,\mu_n| 
\psi_{np}\rangle. \nonumber\\
\end{eqnarray}
The matrix elements of the sigma-nucleon vertex are given by
\begin{align}
&\langle v';\vec k_n', \mu_n';\vec k_p', \mu_p'; \vec k_\sigma'|\hat K^\dagger_\sigma 
|v;\vec k_p, \mu_p;\vec k_n ,\mu_n\rangle \nonumber\\
&= v_0 \delta^3(\vec v-\vec v')\frac{ (2\pi)^3}{(\omega_{k_n'}+\omega_{k_p'}+\omega_{k_\sigma'})^{3/2}
(\omega_{k_p}+\omega_{k_n})^{3/2}} g_\sigma(-1)\nonumber\\
 &\qquad \times \Big\{\bar u_{\mu_n'}(\vec k'_n)u_{\mu_n}(\vec k_n) (2\pi)^3 2\omega_{k_p}
 \delta^3(\vec k_p'-\vec k_p)\delta_{\mu_p' \mu_p} +\nonumber\\
& \qquad \qquad \qquad +\,  \bar u_{\mu_p'}(\vec k'_p)u_{\mu_p}(\vec k_p) (2\pi)^3 2\omega_{k_n}
 \delta^3(\vec k_n'-\vec k_n)\delta_{\mu_n' \mu_n}\Big\}.
\end{align}
It is useful to take into account that in this case
 $\omega_{k'_p}= \omega_{k_n}$.
The final result is\footnote{The Pauli-Villar particles have been ignored for simplicity; they contribute 
as an identical term with the opposite sign and $m_\sigma$ replaced by the cutoff mass $\Lambda_\sigma$.}
\begin{align}
\langle v´;\vec k_n', \mu_n';\vec k_p', \mu_p'; \vec k_\sigma'| \psi_{np\sigma}\rangle
=&\frac{(-g_\sigma)}{(\omega_{k'_n}+\omega_{k'_p}+\omega_{k'_\sigma})^{3/2}}(m_D-\omega_{k_n}-\omega_{k'_p}-\omega_{k'_\sigma})^{-1}\nonumber\\
&\times 2\sqrt{2\omega_{k_p}}\sum_{\mu_n}\bar u_{\mu'_n}(\vec k'_n) u_{\mu_n}(\vec k_n)\nonumber\\
& \times \langle v;\vec k_p, \mu_p;\vec k_n ,\mu_n| \psi_{np}\rangle.
\end{align}
We have thus expressed the wave function of the $np\sigma$ component 
of the deuteron in terms of the wave function for the $np$ component.

\newpage
\section{The electron-deuteron scattering problem}
\subsection{Matrix elements of the optical potential}\label{AppMatrixElementsandCurrents}
The matrix elements of the third term in Eq.~(\ref{Feshbach3}) of the optical have the following structure:
{\allowdisplaybreaks
\begin{align}\label{EqExchangeCurrentskp}
&\langle \ul{v'}; \vec{\ul k}'_e, \ul \mu'_e; \vec{\ul k}'_D, \ul \mu'_D \vert \hat K_\gamma (m-\hat M_{eD\gamma})^{-1} \hat K_\sigma (m-\hat M_{enp\sigma \gamma})^{-1}\nonumber\\
&\quad \times \hat K^\dagger_\gamma (m-\hat M_{enp\sigma})^{-1}\hat K_\sigma^\dagger \vert \ul{v}; \vec{\ul k}_e, \ul \mu_e; \vec{\ul k}_D ,\ul \mu_D\rangle\nonumber \\
&\quad=\ul v_0 \delta(\vec{\ul v}'-\vec{\ul v}) \frac{(2\pi)^3}{\sqrt{(\omega_{k'_e}+\omega_{\ul k'_D})^3(\omega_{k_e} + \omega_{\ul k_D})^3}} e\,\underbrace{\bar{u}_{\mu_e'}(\vec k_e')\gamma^\mu u_{\mu_e}(\vec k_e)}_{j_e^\mu}\nonumber\\
&\quad\times  |e|\; \left(\frac{1}{m-\omega_{k'_D}-\omega_{k_e}-\omega_{k_\gamma}}\right)\nonumber\\
&\quad\times  \sqrt{2\omega_{k_D}2\omega_{k'_D}}\int d^3\tilde k'_p \sqrt{\frac{\omega_{\tilde k'_n}+\omega_{\tilde k'_p}}{\omega_{k'_n}+\omega_{k'_p}}}\frac{1}{2\sqrt{\omega_{\tilde k'_n}\omega_{\tilde k'_p}}}\nonumber\\
&\quad\times \int d^3 \tilde k_p \sqrt{\frac{\omega_{\tilde k_n}+\omega_{\tilde k_p}}{\omega_{k_n}+\omega_{k_p}}}\frac{1}{2\sqrt{\omega_{\tilde k_n}\omega_{\tilde k_p}}}\;  \frac{1}{2\omega_{k_\gamma}}\frac{1}{2\omega_{k''_p}}\frac{1}{2\omega_{k_\sigma}}\nonumber\\
&\quad\times   \left( \frac{1}{m-\omega_{k_e}-\omega_{k_p}-\omega_{k'_n}-\omega_{k_\sigma}}\right) \left(\frac{g_{\mu\nu}}{m-\omega_{k_e}-\omega_{k''_p}-\omega_{k'_n}-\omega_{k_\sigma}-\omega_{k_\gamma}}\right) \nonumber\\
&\quad\times \sum_{\mu_p'\mu_n\mu_p}\sum_{\tilde \mu_n\tilde \mu_p \tilde \mu_n'\tilde \mu_p'} \bar u_{\mu'_p}(\vec k''_p)\Gamma^\nu u_{\mu_p}(\vec k_p) g^2_\sigma  \; \bar u_{\mu'_p}(\vec k'_p) u_{\mu'_p}(\vec k''_p) \; \bar u_{\mu_n}(\vec k'_n) u_{\mu_n}(\vec k_n) \nonumber\\
&\quad\times  C^{1\mu_D'}_{\frac{1}{2}\tilde \mu_p'\frac{1}{2}\tilde \mu_n'}\; D^{1/2}_{\tilde\mu_{p}'\mu_{p}'}\left[ R_W^{-1}\left( \frac{\tilde k_p'}{m_p},B_c\left( v_{np}\right)\right)\right] D^{1/2}_{\mu_{p}\tilde\mu_{p}}
\left[ R_W\left( \frac{\tilde k_p}{m_p},B_c\left( v_{np}\right)\right)\right] \nonumber\\
&\quad\times C^{1\mu_D}_{\frac{1}{2}\tilde \mu_p\frac{1}{2}\tilde \mu_n}
D^{1/2}_{\tilde\mu_{n}'\tilde\mu_{n}}\left[ R_W^{-1}\left( \frac{\tilde k_n'}{m_n},B_c\left( v_{np}'\right)\right)
 R_W\left( \frac{\tilde k_n}{m_n},B_c\left( v_{np}\right)\right)\right]\nonumber\\
&\quad\times  \text{u}^*_{D}(|\vec{\tilde k}_p'|)Y^*_{00}(\hat{\vec{\tilde{k}}}_p') \text{u}_{D}(|\vec{\tilde k}_p|)Y_{00}(\hat{\vec{\tilde{k}}}_p)\;\; +\;\; (p\leftrightarrow n).
\end{align}}
from which it is easy to identify the corresponding contribution to the deuteron 
current (cf.~Chap.~\ref{ChCoupledChannel}).
In addition to the integration over $\tilde k'_p$,
 a second integral appears, which 
runs over the intermediate state $\tilde k_p$ (in Sec.~\ref{SubSecDeutExch} $\tilde k_p''$ was chosen),
 and accounts for the momentum that is transferred
by the $\sigma$-meson from one nucleon to the other. 
By a change of variables one can 
express this as an integration over this momentum transfer, which we call $\vec{\tilde q}$.
 The change requires some work, because $\vec{\tilde k}_p$ and $\vec{\tilde q}$ are defined in 
different reference frames. One has to transform $\vec{\tilde k}_p$ to $\vec k_p$ 
(see Eq.~(\ref{EqChangeTildeNoTilde})), which
is related to $\vec{\tilde q}$  by a simple translation:

{\allowdisplaybreaks
\begin{align}\label{EqExchangeCurrentsq}
&\langle \ul{v'}; \vec{\ul k}'_e, \ul \mu'_e; \vec{\ul k}'_D, \ul \mu'_D \vert \hat K_\gamma (m-\hat M_{eD\gamma})^{-1} \hat K_\sigma (m-\hat M_{enp\sigma \gamma})^{-1}\nonumber\\
&\quad \times \hat K^\dagger_\gamma (m-\hat M_{enp\sigma})^{-1}\hat K_\sigma^\dagger \vert \ul{v}; \vec{\ul k}_e, \ul \mu_e; \vec{\ul k}_D ,\ul \mu_D\rangle\nonumber \\
&\quad =\ul v_0 \delta(\vec{\ul v}'-\vec{\ul v}) \frac{(2\pi)^3}{\sqrt{(\omega_{k'_e}+\omega_{\ul k'_D})^3(\omega_{k_e} + \omega_{\ul k_D})^3}}\underbrace{e\bar{u}_{\mu_e'}(\vec k_e')\gamma^\mu u_{\mu_e}(\vec k_e)}_{j_e^\mu}\nonumber\\
&\quad\times |e|\; \left(\frac{1}{m-\omega_{k'_D}-\omega_{k_e}-\omega_{k_\gamma}}\right) \nonumber\\
&\quad\times \sqrt{2\omega_{k_D}2\omega_{k'_D}} \int d^3\tilde k'_p \sqrt{\frac{\omega_{\tilde k'_n}+\omega_{\tilde k'_p}}{\omega_{k'_n}+\omega_{k'_p}}}\frac{1}{2\sqrt{\omega_{\tilde k'_n}\omega_{\tilde k'_p}}}\nonumber\\
&\quad\times\int d^3 \tilde q \frac{\sqrt{2\omega_{\tilde k_p}2\omega_{\tilde k_n}}}{2\omega_{k_n}2\omega_{k_p}}  \sqrt{\frac{\omega_{k_n}+\omega_{k_p}}{\omega_{\tilde k_n}+\omega_{\tilde k_p}}} \frac{1}{2\omega_{k_\gamma}}\frac{1}{2\omega_{k''_p}}\frac{1}{2\omega_{k_\sigma}}  \nonumber\\
&\quad\times \left( \frac{1}{m-\omega_{k_e}-\omega_{k_p}-\omega_{k'_n}-\omega_{k_\sigma}}\right) \left(\frac{g_{\mu\nu}}{m-\omega_{k_e}-\omega_{k''_p}-\omega_{k'_n}-\omega_{k_\sigma}-\omega_{k_\gamma}}\right)\nonumber\\
&\quad\times \sum_{\mu_p'\mu_n\mu_p}\sum_{\tilde \mu_n\tilde \mu_p \tilde \mu_n'\tilde \mu_p'} \bar u_{\mu'_p}(\vec k''_p)\Gamma^\nu u_{\mu_p}(\vec k_p) g^2_\sigma  \; \bar u_{\mu'_p}(\vec k'_p) u_{\mu'_p}(\vec k''_p) \; \bar u_{\mu_n}(\vec k'_n) u_{\mu_n}(\vec k_n) \nonumber\\
&\quad\times  C^{1\mu_D'}_{\frac{1}{2}\tilde \mu_p'\frac{1}{2}\tilde \mu_n'}\; D^{1/2}_{\tilde\mu_{p}'\mu_{p}'}\left[ R_W^{-1}\left( \frac{\tilde k_p'}{m_p},B_c\left( v_{np}\right)\right)\right] D^{1/2}_{\mu_{p}\tilde\mu_{p}}
\left[ R_W\left( \frac{\tilde k_p}{m_p},B_c\left( v_{np}\right)\right)\right] \nonumber\\
&\quad\times C^{1\mu_D}_{\frac{1}{2}\tilde \mu_p\frac{1}{2}\tilde \mu_n}
D^{1/2}_{\tilde\mu_{n}'\tilde\mu_{n}}\left[ R_W^{-1}\left( \frac{\tilde k_n'}{m_n},B_c\left( v_{np}'\right)\right)
 R_W\left( \frac{\tilde k_n}{m_n},B_c\left( v_{np}\right)\right)\right] \nonumber\\
&\quad\times  \text{u}^*_{D}(|\vec{\tilde k}_p'|)Y^*_{00}(\hat{\vec{\tilde{k}}}_p') \text{u}_{D}(|\vec{\tilde k}_p|)Y_{00}(\hat{\vec{\tilde{k}}}_p)
\;\; +\;\; (p\leftrightarrow n).
\end{align}}
\newpage
It is now easy to identify the exchange current contribution, where the photon couples to the proton:
{\allowdisplaybreaks
\begin{eqnarray}\label{ApExCurrent}
&&J_p^{\mu,\text{ex}}(\vec k_D',\vec k_D; \mu'_D,\mu_D)  = \sqrt{2\omega_{k_D}2\omega_{k'_D}} \int d^3\tilde k'_p \sqrt{\frac{\omega_{\tilde k'_n}+\omega_{\tilde k'_p}}{\omega_{k'_n}+\omega_{k'_p}}}\frac{1}{2\sqrt{\omega_{\tilde k'_n}\omega_{\tilde k'_p}}}\nonumber\\
&&\quad\times \int d^3 \tilde q \frac{\sqrt{2\omega_{\tilde k_p}2\omega_{\tilde k_n}}}{2\omega_{k_n}2\omega_{k_p}}  \sqrt{\frac{\omega_{k_n}+\omega_{k_p}}{\omega_{\tilde k_n}+\omega_{\tilde k_p}}} \frac{1}{2\omega_{k_\gamma}}\frac{1}{2\omega_{k''_p}}\frac{1}{2\omega_{k_\sigma}}  \nonumber\\
&&\quad\times \left(\frac{1}{m-\omega_{k_e}-\omega_{k''_p}-\omega_{k'_n}-\omega_{k_\sigma}-\omega_{k_\gamma}}\right)
\left( \frac{1}{m-\omega_{k_e}-\omega_{k_p}-\omega_{k'_n}-\omega_{k_\sigma}}\right) \nonumber\\
&&\quad\times \sum_{\mu_p'\mu_n\mu_p}\sum_{\tilde \mu_n\tilde \mu_p \tilde \mu_n'\tilde \mu_p'} \bar u_{\mu'_p}(\vec k''_p)\Gamma^\nu u_{\mu_p}(\vec k_p) g^2_\sigma  \; \bar u_{\mu'_p}(\vec k'_p) u_{\mu'_p}(\vec k''_p) \; \bar u_{\mu_n}(\vec k'_n) u_{\mu_n}(\vec k_n) \nonumber\\
&&\quad\times  C^{1\mu_D'}_{\frac{1}{2}\tilde \mu_p'\frac{1}{2}\tilde \mu_n'}\; D^{1/2}_{\tilde\mu_{p}'\mu_{p}'}\left[ R_W^{-1}\left( \frac{\tilde k_p'}{m_p},B_c\left( v_{np}\right)\right)\right] D^{1/2}_{\mu_{p}\tilde\mu_{p}}
\left[ R_W\left( \frac{\tilde k_p}{m_p},B_c\left( v_{np}\right)\right)\right] \nonumber\\
&&\quad\times C^{1\mu_D}_{\frac{1}{2}\tilde \mu_p\frac{1}{2}\tilde \mu_n}
D^{1/2}_{\tilde\mu_{n}'\tilde\mu_{n}}\left[ R_W^{-1}\left( \frac{\tilde k_n'}{m_n},B_c\left( v_{np}'\right)\right)
 R_W\left( \frac{\tilde k_n}{m_n},B_c\left( v_{np}\right)\right)\right] \nonumber\\
&&\quad\times  \text{u}^*_{D}(|\vec{\tilde k}_p'|)Y^*_{00}(\hat{\vec{\tilde{k}}}_p') \text{u}_{D}(|\vec{\tilde k}_p|)Y_{00}(\hat{\vec{\tilde{k}}}_p).
\end{eqnarray}}
The other terms in Eq.~(\ref{Feshbach3}) can be treated in an analogous way.

\subsection{The deuteron exchange currents in the infinite-momentum frame 
$\kappa_D\to\infty$}\label{LimitsInfMomentumFrame}
In the infinite-momentum frame the current matrix elements simplify further.
The spinor product $\bar u_{\mu'_p}(\vec k''_p)\gamma^\nu u_{\mu_p}(\vec k_p)$, 
e.g., becomes in this limit:

\begin{equation}
\bar u_{\frac{1}{2}}(\vec k''_p)\gamma^\mu u_{\frac{1}{2}}(\vec k_p)\rightarrow
\left( \begin{array}{c}
\kappa_D\frac{m'_{np}+2\tilde k'_3}{m'_{np}}\\
2\tilde k'_1+\left(-\frac{1}{2}+\frac{\tilde k'_3}{m'_{np}}\right)\text{Q} -2\tilde q_1\\
2\tilde k'_2+i\text{Q}-2\tilde q_2 \\
\kappa_D\frac{m'_{np}+2\tilde k'_3}{m'_{np}}
 \end{array} \right);
\end{equation}

\begin{equation}
\bar u_{\frac{1}{2}}(\vec k''_p)\gamma^\mu u_{-\frac{1}{2}}(\vec k_p)\rightarrow
\left( \begin{array}{c}
-\text{Q}\\
0\\
0\\
-\text{Q}
 \end{array} \right); \quad \bar u_{-\frac{1}{2}}(\vec k''_p)\gamma^\mu u_{\frac{1}{2}}(\vec k_p)\rightarrow
\left( \begin{array}{c}
\text{Q}\\
0\\
0\\
\text{Q}
 \end{array} \right);
\end{equation}

\begin{equation}
\bar u_{-\frac{1}{2}}(\vec k''_p)\gamma^\mu u_{-\frac{1}{2}}(\vec k_p)\rightarrow
\left( \begin{array}{c}
\kappa_D\frac{m'_{np}+2\tilde k'_3}{m'_{np}}\\
2\tilde k'_1+\left(-\frac{1}{2}+\frac{\tilde k'_3}{m'_{np}}\right)\text{Q} -2\tilde q_1\\
2\tilde k'_2-i\text{Q}-2\tilde q_2 \\
\kappa_D\frac{m'_{np}+2\tilde k'_3}{m'_{np}}
 \end{array} \right).
\end{equation}

\begin{equation}
\omega_{k_D}\rightarrow \kappa_D,\quad
\omega_{k'_P},\,\omega_{k''_P},\,\omega_{k_P} \rightarrow \kappa_D\left( \frac{1}{2}+\frac{\tilde k'_3}{m'_{12}}\right),\quad
\omega_{k'_N}\rightarrow \kappa_D\left( \frac{1}{2}-\frac{\tilde k'_3}{m'_{12}}\right)
\end{equation}
And 
\begin{align}
&\bar u_{\mu'_p}(\vec k'_p)u_{\mu'_p}(\vec k''_p)\rightarrow \kappa_D\frac{m'_{np}+2\tilde k'_3}{m'_{np}},
\quad
\bar u_{\mu_n}(\vec k'_n) u_{\mu_n}(\vec k_n)\to  \kappa_D\frac{m'_{np}-2\tilde k'_3}{m'_{np}},\nonumber\\
& \frac{\sqrt{\omega_{k'_D}\omega_{k_D}}}{2\omega_{k_p}}\to \frac{m_{pn}'}{m_{pn}'+2\tilde k_3}.
\end{align}
For point-like nucleons, i.e. $\Gamma_p^\mu=\gamma^\mu$ and $\Gamma_n^\mu=0$, 
Eq.~(\ref{ApExCurrent}) in the limit
leads to:
\begin{align}
&J_D^{\mu,\text{ex}}(k'_D,\mu'_D;k_D,\mu_D) \nonumber\\ 
&\quad= 
\int d^3 \tilde k'_n \int d^3\tilde q \sqrt{\frac{m_{np}}{m'_{np}}}g^2_\sigma \left(\frac{1}{\omega_\sigma}\right)^2 \frac{1}{\text{Q}+\omega_\sigma}
\frac{2m_{pn}'}{m_{pn}'+2\tilde k_3}
\nonumber \\
&\quad\times\sum_{\mu_p'}\sum_{\tilde \mu_n\tilde \mu_p \tilde \mu_n'\tilde \mu_p'} 
\text{u}^*_{D}(|\vec{\tilde k}_p'|)Y^*_{00}(\hat{\vec{\tilde{k}}}_p') \text{u}_{D}(|\vec{\tilde k}_p|)Y_{00}(\hat{\vec{\tilde{k}}}_p)
\;\,\bar u_{\mu_p'}(\vec k''_p)\gamma^\mu u_{\mu_p}(\vec k_p)
\nonumber\\
&\quad\times  C^{1\mu_D'}_{\frac{1}{2}\tilde \mu_p'\frac{1}{2}\tilde \mu_n'}\; D^{1/2}_{\tilde\mu_{p}'\mu_{p}'}\left[ R_W^{-1}\left( \frac{\tilde k_p'}{m_p},B_c\left( v_{np}\right)\right)
 R_W\left( \frac{\tilde k_p}{m_p},B_c\left( v_{np}\right)\right)\right] \nonumber\\
&\quad\times C^{1\mu_D}_{\frac{1}{2}\tilde \mu_P\frac{1}{2}\tilde \mu_n}
D^{1/2}_{\tilde\mu_{n}'\tilde\mu_{n}}\left[ R_W^{-1}\left( \frac{\tilde k_n'}{m_n},B_c\left( v_{np}'\right)\right)
 R_W\left( \frac{\tilde k_n}{m_n},B_c\left( v_{np}\right)\right)\right]. 
\end{align}

%% file: Ackn.tex
\chapter*{Acknowledgements}

Primarily, I want to thank my advisor, Prof. Wolfgang Schweiger, for the opportunity
to work with him and learn from him, for his constant
guidance and support during my doctoral studies and for his confidence in me. 
  
I also would like to thank Elmar P. Biernat for many helpful and elucidating discussions, 
and for being always willing to explain his own calculations. I am very grateful to 
Prof. Leonid Ya Glozman for helpful discussions concerning the last chapter of the thesis. 
I consider myself lucky for having taken part at the program
``Hadrons in Vacuum nuclei and stars", and I thank all professors and
students who made it possible; for the extraordinary and open-minded atmosphere allowing 
many different directions of research in hadron physics.
I am especially grateful to my mentor Prof.~W. Plessas for his excellent advisory and for 
his constructive complaints on my seminars. My thanks go also to the Iowa group: Prof. William H.~Klink, 
Prof. Wayne N.~Polyzou and Prof. Fritz Coester for the many helpful and interesting
discussions and for their hospitality during my visit.
Finally, I would like to mention Prof. Felipe J. Llanes-Estrada, 
who encouraged me from the beginning in my decision to initiate a scientific career and from whom 
I learned first to work on difficult problems in theoretical hadron
physics. My debt to him is very great. 
 
There are many people -- physicists and not physicists -- 
to whom I could dedicate some lines in this 
Acknowledgements, since they have certainly helped me to become a theoretical
physicist: first my parents, my brothers, In\'es, Isolde, Marisol, Christa, Helga, 
Christina, Mercedes, Suelie, Ydalia, H\`elios, Regina, Martin, Joe, Valentina, Tina, Ki-Seok, Daniel, Alex, 
Vasily...

\vspace{0.5cm}

\noindent\textbf{Financial support}: this work was supported by the ``Fond zur F\"orderung 
der wissenschaftlichen
Forschung in \"Osterreich'' (FWF DK W1203-N16).

\vspace{1cm}
  \begin{flushright}
Mar\'ia G\'omez-Rocha,\\
Graz, December 2012\\
\small{D.O.G.}
\end{flushright}

\newpage
\thispagestyle{empty}

%% file: Bibl.tex
 %